\documentclass[final,leqno,onefignum,onetabnum]{siamltex}

\usepackage{fullpage}
\usepackage{comment}
\usepackage{upgreek}
\usepackage{cite}
\usepackage[mathscr]{eucal}
\usepackage{color}
\usepackage[usenames,dvipsnames,svgnames,table]{xcolor}
\usepackage{amsmath,amssymb,amsopn,mathtools}
\usepackage{graphicx}
\usepackage{hyperref}       
\hypersetup{
	colorlinks=true,   
	citecolor=blue,     
	filecolor=blue,     
	linkcolor=red,    
	urlcolor=blue}      

\usepackage{relsize}
\numberwithin{equation}{section}
\usepackage{enumitem}
\usepackage{epstopdf}

\newlist{Assumption}{enumerate}{1}
\setlist[Assumption]{label=A\arabic*}

\definecolor{Blue}{rgb}{0,0,1}
\definecolor{Red}{rgb}{1,0,0}
\definecolor{Green}{rgb}{0,1,0}
\definecolor{Cyan}{rgb}{0,0.72,0.92}
\definecolor{Amethyst}{rgb}{0.6,0.4,0.8}
\definecolor{Bronze}{rgb}{0.8,0.5,0.2}
\definecolor{Violet}{rgb}{0.54,0.17,0.89}

\newcommand{\solDomainSymbol}{\Omega}

\newcommand{\timeSymbol}{t}

\newcommand{\domain}[1]{\mathsf{#1}}
\newcommand{\paramSymbol}{P}

\newcommand{\sizeFOMsymbol}{N_x}
\newcommand{\sizeROMsymbol}{r}

\newcommand{\timeIndex}{k}
\newcommand{\paramIndex}{i}
\newcommand{\timeWindow}{\tau}
\newcommand{\windowIndex}{j}

\newcommand{\paramDomainSymbol}{D}

\newcommand{\relErrorSymbol}{\varepsilon}

\newcommand{\residualSymbol}{R}

\newcommand{\fullNorm}[1]{{\left\vert\kern-0.25ex\left\vert\kern-0.25ex\left\vert #1 
    \right\vert\kern-0.25ex\right\vert\kern-0.25ex\right\vert}}
\newcommand{\stateSymbol}{T}

\newcommand{\paramDomain}{\domain{\paramDomainSymbol}}

\newcommand{\state}{\boldsymbol{\stateSymbol}}

\newcommand{\relError}{\relErrorSymbol}

\newcommand{\residual}{\boldsymbol{\residualSymbol}}

\newcommand{\param}{\paramSymbol}

\newcommand{\ntimestep}{m}
\newcommand{\timestep}{\Delta\timeSymbol}

\newcommand{\windowIndicator}{\Psi}
\newcommand{\snapshotGroup}{\mathcal{G}}
\newcommand{\nwindow}{J}

\newcommand{\nparam}{N_\paramSymbol}

\DeclareMathOperator*{\argmin}{arg\,min}

\usepackage{amsmath,amssymb}
\usepackage{epsfig, algorithmic, algorithm}
\usepackage{multirow, booktabs}
\usepackage{siunitx}
\usepackage{xr}
\mathtoolsset{showonlyrefs=true}

\newlist{steps}{enumerate}{1}
\setlist[steps, 1]{label = Step \arabic*:}

\begin{document}

\title{Data-scarce surrogate modeling of
shock-induced pore collapse process}

\author{
  Siu Wun Cheung\thanks{Center for Applied Scientific Computing,
  Lawrence Livermore National Laboratory, Livermore, CA 94550 (cheung26@llnl.gov)}
  \and
  Youngsoo Choi\thanks{Center for Applied Scientific Computing,
  Lawrence Livermore National Laboratory, Livermore, CA 94550 (choi15@llnl.gov)}
  \and
  H. Keo Springer \thanks{Material Science Division, Energetic Materials Center,
  Lawrence Livermore National Laboratory, Livermore, CA 94550 (springer12@llnl.gov)}
  \and
  Teeratorn Kadeethum \thanks{Sandia National Laboratories, Albuquerque, NM (tkadeet@sandia.gov)}
}

\date{\today}

\maketitle

\begin{abstract}
Understanding the mechanisms of shock-induced pore collapse is of
great interest in various disciplines in sciences and engineering,
including materials science, biological sciences, and geophysics.
However, numerical modeling of the complex pore collapse processes can be costly.
To this end, a strong need exists to develop surrogate models for
generating economic predictions of pore collapse processes.
In this work, we study the use of a data-driven reduced order model, namely dynamic mode decomposition,
and a deep generative model, namely conditional generative adversarial networks, to resemble the numerical simulations of the pore collapse process at representative training shock pressures.
Since the simulations are expensive, the training data are scarce, which makes training an accurate surrogate model challenging.
To overcome the difficulties posed by the complex physics phenomena,
we make several crucial treatments to the plain original form of the methods
to increase the capability of approximating and predicting the dynamics.
In particular, physics information is used as indicators or conditional inputs to guide the prediction.
In realizing these methods,
the training of each dynamic mode composition model takes only around 30 seconds on CPU. 
In contrast, training a generative adversarial network model takes 8 hours on GPU.
Moreover, using dynamic mode decomposition,
the final-time relative error is around 0.3\% in the reproductive cases.
We also demonstrate the predictive power of the methods at unseen testing shock pressures,
where the error ranges from 1.3\% to 5\% in the interpolatory cases
and 8\% to 9\% in extrapolatory cases.

\end{abstract}

\section{Introduction}

\begin{figure}[h]
\centering
\includegraphics[width=\linewidth]{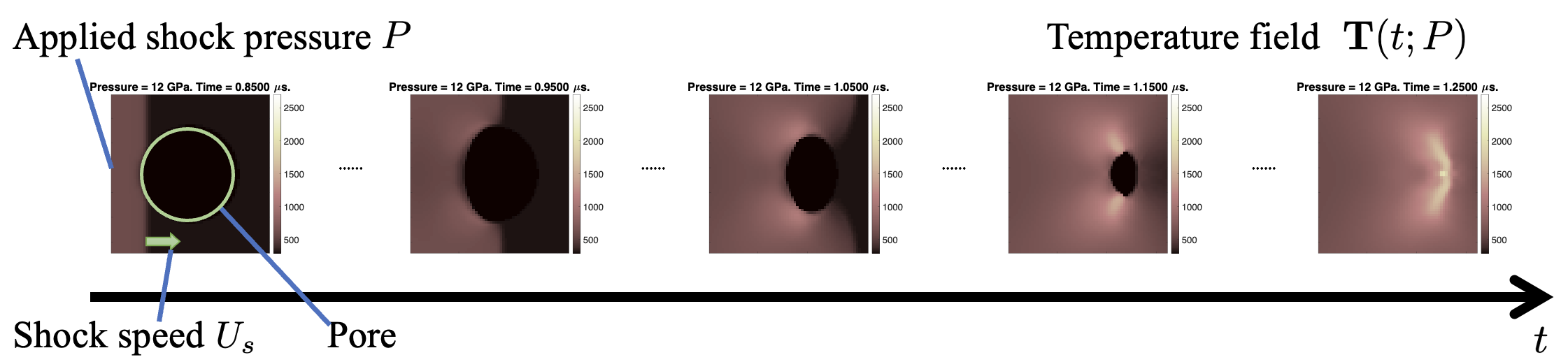}
  \caption{Schematic diagram for illustration of shock-induced pore collapse process.
  At first, the shock approaches and travels through the pore.
  The pore eventually deforms and develops into a high-temperature profile after the interaction with the shock.
  }
  \label{fig:pore_coll_fig}
\end{figure}

Shock-induced pore collapse is a phenomenon that occurs when a shock wave passes through a porous material,
causing the pores to collapse or deform.
Figure \ref{fig:pore_coll_fig} illustrates a shock-induced pore collapse process.
At first, the shock approaches and travels through the pore.
The pore eventually deforms and develops into a high-temperature profile after the interaction with the shock.
This phenomenon has been observed and studied in a variety of materials,
including viscoelastic materials \cite{tong1993dynamic},
nanoporous metals \cite{erhart2005atomistic},
sedimentary rocks \cite{schade2007numerical},
biological cells \cite{adhikari2015mechanism}, and
polymers \cite{dattelbaum2019shock}.
The collapse of pores can have a significant impact on the mechanical properties of the material,
including its strength, stiffness, and ductility.
For example, in metals, shock-induced pore collapse can lead to a reduction in ductility and toughness,
which can make the material more prone to brittle failure.
In geological materials, pore collapse can affect the permeability and porosity of the material,
which can have implications for groundwater flow and oil recovery.
Understanding the mechanisms of shock-induced pore collapse is therefore of great interest
in various disciplines in sciences and engineering,
including materials science, biological sciences and geophysics.

However, accurately analyzing the pore collapse dynamics is challenging
due to the complex and nonlinear nature of the deformation process.
Traditional analytical models, which rely on simplified assumptions
about the material properties and pore geometry,
often fail to capture the true behavior of the system.
Numerical methods is a powerful alternative to obtain approximate
solutions through computer simulation in this scenario.
For instance, the pore collapse processes can be accurately
simulated by the multi-physics hydro-code, ALE3D \cite{noble2017ale3d}.
However, a single simulation takes up to 1 week on 1024 cores.
It is therefore desirable to develop efficient techniques for
resembling the dynamics in these computationally expensive simulations
and predicting the dynamics in unseen generic shock pressures.

Obtaining computationally economical prediction of complex physics phenomena
remains a demanding and challenging task
in many applications in engineering and science.
In recent years, numerous research efforts have been devoted to
develop surrogate models, which work as simplified representation of the underlying physical process
and reduce the computational cost of simulating or analyzing the original system.
One important class of these surrogate models is the projection-based reduced order models (ROMs),
which aims to reduce the dimensionality by projecting high-fidelity physics-based models onto
low-dimensional structures, which are constructed from compression of the representative snapshot solution data.
The data compression techniques include linear approaches such as
proper orthogonal decomposition (POD) \cite{berkooz1993proper},
balanced truncation \cite{safonov1989schur}, and
reduced basis method \cite{rozza2008reduced}, or
nonlinear compression approaches such as autoencoders (AE)
\cite{lee2020model,maulik2021reduced,kim2022fast}.
Projection-based ROMs are intrusive in the sense that
involve incorporating the reduced solution representation
into the governing equations, physics laws, and numerical discretization methods,
such as finite element, finite volume, and finite difference methods.
As a result, these approaches are data-driven but also constrained by physics,
requiring less data to achieve the same level of accuracy.
Linear subspace ROMs had been applied to different applications with great success, including
nonlinear diffusion equations \cite{hoang2021domain, fritzen2018algorithmic},
Burgers equation and Euler equations in small-scale \cite{choi2019space,
choi2020sns, carlberg2018conservative},
convection--diffusion equations \cite{mojgani2017lagrangian, kim2021efficientII},
Navier--Stokes equations \cite{xiao2014non, burkardt2006pod},
Lagrangian hydrodynamics \cite{copeland2022reduced, cheung2023local}.
porous media flow \cite{ghasemi2015localized, cheung2020constraint},
reservoir simulations \cite{jiang2019implementation,yang2016fast},
computational electro-cardiology \cite{yang2017efficient},
shallow water equations \cite{zhao2014pod,cstefuanescu2013pod},
Boltzmann transport problems \cite{choi2021space},
wave equations \cite{fares2011reduced,cheng2016reduced,cheung2021explicit},
computing electromyography \cite{mordhorst2017pod},
spatio-temporal dynamics of a predator--prey system \cite{dimitriu2013application},
acoustic wave-driven microfluidic biochips \cite{antil2012reduced},
rocket nozzle shape design \cite{amsallem2015design},
flutter avoidance wing shape optimization \cite{choi2020gradient},
topology optimization of wind turbine blades \cite{choi2019accelerating},
and lattice structure design \cite{mcbane2021component, mcbane2022stress}.
Survey papers for the projection-based
ROMs can be found in \cite{gugercin2004survey, benner2015survey}.
It is noteworthy that in spite of the successes of the classical linear subspace projection-based ROMs
in many applications, these approaches are limited to the assumption that the intrinsic solution space falls
into a subspace with a small dimension, i.e., the solution space with a
Kolmogorov $n$-width decays fast.  This assumption is violated in advection-dominated
problems, due to features such as sharp gradients, moving shock fronts, and turbulence, which
prevent these model reduction schemes from being practical.
A way to overcome this challenge is to build small and accurate projection-based reduced-order models
by decomposing the solution manifold into submanifolds.
These reduced-order models are local in the sense that
each of them are valid only over a certain subset of the parameter-time domain.
The appropriate local reduced order model is chosen based on the current state of the system,
and all the local reduced order models cover the whole time marching in the online phase.
The concept of a local reduced order model was introduced in \cite{washabaugh2012nonlinear,
amsallem2012nonlinear}, where unsupervised clustering is used for the solution
manifold decomposition.
In \cite{parish2019windowed,shimizu2020windowed}, windowed ROM apporaches
were introduced to construct temporally-local ROMs which are small but accurate within a
short period in advection-dominated problems.
In \cite{copeland2022reduced,cheung2023local}, windowed ROM approaches
were developed for Lagrangian hydrodynamics by decomposing the solution manifold decomposition
based on physical time or more generally a suitably defined physics-based indicator.

A drawback of projection-based ROMs is that the implementation requires knowledge of
the underlying numerical methods used in the high-fidelity simulation.
Conversely, the class of non-intrusive surrogate models do not require access to the source code
of the high-fidelity physics solver, and they are solely based on data.
With the growing availability of data,
there has been extensive research on non-intrusive surrogate models of discrete dynamics,
using different dimensionality reduction and machine learning techniques.
Similar to the projection-based ROMs, many non-intrusive surrogate models
construct low-dimensional structure for approximating the solution manifold and
approximate the dynamics in the low-dimensional latent code.
While the projection-based ROMs use the governing equations to derive the dynamics in the low-dimensional latent space,
non-intrusive surrogate models are purely data-driven.
For example, several approaches use linear compression techniques,
to construct a reduced subspace from snapshots,
such as dynamic mode decomposition (DMD)
\cite{schmid2010dynamic,rowley2009spectral,tu2014dynamic,proctor2016dynamic}
which seeks the best-fit linear model,
operator inference (OpInf)
\cite{peherstorfer2016data,mcquarrie2021data,mcquarrie2021non}
which seeks the best-fit polynomial model, and
sparse identification of nonlinear dynamics (SINDy)
\cite{brunton2016discovering,messenger2021weak}
which seeks the best-fit sparse regression.
The idea of identifying the best reduced discrete dynamic model within a certain family of functions
can be extended to nonlinear compression techniques by AE,
for example, using SINDy \cite{champion2019data},
parametric Latent Space Dynamics Identification
(LaSDI) \cite{fries2022lasdi,he2022glasdi},
and DeepFluids \cite{kim2019deep}.
Besides dimensionality reduction techniques,
neural networks can also be used in approximate the nonlinear operator
in the dynamical system as non-intrusive surrogate models,
such as Fourier neural operator (FNO) \cite{li2020fourier,kovachki2021neural},
deep operator network (DeepONet) \cite{lu2021learning},
and other relevant works \cite{wang2020deep,cheung2020deep,
zhang2019deep,wang2020efficient,chen2021generalized}.

\begin{figure}[htp!]
\centering
\includegraphics[width=\linewidth]{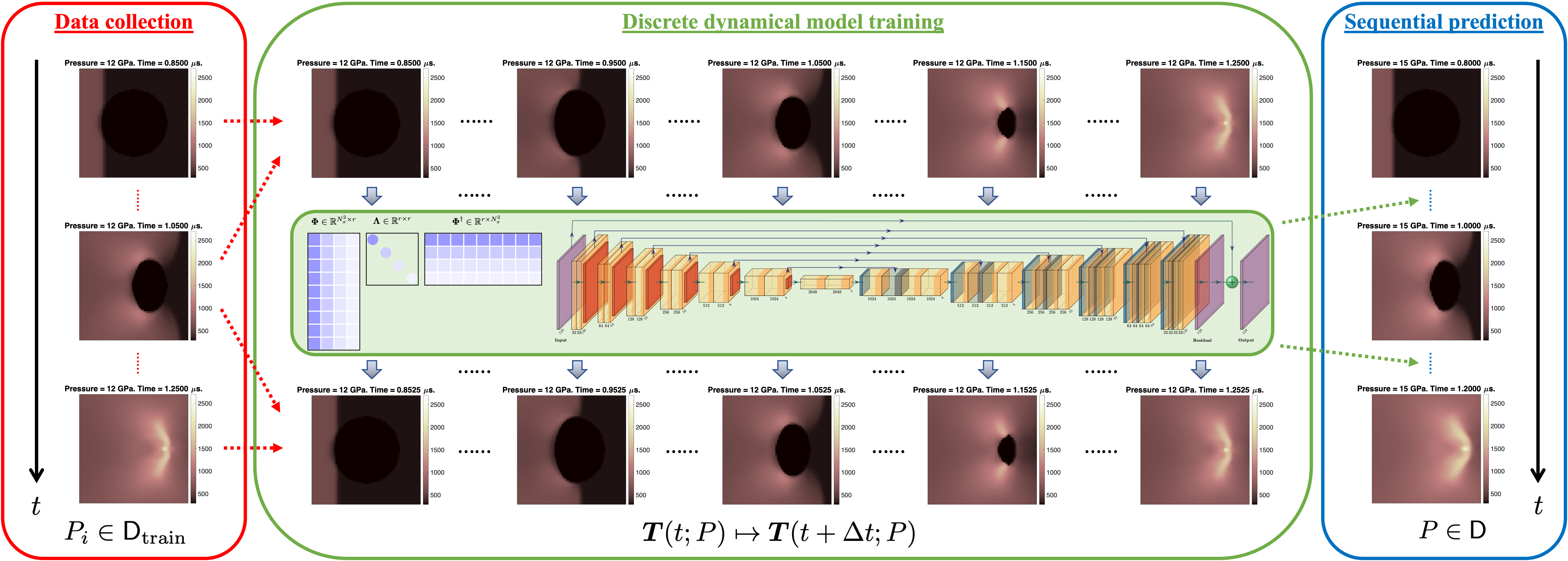}
\caption{
Schematics of non-intrusive surrogate models of the discrete dynamics of pore collapse.
In the offline phase, the snapshot data from training shock pressures
are used as the input and the output of the recurrence relation in the discrete dynamics, and
dynamic mode decomposition or U-Net generative adversarial networks are employed as
functional approximation to model the relation.
In the online phase, composition of the trained model is used to perform sequential prediction of the discrete
dynamics of pore collapse process at a general shock pressure.
}
\label{fig:schematic}
\end{figure}

In this work, we employ and compare two
data-driven and machine-learning based methods, namely
dynamic mode decomposition (DMD) and
U-Net generative adversarial networks (GAN),
to serve as efficient non-intrusive surrogate models of the discrete dynamics.
As illustrated in Figure~\ref{fig:schematic},
these methods are used to model the discrete dynamics and
snapshot data from selected training shock pressure are used to train the model.
Composition of the trained model is used to perform sequential prediction of the discrete
dynamics of the pore collapse process at a general shock pressure.
We remark that, since the simulations are expensive,
the training data are scarce.
To the best of our knowledge,
this is the first work in using data-driven non-intrusive surrogate modeling
methods for the pore collapse process.
We make several crucial treatments to the plain original form of the methods
in order to increase the capability of approximating and predicting the dynamics.
For enhancing DMD, we combine the idea of physics-indicated local ROM in \cite{copeland2022reduced,cheung2023local}
and parametric DMD with matrix manifold interpolation in \cite{amsallem2008interpolation,amsallem2011online,choi2020gradient}.
On the other hand, for enhancing GAN, we combine the improved architecture
with conditional continuous input in \cite{kadeethum2022continuous}
and the residual network structure for approximating discrete dynamics (c.f. \cite{chen2021generalized}).

The rest of the paper is organized as follows.
In Section~\ref{sec:pore-coll}, we describe the phenomenon of pore collapse process
and the physics-based high-fidelity simulations.
Next, in Section~\ref{sec:dmd} and Section~\ref{sec:gan}, we discuss the details
of surrogate modeling by DMD and GAN, respectively.
In Section~\ref{sec:exp}, we present some numerical results to test and compare the performance of
the proposed methods. Finally, a conclusion is given in Section~\ref{sec:conclusion}.

\section{Physics-based simulations of pore collapse}
\label{sec:pore-coll}

\begin{figure}[htp!]
\centering
\includegraphics[width=0.19\linewidth]{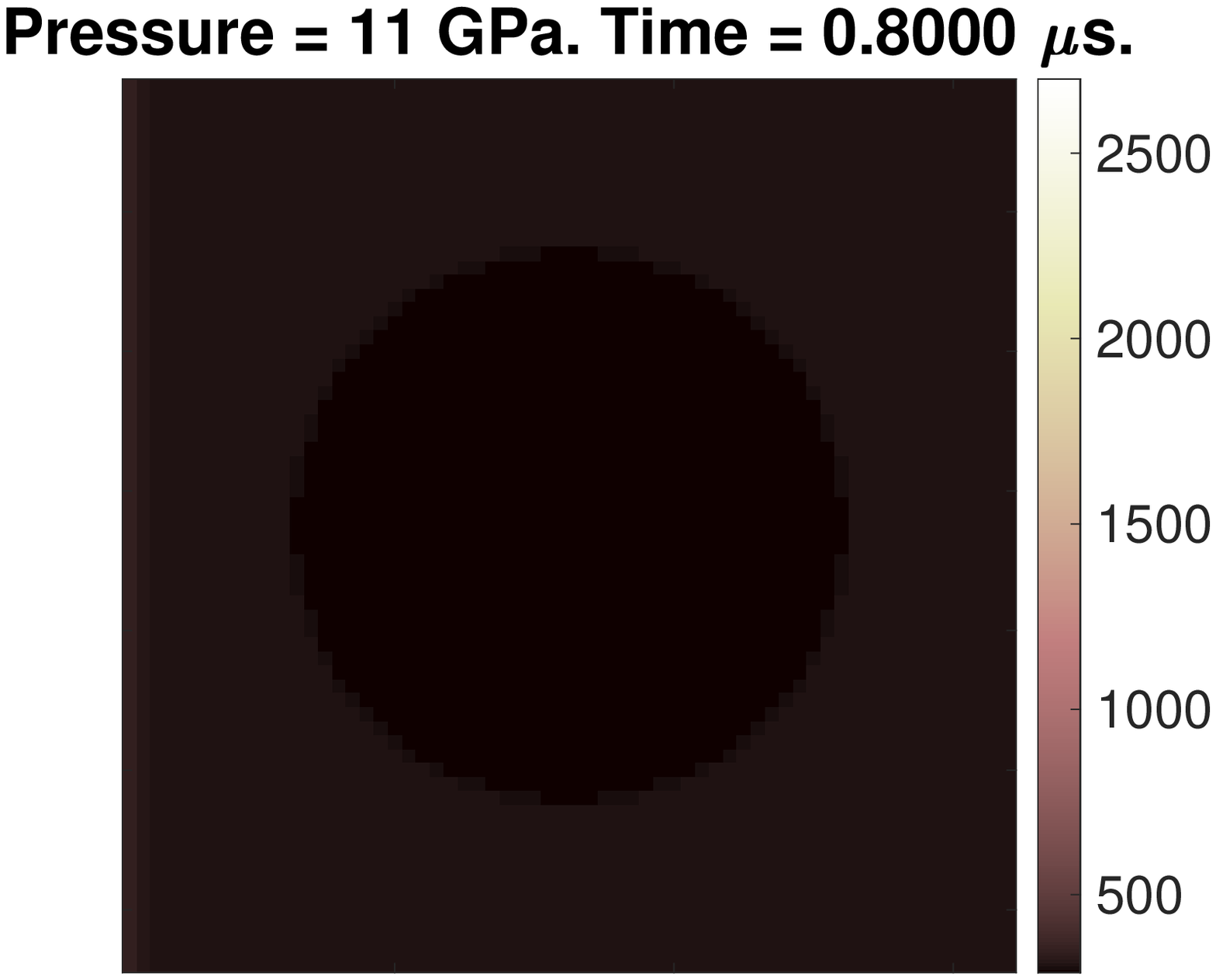}
\includegraphics[width=0.19\linewidth]{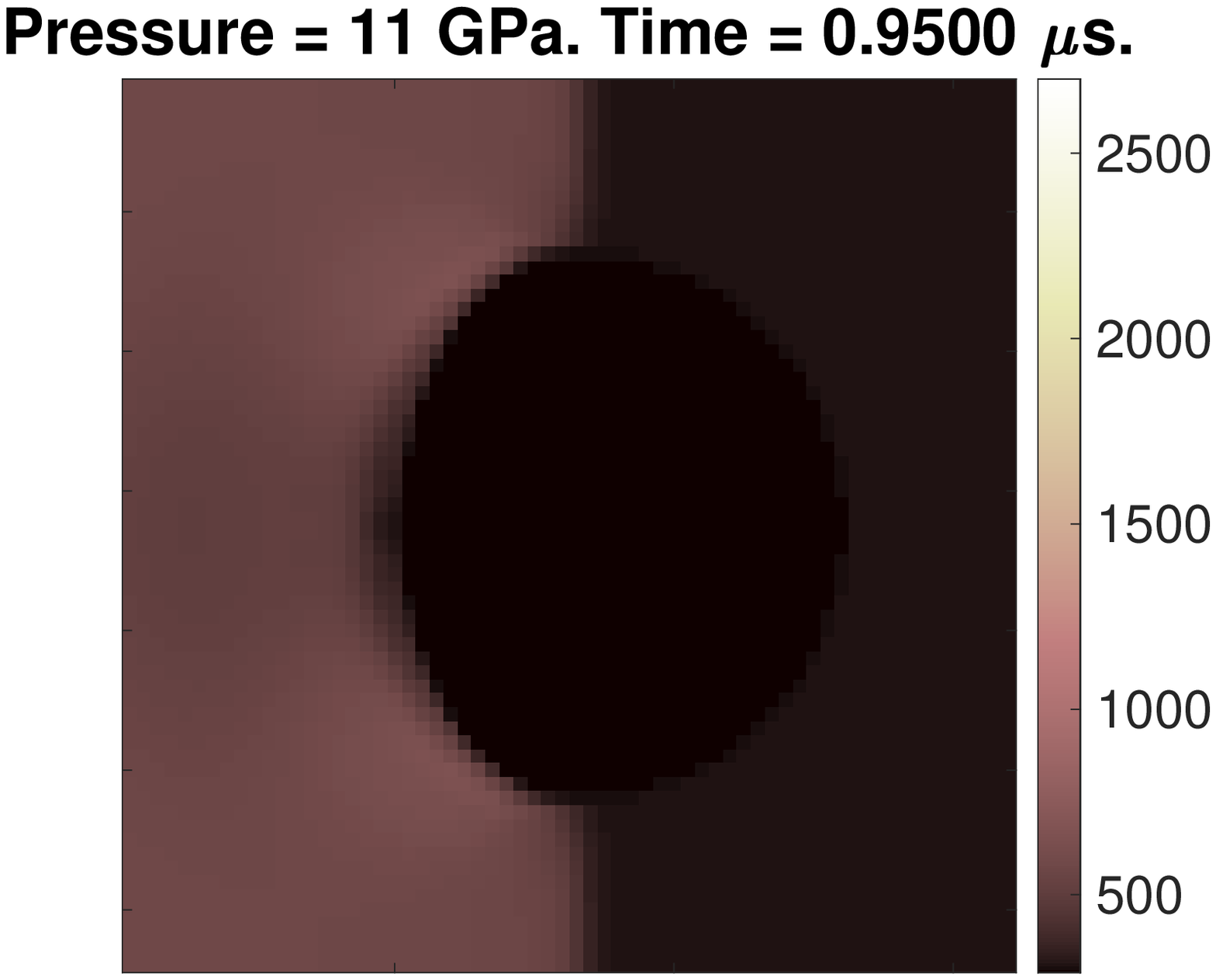}
\includegraphics[width=0.19\linewidth]{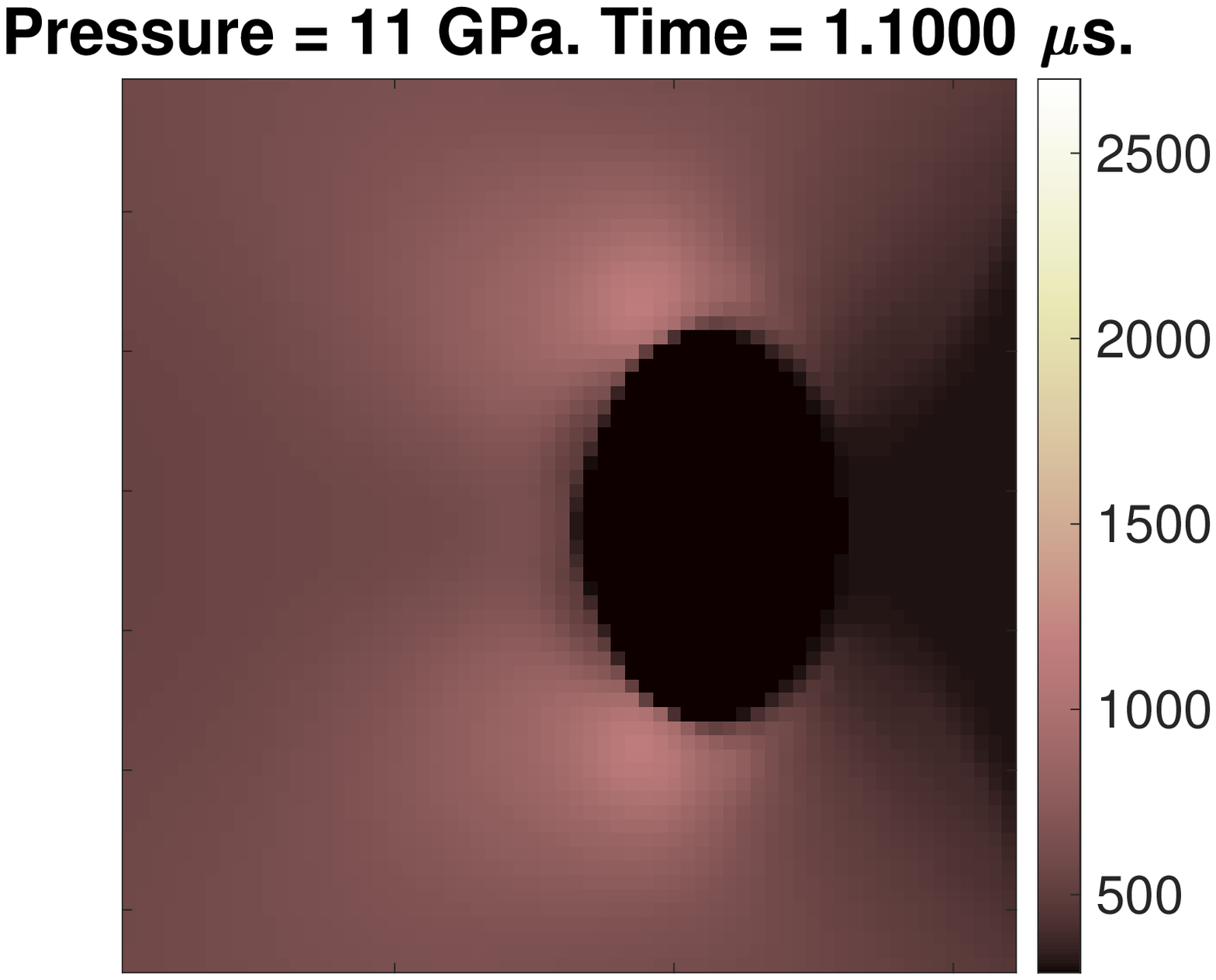}
\includegraphics[width=0.19\linewidth]{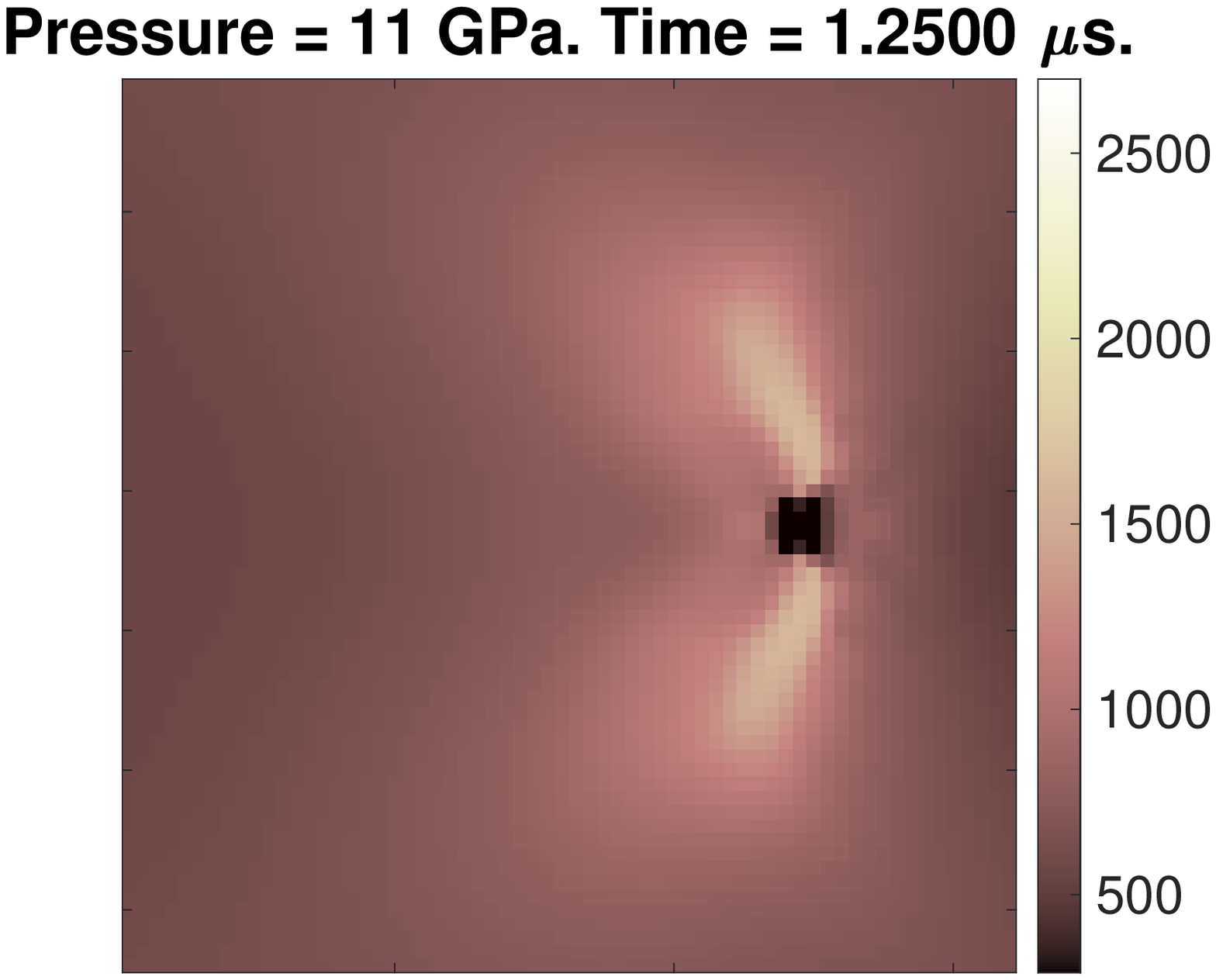}
\includegraphics[width=0.19\linewidth]{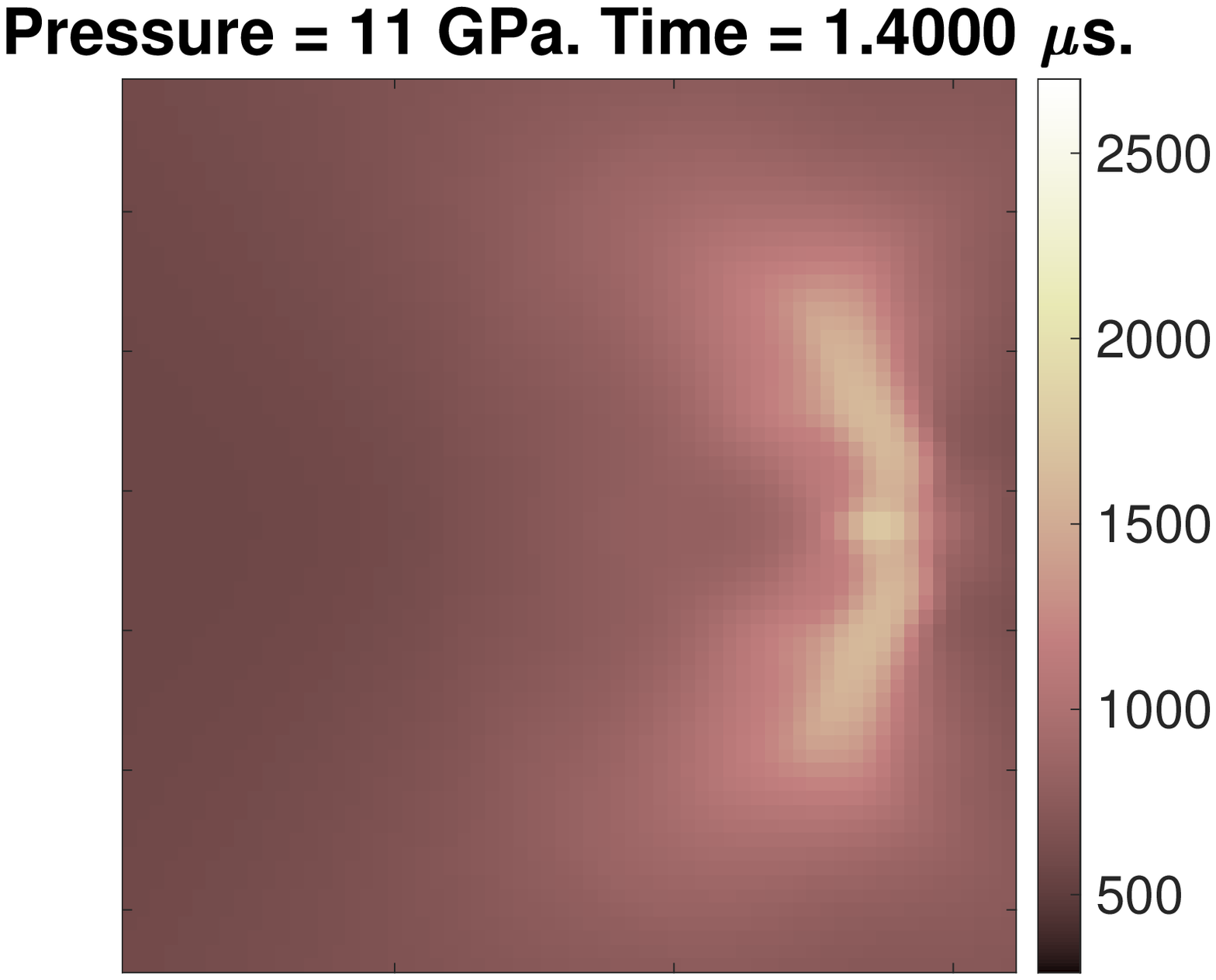}\\
\includegraphics[width=0.19\linewidth]{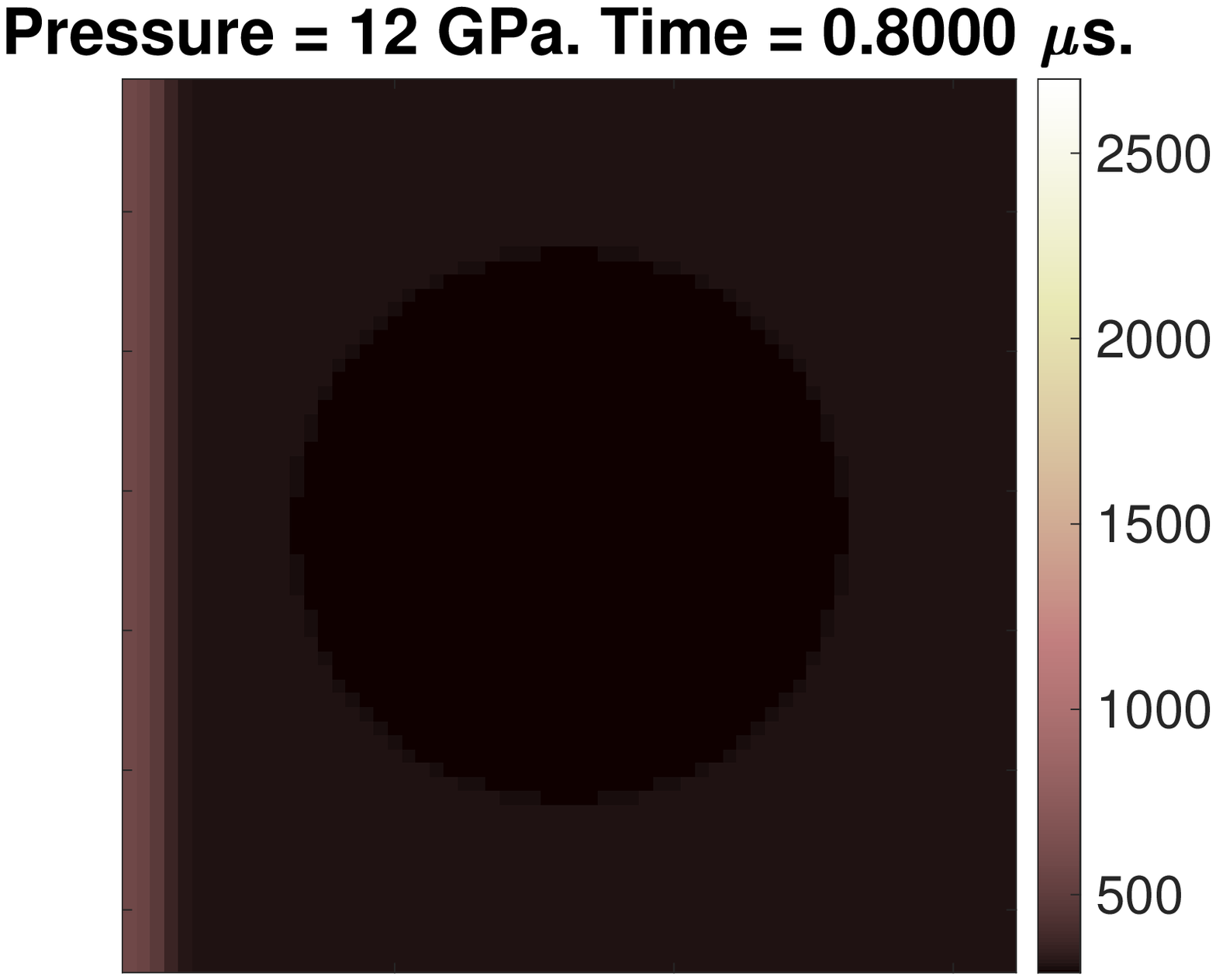}
\includegraphics[width=0.19\linewidth]{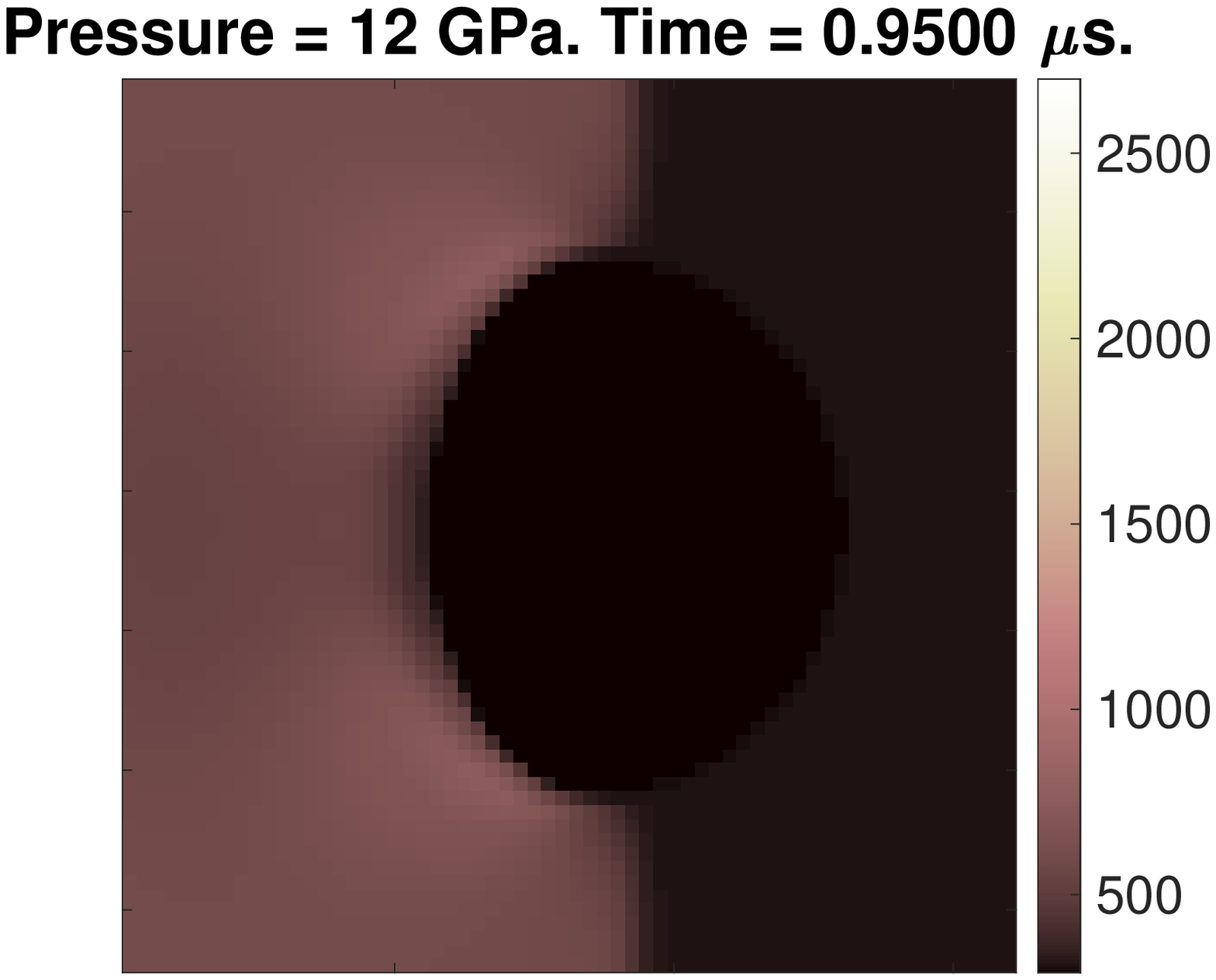}
\includegraphics[width=0.19\linewidth]{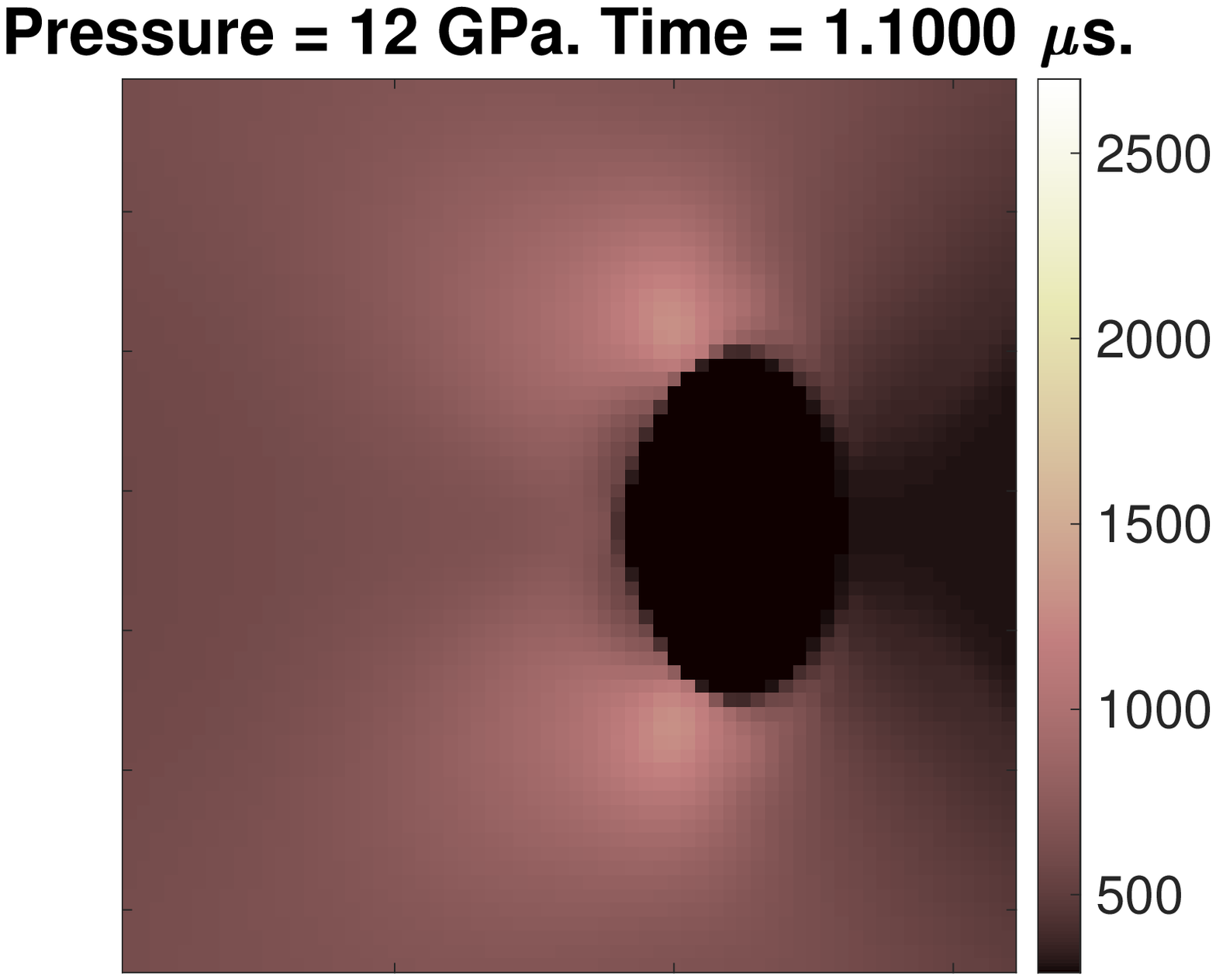}
\includegraphics[width=0.19\linewidth]{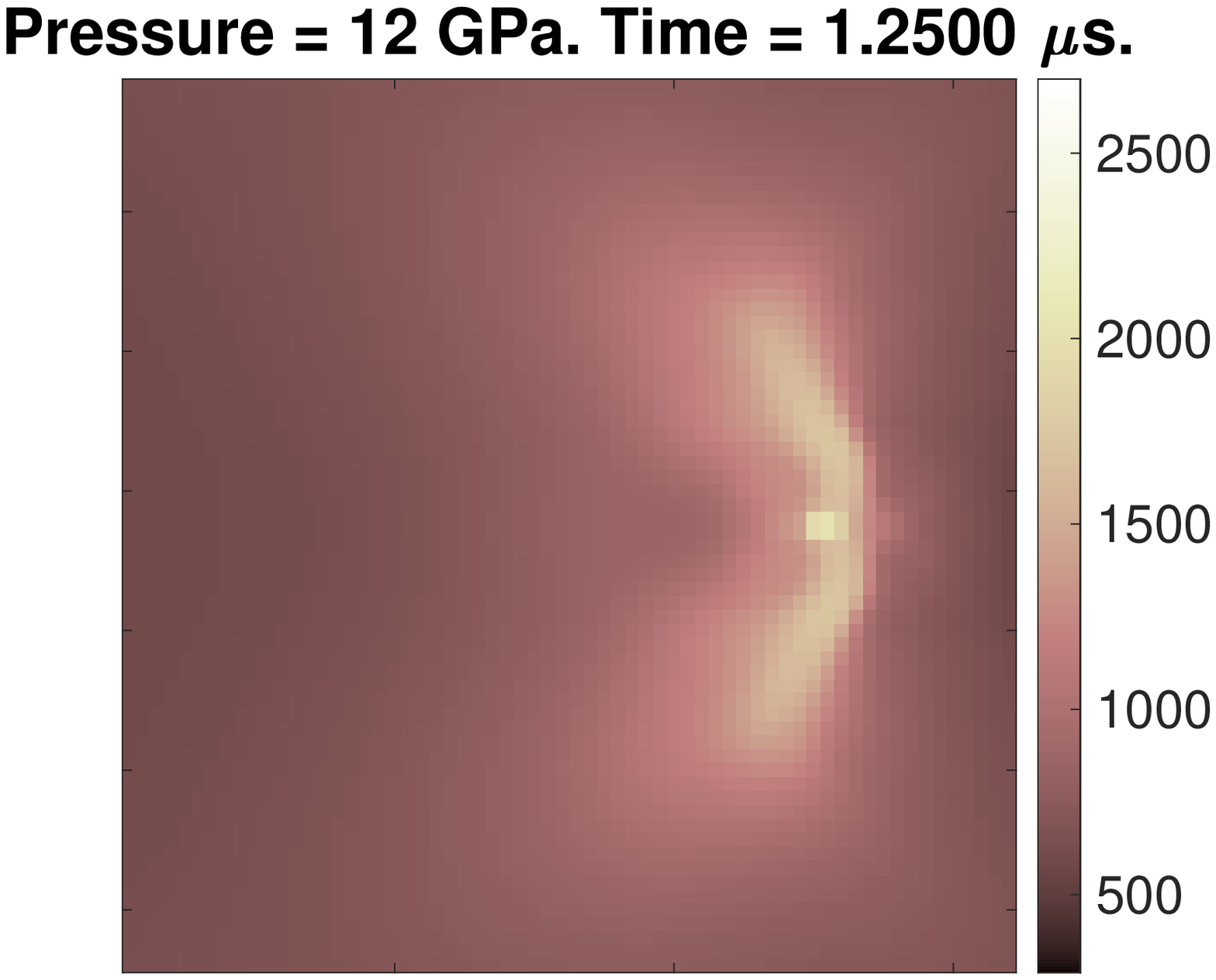}
\includegraphics[width=0.19\linewidth]{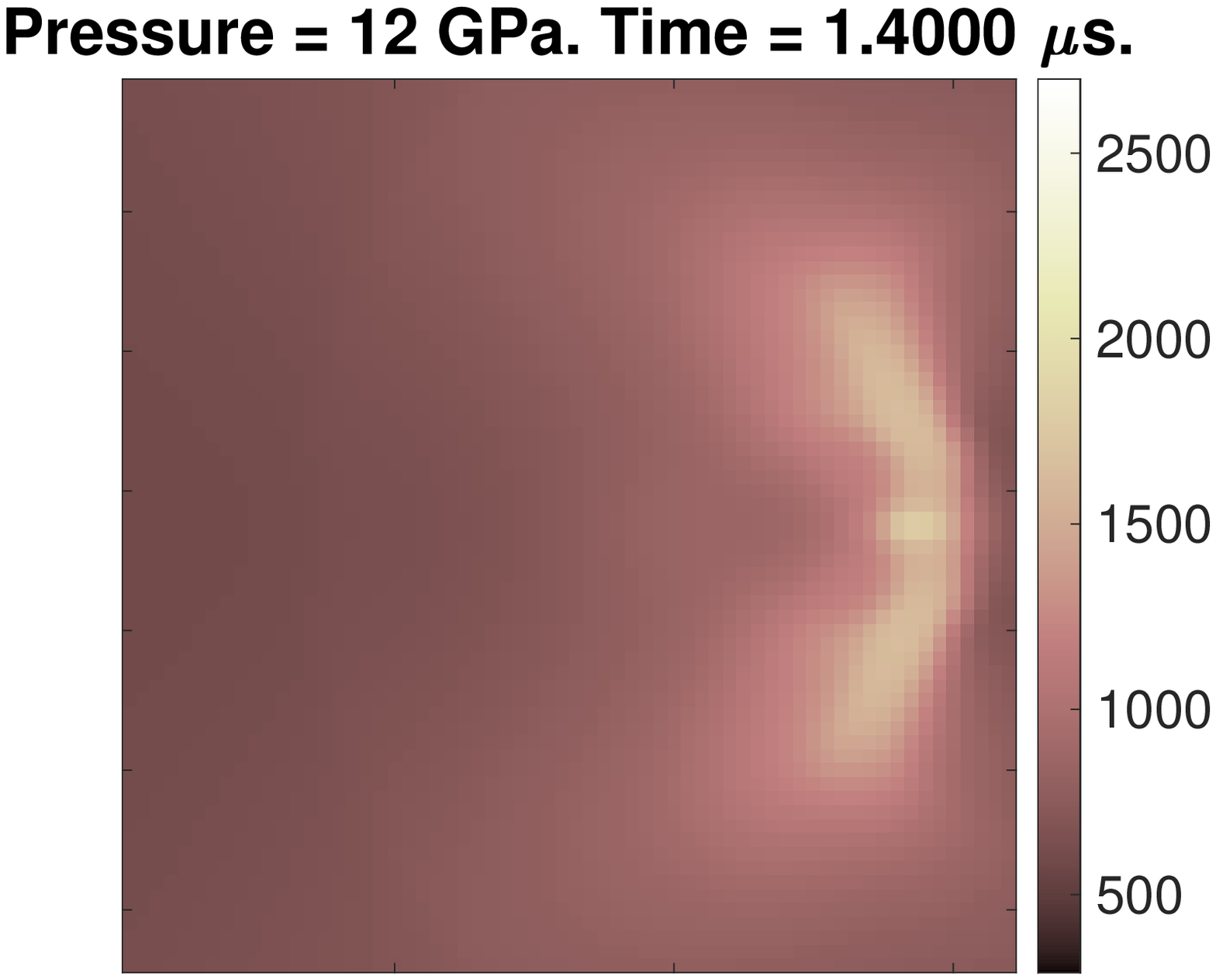}\\
\includegraphics[width=0.19\linewidth]{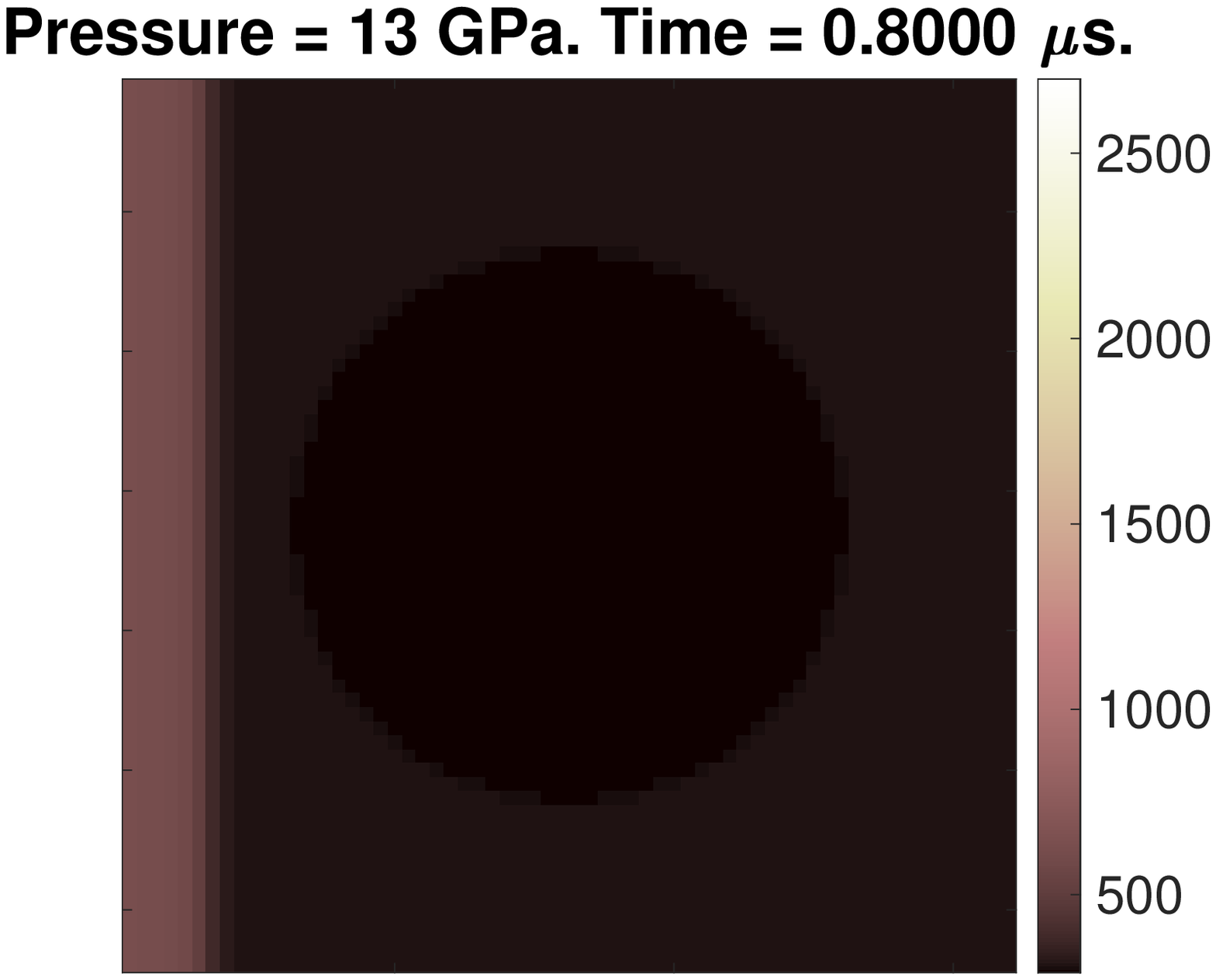}
\includegraphics[width=0.19\linewidth]{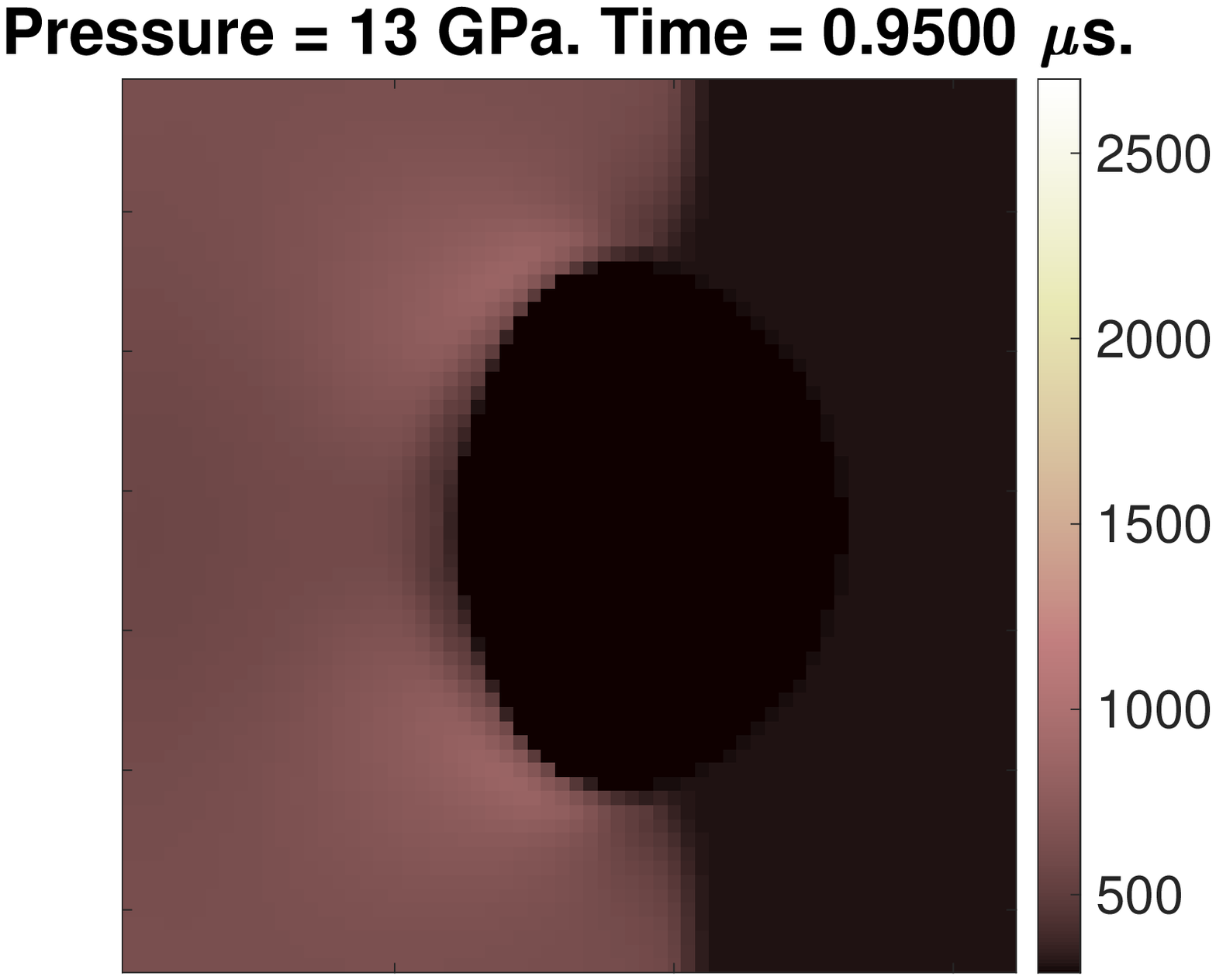}
\includegraphics[width=0.19\linewidth]{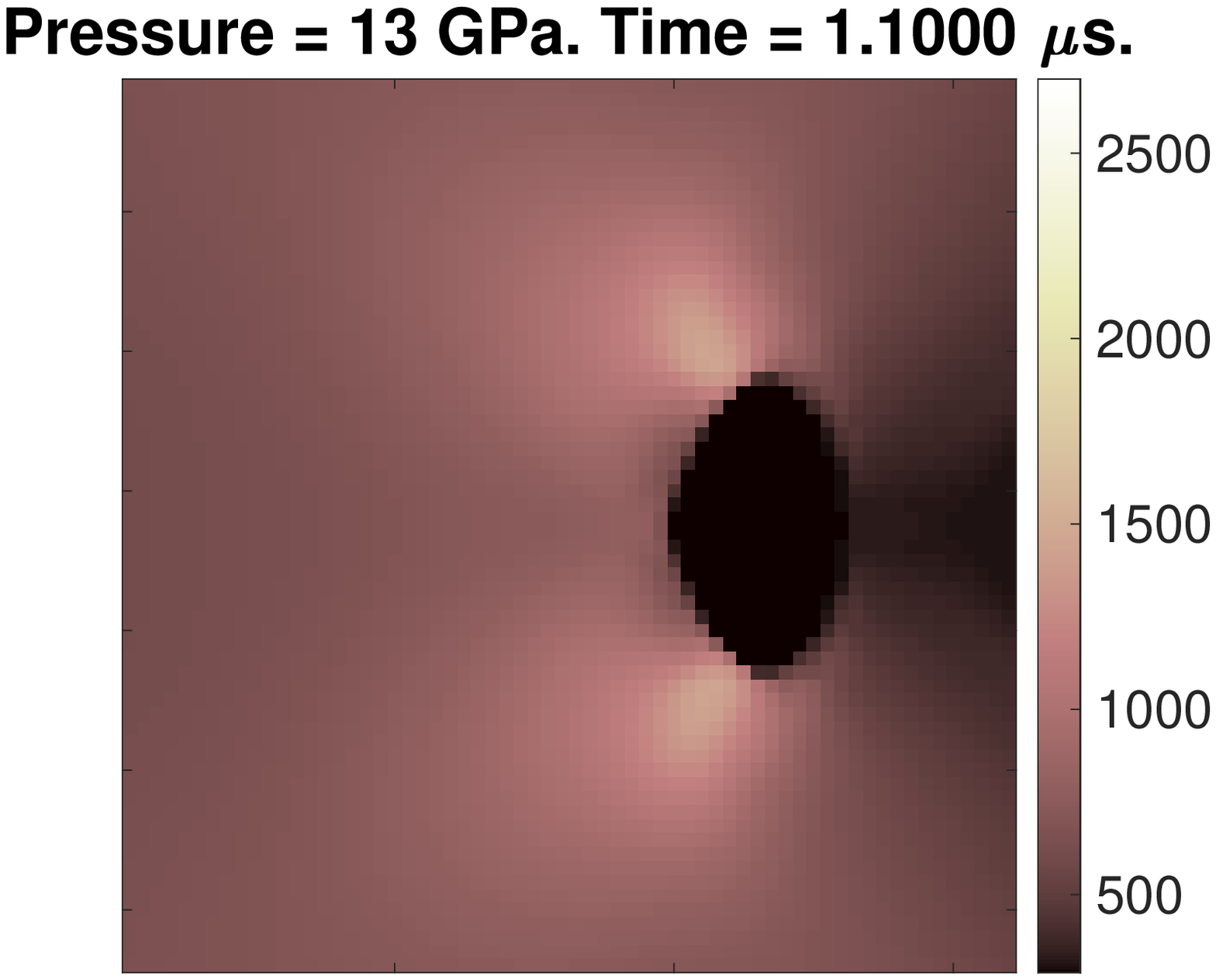}
\includegraphics[width=0.19\linewidth]{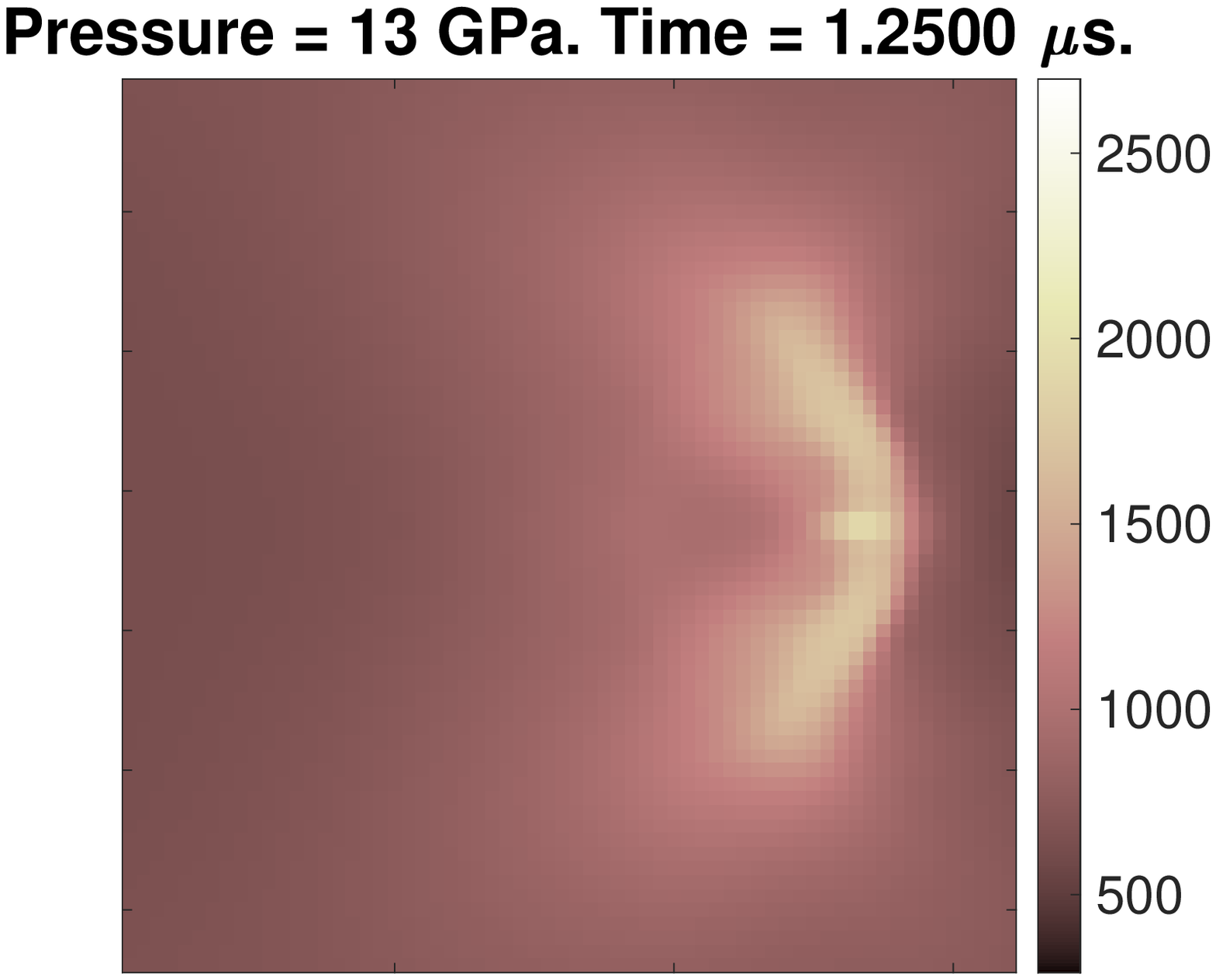}
\includegraphics[width=0.19\linewidth]{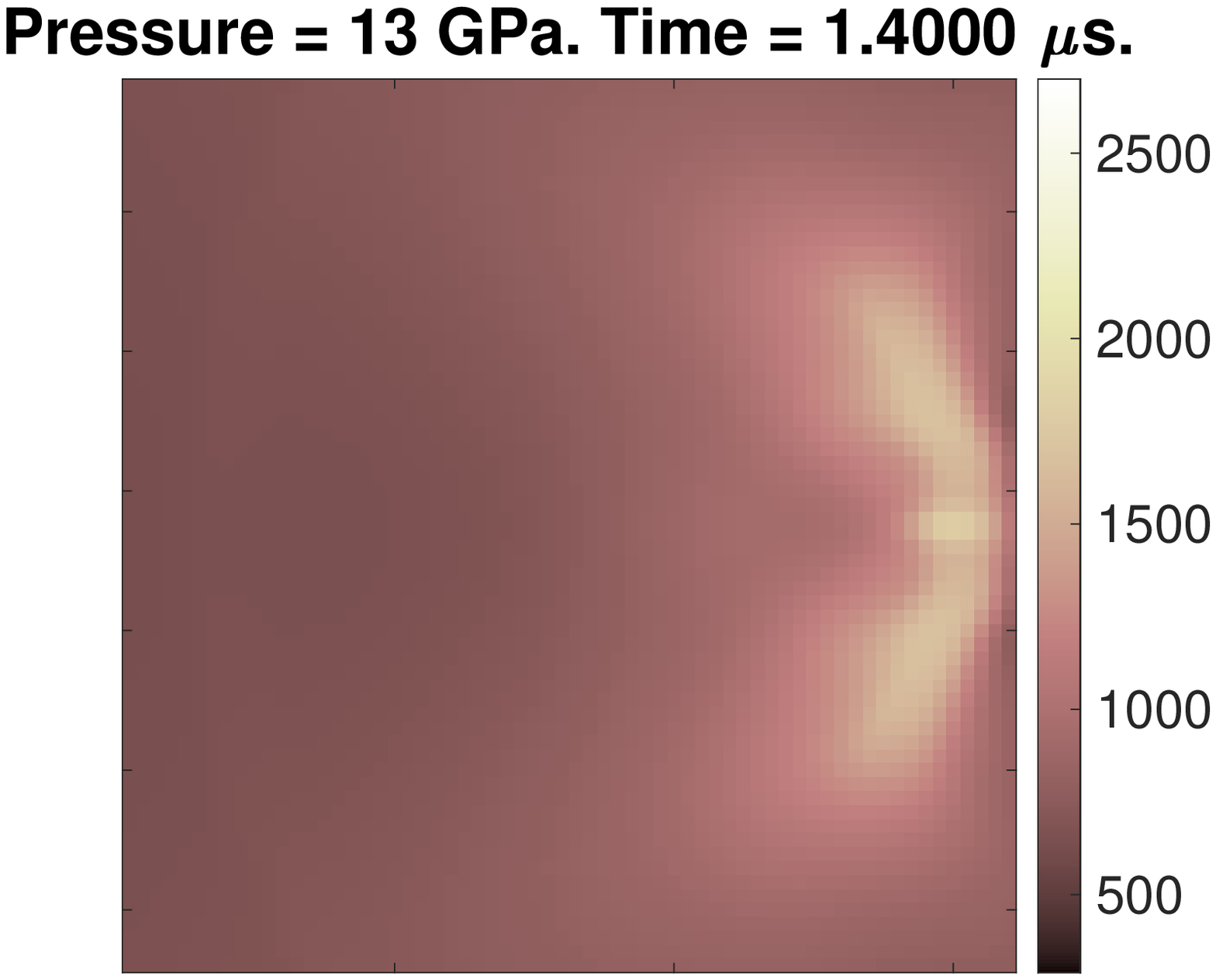}\\
\includegraphics[width=0.19\linewidth]{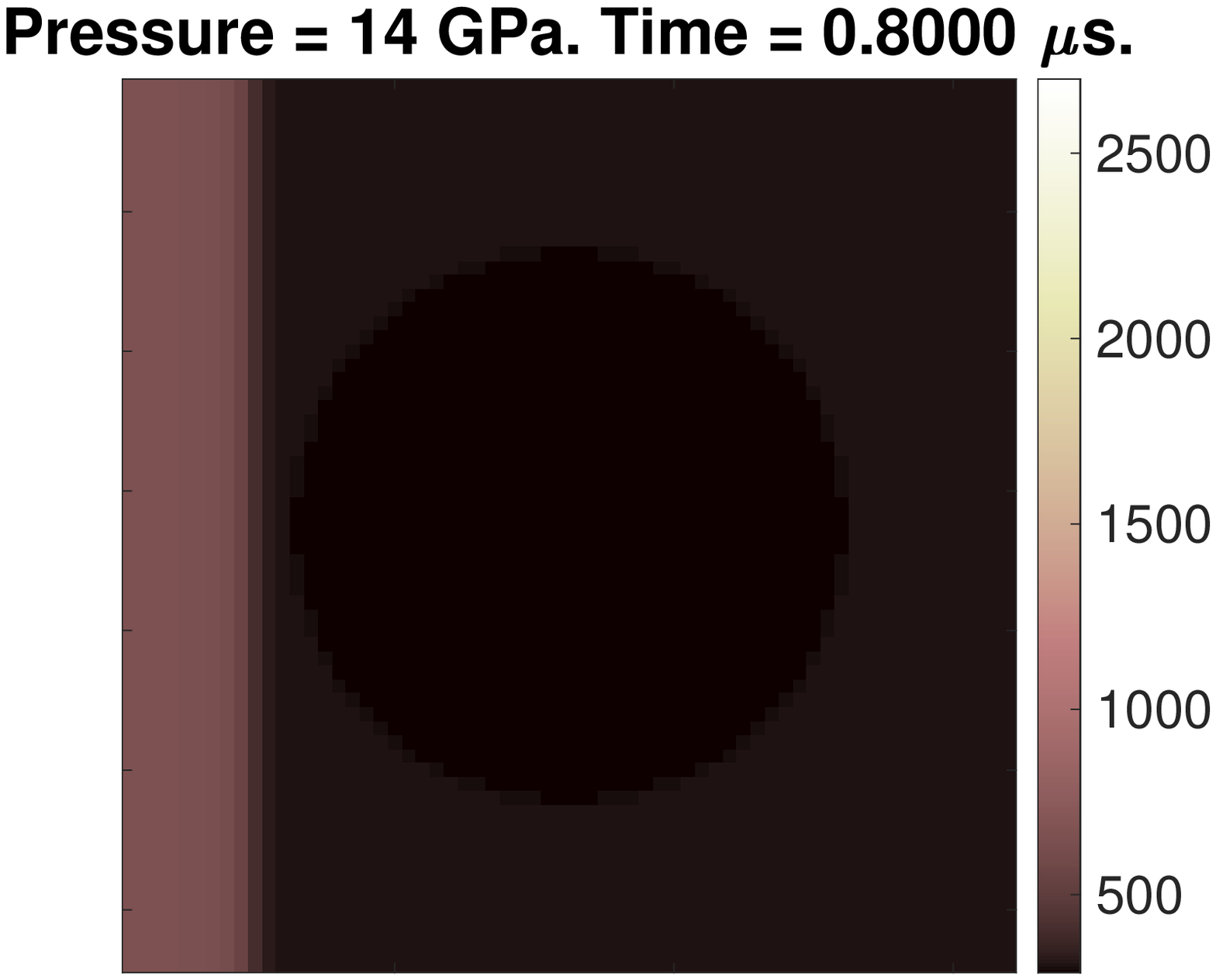}
\includegraphics[width=0.19\linewidth]{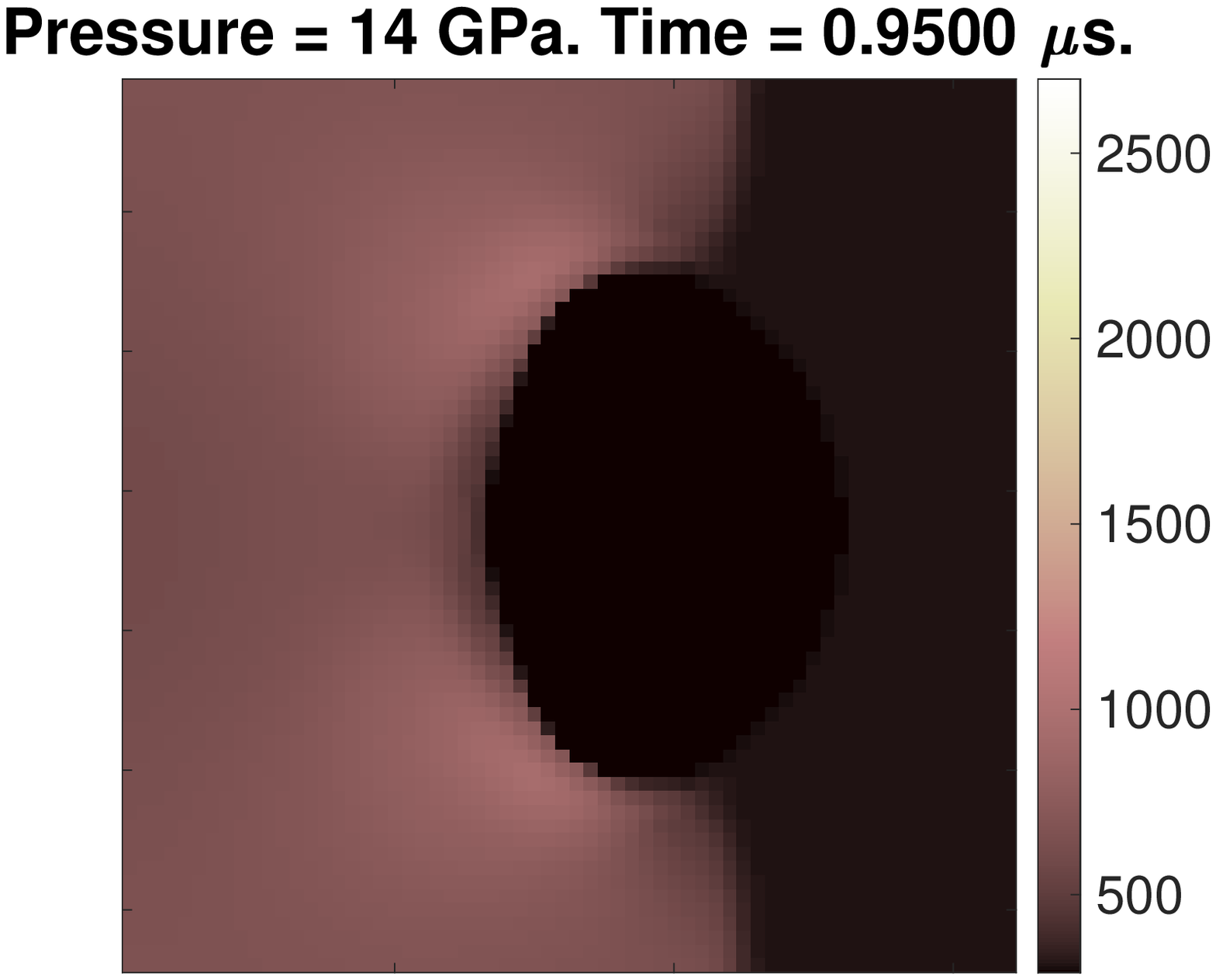}
\includegraphics[width=0.19\linewidth]{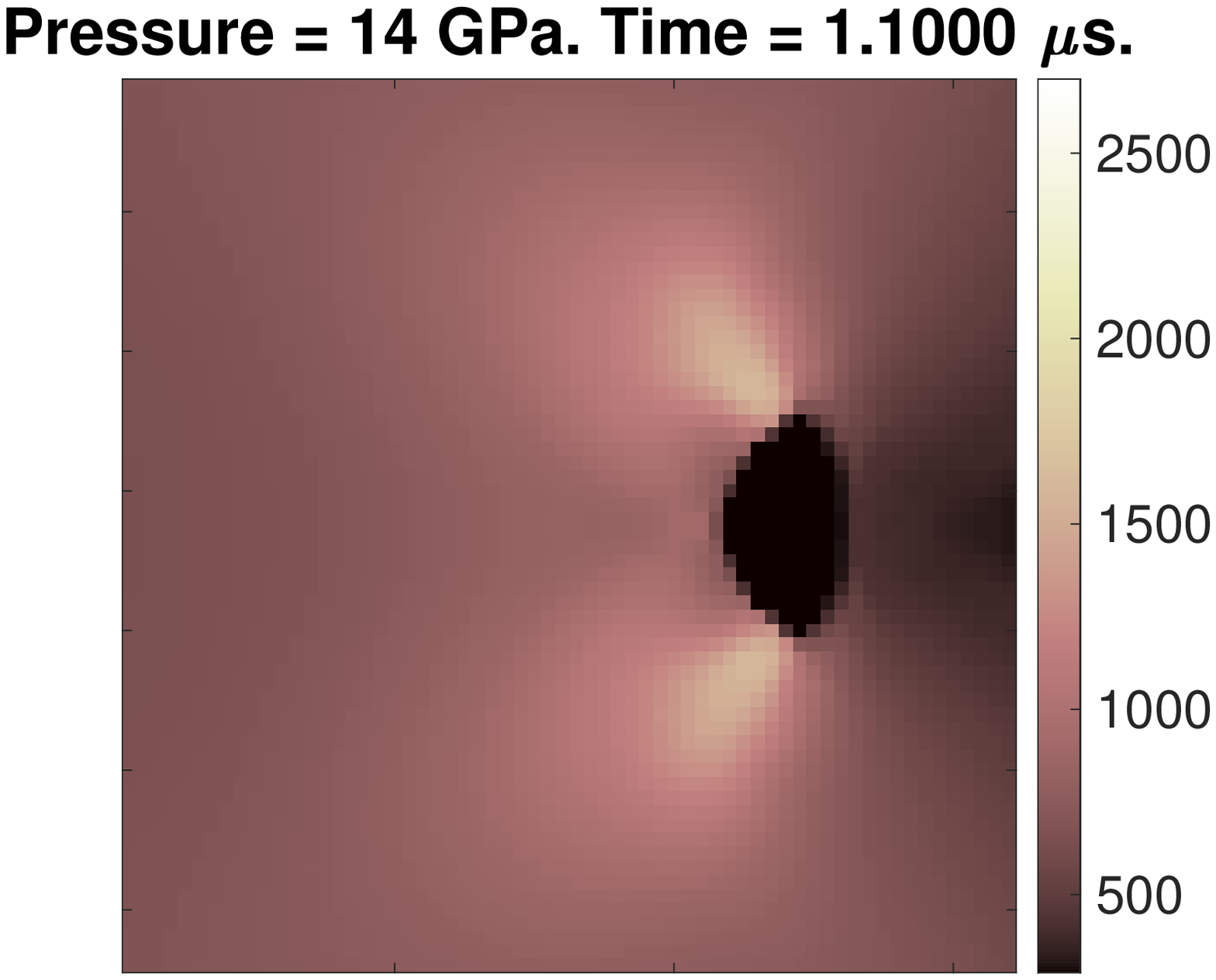}
\includegraphics[width=0.19\linewidth]{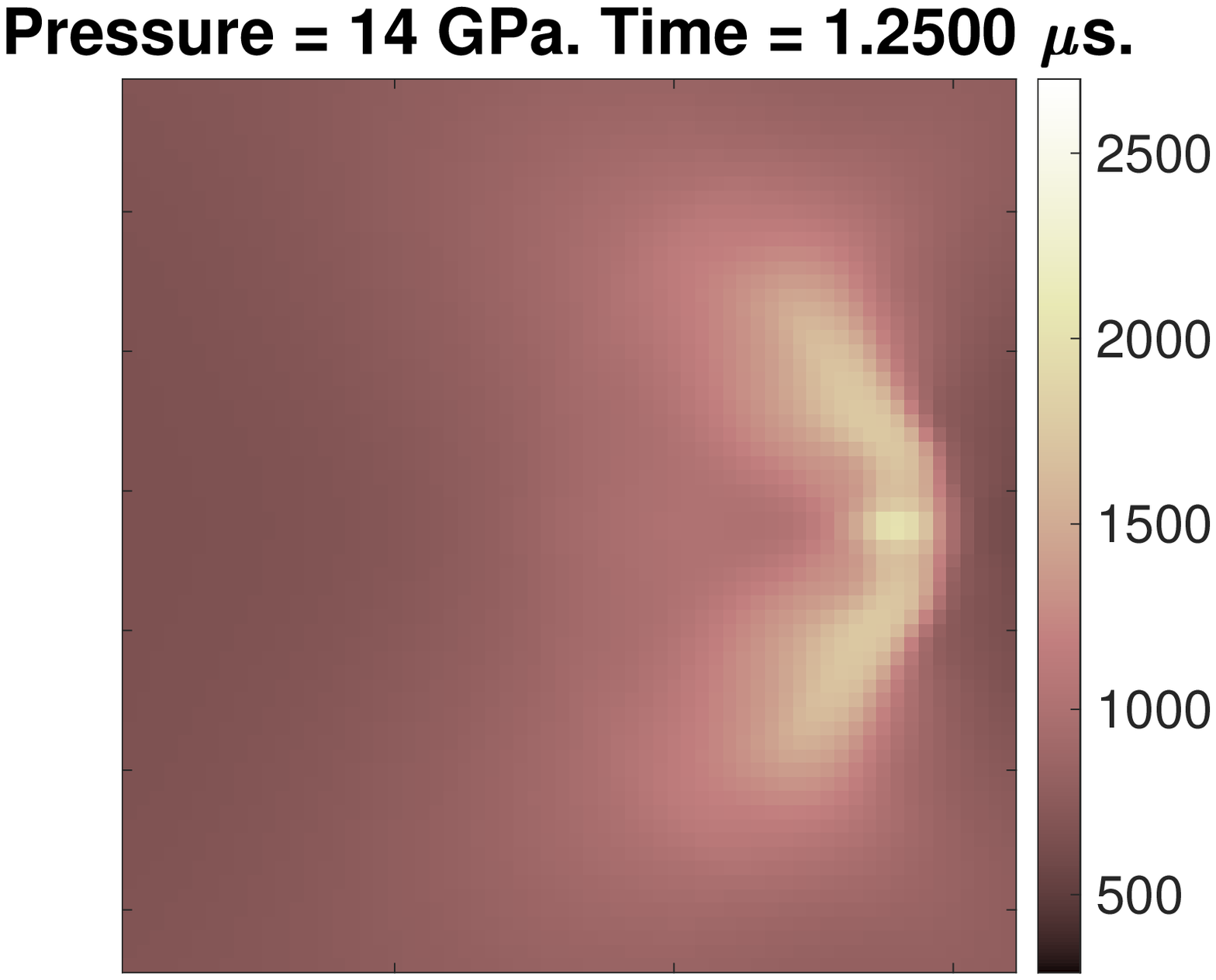}
\includegraphics[width=0.19\linewidth]{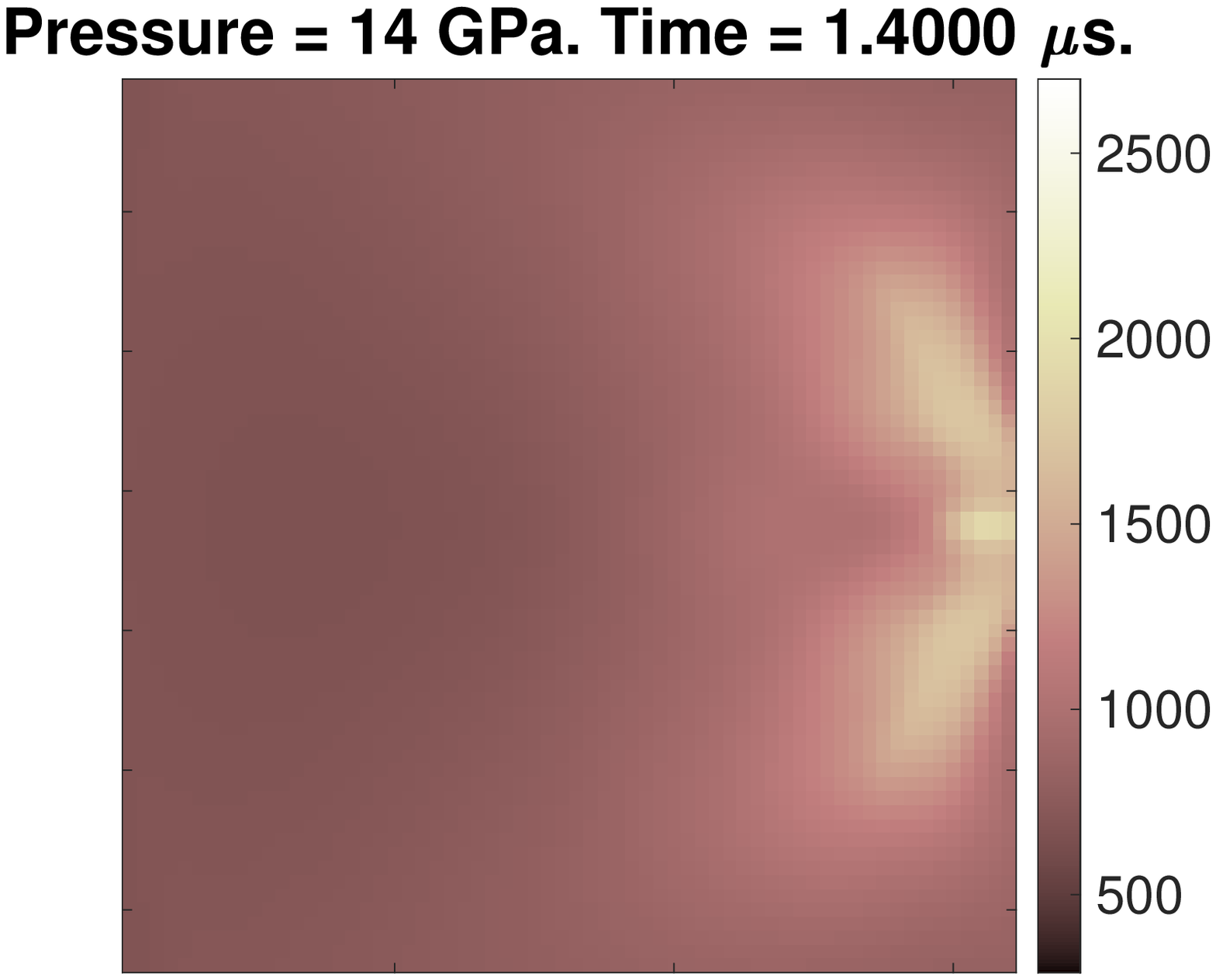}\\
\includegraphics[width=0.19\linewidth]{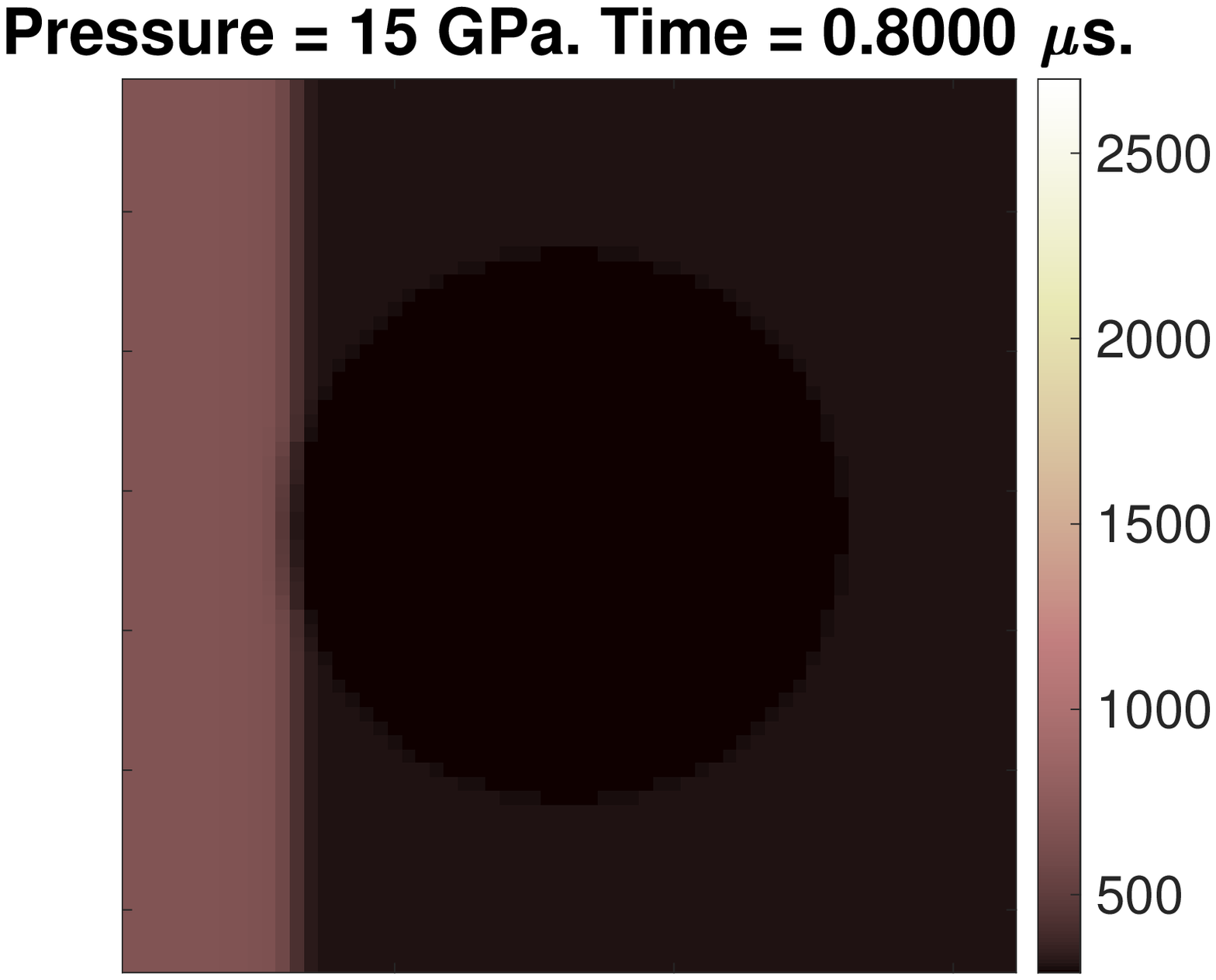}
\includegraphics[width=0.19\linewidth]{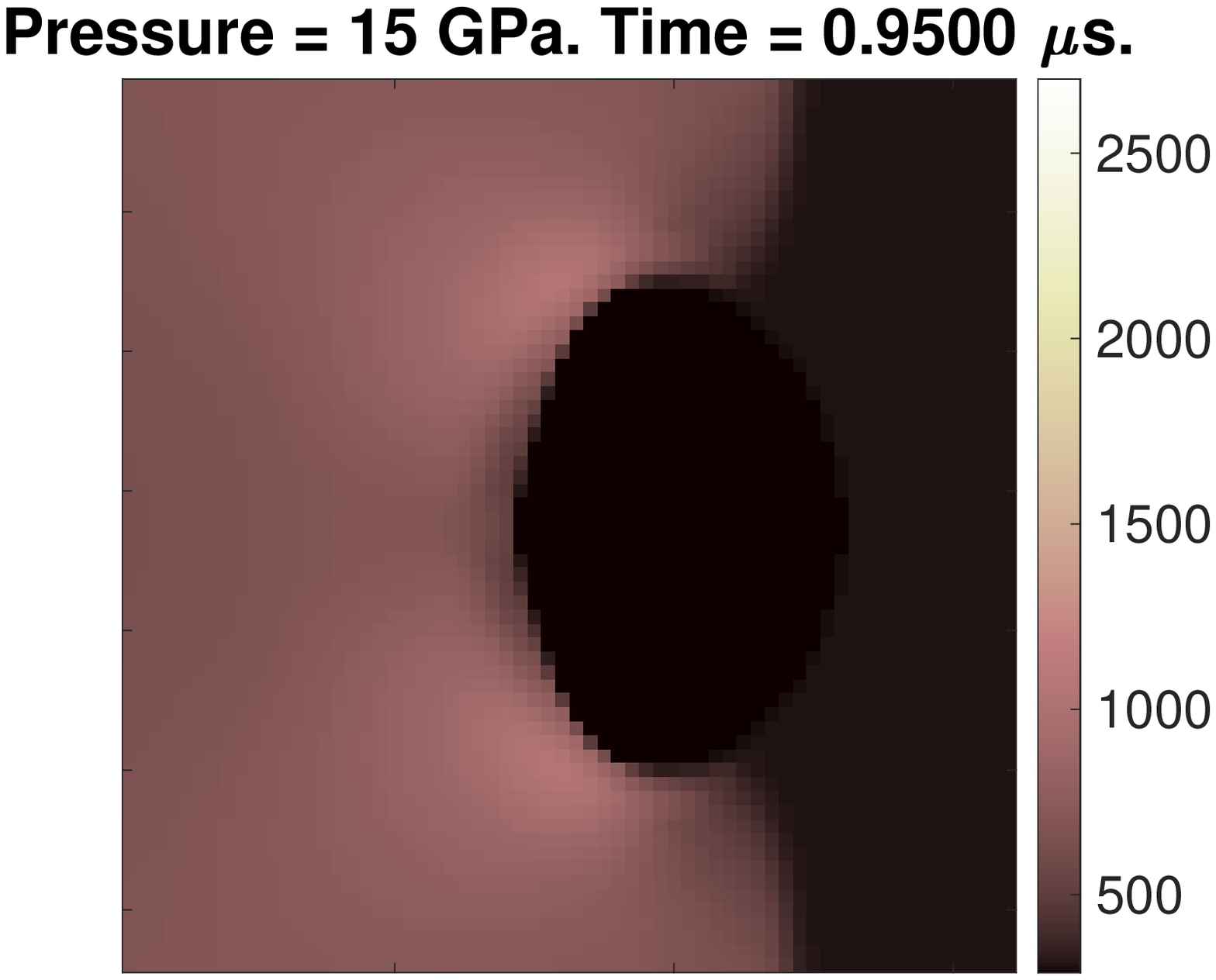}
\includegraphics[width=0.19\linewidth]{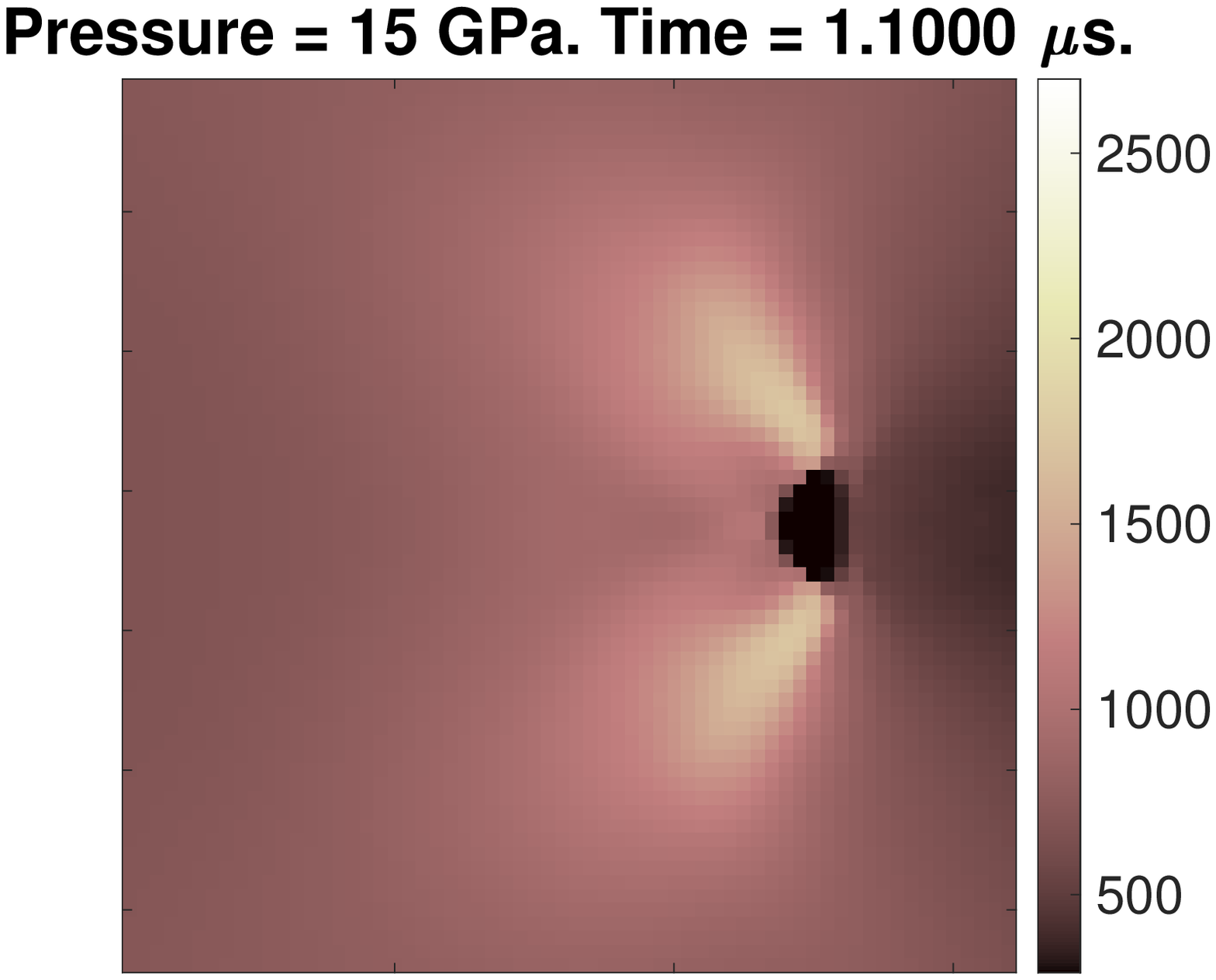}
\includegraphics[width=0.19\linewidth]{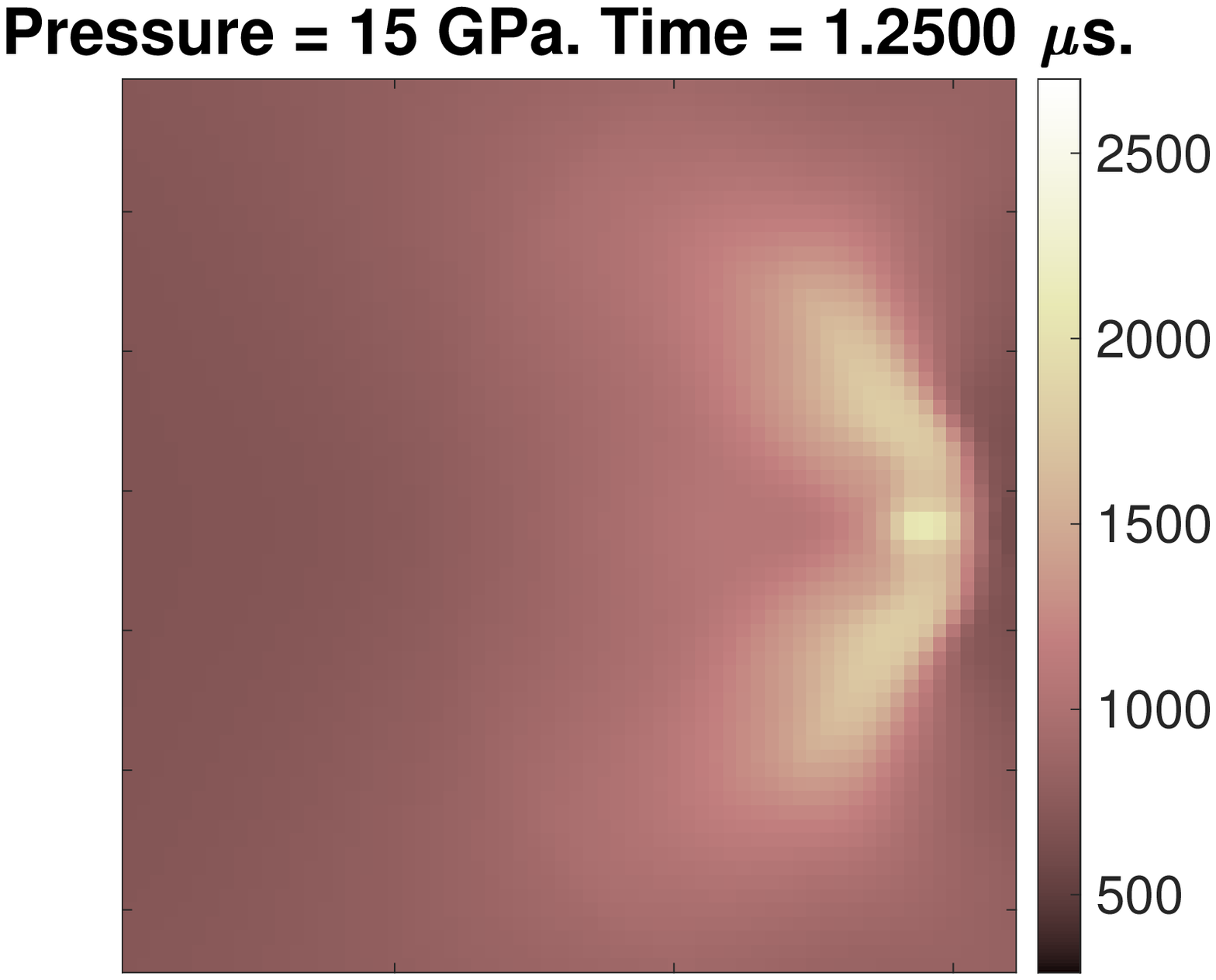}
\includegraphics[width=0.19\linewidth]{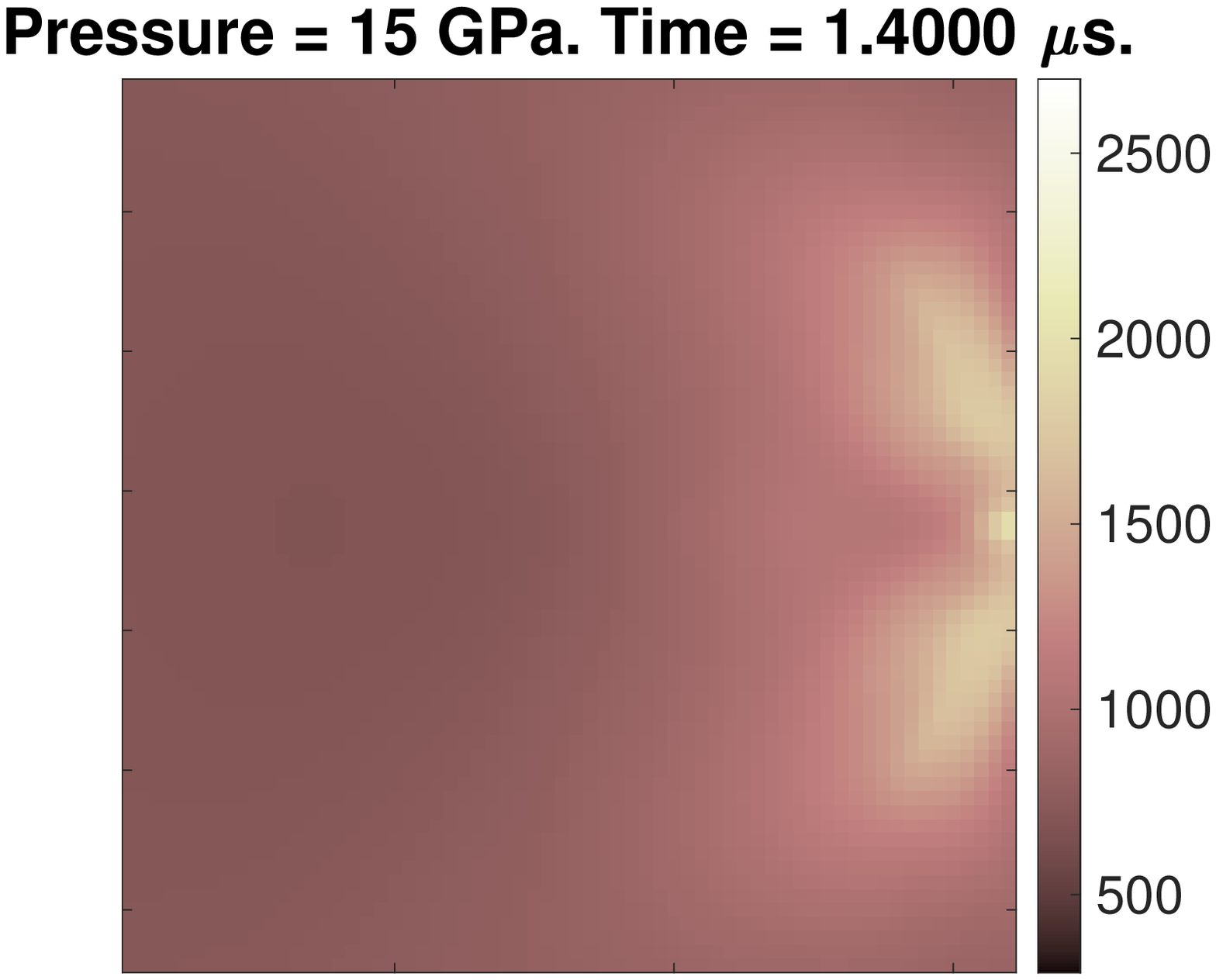}
\caption{Selected representative snapshots of temperature fields at
different shock pressures (11--15 GPa, row-wise)
and time instances (0.8--1.4 $\mu$s, column-wise).
With higher shock pressure, the pore collapse
takes place at an earlier time.}
\label{fig:snap_unif}
\end{figure}

We perform pore collapse simulations using the multi-physics arbitrary Lagrangian Eulerian 
finite element hydrocode, ALE3D \cite{noble2017ale3d}. 
Our simulations consist of a 10 $\mu$m by 10 $\mu$m with a central circular pore whose diameter is 1 $\mu$m. 
The applied shock pressure ranges from 10 to 20 GPa.  Simulations are performed under 2D plane strain conditions. 
Symmetry conditions are imposed on the upper and lower boundaries of the domain. 
Figure~\ref{fig:snap_unif} depicts some selected representative snapshots
of temperature fields at different shock pressures ranging from 11 to 15 GPa,
and time instances ranging from 0.8 to 1.4 $\mu$s.
Each row corresponds to the same shock pressure and
each column corresponds to same time instance.
It can be observed that, with higher shock pressure,
the pore collapse takes place at an earlier time.


Next, we introduce some notations and dimensionless quantities
to simplify the discussion.
Let $\paramDomain = [\param_\text{min}, \param_\text{max}]$
denote the range of applied shock pressure measured in GPa, and
$\solDomainSymbol = [x_\text{min}, x_\text{max}]^2 \subset \mathbb{R}^2$
denote the spatial region of interest with length scale in nm.
The temperature fields are
measured at $\sizeFOMsymbol^2$ square sub-zones
with equal length $h_x$ in $\solDomainSymbol$,
at a uniform sampling rate $\timestep$, and
are represented as matrices
$\state(\timeSymbol; \param) \in \mathbb{R}^{\sizeFOMsymbol \times \sizeFOMsymbol}$ or
vectors $\state(\timeSymbol; \param) \in \mathbb{R}^{\sizeFOMsymbol^2}$,
depending on the surrogate modeling approach under consideration.
For a shock pressure $\param$, the time interval of interest
measured in $\mu$s is denoted by
$\mathcal{T}(\param) = [\timeSymbol^{(0)}(\param), \timeSymbol^{(0)}(\param) +
\ntimestep \timestep]$.
It is important to note that,
since the dynamics is advective and transport in nature and
the traveling speed of the shock varies with shock pressure,
in order to capture the corresponding physics phenonemena,
the initial time $\timeSymbol^{(0)}(\param)$
must be adjusted depending on the shock pressure $\param$.

We end this section by describing the simulation data used
for constructing reduced order models.
The samples of temperature fields $\state_\paramIndex^{k} = \state(\timeSymbol^{(\timeIndex)}_\paramIndex; \param_\paramIndex)$ are measured at
training shock pressures $\paramDomain_{\text{train}} =
\{ \param_\paramIndex \}_{\paramIndex=1}^{\nparam} \subset \paramDomain$
and time instances
$\timeSymbol^{(\timeIndex)}_\paramIndex =
\timeSymbol^{(0)}(\param_\paramIndex) + \timeIndex \timestep$
for $0 \leq \timeIndex \leq \ntimestep$
within the time interval $\mathcal{T}_\paramIndex = \mathcal{T}(\param_\paramIndex)$.
Our goal is to construct reduced order models from the training samples
to resemble the numerical simulations of pore collapse process,
and make predictions of the temperature fields $\widetilde{\state}(\timeSymbol; \param)$
in the time interval of query $\timeSymbol \in \widetilde{\mathcal{T}}(\param)$,
given the initial condition $\state^{(0)}(\param) = \state(\timeSymbol^{(0)}(\param); \param)$,
at generic shock pressures $\param \in \paramDomain \setminus \paramDomain_\text{train}$.
In the rest of this paper, we will introduce techniques to
overcome the difficulties posed to surrogate modeling by
the advective and transport nature of the dynamics.

\section{Dynamic mode decomposition}
\label{sec:dmd}

Dynamic mode decomposition (DMD) was introduced in \cite{schmid2010dynamic}
as a numerical technique for extracting discrete dynamical features
from a sequence of sample data and further studied in
\cite{rowley2009spectral,tu2014dynamic}.
We will given a brief overview of DMD in Section~\ref{sec:serial-dmd}
in the context of numerical simulation data.
Next, in Section~\ref{sec:local-dmd}, we will discuss a specific approach of
modifying DMD to tackle the challenges from the nature of
advective and transport of the shock front.
In Section~\ref{sec:prediction-dmd}, we will introduce the predictive procedure
of DMD on generic shock pressure
$\param \in \paramDomain$,
which is in general unseen in the training samples.

\subsection{Offline stage: serial DMD}
\label{sec:serial-dmd}
We start the offline procedure in DMD
with the sequence of samples
$\{ \state^{(\timeIndex)}_\paramIndex \}_{k=0}^{\ntimestep}$
at a particular training shock pressure
$\param_\paramIndex \in \paramDomain_{\text{train}}$.
The samples $\{ \state_\paramIndex^{(\timeIndex)} \}_{k=0}^{\ntimestep}$
are represented as vectors
in $\mathbb{R}^{\sizeFOMsymbol^2}$.
DMD seeks a linear transformation $\mathbf{A}_\paramIndex \in \mathbb{R}^{\sizeFOMsymbol^2 \times \sizeFOMsymbol^2}$ which approximates the discrete dynamics
\begin{equation}
\state_\paramIndex^{(\timeIndex+1)} \approx \mathbf{A}_\paramIndex \state_\paramIndex^{(\timeIndex)} \text{ for all } 0 \leq \timeIndex < \ntimestep.
\end{equation}
The input snapshot matrix $\mathbf{S}_\paramIndex^-$ and the output snapshot matrix $\mathbf{S}_\paramIndex^+$ of the linear recurrence relation are
\begin{equation}
\label{eq:uniform-snap}
\begin{split}
\mathbf{S}_\paramIndex^- & = \left[ \state_\paramIndex^{(0)}, \state_\paramIndex^{(1)}, \cdots, \state_\paramIndex^{(\ntimestep-1)} \right] \in \mathbb{R}^{\sizeFOMsymbol^2 \times \ntimestep}, \\
\mathbf{S}_\paramIndex^+ & = \left[ \state_\paramIndex^{(1)}, \state_\paramIndex^{(2)}, \cdots, \state_\paramIndex^{(\ntimestep)} \right] \in \mathbb{R}^{\sizeFOMsymbol^2 \times \ntimestep}.
\end{split}
\end{equation}
Performing rank-$\sizeROMsymbol$ truncated
singular value decomposition (SVD) on $\mathbf{S}_\paramIndex^-$ yields
\begin{equation}
\mathbf{S}_\paramIndex^- = \mathbf{U}_\paramIndex\mathbf{\Sigma}_\paramIndex\mathbf{V}_\paramIndex^\top,
\end{equation}
where $\mathbf{U}_\paramIndex \in \mathbb{R}^{\sizeFOMsymbol^2 \times \sizeROMsymbol},
\mathbf{\Sigma}_\paramIndex \in \mathbb{R}^{\sizeROMsymbol \times \sizeROMsymbol},
\mathbf{V}_\paramIndex \in \mathbb{R}^{\ntimestep \times \sizeROMsymbol}$, and
$\sizeROMsymbol \leq \text{rank}(\mathbf{S}^-) \leq \min\{\ntimestep,\sizeFOMsymbol^2\}$.
We remark that the reduced dimension $\sizeROMsymbol$
is assumed to be identical for all training parameters in $\paramDomain_{\text{train}}$.
Then we define the reduced discrete dynamical system by
\begin{equation}
\begin{split}
\widehat{\mathbf{A}}_\paramIndex & = \mathbf{U}_\paramIndex^\top \mathbf{S}_\paramIndex^+ \mathbf{V}_\paramIndex \mathbf{\Sigma}_\paramIndex^{-1} \in \mathbb{R}^{\sizeROMsymbol \times \sizeROMsymbol},
\end{split}
\end{equation}
and perform the spectral decomposition on $\widehat{\mathbf{A}}_\paramIndex$, i.e.
\begin{equation}
\widehat{\mathbf{A}}_\paramIndex \mathbf{X}_\paramIndex = \mathbf{X}_\paramIndex \mathbf{\Lambda}_\paramIndex,
\end{equation}
where $\mathbf{X}_\paramIndex \in \mathbb{C}^{\sizeROMsymbol \times \sizeROMsymbol}$
consists of the eigenvectors of $\widehat{\mathbf{A}}_\paramIndex$ and
$\mathbf{\Lambda}_\paramIndex \in \mathbb{C}^{\sizeROMsymbol \times \sizeROMsymbol}$
is the diagonal matrix containing the DMD eigenvalues.
The DMD basis is then given by $\mathbf{\Phi}_\paramIndex = \mathbf{U}_\paramIndex\mathbf{X}_\paramIndex \in \mathbb{C}^{\sizeFOMsymbol^2 \times \sizeROMsymbol}$.
Then the DMD modes $(\mathbf{\Phi}_\paramIndex, \mathbf{\Lambda}_\paramIndex)$
are used for reproductive approximation
$\widetilde{\state}_{\text{DMD}}(\timeSymbol; \param_\paramIndex)$
of the dynamics at the shock pressure $\param_\paramIndex$, which is given by:
for $\timeSymbol \in \mathcal{T}_\paramIndex = [\timeSymbol_\paramIndex^{(0)}, \timeSymbol_\paramIndex^{(\ntimestep)}]$,
\begin{equation}
\label{eq:dmd-formula}
\widetilde{\state}_{\text{DMD}}(\timeSymbol; \param_\paramIndex) =
\mathbf{\Phi}_\paramIndex \mathbf{\Lambda}_\paramIndex^{\frac{\timeSymbol-\timeSymbol_\paramIndex^{(0)}}{\timestep}} \mathbf{\Phi}_\paramIndex^{\dagger}
\state_\paramIndex^{(0)}.
\end{equation}

\subsection{Offline stage: windowed DMD}
\label{sec:local-dmd}
Section~\ref{sec:serial-dmd} presented a serial DMD, in which the
high-fidelity temperature fields are represented by
ROM subspaces.
However, the advective-dominated nature of the temperature field
implies the weak linear dependence among the
snapshots. As a result, there is no intrinsic low-dimensional subspace
that can universally approximate the solution manifold comprised of
all the solutions over the temporal domain.
In mathematical terms, the solution manifold has slow decay in
Kolmogorov $n$-width.
In order to maintain accuracy with longer simulation time,
the dimension of the reduced subspaces
becomes large if we use the serial DMD.
Furthermore, the large number of high-fidelity snapshot samples also imposes a heavy
burden in storage and computational cost for the SVD computations.

To this end, we employ multiple reduced order models in time
to overcome these difficulties.
The main idea is to construct windowed DMDs in the parameter-time domain
using a suitable indicator for clustering and classification.
In the offline phase, we construct each of these reduced order models
from a small subset of the snapshot samples to ensure low dimension.
In the online phase, each of these reduced order models are used
in a certain subset of the parameter-time domain where they are
supposed to provide good approximation.
Local ROMs have been well studied in the literature
\cite{parish2019windowed,shimizu2020windowed,copeland2022reduced,cheung2023local}.
Following \cite{cheung2023local},
the windowed DMD framework in this paper involves a decomposition of the solution manifold
and relies on an indicator which is used to classify the snapshot samples
and assign the reduced-order models.
The rationale is to decompose the solution manifold into submanifolds
where the Kolmogorov $n$-width decays fast with respect to the subspace dimension,
within which we can collect snapshots with strong linear dependence.
This enables us to build accurate multiple low-dimensional subspaces.

We describe the general framework of indicator-based decomposition of the solution manifold
from which we will derive two practical examples later in this section.
Let $\windowIndicator: \mathbb{R}^{\sizeFOMsymbol^2} \times \mathbb{R}^+
\times \paramDomain \to \mathbb{R}$ be an indicator
which maps the triplet $(\state, \timeSymbol, \param)$ to a real value
in the range $[\windowIndicator_\text{min}, \windowIndicator_\text{max}]$.
For any $\param \in \paramDomain$,
we assume $\windowIndicator(\state^{(0)}(\param), \timeSymbol^{(0)}(\param), \param) = \Psi_\text{min}$,
and $\windowIndicator(\state(\timeSymbol, \param), \timeSymbol, \param)$
is increasing with time $\timeSymbol$.
The range of the indicator is partitioned into $\nwindow$ subintervals, i.e.
\begin{equation}
\label{eq:indicator-partition}
\windowIndicator_\text{min} = \windowIndicator_{0} < \windowIndicator_{1} < \cdots <
\windowIndicator_{\nwindow-1} < \windowIndicator_{\nwindow} = \windowIndicator_\text{max}.
\end{equation}

In the training phase,
at a given training parameter $\param_{i} \in \paramDomain$,
instead of directly assembling all the snapshot samples into huge
snapshot matrices as in \eqref{eq:uniform-snap},
the FOM states are first classified into $\nwindow$ groups.
Given the samples $\{ \state^{(\timeIndex)}_\paramIndex \}_{\timeIndex=0}^{\ntimestep}$
at a shock pressure $\param_\paramIndex \in \paramDomain_{\text{train}}$
and a group index $1 \leq \windowIndex \leq \nwindow$,
we denote by $\snapshotGroup^{(\windowIndex)}_\paramIndex$
the subset of temporal indices
whose corresponding snapshot belongs to the $\windowIndex$-th group, i.e.
\begin{equation}\label{eq:snapsho\timeSymbol_cluster}
\snapshotGroup^{(\windowIndex)}_\paramIndex
= \left\{ 0 \leq \timeIndex < \ntimestep:
\windowIndicator\left(\state^{(\timeIndex)}_\paramIndex, \timeSymbol^{(\timeIndex)}_\paramIndex,
\param_\paramIndex\right) \in [\windowIndicator_{\windowIndex-1}, \windowIndicator_{\windowIndex})
\right\},
\end{equation}
and denote
$K^{(\windowIndex-1)}_\paramIndex = \min \snapshotGroup^{(\windowIndex)}_\paramIndex$ and
$\ntimestep^{(\windowIndex)}_\paramIndex = \vert \snapshotGroup^{(\windowIndex)}_\paramIndex \vert$.
Then $\ntimestep =
\sum_{\windowIndex=1}^{\nwindow} \ntimestep^{(\windowIndex)}$.
Consequently, by
extending $K^{(\nwindow)}_\paramIndex = \ntimestep$
and taking
$\timeWindow^{(\windowIndex)}_\paramIndex = \timeSymbol_{K^{(\windowIndex)}_\paramIndex}$
for $0 \leq j \leq \nwindow$, the time interval
$\mathcal{T}_\paramIndex = [\timeSymbol^{(0)}_\paramIndex,
\timeSymbol^{(\ntimestep)}_\paramIndex]$
at the shock pressure $\param_\paramIndex$
is partitioned into $\nwindow$ subintervals, i.e.
\begin{equation}
\label{eq:temporal-partition}
\timeSymbol^{(0)}_\paramIndex = \timeWindow^{(0)}_\paramIndex
< \timeWindow^{(1)}_\paramIndex < \cdots <
\timeWindow^{(\nwindow-1)}_\paramIndex < \timeWindow^{(\nwindow)}_\paramIndex = \timeSymbol^{(\ntimestep)}_\paramIndex.
\end{equation}
For $1 \leq \paramIndex < \nparam$ and $1 \leq \windowIndex \leq \nwindow$,
we define the snapshot submatrices by
\begin{equation}
\begin{split}
\mathbf{S}^{(\windowIndex), -}_\paramIndex & = \left[ \state^{(\timeIndex)}_\paramIndex
\right]_{k \in \snapshotGroup^{(\windowIndex)}_\paramIndex} \in  \mathbb{R}^{\sizeFOMsymbol^2 \times \ntimestep^{(\windowIndex)}_\paramIndex}, \\
\mathbf{S}^{(\windowIndex), +}_\paramIndex & = \left[ \state^{(\timeIndex+1)}_\paramIndex
\right]_{k \in \snapshotGroup^{(\windowIndex)}_\paramIndex} \in  \mathbb{R}^{\sizeFOMsymbol^2 \times \ntimestep^{(\windowIndex)}_\paramIndex}.
\end{split}
\end{equation}
By carrying out the truncated SVD as discussed in
Section~\ref{sec:serial-dmd} with the pair of snapshot matrices
$(\mathbf{S}^{(\windowIndex), -}_\paramIndex , \mathbf{S}^{(\windowIndex), +}_\paramIndex)$,
we obtain the modal discrete dynamical system
$(\mathbf{U}^{(\windowIndex)}_\paramIndex, \widehat{\mathbf{A}}^{(\windowIndex)}_\paramIndex)
\in \mathbb{R}^{\sizeFOMsymbol^2 \times \sizeROMsymbol_\windowIndex} \times \mathbb{R}^{\sizeROMsymbol_\windowIndex \times \sizeROMsymbol_\windowIndex}$ by
\begin{equation}
\begin{split}
\mathbf{S}^{(\windowIndex), -}_\paramIndex & =
\mathbf{U}^{(\windowIndex)}_\paramIndex
\mathbf{\Sigma}^{(\windowIndex)}_\paramIndex
\left[\mathbf{V}^{(\windowIndex)}_\paramIndex\right]^\top, \\
\widehat{\mathbf{A}}^{(\windowIndex)}_\paramIndex & =
\left[\mathbf{U}^{(\windowIndex)}_\paramIndex\right]^\top
\mathbf{S}^{(\windowIndex), +}_\paramIndex
\mathbf{V}^{(\windowIndex)}_\paramIndex
\left[ \mathbf{\Sigma}^{(\windowIndex)}_\paramIndex \right]^{-1}.
\end{split}
\end{equation}
Again, it is assumed that the reduced dimension $\sizeROMsymbol_\windowIndex$
is identical for all training parameters in $\paramDomain_{\text{train}}$.
Then we perform eigenvalue decomposition
as in Section~\ref{sec:serial-dmd} and obtain the DMD modes
$(\mathbf{\Phi}^{(\windowIndex)}_\paramIndex, \mathbf{\Lambda}^{(\windowIndex)}_\paramIndex)
\in \mathbb{C}^{\sizeFOMsymbol^2 \times \sizeROMsymbol_\windowIndex} \times \mathbb{C}^{\sizeROMsymbol_\windowIndex \times \sizeROMsymbol_\windowIndex}$ by
\begin{equation}
\begin{split}
\widehat{\mathbf{A}}^{(\windowIndex)}_\paramIndex
\mathbf{X}^{(\windowIndex)}_\paramIndex
& = \mathbf{X}^{(\windowIndex)}_\paramIndex
\mathbf{\Lambda}^{(\windowIndex)}_\paramIndex, \\
\mathbf{\Phi}^{(\windowIndex)}_\paramIndex & =
\mathbf{U}^{(\windowIndex)}_\paramIndex
\mathbf{X}^{(\windowIndex)}_\paramIndex,
\end{split}
\end{equation}
which are used for the DMD reproductive approximation
$\widetilde{\state}_{\text{DMD}}(\timeSymbol; \param_\paramIndex)$
at the shock pressure $\param_\paramIndex$ given by:
iteratively for $1 \leq \windowIndex \leq \nwindow$,
for $\timeSymbol \in [\timeWindow^{(\windowIndex-1)}_\paramIndex, \timeWindow^{(\windowIndex)}_\paramIndex]$,
\begin{equation}
\label{eq:DMD-reproductive}
\widetilde{\state}_{\text{DMD}}(t; \param_\paramIndex) = \mathbf{\Phi}_\paramIndex^{(\windowIndex)} \left[\mathbf{\Lambda}_\paramIndex^{(\windowIndex)}\right]^{\frac{\timeSymbol-\timeWindow^{(\windowIndex)}_\paramIndex}{\timestep}} \left[\mathbf{\Phi}_{\paramIndex}^{(\windowIndex)}\right]^{\dagger} \widetilde{\state}_{\text{DMD}}(\timeWindow_\paramIndex^{(\windowIndex-1)}; \param_\paramIndex),
\end{equation}
where $\widetilde{\state}_{\text{DMD}}(\timeWindow_\paramIndex^{(\windowIndex-1)}; \param_\paramIndex)$
is set to be the initial state $\state_\paramIndex^{(0)}$ if $\windowIndex=1$,
and is obtained from DMD approximation in the previous time subinterval for $\windowIndex>1$.
We remark that if $\nwindow = 1$, it reduces
to the serial DMD as discussed in Section~\ref{sec:serial-dmd}.

We end this subsection with two practical choices of the indicator $\windowIndicator$
for the decomposition of solution manifold.
One natural choice is the time windowing (TW) DMD,
where we use the physical time as the indicator, i.e.
$\windowIndicator(\state, \timeSymbol, \param) = (\timeSymbol - \timeSymbol^{(0)}(\param)) / \timestep$.
In this case, $\windowIndicator_\text{min} = 0$ and
$\windowIndicator_\text{max} = \ntimestep$, and
the temporal partition \eqref{eq:temporal-partition} is actually an affine transformation of
indicator range partition \eqref{eq:indicator-partition},
i.e. $\timeWindow^{(\windowIndex)}_\paramIndex = \timeSymbol_\paramIndex^{(0)} + \windowIndicator_{\windowIndex} \timestep$,
for all $1 \leq \windowIndex \leq \nwindow$.

Inspired by \cite{cheung2023local},
another choice of indicator-based decomposition of solution manifold
that is applicable to pore collapse process
is the distance windowing (DW) DMD,
where we use the horizontal translation distance
of the primary shock as the indicator.
Among the $\sizeFOMsymbol^2$ sub-zones,
we select $\sizeFOMsymbol$ sub-zones on the bottom boundary
$x_2 = x_{\text{min}}$ as markers,
and collect their indices into a subset $\mathcal{I}$.
Then the indicator of shock distance is defined as
the number of markers whose temperature values
exceed the temperature threshold $\stateSymbol_{\text{threshold}} = 300$,
which is the critical value distinguishing between the dark background temperature and
the bright hot temperature as illustrated in Figure~\ref{fig:pore_coll_fig}, i.e.
$$\windowIndicator(\state, \timeSymbol, \param) =
\left \vert \{ s \in \mathcal{I}: \mathbf{e}_s^\top \state > \stateSymbol_{\text{threshold}} \} \right \vert. $$
In this case, $\windowIndicator_\text{min} \geq 0$ and
$\windowIndicator_\text{max} = \sizeFOMsymbol$.

\subsection{Prediction stage}
\label{sec:prediction-dmd}

For parametric DMD prediction at a generic shock pressure $\param \in \paramDomain$,
we construct an appropriate temporal partition and use corresponding DMD models
for approximation in each of temporal subintervals.
More precisely, for $1 \leq \windowIndex \leq \nwindow$,
we need to determine the temporal subinterval endpoint
$\timeWindow^{(\windowIndex)}(\param) \in \mathbb{R}$
by scalar-valued interpolation,
and the modal discrete dynamical system
$(\mathbf{U}^{(\windowIndex)}( \param), \widehat{\mathbf{A}}^{(\windowIndex)}( \param))
\in \mathbb{R}^{\sizeFOMsymbol^2 \times \sizeROMsymbol_\windowIndex} \times \mathbb{R}^{\sizeROMsymbol_\windowIndex \times \sizeROMsymbol_\windowIndex}$
by matrix-valued interpolation, with the interpolating points
as the training shock pressures in $\paramDomain_\text{train}$
and the interpolating values in the database obtained at the training shock pressures
as described in Section~\ref{sec:local-dmd}, i.e.
\begin{equation}
\mathcal{DB}^{(\windowIndex)} = \left\{
\left(\param_\paramIndex,
\timeWindow^{(\windowIndex)}_\paramIndex,
\mathbf{U}^{(\windowIndex)}_\paramIndex,
\widehat{\mathbf{A}}^{(\windowIndex)}_\paramIndex\right) \right\}_{\paramIndex=1}^{\nparam}
\subset \paramDomain_\text{train} \times \mathbb{R} \times \mathbb{R}^{\sizeFOMsymbol^2 \times \sizeROMsymbol_\windowIndex} \times \mathbb{R}^{\sizeROMsymbol_\windowIndex \times \sizeROMsymbol_\windowIndex}.
\end{equation}

We adopt the radial basis functions (RBF) interpolation method.
We choose an infinitely smooth radial basis function $\varphi: [0,\infty) \to [0, \infty)$, and
define the interpolation matrix $\mathbf{B} \in \mathbb{R}^{\nparam \times \nparam}$ by
$$ \mathbf{B}_{\paramIndex, \paramIndex^\prime} =
\varphi\left( \left\| \param_\paramIndex - \param_{\paramIndex^\prime} \right\| \right)
\text{ for all } 1 \leq \paramIndex, \paramIndex^\prime \leq \nparam. $$
The scalar-valued interpolant
of the temporal subinterval endpoint
$\timeWindow^{(\windowIndex)}(\param) \in \mathbb{R}$
is given by the linear combination
\begin{equation}
\label{eq:scalar-interpolation}
\timeWindow^{(\windowIndex)}(\param) = \sum_{\paramIndex = 1}^{\nparam}
\omega^{(\windowIndex)}_\paramIndex \varphi\left( \left\| \param - \param_{\paramIndex} \right\| \right),
\end{equation}
where the weights $\boldsymbol{\omega}^{(\windowIndex)} =
(\omega^{(\windowIndex)}_1, \omega^{(\windowIndex)}_2, \ldots,
\omega^{(\windowIndex)}_{\nparam})^\top \in \mathbb{R}^{\nparam}$
are defined by solving
$\mathbf{B} \boldsymbol{\omega}^{(\windowIndex)} = \boldsymbol{\timeWindow}^{(\windowIndex)}
= (\timeWindow^{(\windowIndex)}_1, \timeWindow^{(\windowIndex)}_2, \ldots,
\timeWindow^{(\windowIndex)}_{\nparam})^\top \in \mathbb{R}^{\nparam}$,
which is derived from
\begin{equation}
\label{eq:interpolating-time-window}
\timeWindow^{(\windowIndex)}(\param_\paramIndex) = \timeWindow^{(\windowIndex)}_\paramIndex
\text{ for all } 1 \leq \paramIndex \leq \nparam.
\end{equation}
The interpolated values form a partition for the time interval of query
$\widetilde{\mathcal{T}}(\param) = [\timeWindow^{(0)}(\param),
\timeWindow^{(\nwindow)}(\param)] \subseteq \mathcal{T}(\param)$, i.e.
\begin{equation}
\label{eq:temporal-partition-query}
\timeWindow^{(0)}(\param)
< \timeWindow^{(1)}(\param) < \cdots <
\timeWindow^{(\nwindow-1)}(\param) < \timeWindow^{(\nwindow)}(\param).
\end{equation}

It remains to describe the matrix-valued interpolation.
For a comprehensive discussion on the theory and practice of
interpolation on a matrix manifold in the context of linear subspace
reduced order models, the reader is referred to
\cite{amsallem2008interpolation,amsallem2011online,choi2020gradient}.
Here, we present only the necessary details of RBF interpolation of DMD matrix components
at a generic shock pressure $\param \in \paramDomain$.
The first step is identify a reference training shock pressure index
$1 \leq \paramIndex_\text{ref}(\param) \leq \nparam$ by
\begin{equation}
\paramIndex_\text{ref}(\param) = \argmin_{1 \leq \paramIndex \leq \nparam}
\vert \param - \param_\paramIndex \vert.
\end{equation}
Next, we rotate the reduced order operator to enforce the
consistency in the generalized coordinate system.
For $1 \leq \paramIndex \leq \nparam$,
we perform SVD of the matrix product
$\left[\mathbf{U}^{(\windowIndex)}_\paramIndex\right]^\top
\mathbf{U}^{(\windowIndex)}_{\paramIndex_\text{ref}(\param)}$, i.e.
\begin{equation}
\left[\mathbf{U}^{(\windowIndex)}_\paramIndex\right]^\top
\mathbf{U}^{(\windowIndex)}_{\paramIndex_\text{ref}(\param)}
= \left[\mathbf{Y}^{(\windowIndex)}_{\paramIndex}(\param)\right]^\top
\mathbf{\Gamma}^{(\windowIndex)}_{\paramIndex} (\param)
\mathbf{Z}^{(\windowIndex)}_{\paramIndex} (\param).
\end{equation}
Then we define the rotation matrix
$\mathbf{Q}^{(\windowIndex)}_{\paramIndex}(\param)
\in \mathbb{R}^{\sizeROMsymbol_\windowIndex
\times \sizeROMsymbol_\windowIndex}$ by
\begin{equation}
\mathbf{Q}^{(\windowIndex)}_{\paramIndex}(\param)
= \left[\mathbf{Y}^{(\windowIndex)}_{\paramIndex}(\param)\right]^\top
\mathbf{Z}^{(\windowIndex)}_{\paramIndex} (\param),
\end{equation}
which is the solution to the classical orthogonal Procrustes problem.
The matrix-valued interpolant of the modal discrete dynamical system
$(\mathbf{U}^{(\windowIndex)}(\param), \widehat{\mathbf{A}}^{(\windowIndex)}(\param))
\in \mathbb{R}^{\sizeFOMsymbol^2 \times \sizeROMsymbol_\windowIndex} \times \mathbb{R}^{\sizeROMsymbol_\windowIndex \times \sizeROMsymbol_\windowIndex}$
is then given by the linear combination
\begin{equation}
\begin{split}
\mathbf{U}^{(\windowIndex)}(\param)
& = \mathbf{U}^{(\windowIndex)}_{\paramIndex_{\text{ref}}(\param)} +
\sum_{\paramIndex = 1}^{\nparam}
\mathbf{F}^{(\windowIndex)}_\paramIndex(\param)
\varphi\left( \left\| \param - \param_{\paramIndex} \right\| \right), \\
\widehat{\mathbf{A}}^{(\windowIndex)}(\param)
& = \widehat{\mathbf{A}}^{(\windowIndex)}_{\paramIndex_{\text{ref}}(\param)} +
\sum_{\paramIndex = 1}^{\nparam}
\mathbf{G}^{(\windowIndex)}_\paramIndex(\param)
\varphi\left( \left\| \param - \param_{\paramIndex} \right\| \right).
\end{split}
\end{equation}
Here, for $1 \leq \ell_1 \leq \sizeFOMsymbol^2$ and $1 \leq \ell_2 \leq \sizeROMsymbol_\windowIndex$,
the $(\ell_1, \ell_2)$-entry of the weights $\mathbf{F}^{(\windowIndex)}_\paramIndex(\param) \in
\mathbb{R}^{\sizeFOMsymbol^2 \times \sizeROMsymbol_\windowIndex}$,
denoted by $\mathbf{f}^{(\windowIndex)}_{\ell_1, \ell_2}(\param)
= \left([\mathbf{F}^{(\windowIndex)}_\paramIndex(\param)]_{\ell_1, \ell_2}\right)_{\paramIndex=1}^{\nparam}
\in \mathbb{R}^{\nparam}$,
are defined by solving
\begin{equation}
\mathbf{B} \mathbf{f}^{(\windowIndex)}_{\ell_1, \ell_2}(\param)
= \left( \left[
\mathbf{U}^{(\windowIndex)}_\paramIndex
\mathbf{Q}^{(\windowIndex)}_{\paramIndex}(\param) -
\mathbf{U}^{(\windowIndex)}_{\paramIndex_\text{ref}(\param)}
\right]_{\ell_1, \ell_2} \right)_{\paramIndex=1}^{\nparam}
\in \mathbb{R}^{\nparam}.
\end{equation}
Similarly, for $1 \leq \ell_1, \ell_2 \leq \sizeROMsymbol_\windowIndex$,
the $(\ell_1, \ell_2)$-entry of the weights $\mathbf{G}^{(\windowIndex)}_\paramIndex(\param) \in
\mathbb{R}^{\sizeROMsymbol_\windowIndex \times \sizeROMsymbol_\windowIndex}$,
denoted by $\mathbf{g}^{(\windowIndex)}_{\ell_1, \ell_2}(\param)
= \left([\mathbf{G}^{(\windowIndex)}_\paramIndex(\param)]_{\ell_1, \ell_2}\right)_{\paramIndex=1}^{\nparam}
\in \mathbb{R}^{\nparam}$,
are defined by solving
\begin{equation}
\mathbf{B} \mathbf{g}^{(\windowIndex)}_{\ell_1, \ell_2}(\param)
= \left( \left[
\mathbf{Q}^{(\windowIndex)}_{\paramIndex}(\param)^\top
\widehat{\mathbf{A}}^{(\windowIndex)}_\paramIndex
\mathbf{Q}^{(\windowIndex)}_{\paramIndex}(\param) -
\widehat{\mathbf{A}}^{(\windowIndex)}_{\paramIndex_\text{ref}(\param)}
\right]_{\ell_1, \ell_2} \right)_{\paramIndex=1}^{\nparam}
\in \mathbb{R}^{\nparam}.
\end{equation}

As in Section~\ref{sec:serial-dmd},
we perform eigenvalue decomposition
and obtain the DMD modes
$(\mathbf{\Phi}^{(\windowIndex)}(\param), \mathbf{\Lambda}^{(\windowIndex)}(\param)
\in \mathbb{C}^{\sizeFOMsymbol^2 \times \sizeROMsymbol_\windowIndex} \times \mathbb{C}^{\sizeROMsymbol_\windowIndex \times \sizeROMsymbol_\windowIndex}$ by
\begin{equation}
\begin{split}
\widehat{\mathbf{A}}^{(\windowIndex)}(\param)
\mathbf{X}^{(\windowIndex)}(\param)
& = \mathbf{X}^{(\windowIndex)}(\param)
\mathbf{\Lambda}^{(\windowIndex)}(\param), \\
\mathbf{\Phi}^{(\windowIndex)}(\param) & =
\mathbf{U}^{(\windowIndex)}(\param)
\mathbf{X}^{(\windowIndex)}(\param).
\end{split}
\end{equation}
With the initial condition
$\widetilde{\state}_{\text{DMD}}(\timeSymbol^{(0)}(\param); \param) = \state^{(0)}(\param)$,
the DMD prediction $\widetilde{\state}_{\text{DMD}}(\timeSymbol; \param)$
is then given by:
iteratively for $1 \leq \windowIndex \leq \nwindow$,
for $\timeSymbol \in [\timeWindow^{(\windowIndex-1)}(\param),
\timeWindow^{(\windowIndex)}(\param)]$,
\begin{equation}
\label{eq:DMD-prediction}
\widetilde{\state}_{\text{DMD}}(\timeSymbol; \param) = \mathbf{\Phi}^{(\windowIndex)}(\param) \left[\mathbf{\Lambda}^{(\windowIndex)}(\param)\right]^{\frac{\timeSymbol-\timeWindow^{(\windowIndex)}(\param)}{\timestep}} \left[\mathbf{\Phi}^{(\windowIndex)}(\param)\right]^{\dagger} \widetilde{\state}_{\text{DMD}}(\timeWindow^{(\windowIndex-1)}(\param); \param),
\end{equation}
where $\widetilde{\state}_{\text{DMD}}(\timeWindow^{(\windowIndex-1)}; \param)$
is set to be the initial state $\state^{(0)}(\param)$ if $\windowIndex=1$,
and is obtained from DMD approximation in the previous time subinterval for $\windowIndex>1$.

As a final remark, for all $1 \leq \paramIndex \leq \nparam$,
we have $\paramIndex_\text{ref}(\param_\paramIndex) = \paramIndex$,
which implies $\mathbf{Q}^{(\windowIndex)}_{\paramIndex}(\param_\paramIndex)
= \mathbf{I}_{\sizeROMsymbol_\windowIndex}$.
Thanks to \eqref{eq:interpolating-time-window}, we have
$\timeWindow^{(\windowIndex)}(\param_\paramIndex) =
\timeWindow^{(\windowIndex)}_\paramIndex$
for all $0 \leq \windowIndex \leq \nwindow$, and
 $\mathbf{U}^{(\windowIndex)}(\param_\paramIndex) =
\mathbf{U}^{(\windowIndex)}_{\paramIndex}$ and
$\widehat{\mathbf{A}}^{(\windowIndex)}(\param_\paramIndex)
= \widehat{\mathbf{A}}^{(\windowIndex)}_{\paramIndex}$
for all $1 \leq \windowIndex \leq \nwindow$.
Therefore, \eqref{eq:DMD-prediction} actually reproduces
\eqref{eq:DMD-reproductive} at the training shock pressures
$\param_\paramIndex \in \paramDomain_\text{train}$.

\section{Continuous conditional generative adversarial network}
\label{sec:gan}

Generative adversarial network (GAN) was introduced in \cite{goodfellow2014generative}
as a deep learning method that learns a parametrized representation
by random latent codes for a set of training data in an unsupervised manner and
allows fast sampling from the distribution represented by the dataset.
In the original work \cite{goodfellow2014generative}, GAN formulates
a two-player minimax game with a binary classification score as an optimization problem,
and trains two artificial neural networks, the discriminator and the generator,
simultaneously to optimize the objective function in opposing ways.
These networks compete with each other, with one aiming to maximize the objective function
and the other aiming to minimize it.
In \cite{radford2015unsupervised}, deep convolutional generative adversarial network (DCGAN) is
developed by for image generation tasks by utilizing deep convolutional architectures in GAN.

In this section, we introduce a GAN-based
dynamical prediction scheme for the numerical simulation data.
Our method is based on residual network structure and
modified from \cite{kadeethum2022continuous}
which adopts several recent improvements to GAN, including
batch-based critic architecture \cite{demir2018patch} and
U-Net generator architecture \cite{ronneberger2015unet}
in \texttt{pix2pix} \cite{isola2017image} for the image-to-image translation task,
earth mover distance as loss function in Wasserstein GAN \cite{arjovsky2017wasserstein},
and continuous conditional generator input \cite{ding2020continuous}.
In Section~\ref{sec:offline-gan}, we will discuss the details of the neural network.
In Section~\ref{sec:prediction-gan}, we will introduce the predictive procedure
of GAN on generic shock pressure
$\param \in \paramDomain$, which is in general unseen in the training samples.

\subsection{Offline stage}
\label{sec:offline-gan}
We begin the discussion of the offline procedure
in the continuous conditional generative adversarial network (CcGAN)
approach with data preprocessing.
We represent the sampled data of the temperature fields
$\state_\paramIndex^{(\timeIndex)}$ as matrices in $\mathbb{R}^{\sizeFOMsymbol \times \sizeFOMsymbol}$,
and define the residual as
\begin{equation}
{\residual}_\paramIndex^{(\timeIndex)}
= {\state}_\paramIndex^{(\timeIndex+1)} - {\state}_\paramIndex^{(\timeIndex)}
\in \mathbb{R}^{\sizeFOMsymbol \times \sizeFOMsymbol}.
\end{equation}
The training data are normalized by:
for $1 \leq \paramIndex \leq \nparam$ and $0 \leq \timeIndex < \ntimestep$,
\begin{equation}
\begin{split}
\overline{\timeSymbol}^{(\timeIndex)} & = \timeIndex / (\ntimestep - 1) \in [0,1], \\
\overline{\param}_\paramIndex & = (\param_\paramIndex - \param_\text{min})/(\param_\text{max} - \param_\text{min}) \in [0,1], \\
\overline{\state}_\paramIndex^{(\timeIndex)} & = {\state}_\paramIndex^{(\timeIndex)}/(\stateSymbol_\text{max} - \stateSymbol_\text{min}) \in [0,1]^{\sizeFOMsymbol \times \sizeFOMsymbol}, \\
\overline{\residual}_\paramIndex^{(\timeIndex)} & = {\residual}_\paramIndex^{(\timeIndex)}/(\residualSymbol_\text{max} - \residualSymbol_\text{min}) \in [0,1]^{\sizeFOMsymbol \times \sizeFOMsymbol},
\end{split}
\end{equation}
where
\begin{equation}
\begin{split}
\stateSymbol_\text{max} & = \max_{1 \leq \paramIndex \leq \nparam, 0 \leq \timeIndex < \ntimestep}
\state_\paramIndex^{(\timeIndex)}, \\
\stateSymbol_\text{in} & = \max_{1 \leq \paramIndex \leq \nparam, 0 \leq \timeIndex < \ntimestep}
\state_\paramIndex^{(\timeIndex)}, \\
\residualSymbol_\text{max} & = \max_{1 \leq \paramIndex \leq \nparam, 0 \leq \timeIndex < \ntimestep}
\residual_\paramIndex^{(\timeIndex)}, \\
\residualSymbol_\text{in} & = \max_{1 \leq \paramIndex \leq \nparam, 0 \leq \timeIndex < \ntimestep}
\residual_\paramIndex^{(\timeIndex)}.
\end{split}
\end{equation}
Then the labelled paired training dataset is given by
\begin{equation}
\begin{split}
\mathcal{S}_\text{in} & = \left\{ \left(\overline{\timeSymbol}^{(\timeIndex)},
\overline{\param}_\paramIndex, \overline{\state}_\paramIndex^{(\timeIndex)}\right) :
1 \leq \paramIndex \leq \nparam \text{ and } 0 \leq \timeIndex < \ntimestep \right\}
\subset [0,1] \times [0,1] \times [0,1]^{\sizeFOMsymbol \times \sizeFOMsymbol}, \\
\mathcal{S}_\text{out} & = \left\{ \overline{\residual}_\paramIndex^{(\timeIndex)} :
1 \leq \paramIndex \leq \nparam \text{ and } 0 \leq \timeIndex < \ntimestep \right\}
\subset [0,1]^{\sizeFOMsymbol \times \sizeFOMsymbol},
\end{split}
\end{equation}

Given the normalized datasets $(\mathcal{S}_{\text{in}}, \mathcal{S}_{\text{out}})$,
the goal is to learn a generator
$G^\star: \mathbb{R} \times \mathbb{R} \times \mathbb{R}^{\sizeFOMsymbol \times \sizeFOMsymbol} \to [0,1]^{\sizeFOMsymbol \times \sizeFOMsymbol}$
which approximates the discrete dynamics
$$
\overline{\residual}_\paramIndex^{(\timeIndex)} \approx
G^\star \left(\overline{\timeSymbol}^{(\timeIndex)},
\overline{\param}_\paramIndex, \overline{\state}_\paramIndex^{(\timeIndex)}\right)
\text{ for all } 1 \leq \paramIndex \leq \nparam \text{ and } 0 \leq \timeIndex < \ntimestep.
$$
In the GAN framework, the generator $G^\star$ is learnt
through optimizing the function $G$ to minimize a objective functional which
measures the distance of the generator distribution
and the groundtruth distribution in a certain metric.
The generator $G$ is set to compete with another neural network
$D: \mathbb{R} \times \mathbb{R} \times \mathbb{R}^{\sizeFOMsymbol \times \sizeFOMsymbol} \to \mathbb{R}$, called the critic.
The two functions have opposite objectives, as the critic aims to distinguish
the generator distribution from the groundtruth distribution,
while the generator aims to fool the discriminator.
In our work, the overall objective has three components.
First, we use the earth mover distance in \cite{arjovsky2017wasserstein},
as the competing objective, which is formally defined as
\begin{equation}\label{eq:W-loss}
\mathcal{L}_{\text{WGAN}}(D,G) =
\sum_{\paramIndex=1}^{\nparam} \sum_{\timeIndex=0}^{\ntimestep-1}
D\left(\overline{\timeSymbol}^{(\timeIndex)}, \overline{\param}_\paramIndex,
\overline{\residual}_\paramIndex^{(\timeIndex)}\right) -
D\left(\overline{\timeSymbol}^{(\timeIndex)}, \overline{\param}_\paramIndex,
G \left(\overline{\timeSymbol}^{(\timeIndex)},
\overline{\param}_\paramIndex, \overline{\state}_\paramIndex^{(\timeIndex)}\right)\right).
\end{equation}
Second, we use the gradient penalty in \cite{gulrajani2017improved}
as a regularizer to weakly enforce the 1-Lipschitz continuity in the critic, which is given by
\begin{equation}\label{eq:Lipschitz-loss}
\mathcal{L}_{\text{Lip}}(D) =
\sum_{\paramIndex=1}^{\nparam} \sum_{\timeIndex=0}^{\ntimestep-1}
\left(\left\| \nabla_{\overline{\state}} D\left(\overline{\timeSymbol}^{(\timeIndex)}, \overline{\param}_\paramIndex, \varepsilon_\paramIndex^{(\timeIndex)} \overline{\residual}_\paramIndex^{(\timeIndex)} + (1-\varepsilon_\paramIndex^{(\timeIndex)}) G \left(\overline{\timeSymbol}^{(\timeIndex)},
\overline{\param}_\paramIndex, \overline{\state}_\paramIndex^{(\timeIndex)}\right) \right) \right\|_2 - 1 \right)^2,
\end{equation}
where $\varepsilon_\paramIndex^{(\timeIndex)} \sim \mathcal{U}(0,1)$ is independent and identically distributed. Third, we use the absolute distance as the reconstruction objective, which is defined as
\begin{equation}\label{eq:reconstrunction-loss}
\mathcal{L}_{\text{recon}}(G) =
\sum_{\paramIndex=1}^{\nparam} \sum_{\timeIndex=0}^{\ntimestep-1}
\left\vert \overline{\residual}_\paramIndex^{(\timeIndex)} -
G \left(\overline{\timeSymbol}^{(\timeIndex)},
\overline{\param}_\paramIndex, \overline{\state}_\paramIndex^{(\timeIndex)}\right)\right\vert.
\end{equation}
The optimization problem is then formulated as
\begin{equation}
\min_{G \in \mathcal{G}} \max_{D \in \mathcal{D}}
\mathcal{L}_{\text{WGAN}}(D,G) +
\mu_{\text{Lip}} \mathcal{L}_{\text{Lip}}(D) +
\mu_{\text{recon}} \mathcal{L}_{\text{recon}}(G),
\label{eq:total-gan-loss}
\end{equation}
where $\mu_{\text{Lip}} > 0$ and $\mu_{\text{recon}} > 0$ are regularization parameters
which control the tradeoff between the three components in the overall objective,
$\mathcal{G}$ is a class of neural networks with the U-Net architecture,
and $\mathcal{D}$ is a class of convolutional neural networks.
The generator and the critic are trained simultaneously
and the objective functional is dynamic to each of them in the training process.
In our work, we use the adaptive moment estimation (ADAM) method \cite{kingma2014adam}
to update the critic $D$ and the generator $G$ in alternating direction.

\subsection{Prediction stage}
\label{sec:prediction-gan}
After sufficient training, the generator $G^\star$ can serve as a
global surrogate model for predicting the temperature field.
At a generic shock pressure $\param \in \paramDomain$,
with the initial condition
$\widetilde{\state}_{\text{GAN}}(\timeSymbol^{(0)}(\param); \param) = \state^{(0)}(\param)$,
for $0 \leq \timeIndex < \ntimestep$,
the GAN prediction $\widetilde{\state}_{\text{GAN}}(\timeSymbol^{(\timeIndex+1)}(\param); \param)
\in \mathbb{R}^{\sizeFOMsymbol \times \sizeFOMsymbol}$ is iteratively given by
\begin{equation}
\widetilde{\state}_{\text{GAN}}({\timeSymbol}^{(\timeIndex+1)}(\param); {\param}) =
\widetilde{\state}_{\text{GAN}}({\timeSymbol}^{(\timeIndex)}(\param); {\param}) +
(R_\text{max} - R_\text{min}) G^\star \left(\overline{\timeSymbol}^{(\timeIndex)},
\overline{\param}, \overline{\state}_{\text{GAN}}^{(\timeIndex)}(\param) \right),
\end{equation}
where
\begin{equation}
\begin{split}
\overline{\param} & = (\param - \param_\text{min})/(\param_\text{max} - \param_\text{min}), \\
\overline{\state}_{\text{GAN}}^{(\timeIndex)}(\param) & = \widetilde{\state}_{\text{GAN}}({\timeSymbol}^{(\timeIndex)}(\param); \param)/(\stateSymbol_\text{max} - \stateSymbol_\text{min}).
\end{split}
\end{equation}

\section{Numerical experiments}
\label{sec:exp}

In this section, we present some numerical results to test the performance of
our proposed methods when applied to the numerical simulation data
for the pore collapse process.

\subsection{Problem specification}
In our numerical experiments,
the bounds of the range $\paramDomain$
of applied shock pressure are
$\param_\text{min} = 11$
and $\param_\text{max} = 15$, and
the spatial region of interest
$\solDomainSymbol$ is a square
which is partitioned into
$\sizeFOMsymbol^2 = 128^2$ square sub-zones
with equal length $h_x = 250$.
In order to depict the pore collapse process,
explained in Figure~\ref{fig:pore_coll_fig},
we choose $\ntimestep = 180$, $\timestep = 0.0025$, and
$\timeSymbol^{(0)}(\param) = 0.9875 - 0.0125\param$,
as the initial time of the time interval of interest $\mathcal{T}(\param)$
for the shock pressure $\param \in \paramDomain$.
Figure~\ref{fig:snap_unif} depicts some selected representative snapshots
of temperature fields at different shock pressures ranging from 11 to 15 GPa,
in the corresponding time interval of interest.
Each row corresponds to the same shock pressure.
Unlike Figure~\ref{fig:snap_unif},
the snapshots in the same column do not correspond to the same time instance.

\begin{figure}[htp!]
\centering
\includegraphics[width=0.19\linewidth]{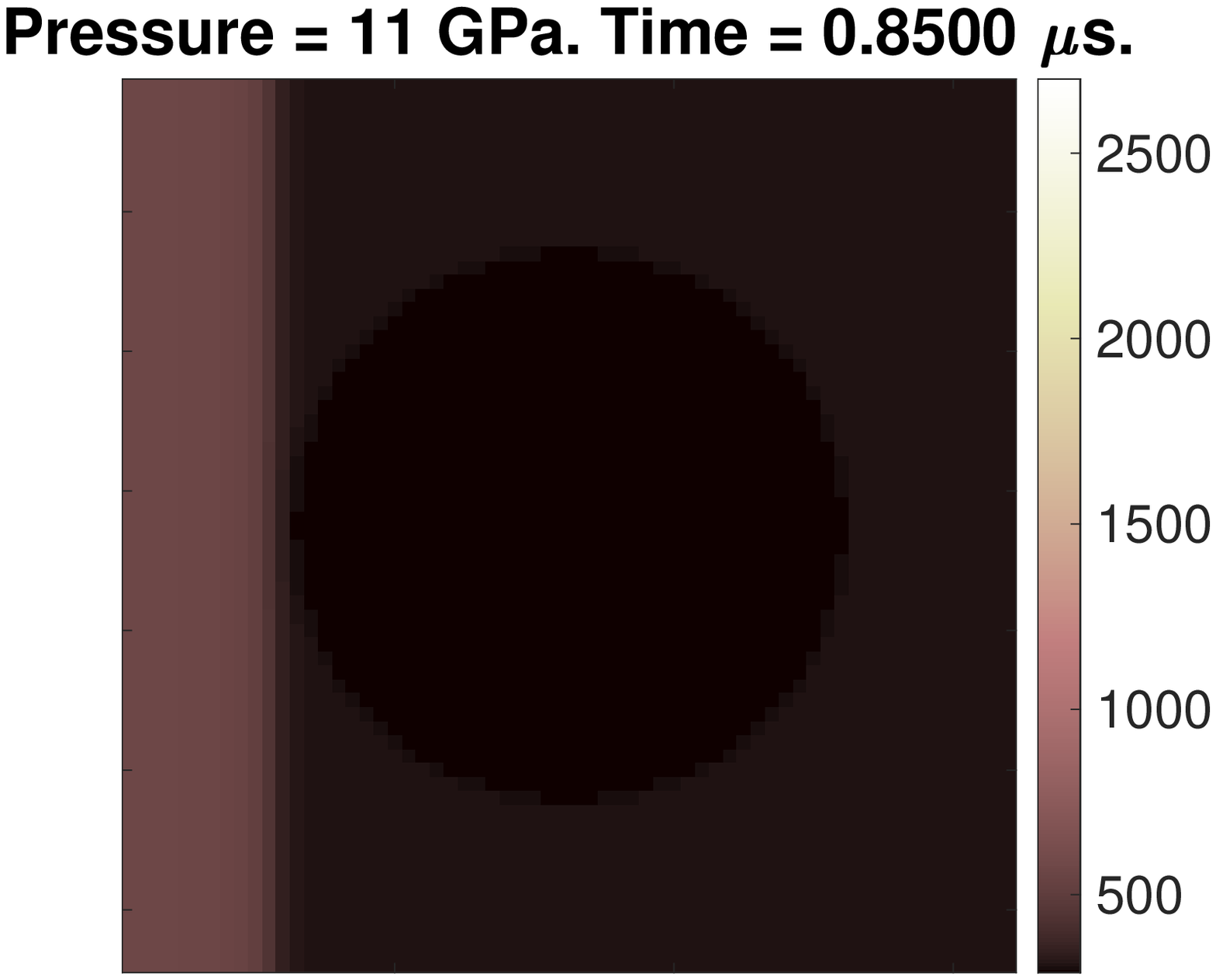}
\includegraphics[width=0.19\linewidth]{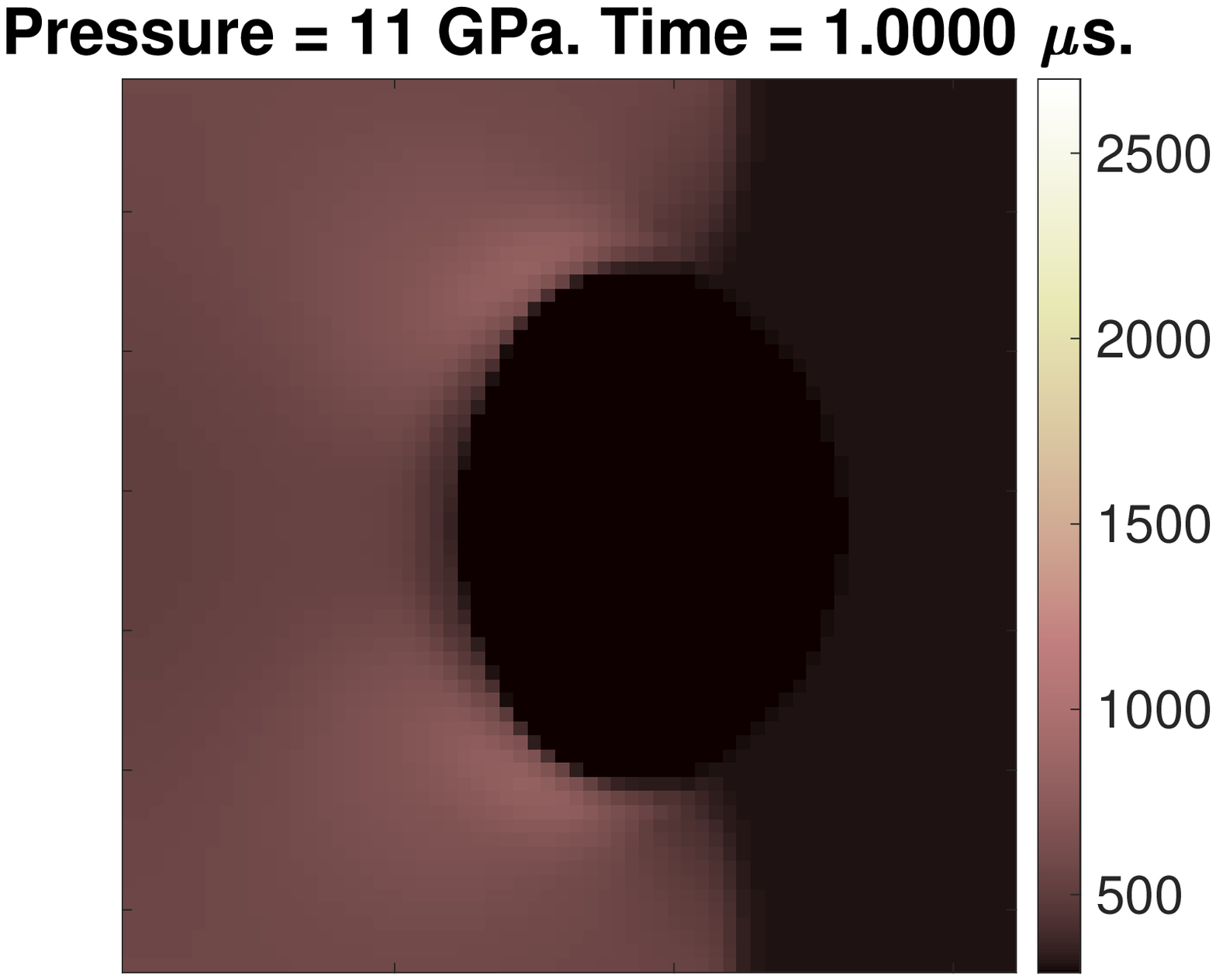}
\includegraphics[width=0.19\linewidth]{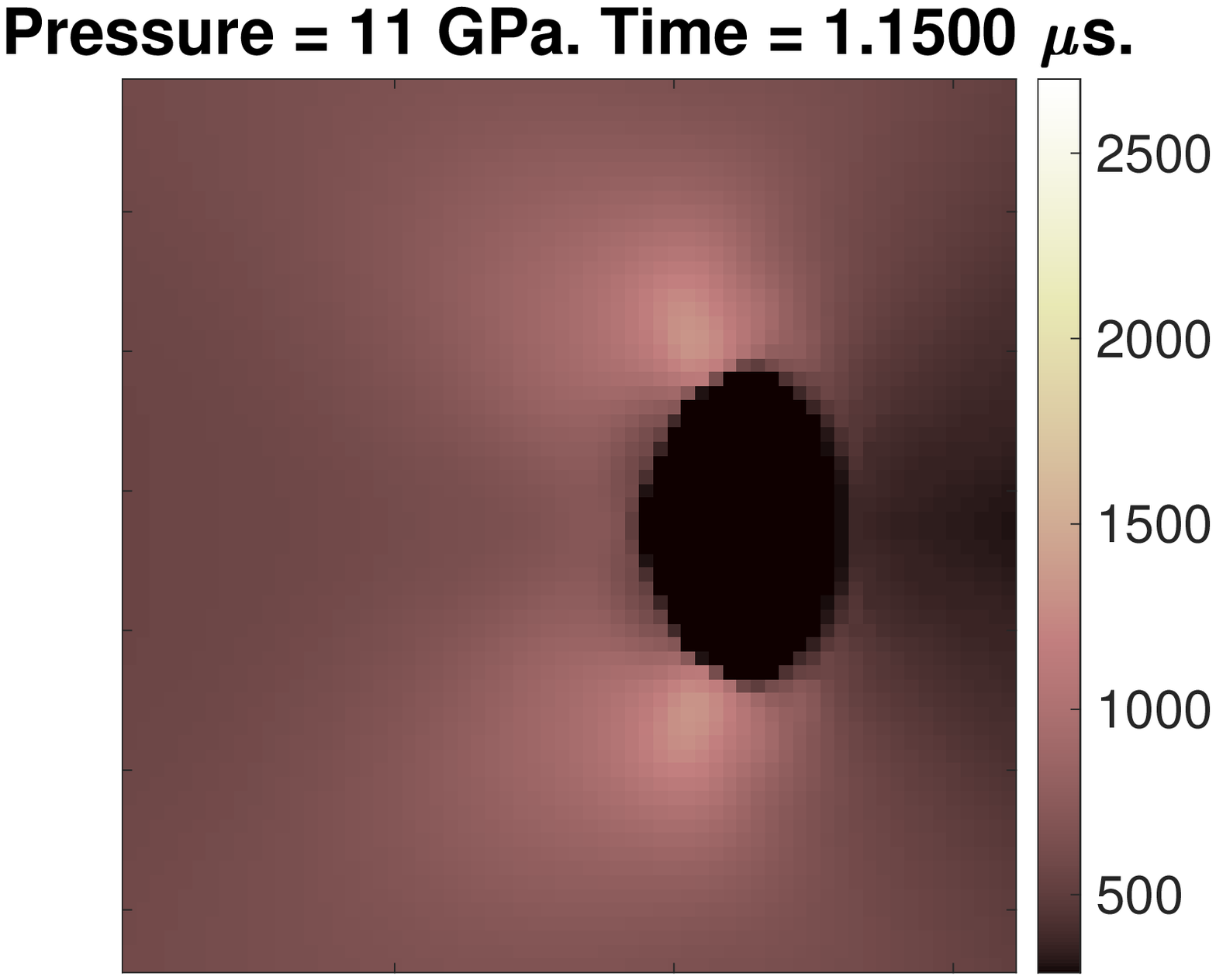}
\includegraphics[width=0.19\linewidth]{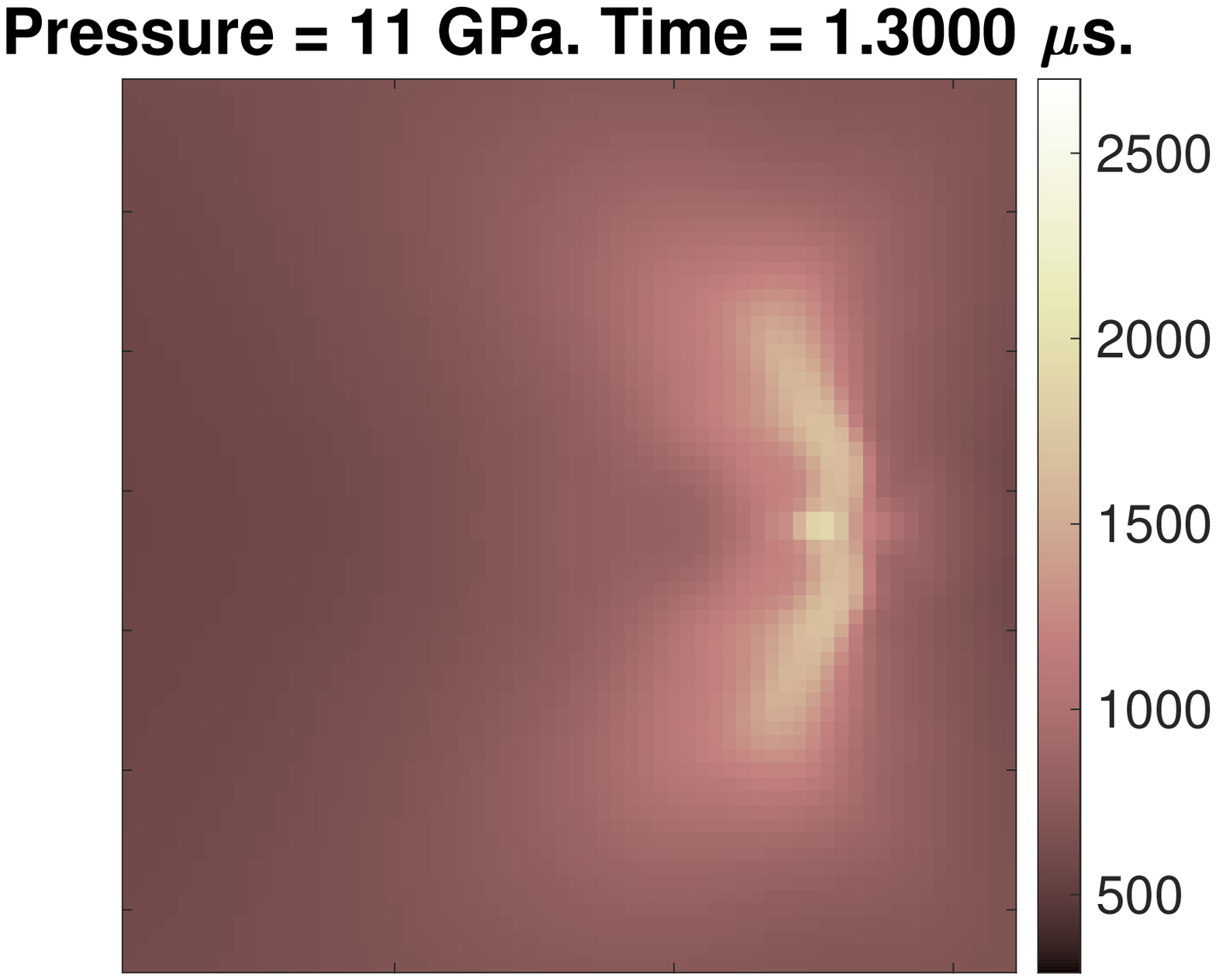}
\includegraphics[width=0.19\linewidth]{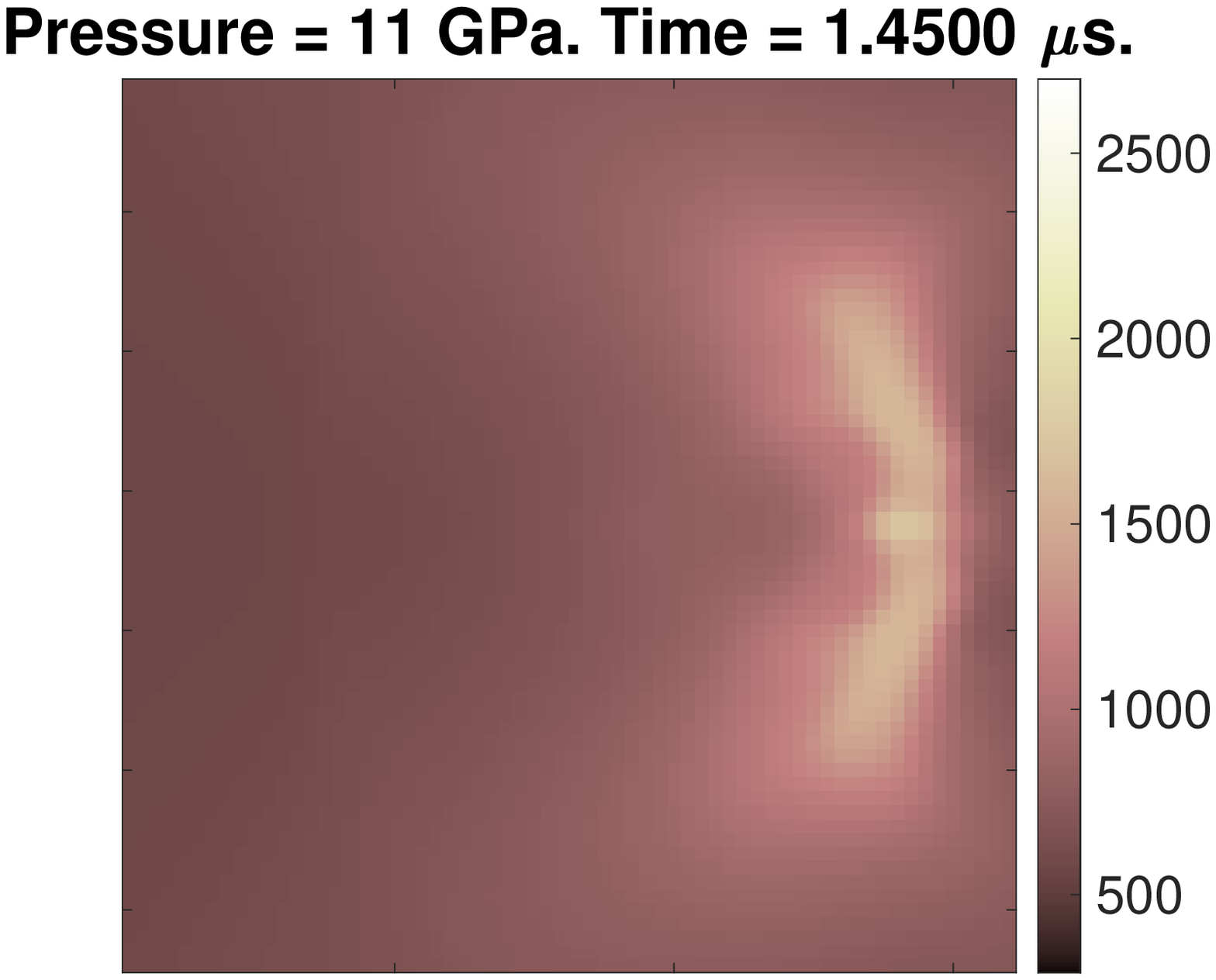}\\
\includegraphics[width=0.19\linewidth]{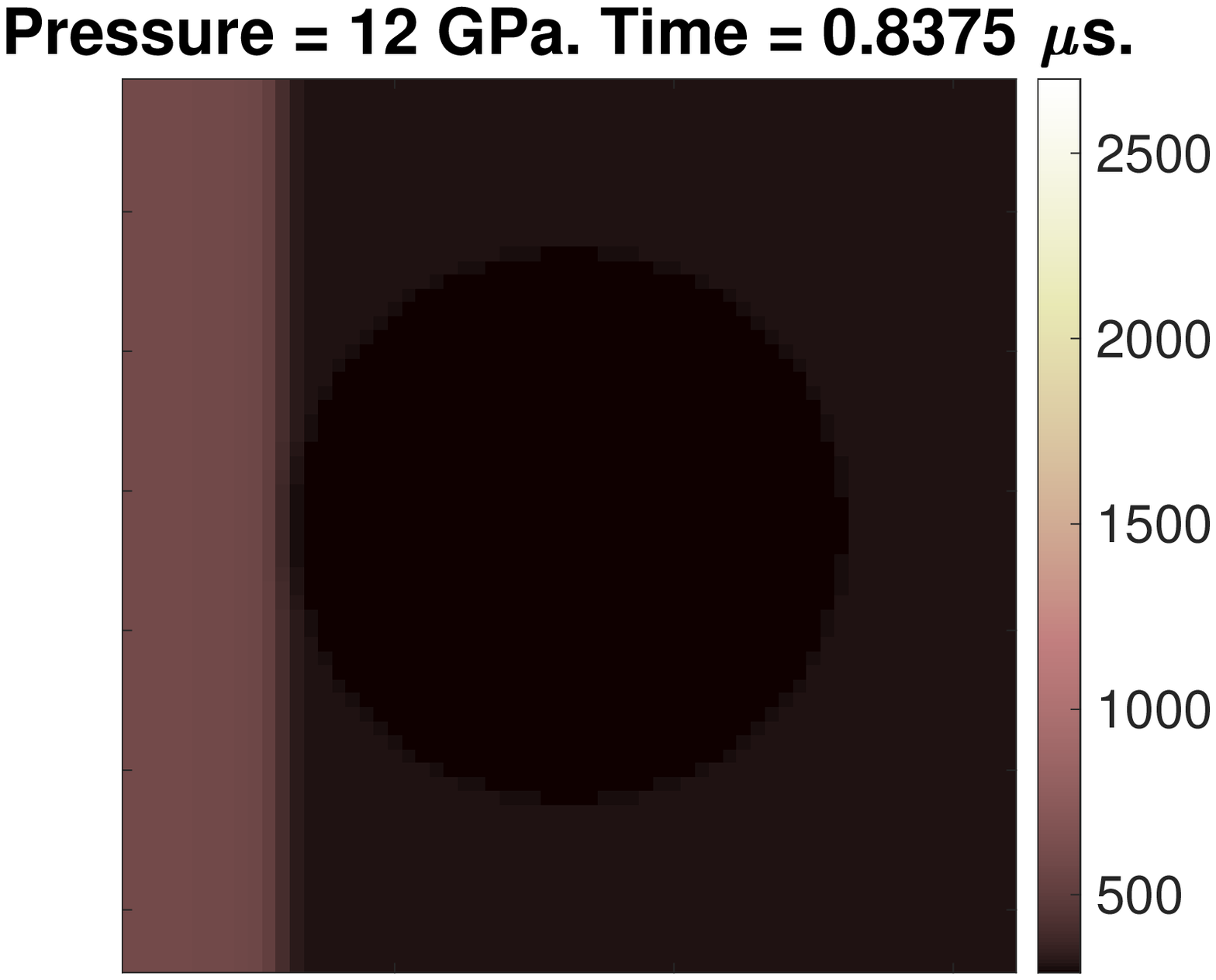}
\includegraphics[width=0.19\linewidth]{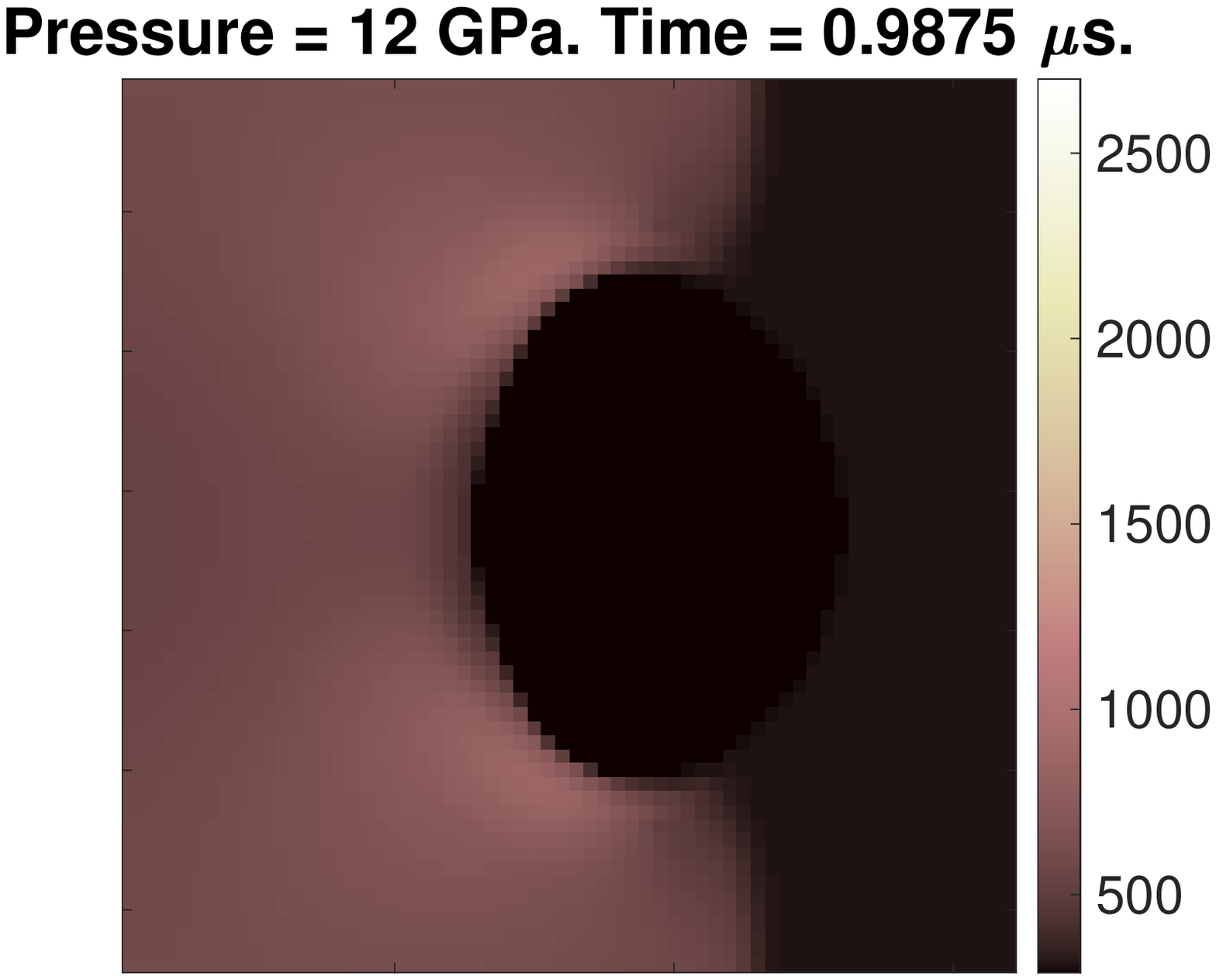}
\includegraphics[width=0.19\linewidth]{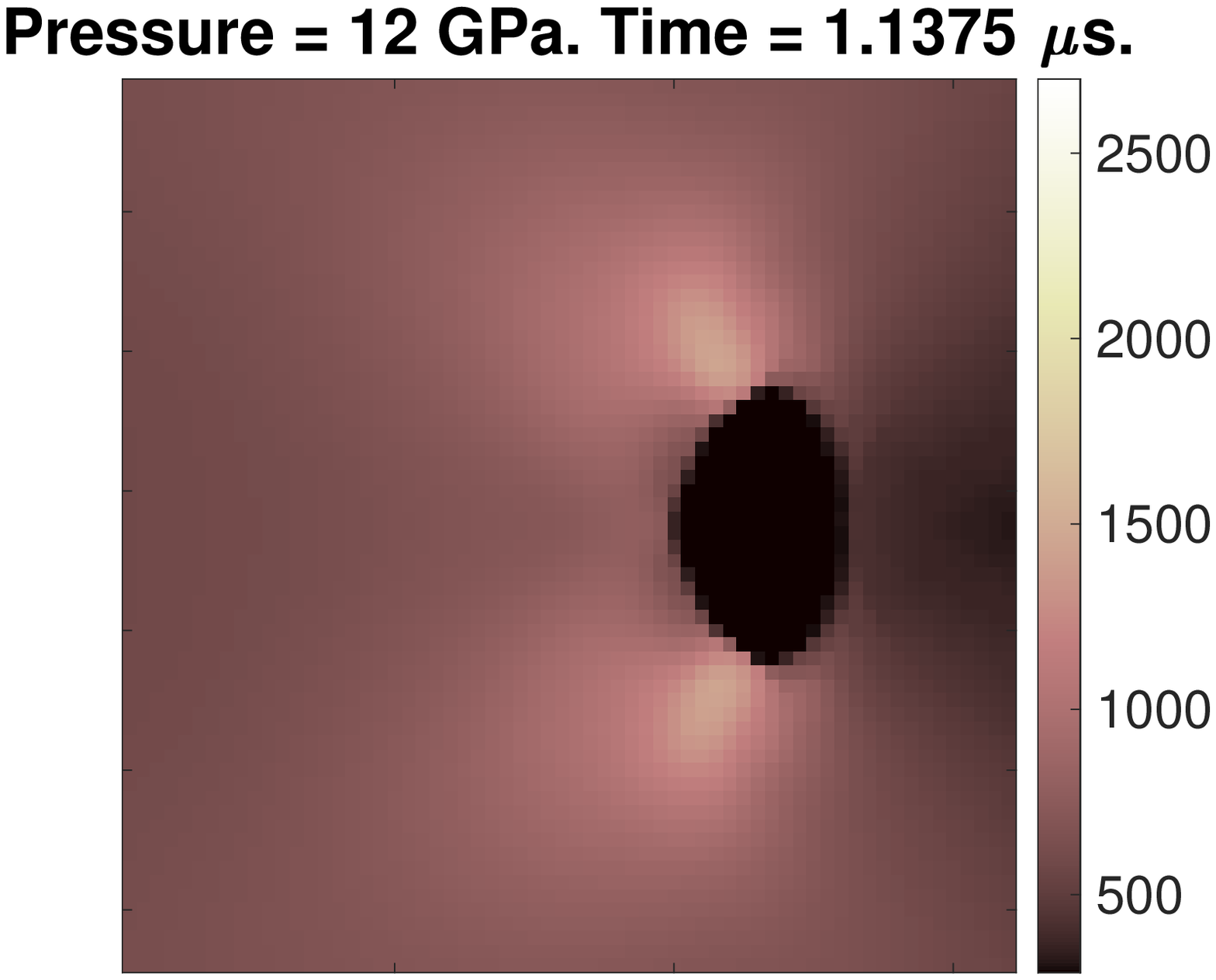}
\includegraphics[width=0.19\linewidth]{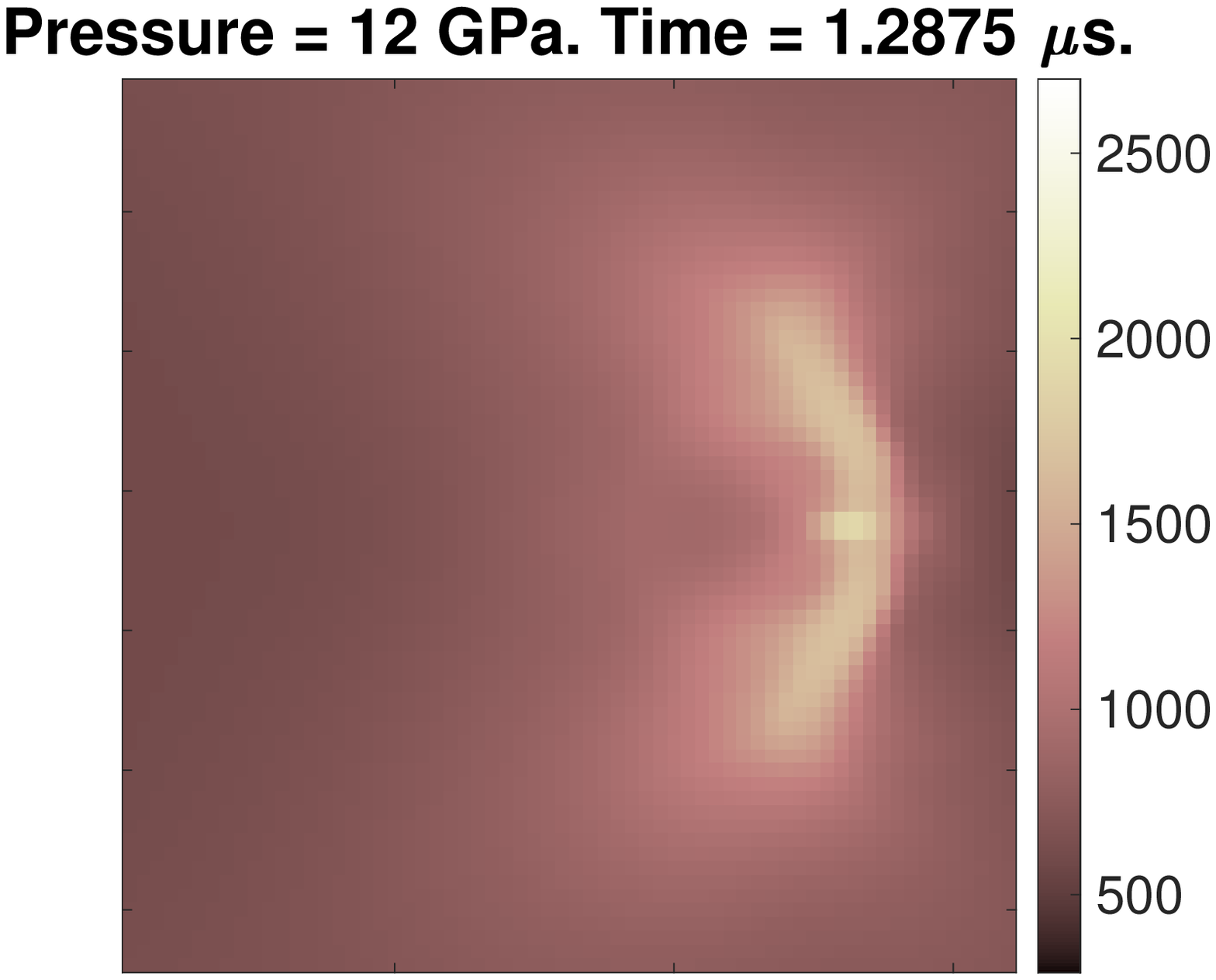}
\includegraphics[width=0.19\linewidth]{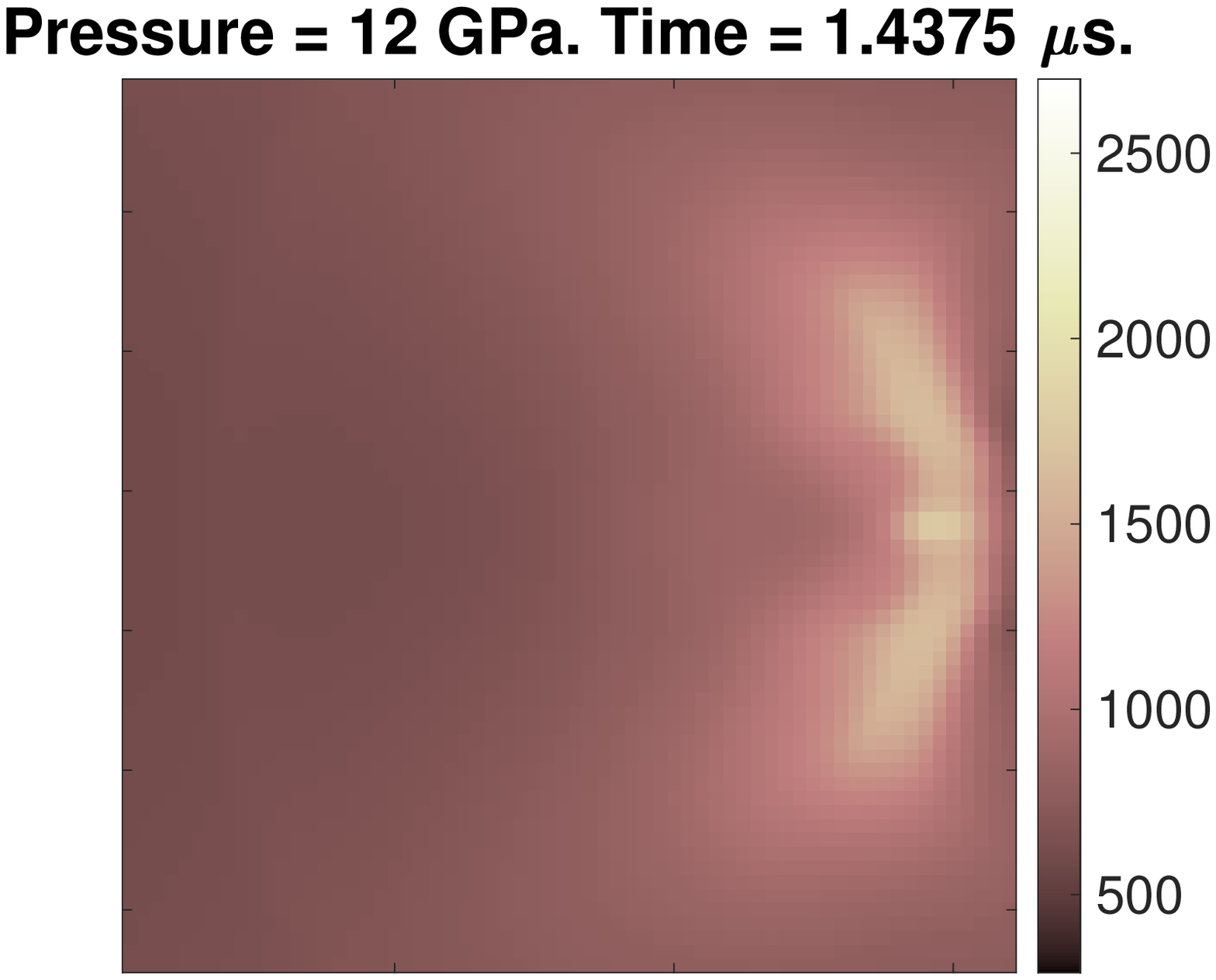}\\
\includegraphics[width=0.19\linewidth]{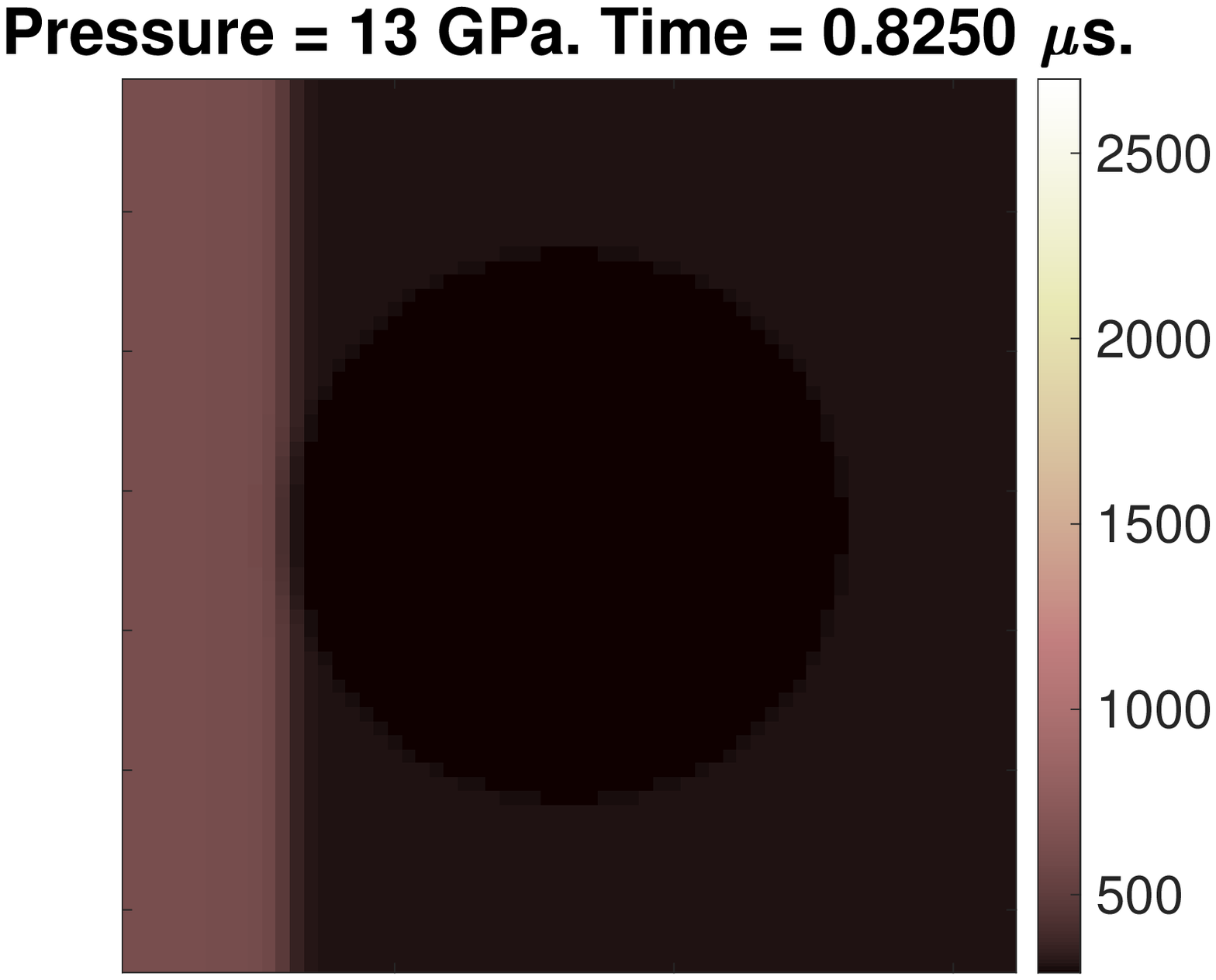}
\includegraphics[width=0.19\linewidth]{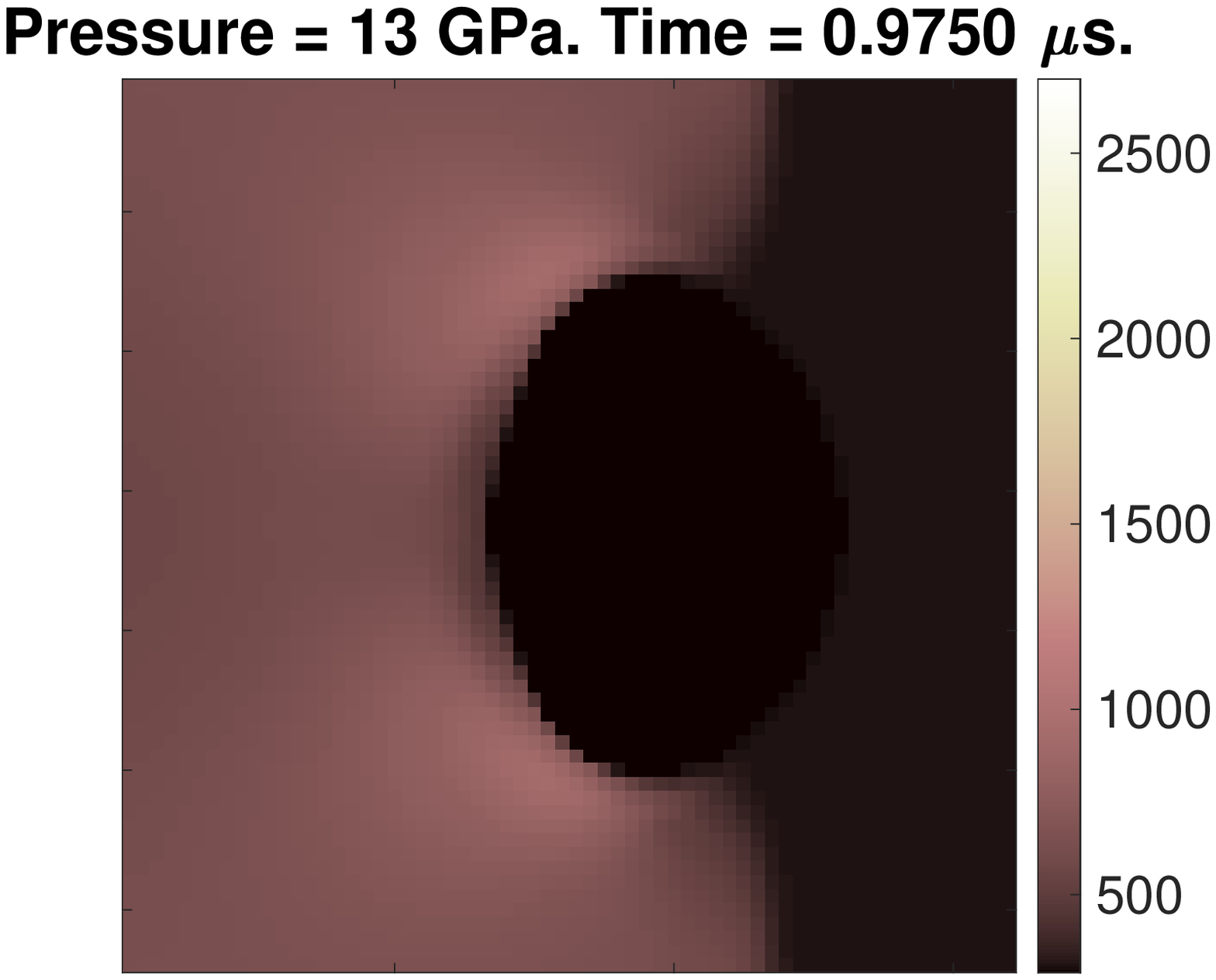}
\includegraphics[width=0.19\linewidth]{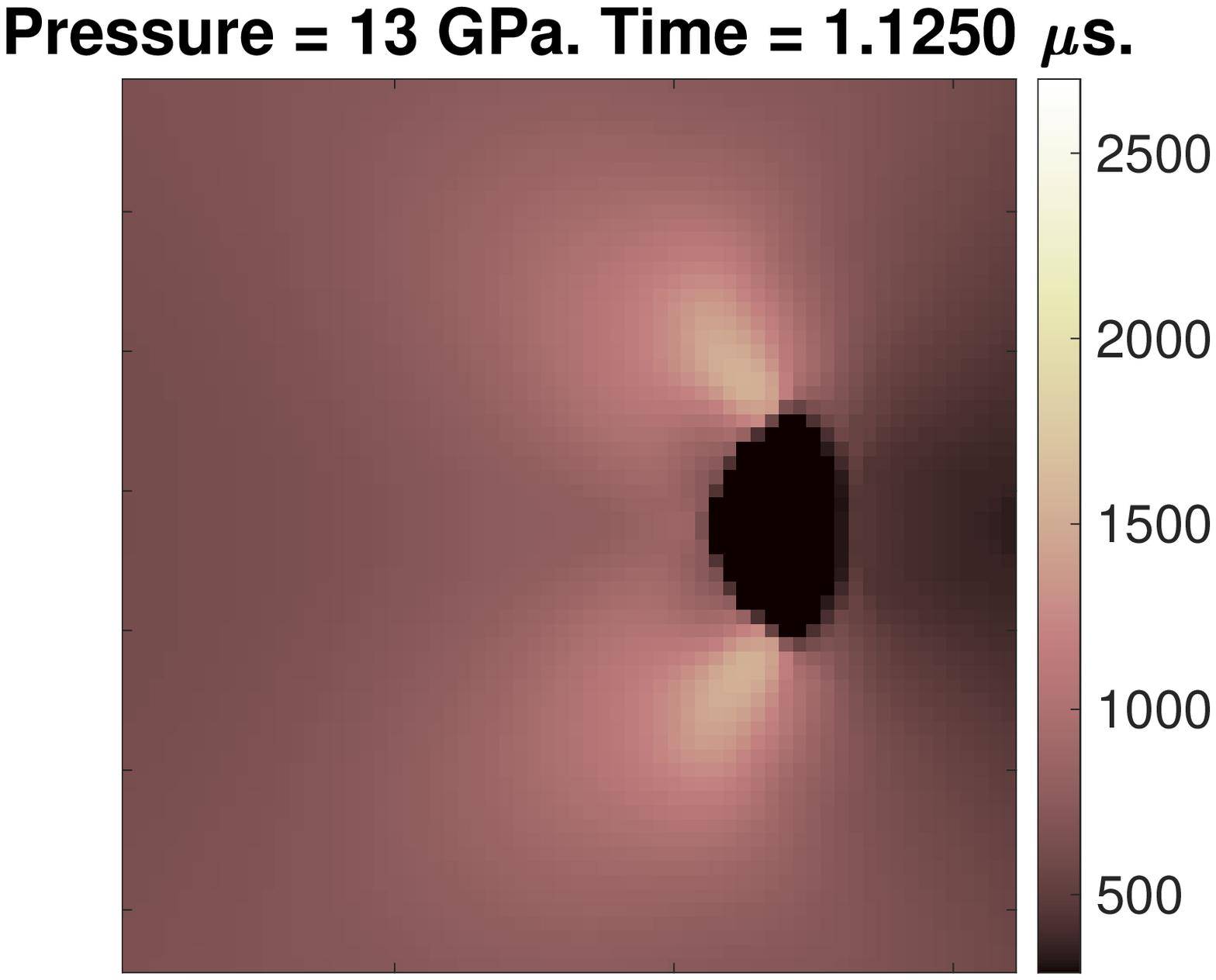}
\includegraphics[width=0.19\linewidth]{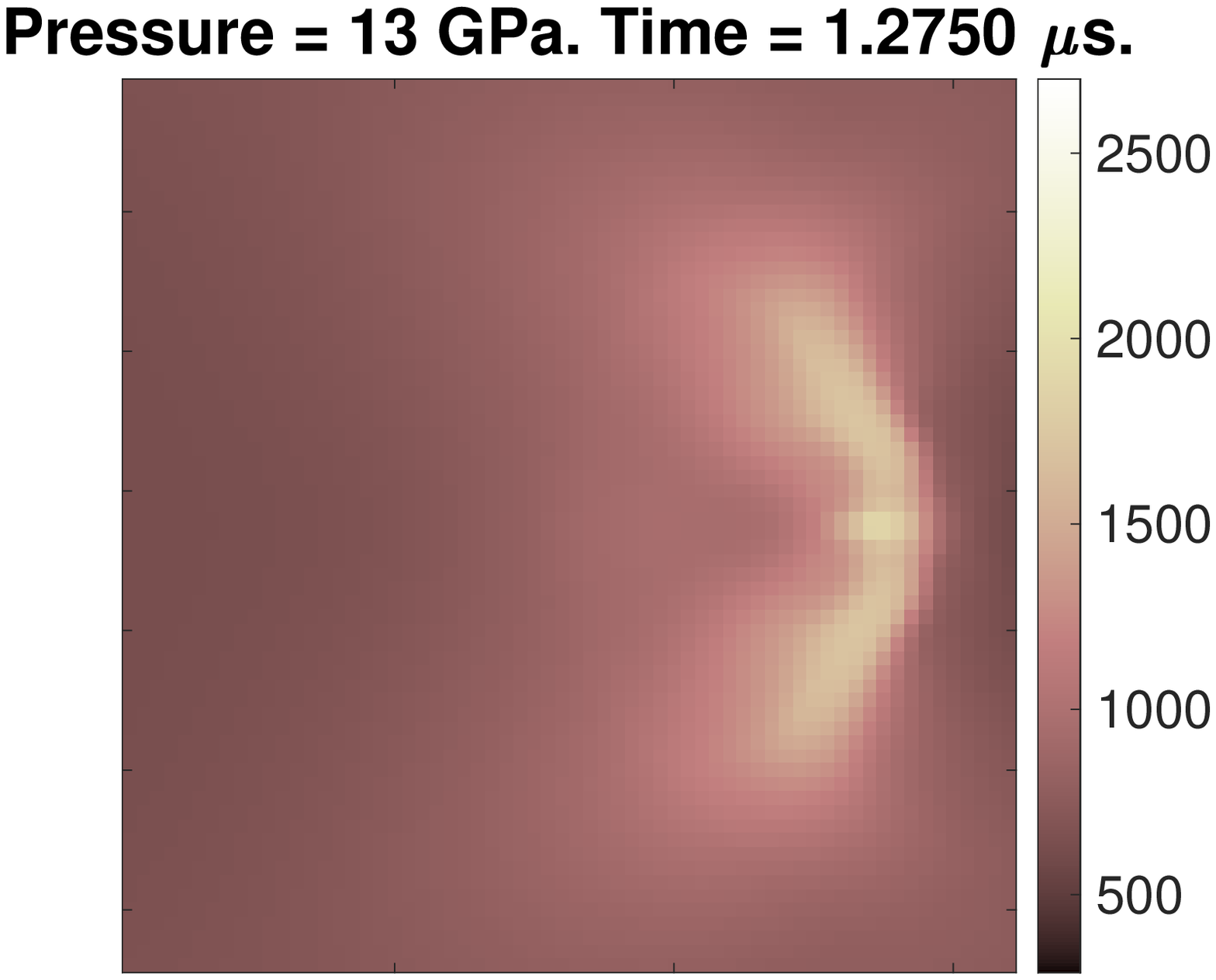}
\includegraphics[width=0.19\linewidth]{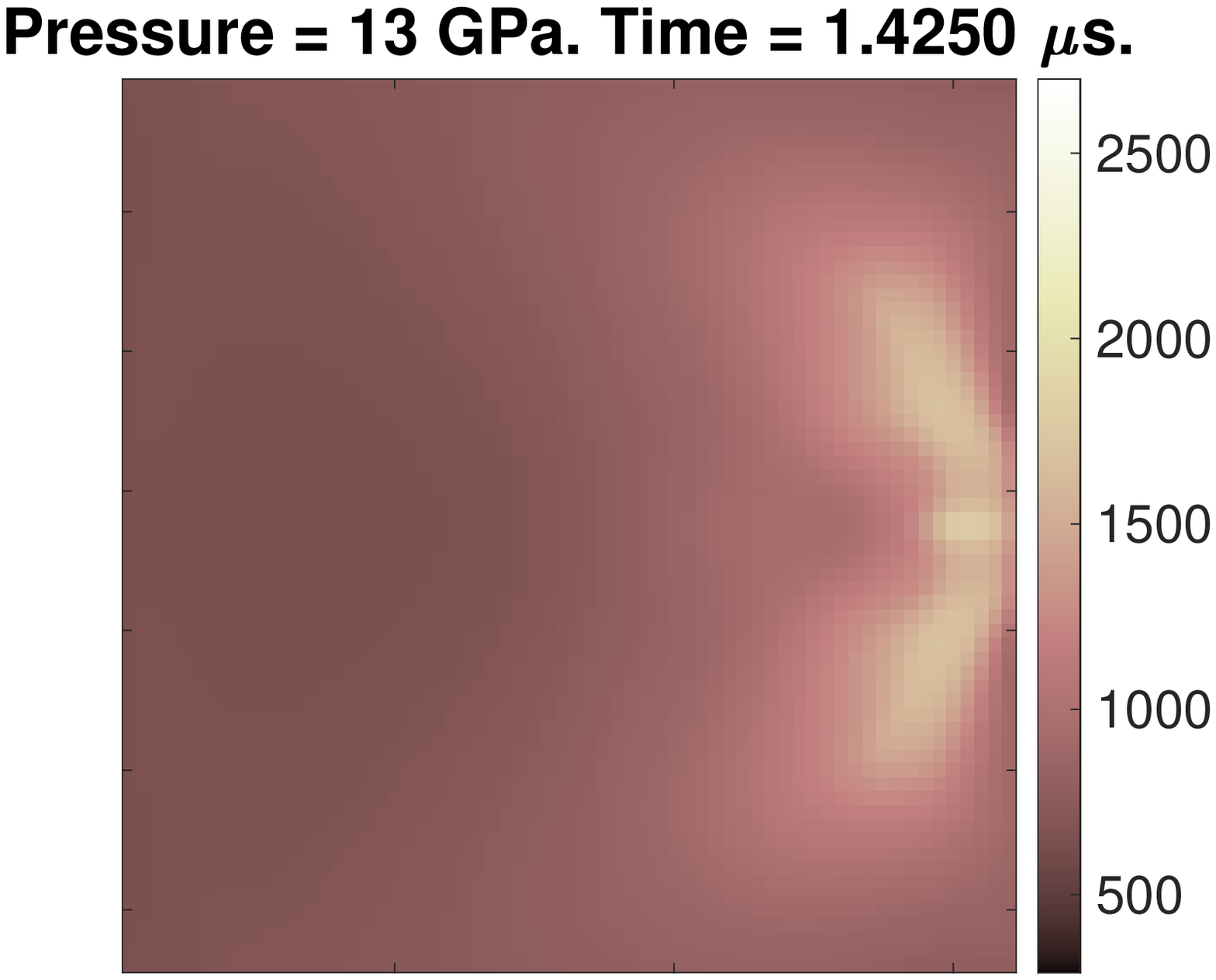}\\
\includegraphics[width=0.19\linewidth]{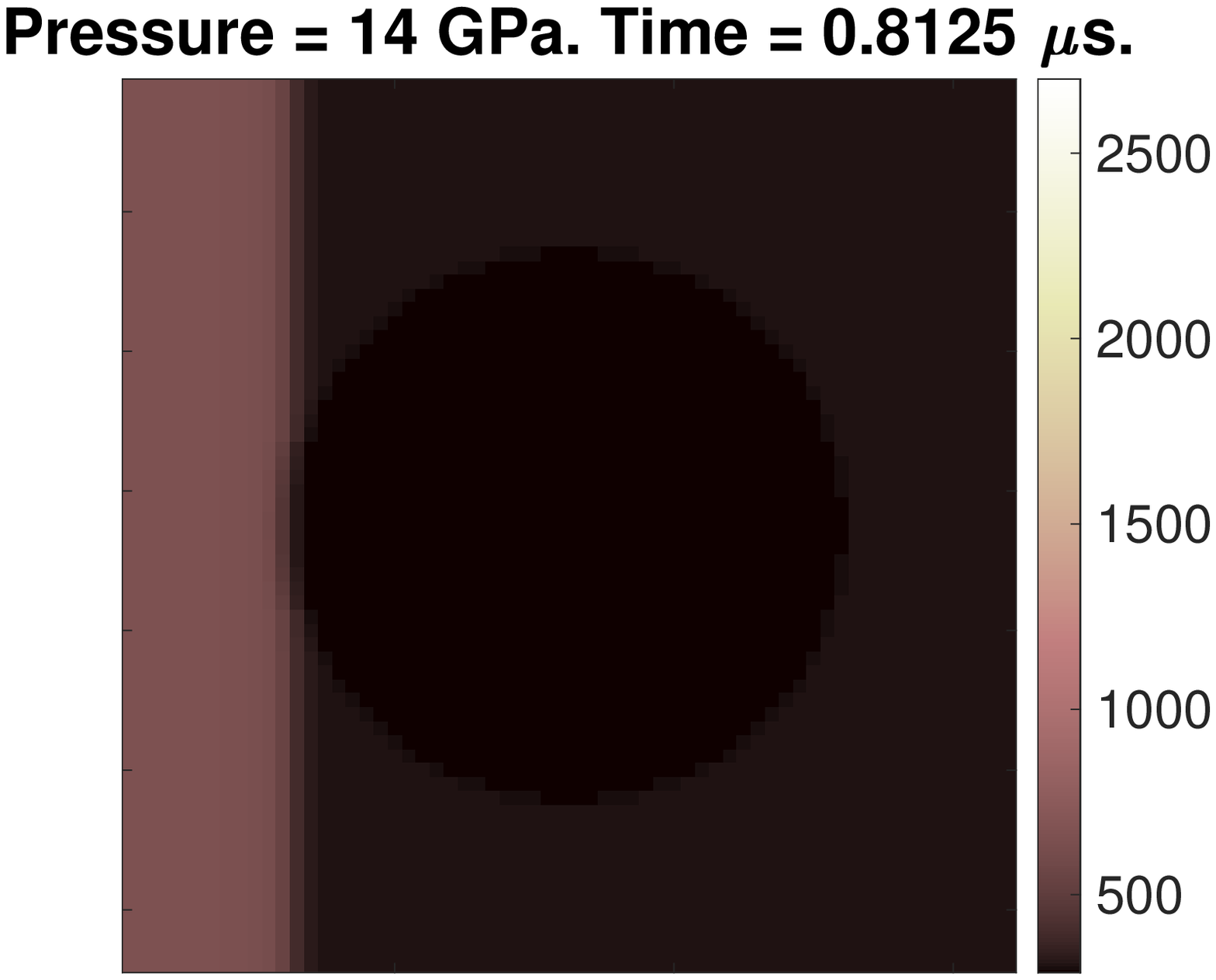}
\includegraphics[width=0.19\linewidth]{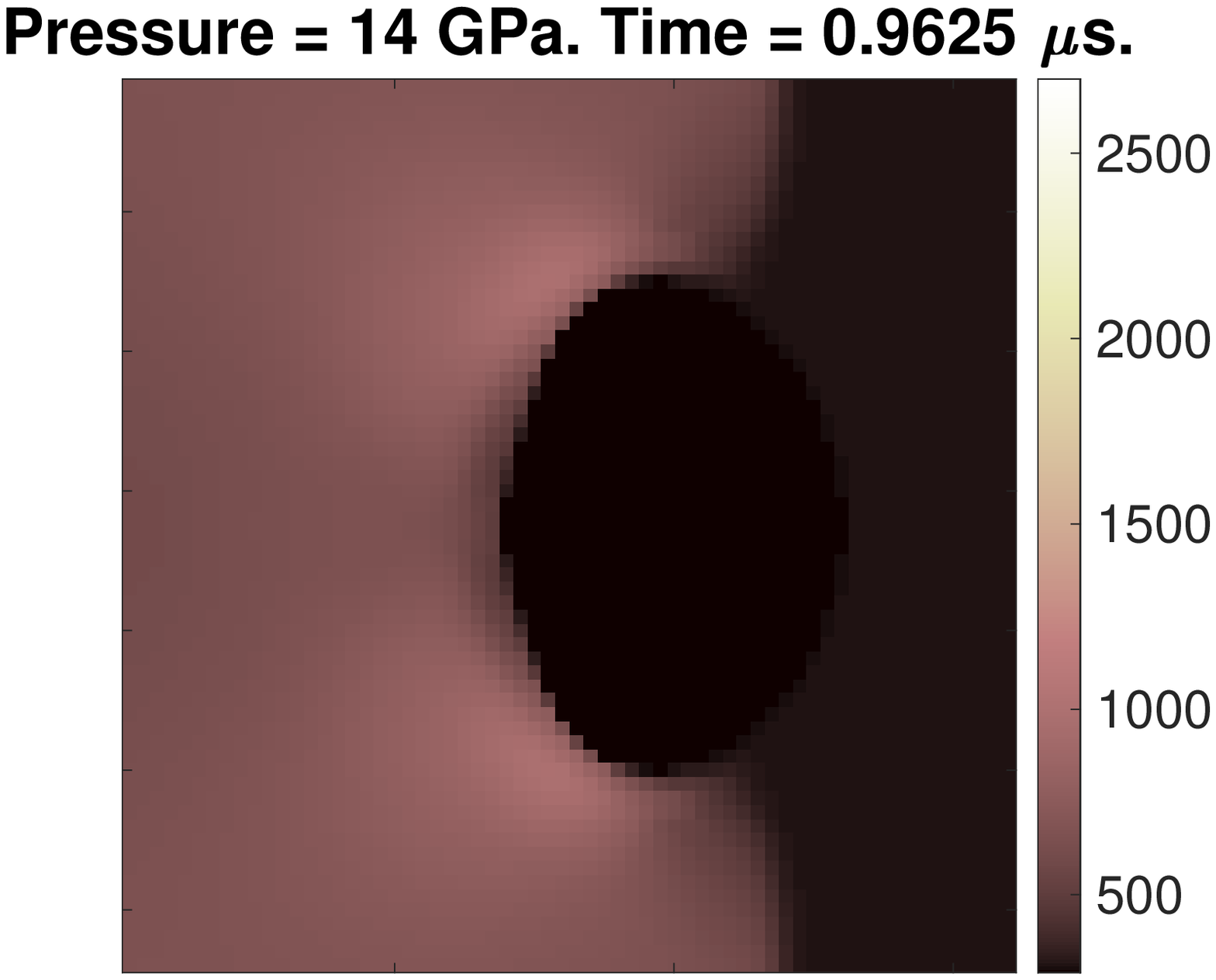}
\includegraphics[width=0.19\linewidth]{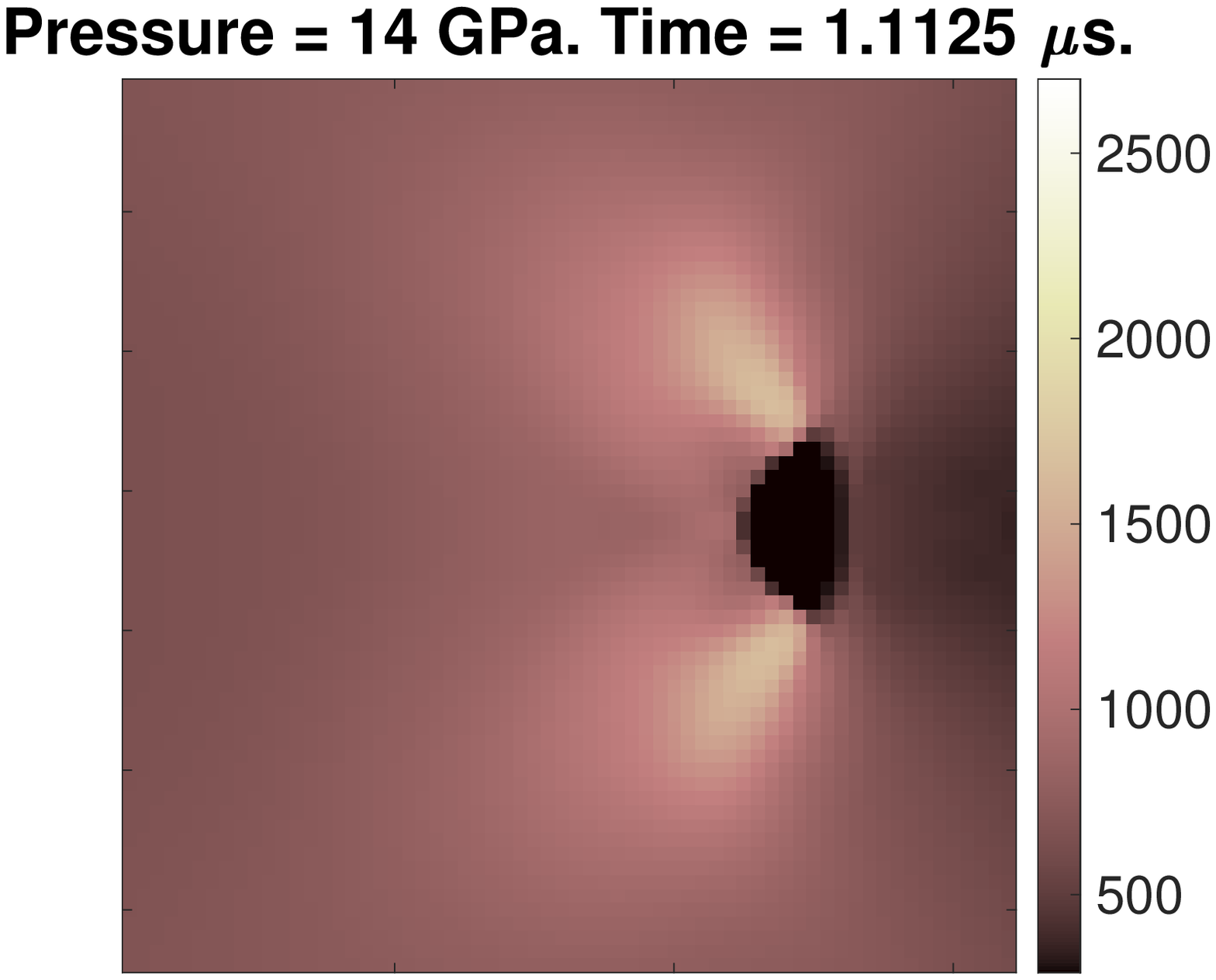}
\includegraphics[width=0.19\linewidth]{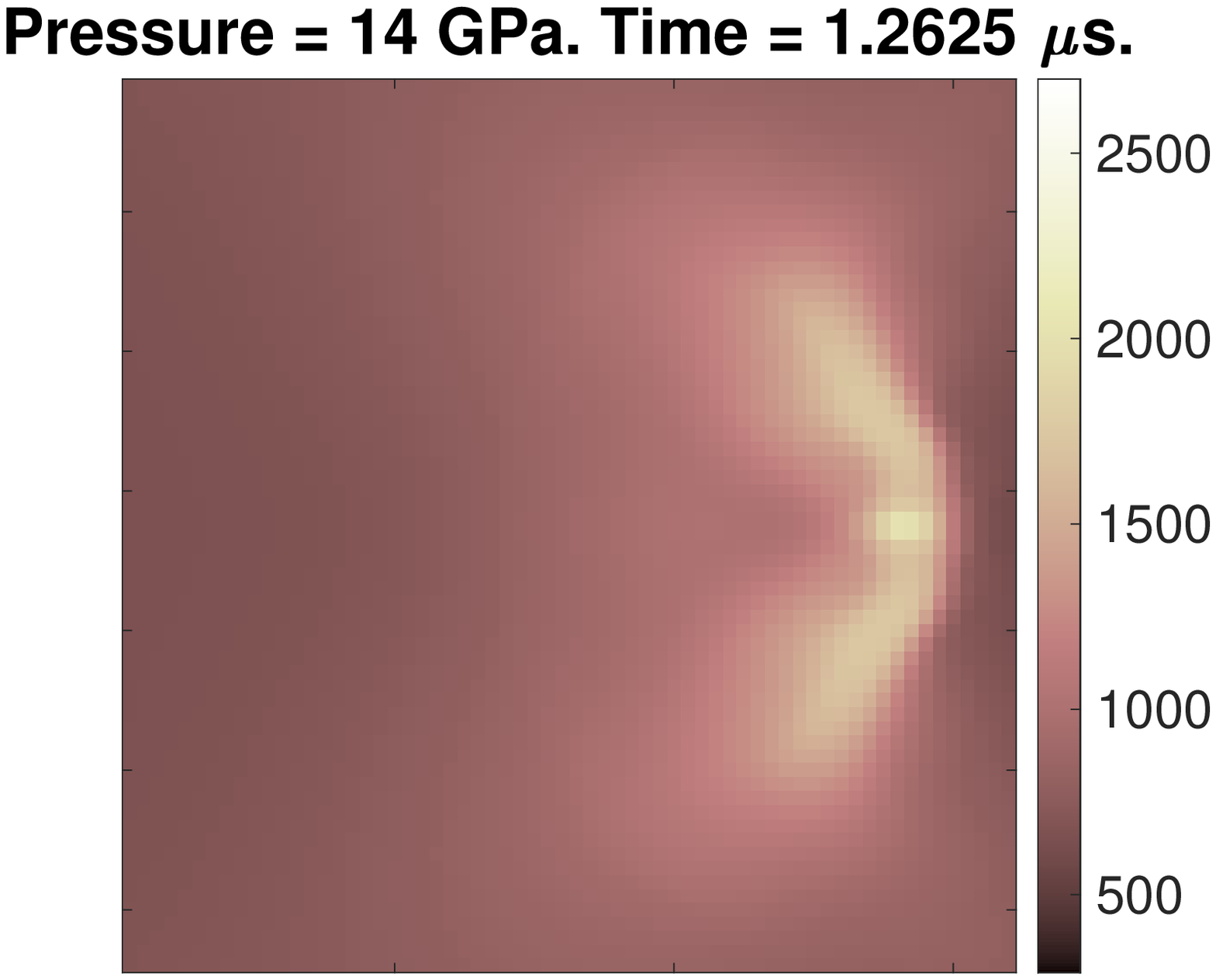}
\includegraphics[width=0.19\linewidth]{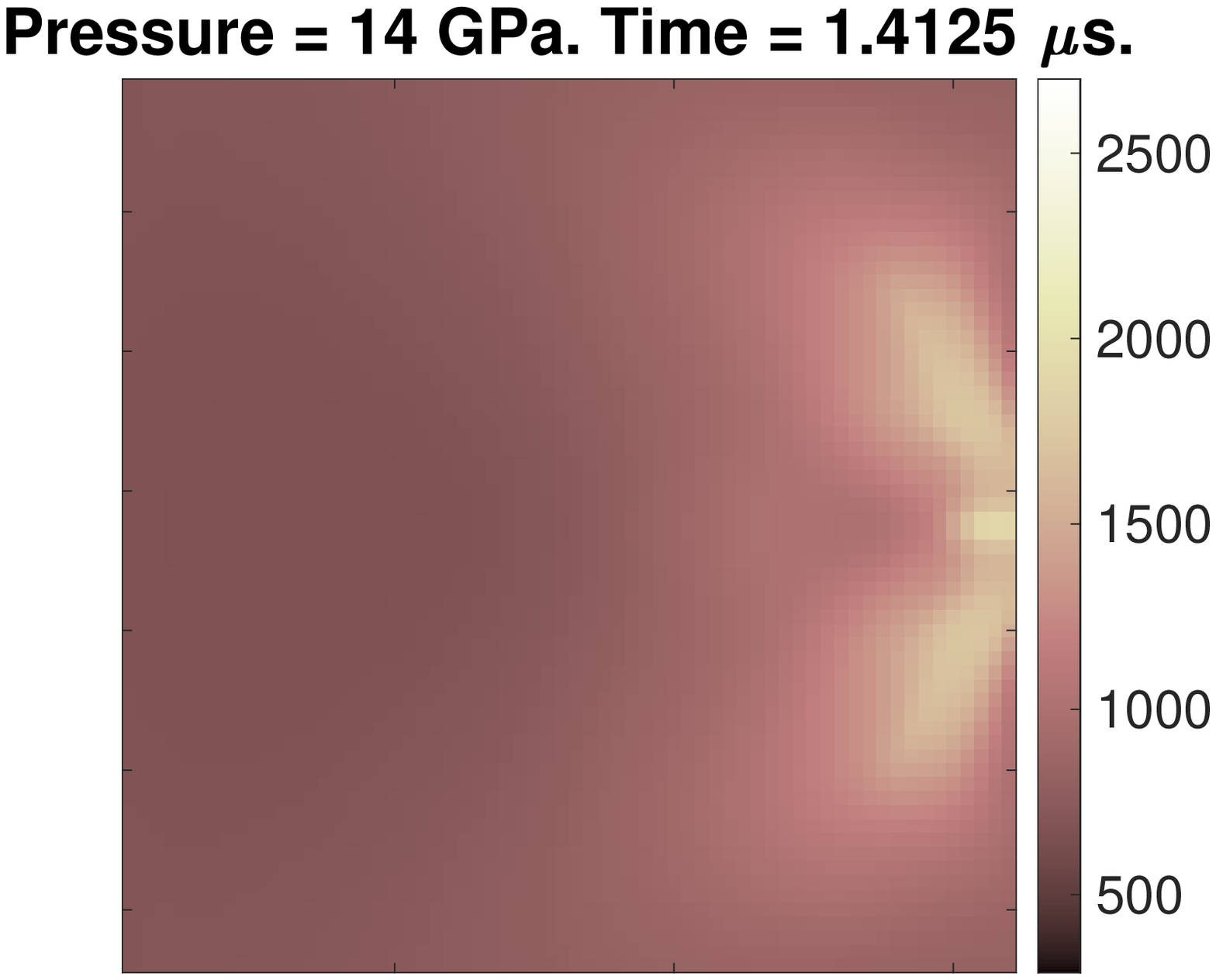}\\
\includegraphics[width=0.19\linewidth]{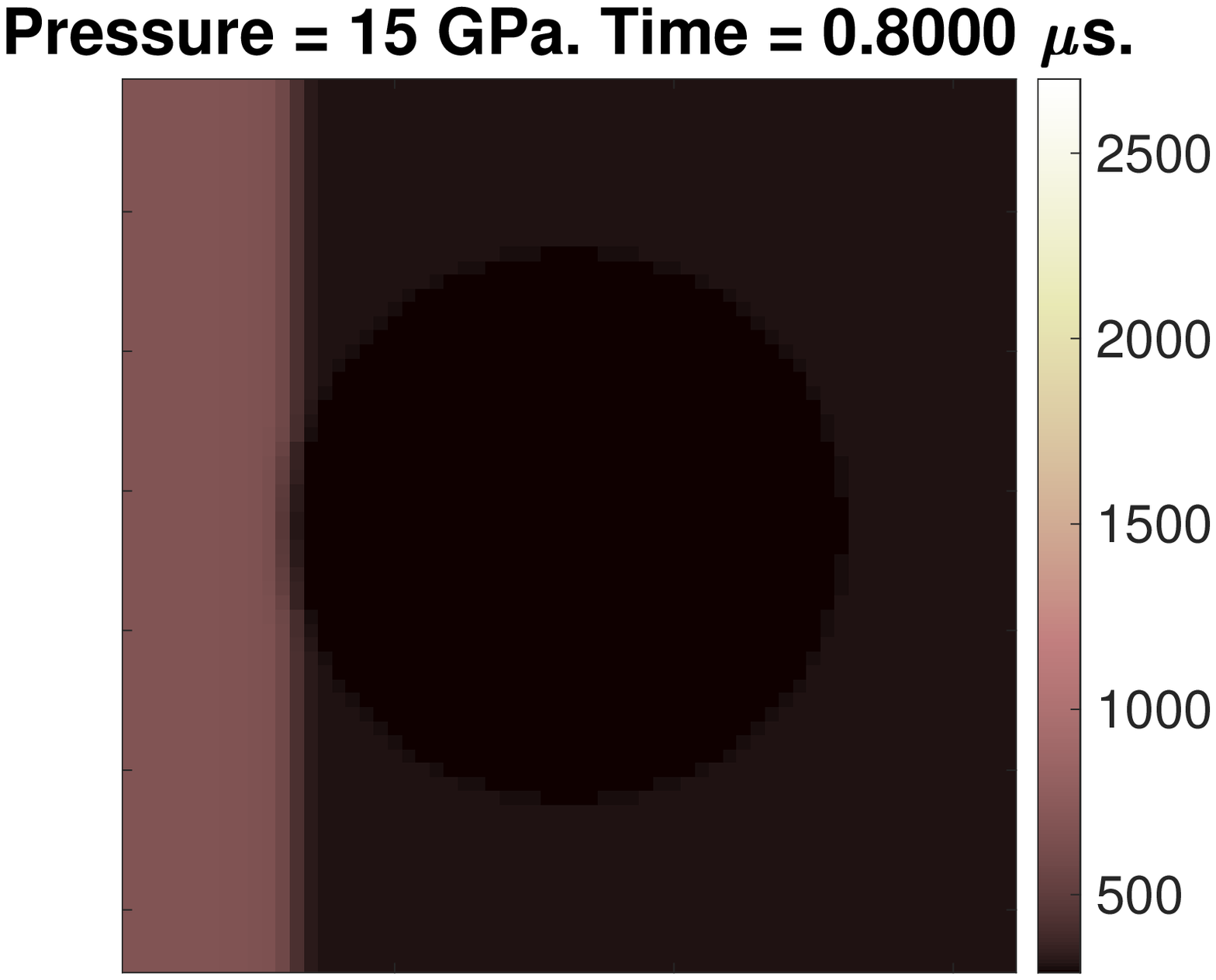}
\includegraphics[width=0.19\linewidth]{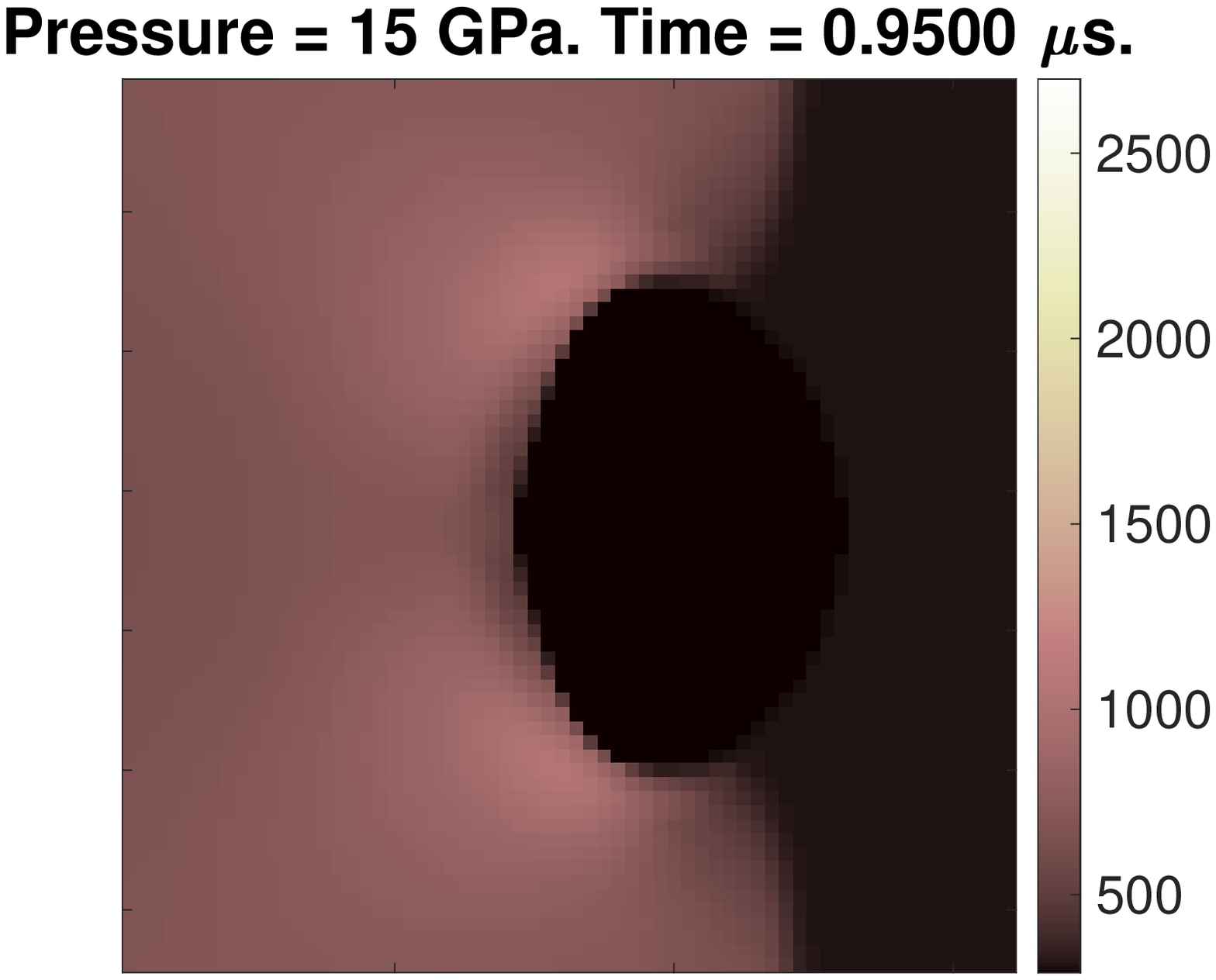}
\includegraphics[width=0.19\linewidth]{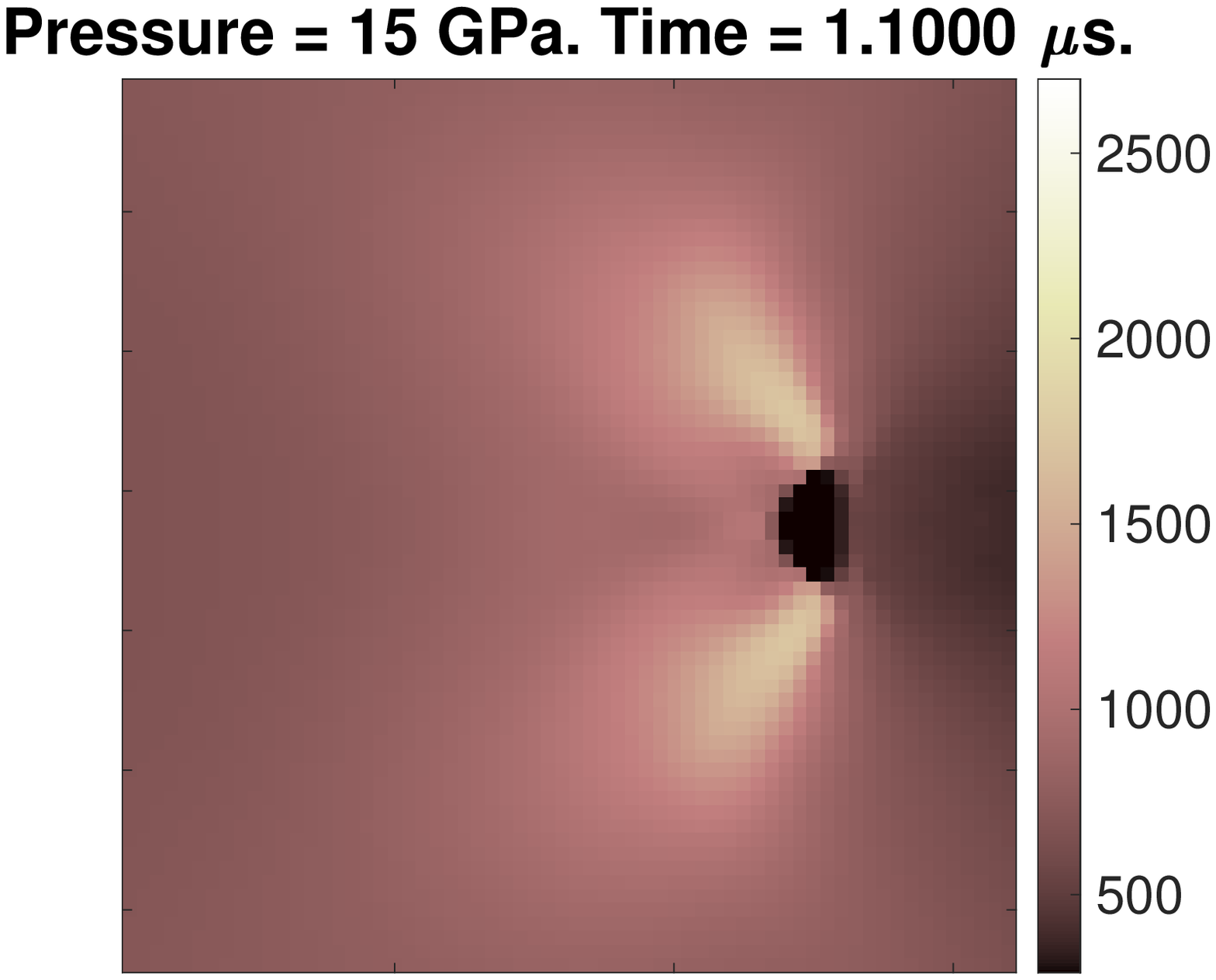}
\includegraphics[width=0.19\linewidth]{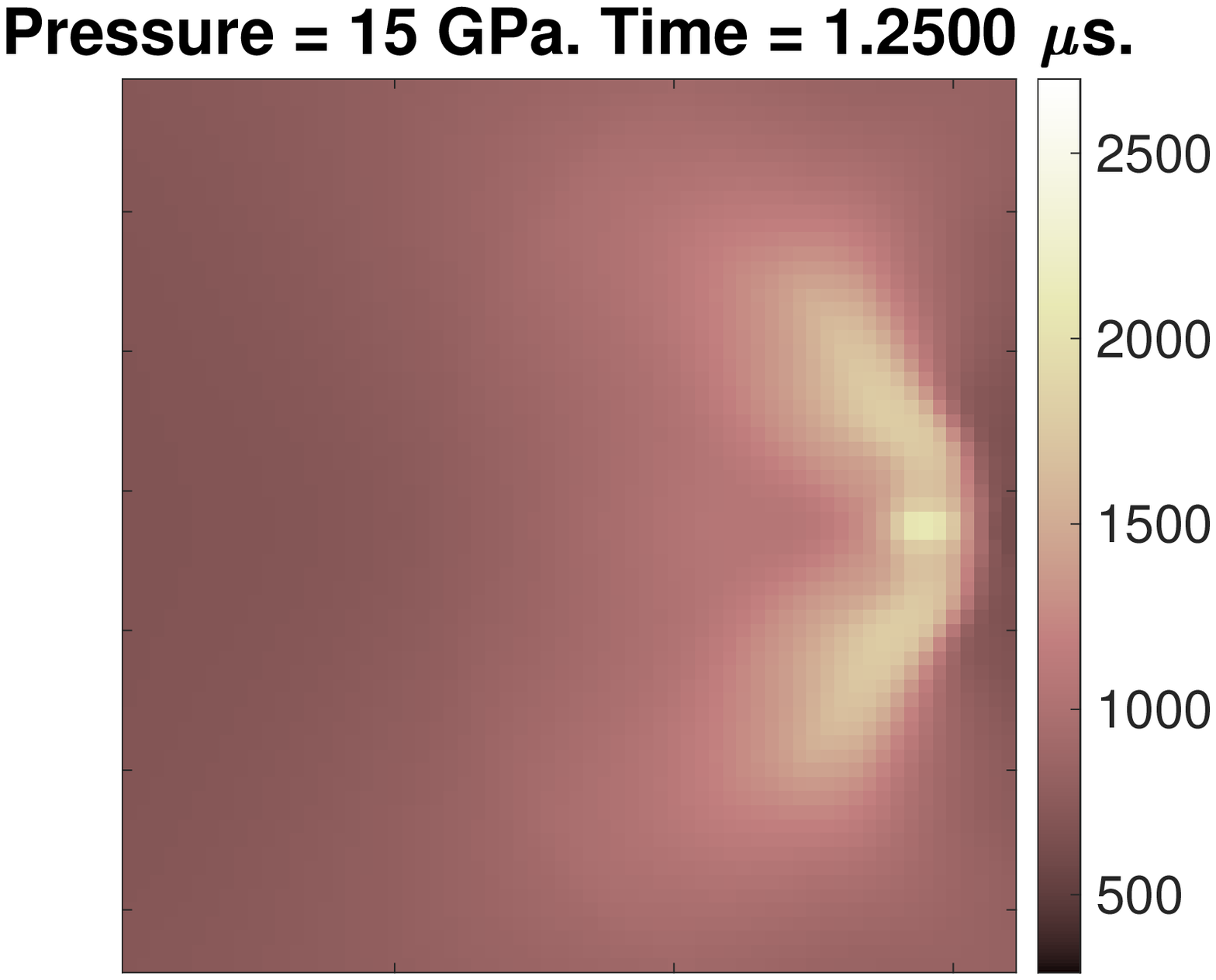}
\includegraphics[width=0.19\linewidth]{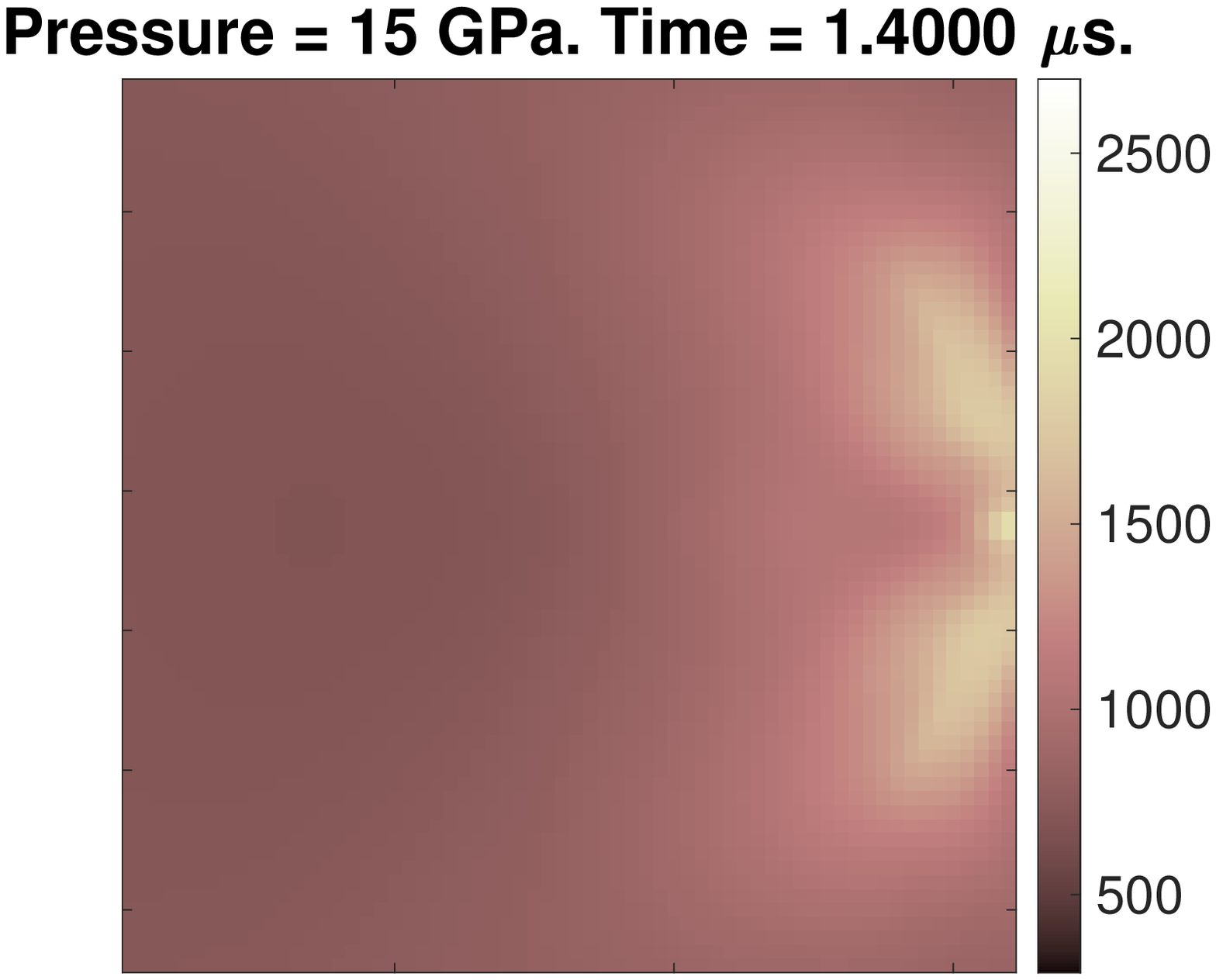}
\caption{Selected representative snapshots of temperature fields at
different shock pressures (11--15 GPa, row-wise)
in the corresponding time interval of interest,
which is adjusted to depict the pore collapse process.
Unlike Figure~\ref{fig:snap_unif},
the snapshots in the same column do not correspond to same time instance.}
\label{fig:snap_shift}
\end{figure}

\subsection{Methodology specification}

In this subsection, we discuss the details of
the surrogate modeling approaches
in performing the numerical experiments.

We first discuss the details about DMD in Section~\ref{sec:dmd}.
We use the DW-DMD approach described in Section~\ref{sec:local-dmd}.
with $\nwindow = 20$ and $\sizeROMsymbol_\windowIndex \equiv 9$,
and we use Gaussian functions in RBF interpolation.
All the DMD results are generated using
the implementation in libROM \footnote{GitHub page, {\it https://github.com/LLNL/libROM}.}
on Quartz in Livermore Computing
Center\footnote{High performance computing at LLNL, https://hpc.llnl.gov/hardware/platforms/quartz},
on Intel Xeon CPUs with 128 GB memory, peak TFLOPS of 3251.4, and peak single
CPU memory bandwidth of 77 GB/s.
The training of each local DMD model takes around 30 seconds on CPU.

\begin{figure}[htp!]
\centering
\includegraphics[width=\linewidth]{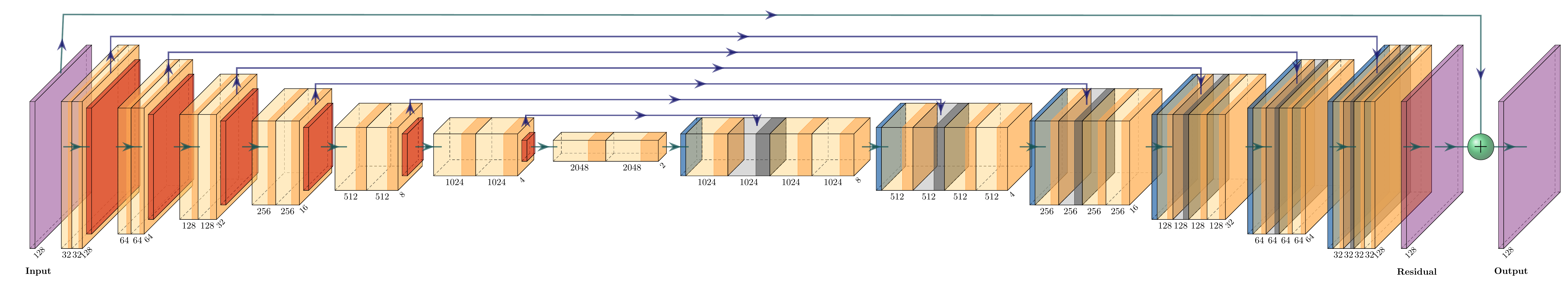}
\caption{
U-Net generator architecture used in the examples presented in Section~\ref{sec:exp}.
}
\label{fig:architecture}
\end{figure}

Next, we discuss the details about CcGAN in Section~\ref{sec:gan}.
The U-Net generator architecture is presented in Figure~\ref{fig:architecture}.
Following \cite{kadeethum2022continuous},
we take $\mu_{\text{Lip}} = 10$ and $\mu_{\text{recon}} = 500$ in the
objective \eqref{eq:total-gan-loss}.
All the CcGAN results are generated on
Lassen in Livermore Computing Center\footnote{High
performance computing at LLNL, https://hpc.llnl.gov/hardware/platforms/lassen},
on Intel Power9 CPUs with 256 GB memory and NVIDIA V100 GPUs,
peak TFLOPS of 23,047.20, and peak single
CPU memory bandwidth of 170 GB/s.
With a batch size 6 and 2000 epoches,
the training of global CcGAN model takes 8 hours on GPU.

\subsection{Prediction and performance evaluation}

\begin{figure}[htp!]
\centering
\includegraphics[width=0.19\linewidth]{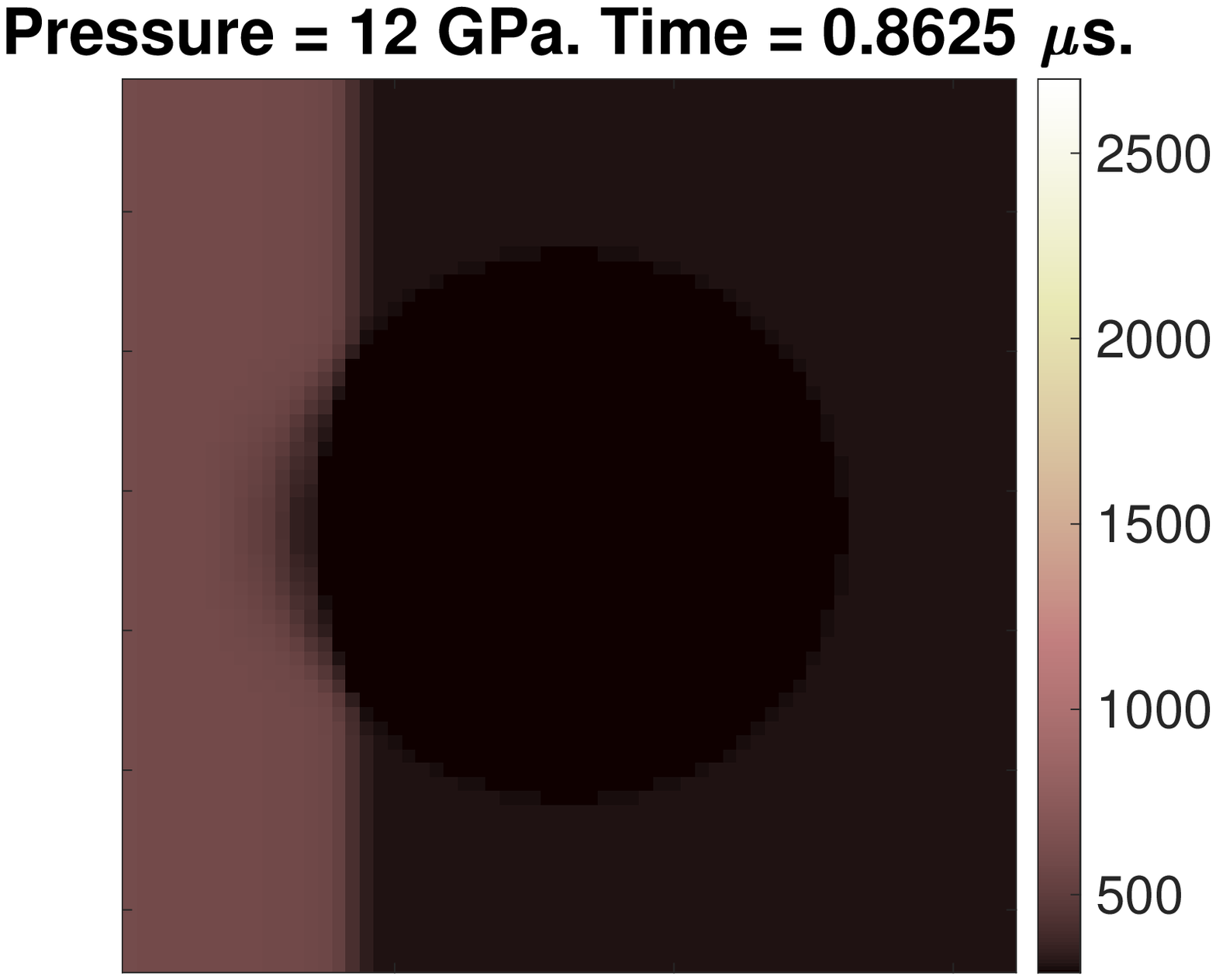}
\includegraphics[width=0.19\linewidth]{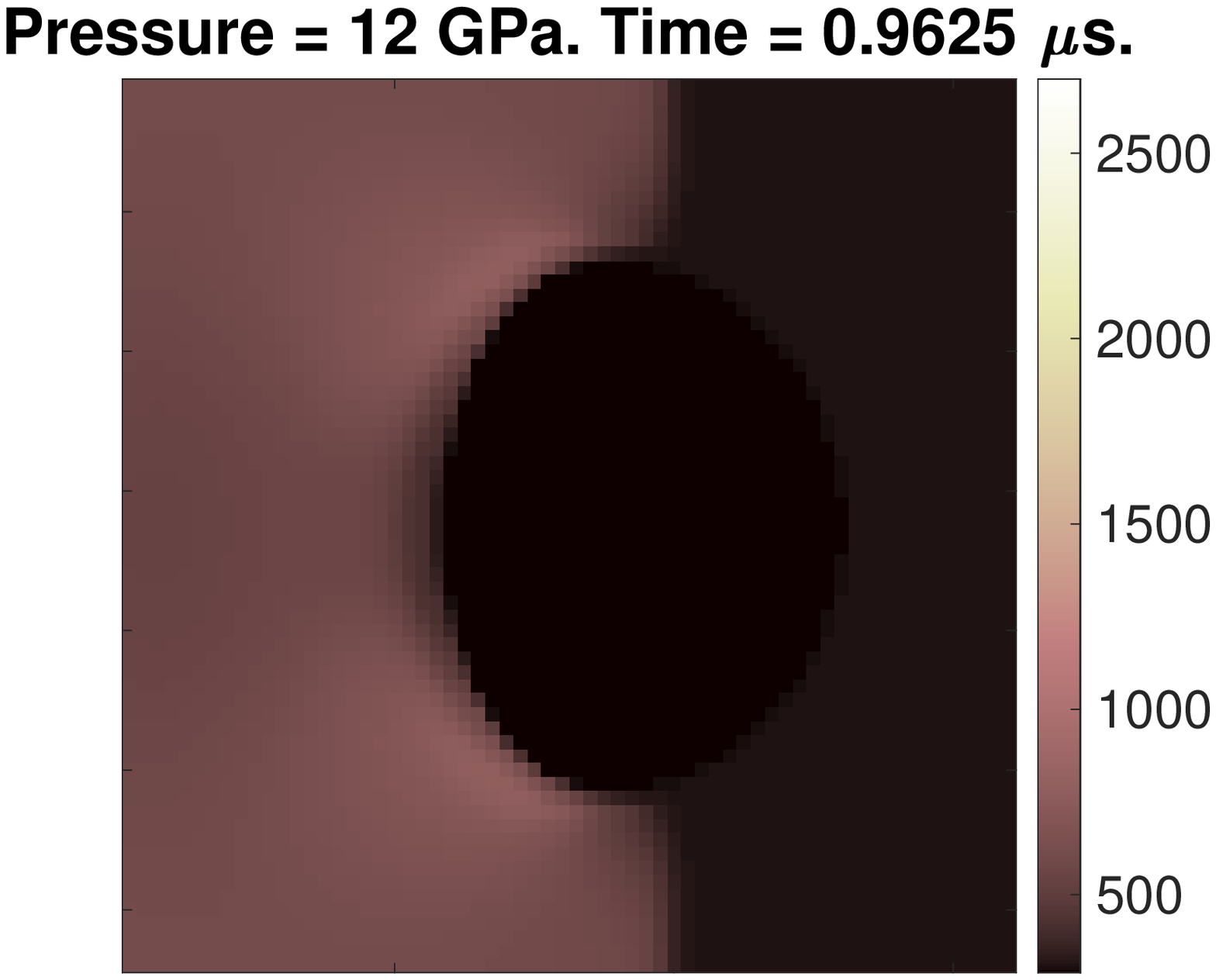}
\includegraphics[width=0.19\linewidth]{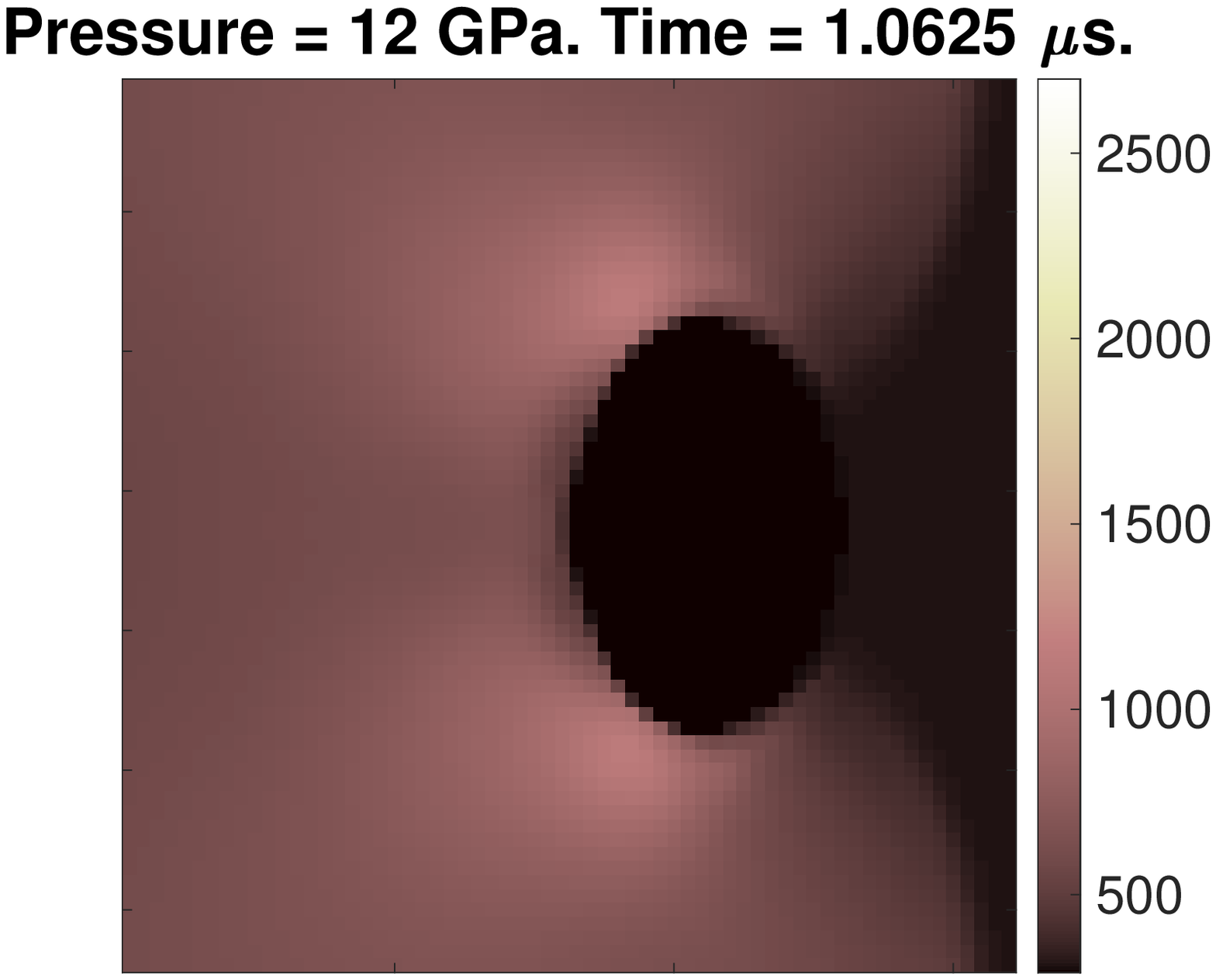}
\includegraphics[width=0.19\linewidth]{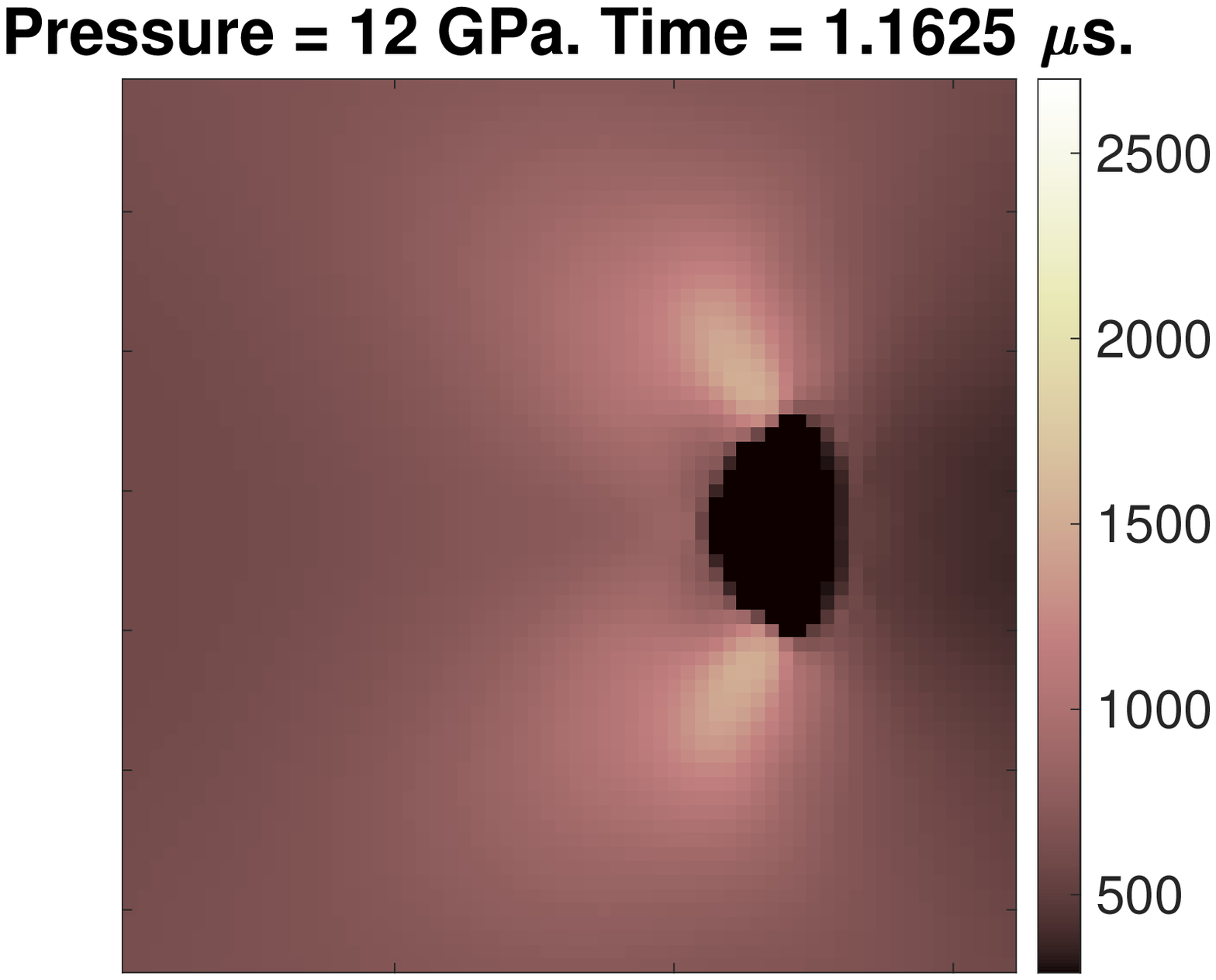}
\includegraphics[width=0.19\linewidth]{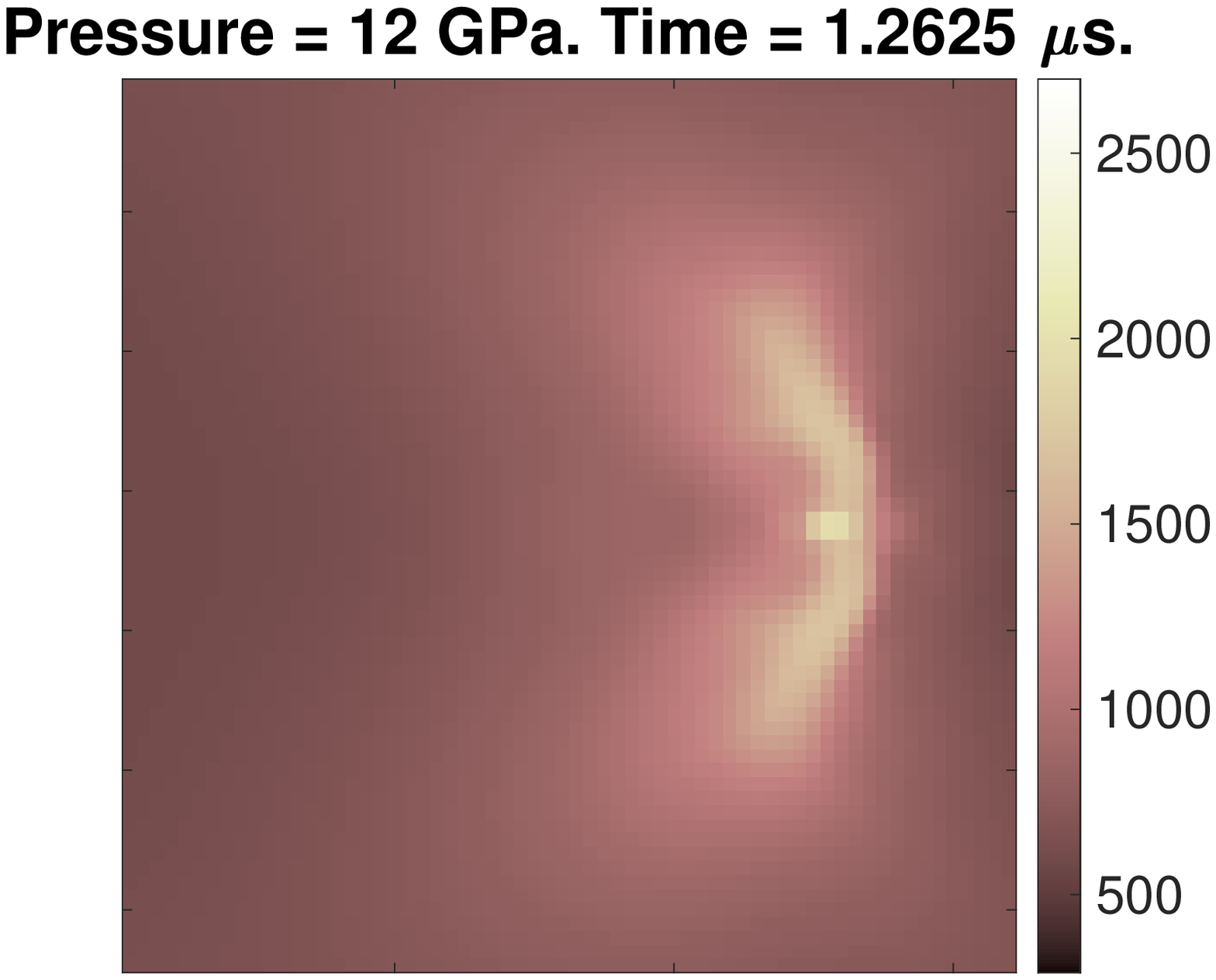}\\
\includegraphics[width=0.19\linewidth]{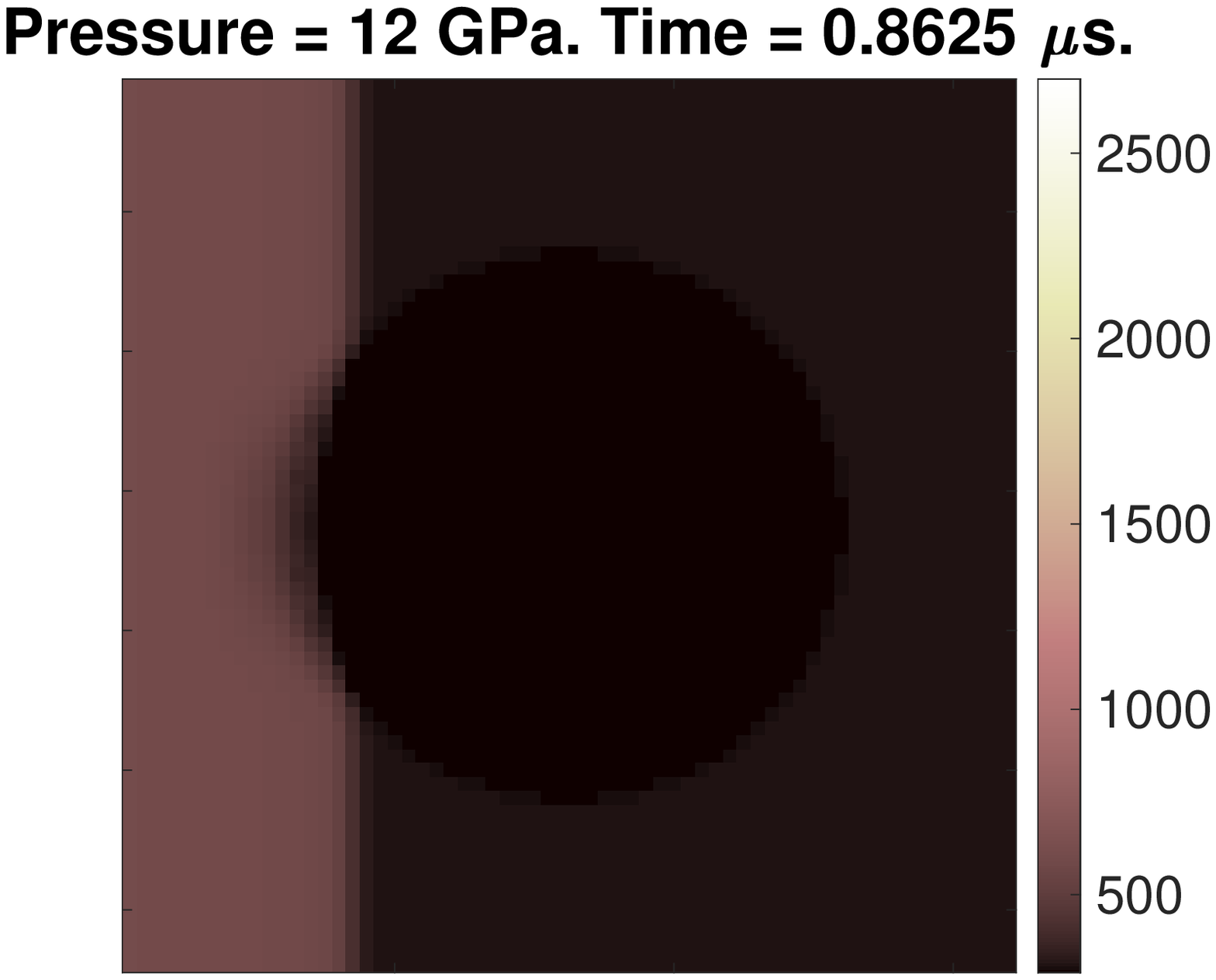}
\includegraphics[width=0.19\linewidth]{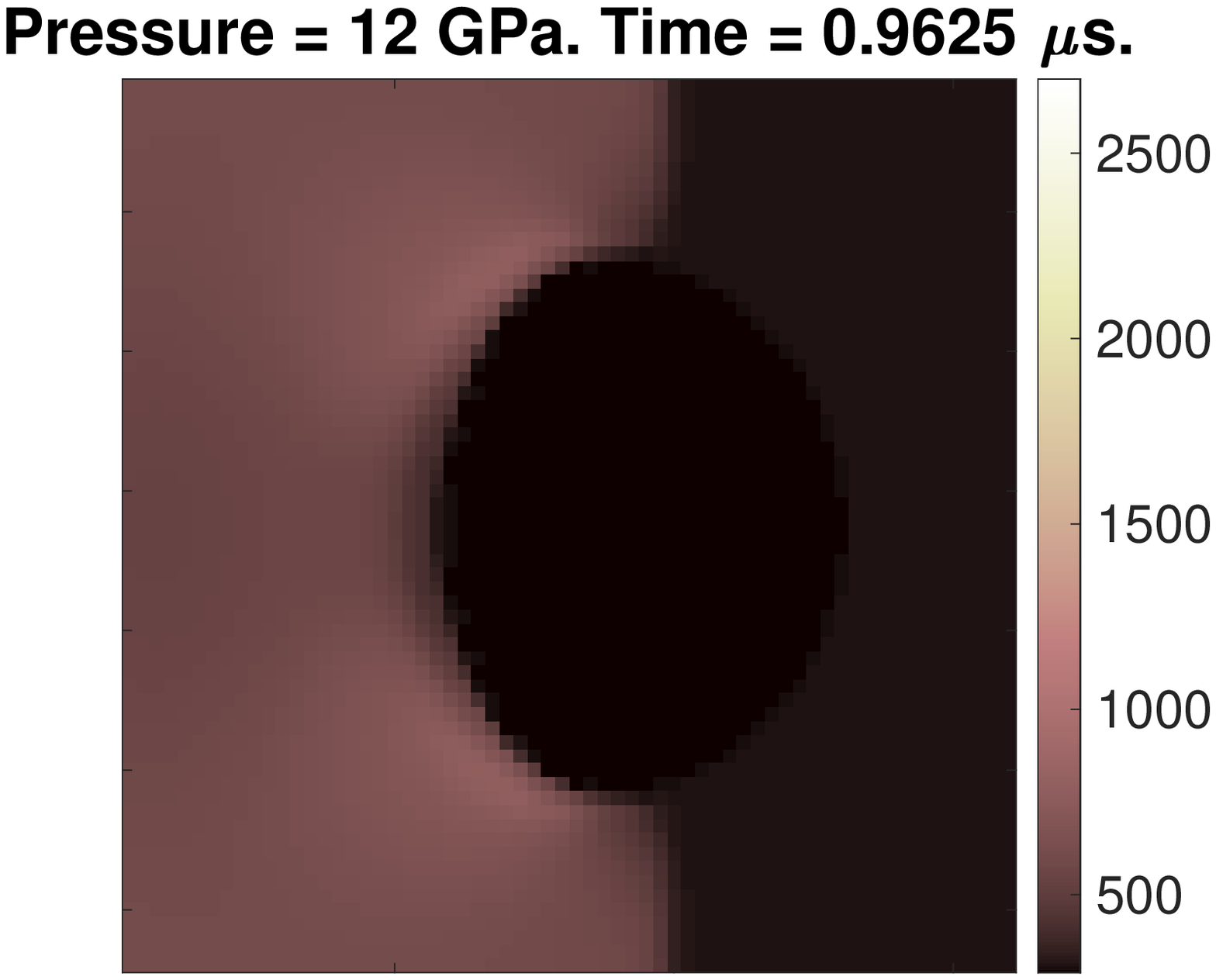}
\includegraphics[width=0.19\linewidth]{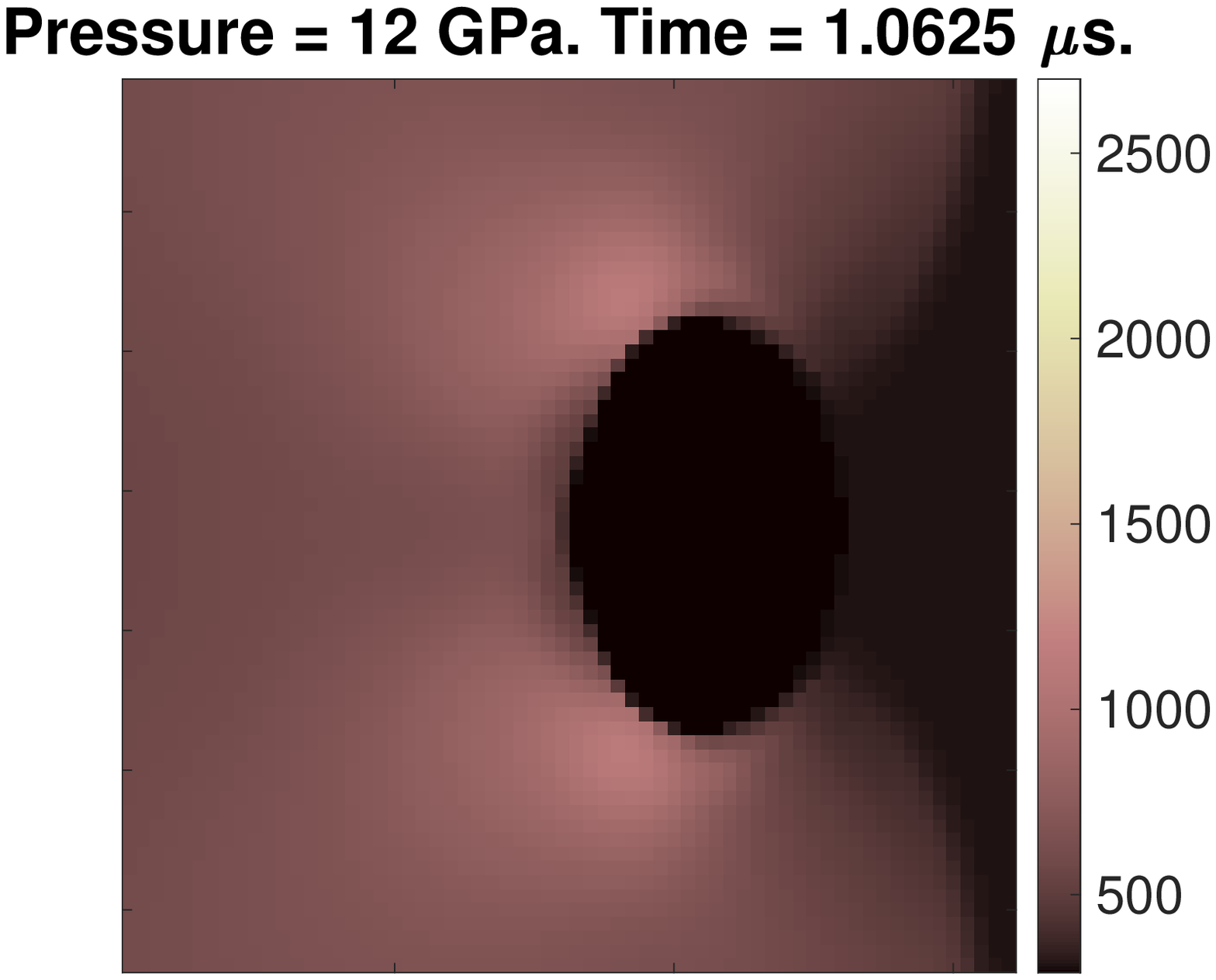}
\includegraphics[width=0.19\linewidth]{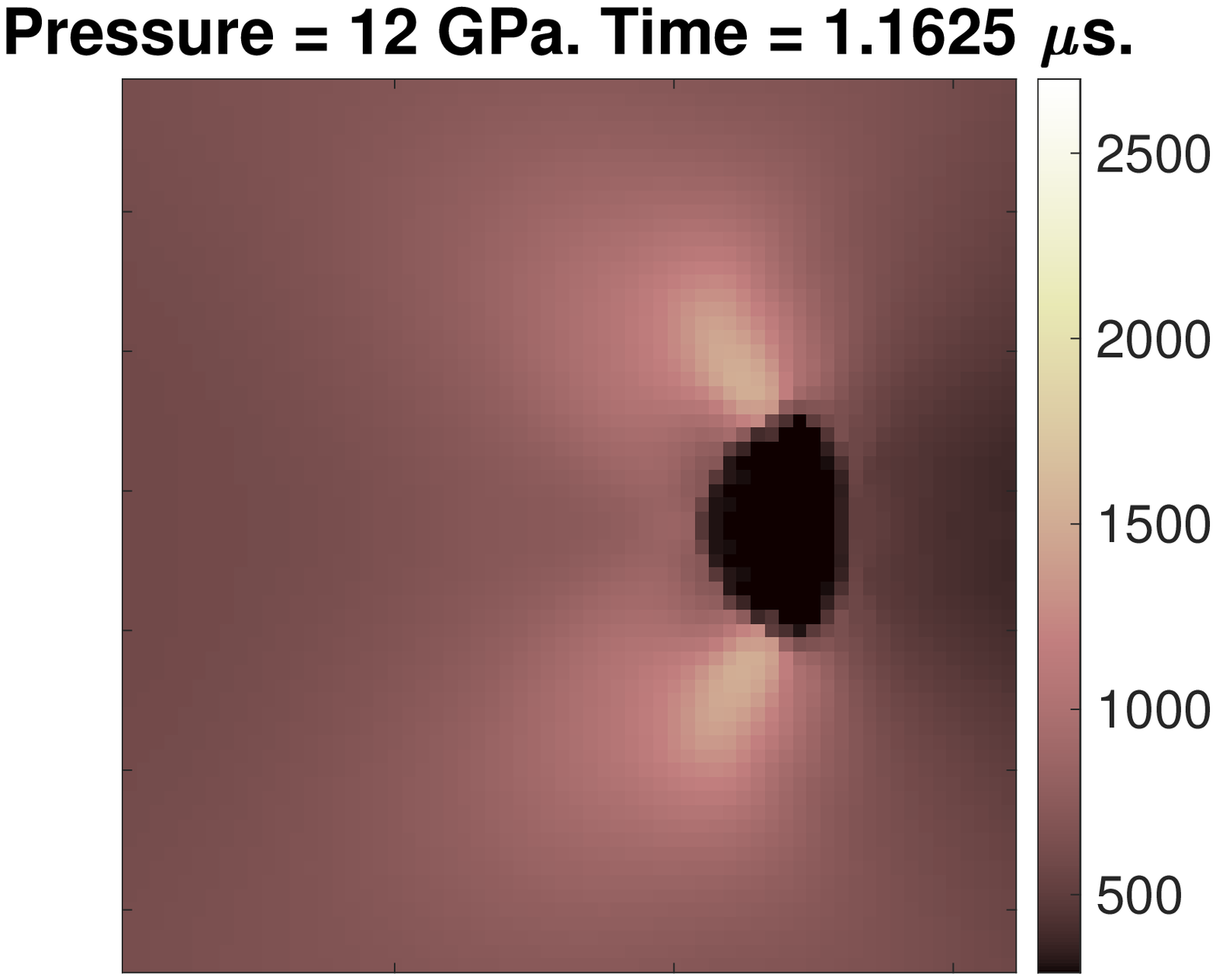}
\includegraphics[width=0.19\linewidth]{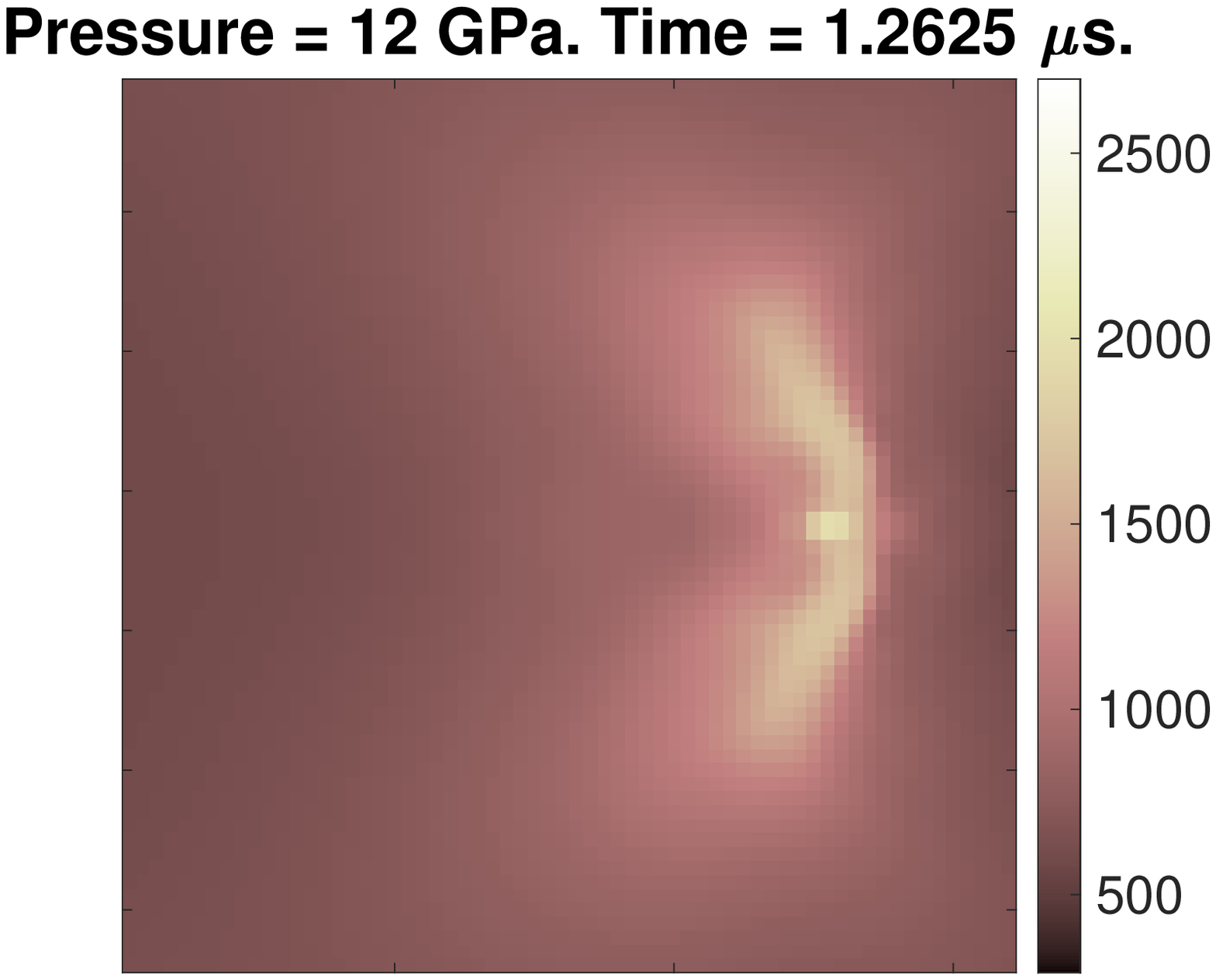}\\
\includegraphics[width=0.19\linewidth]{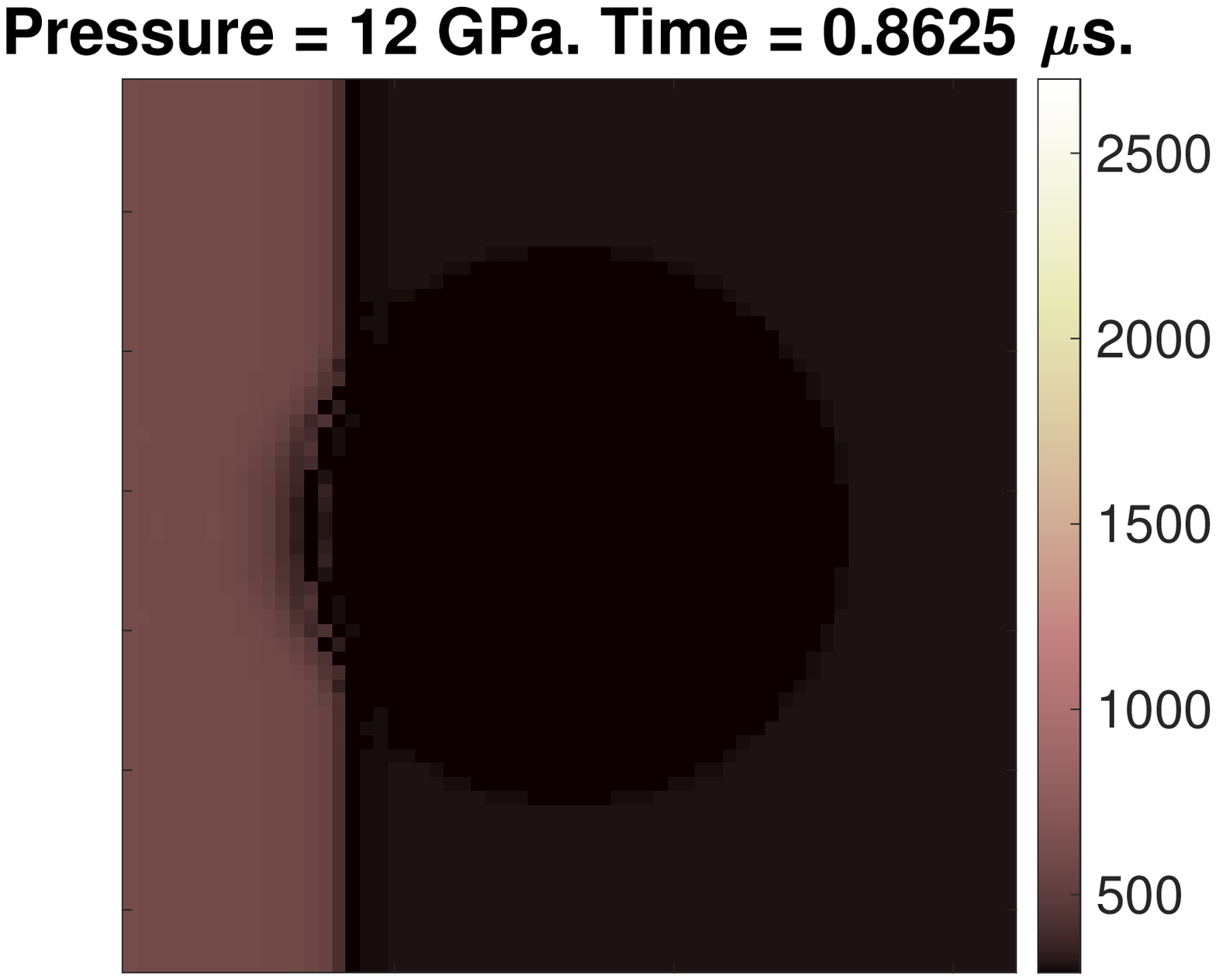}
\includegraphics[width=0.19\linewidth]{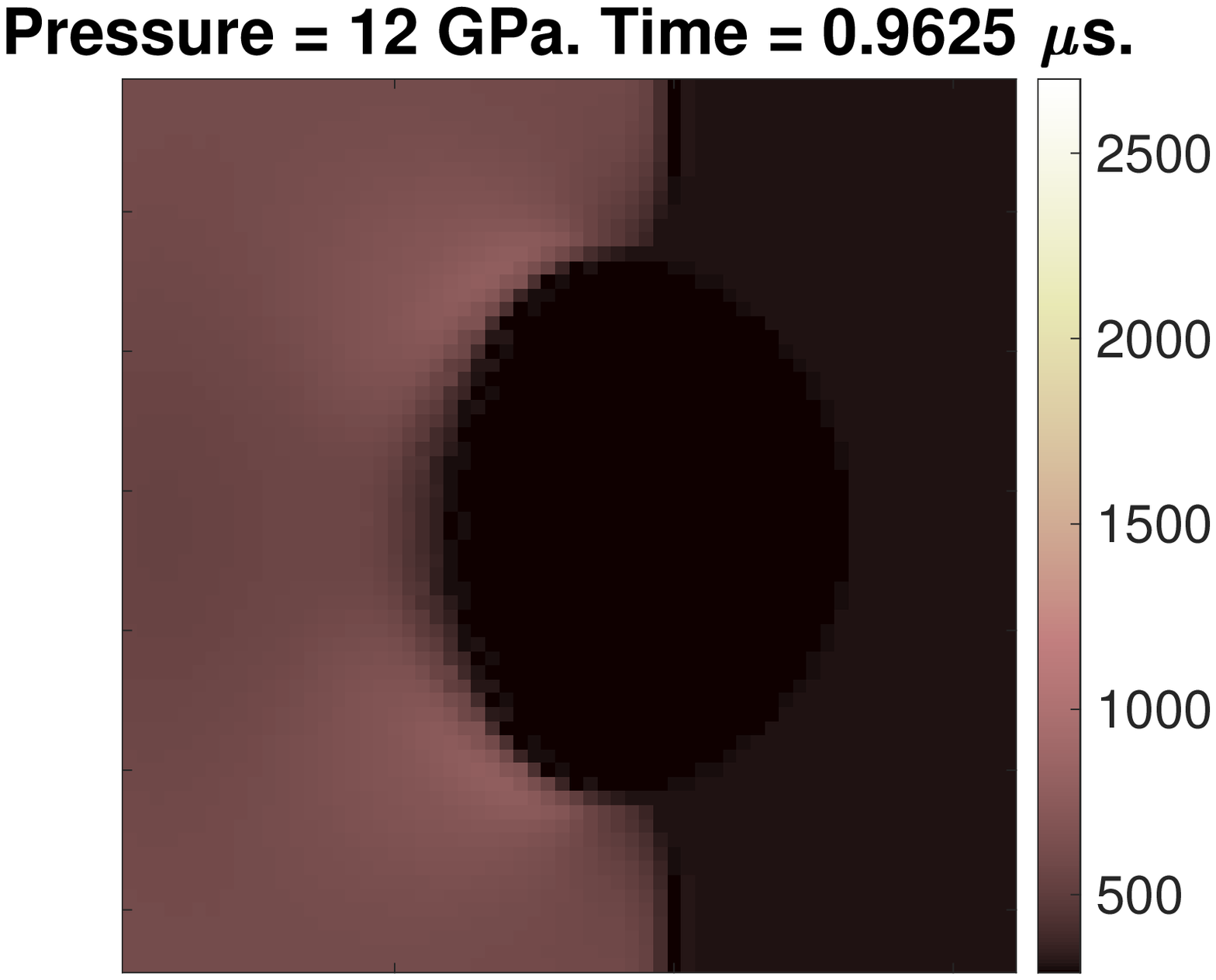}
\includegraphics[width=0.19\linewidth]{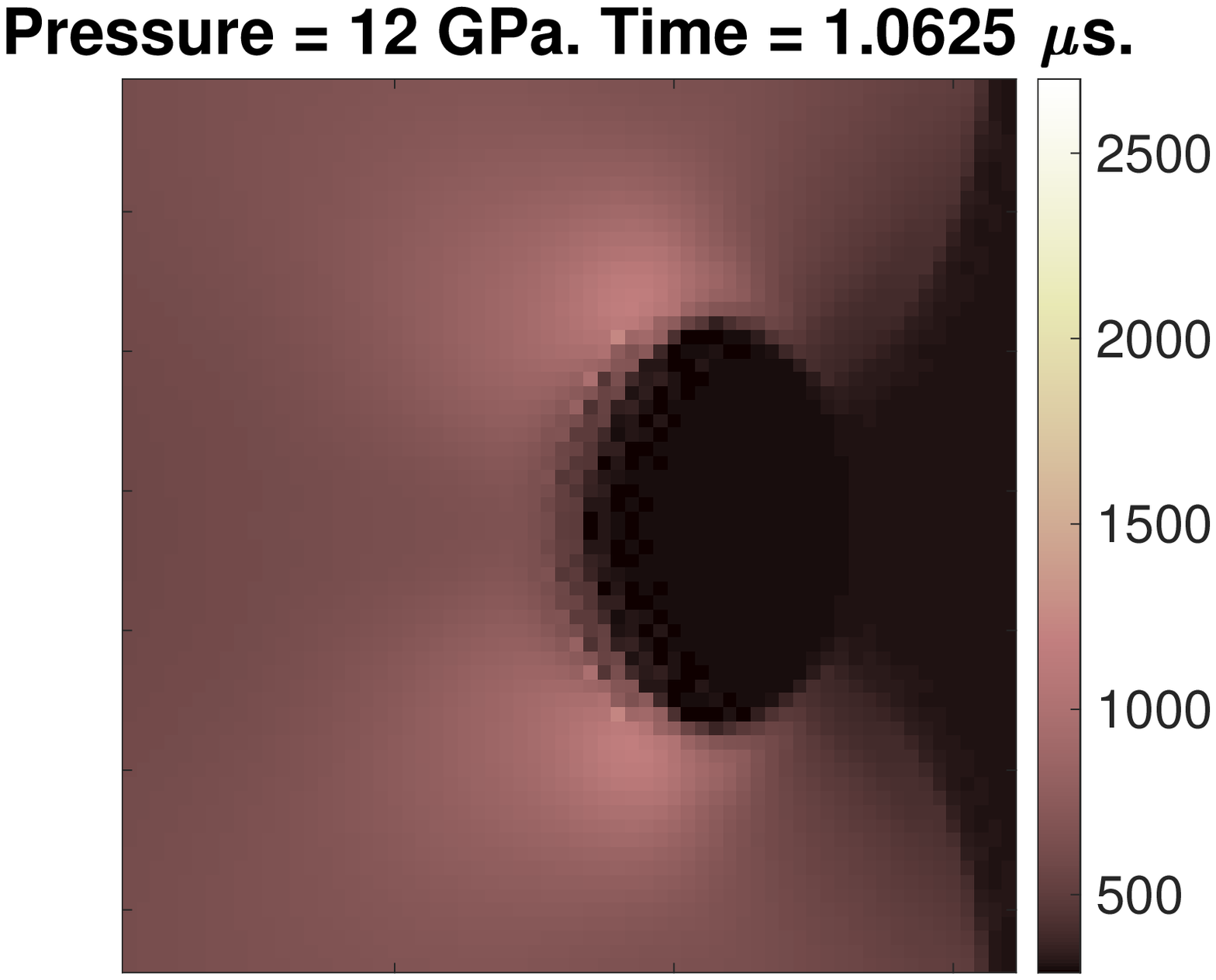}
\includegraphics[width=0.19\linewidth]{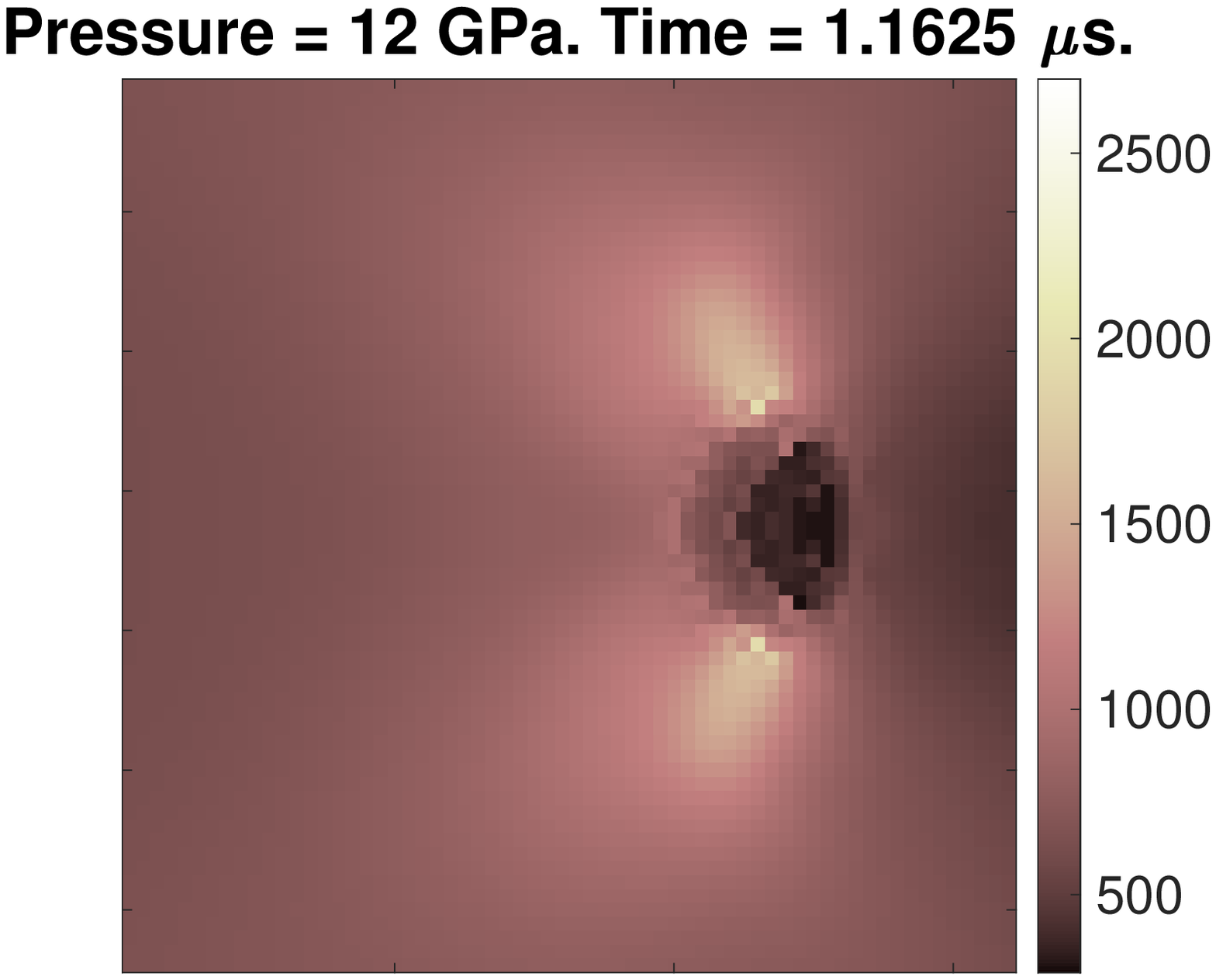}
\includegraphics[width=0.19\linewidth]{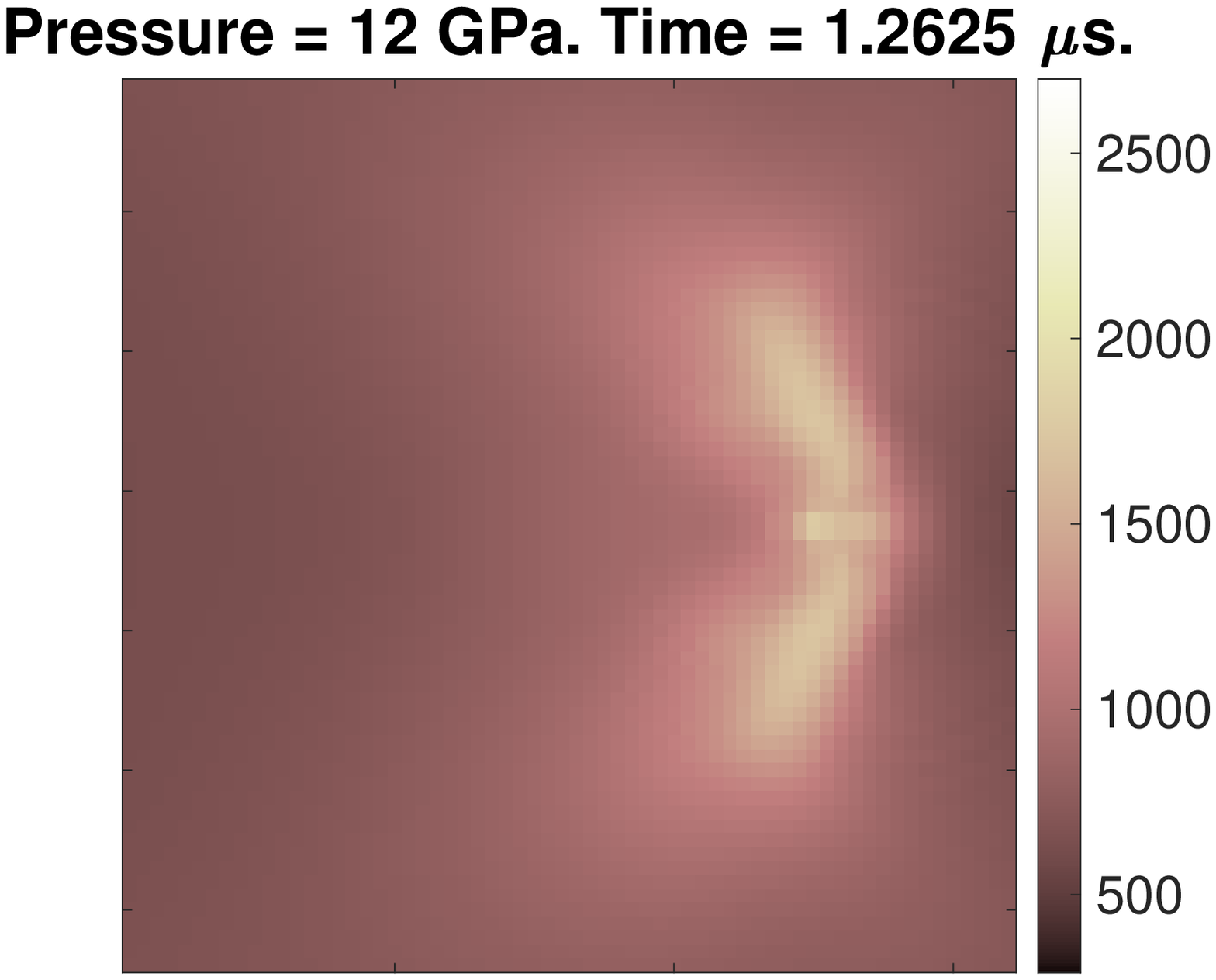}\\
\includegraphics[width=0.19\linewidth]{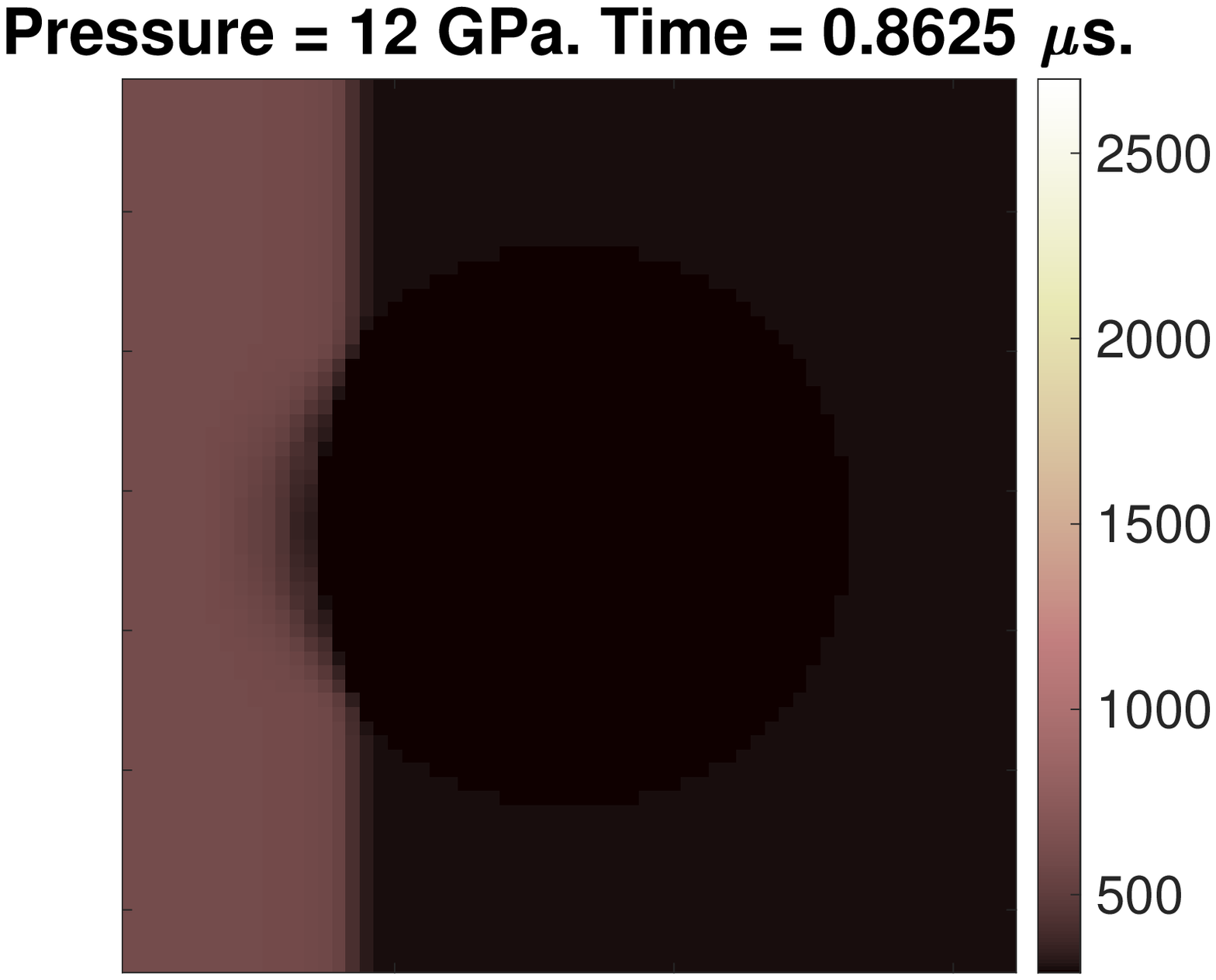}
\includegraphics[width=0.19\linewidth]{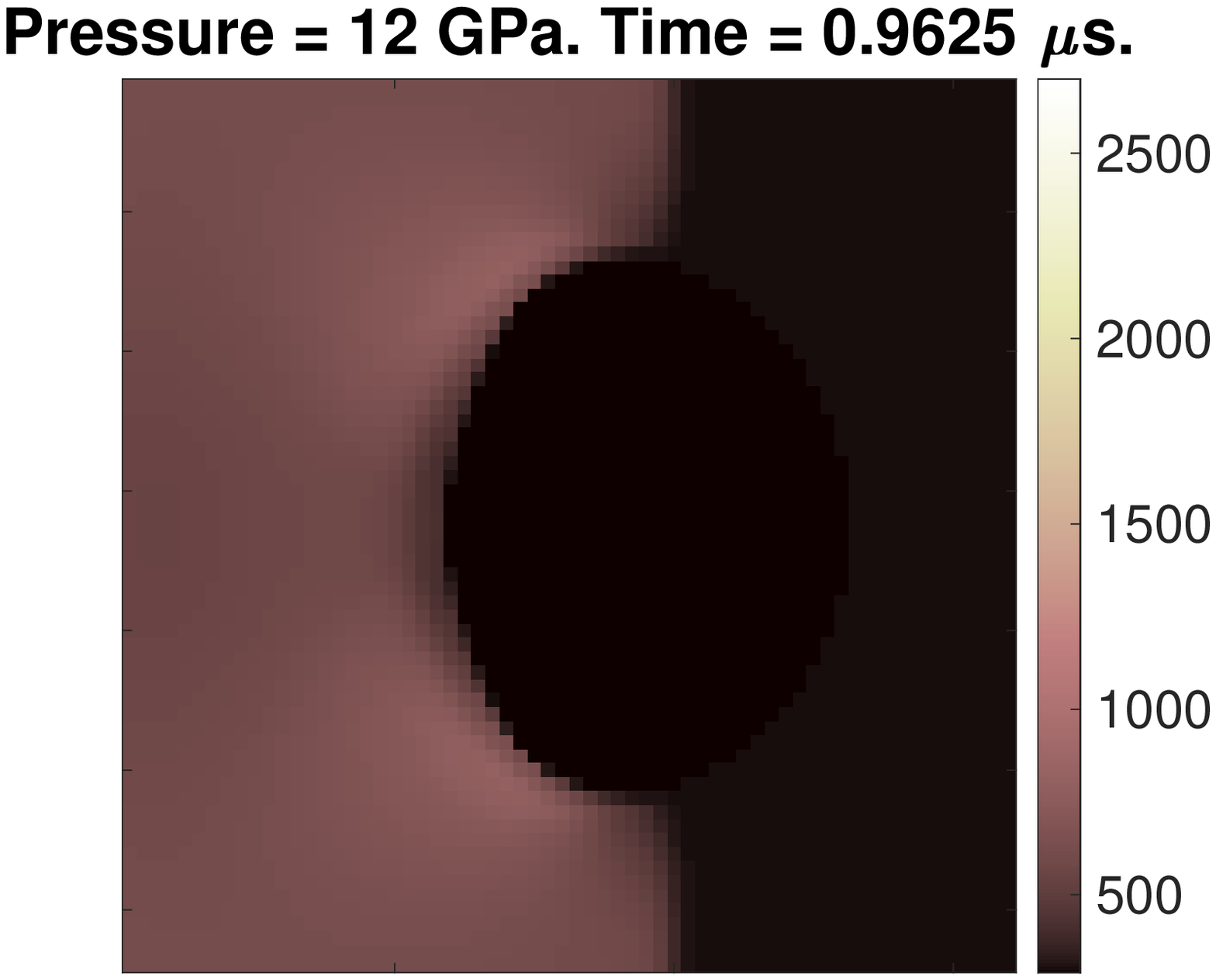}
\includegraphics[width=0.19\linewidth]{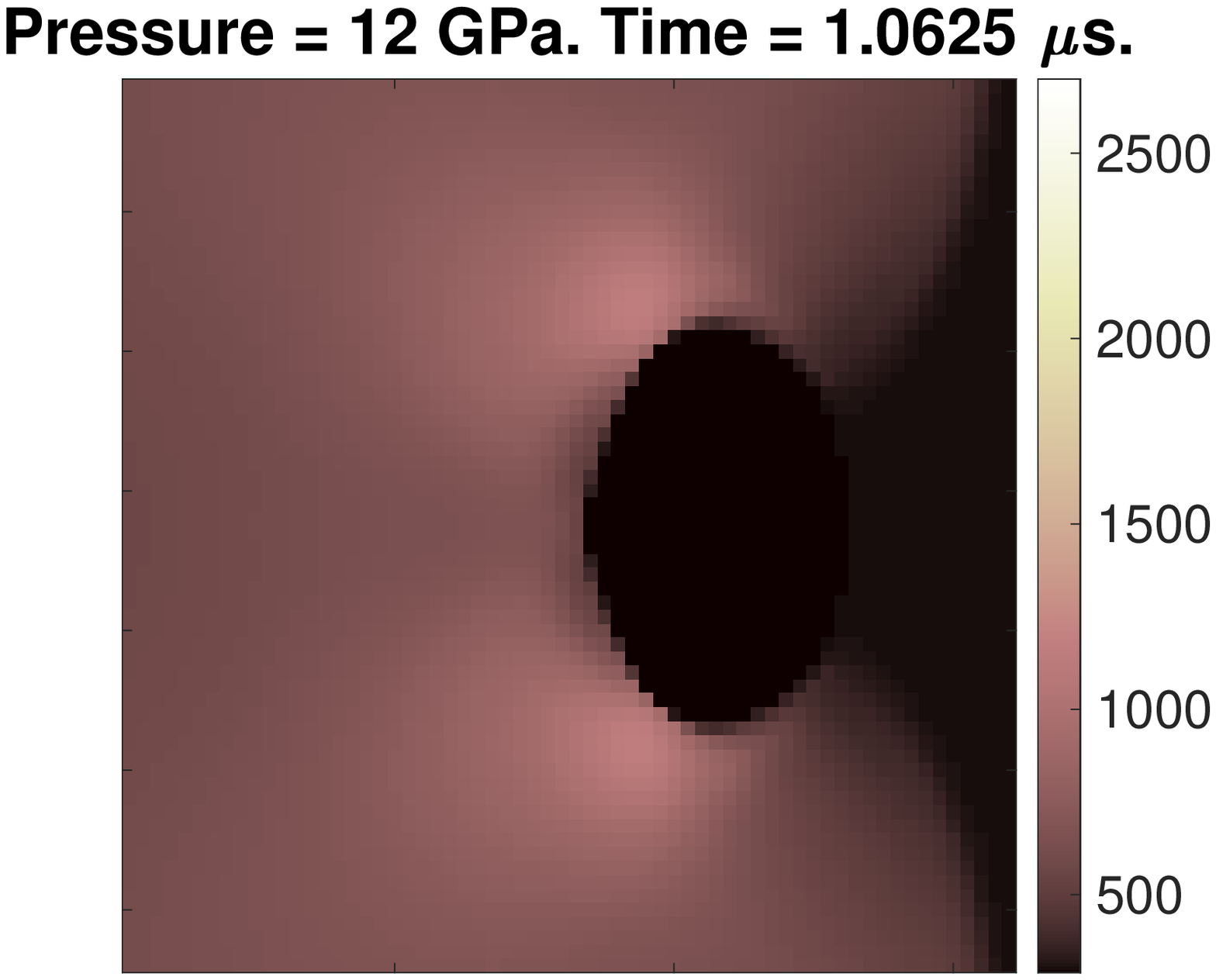}
\includegraphics[width=0.19\linewidth]{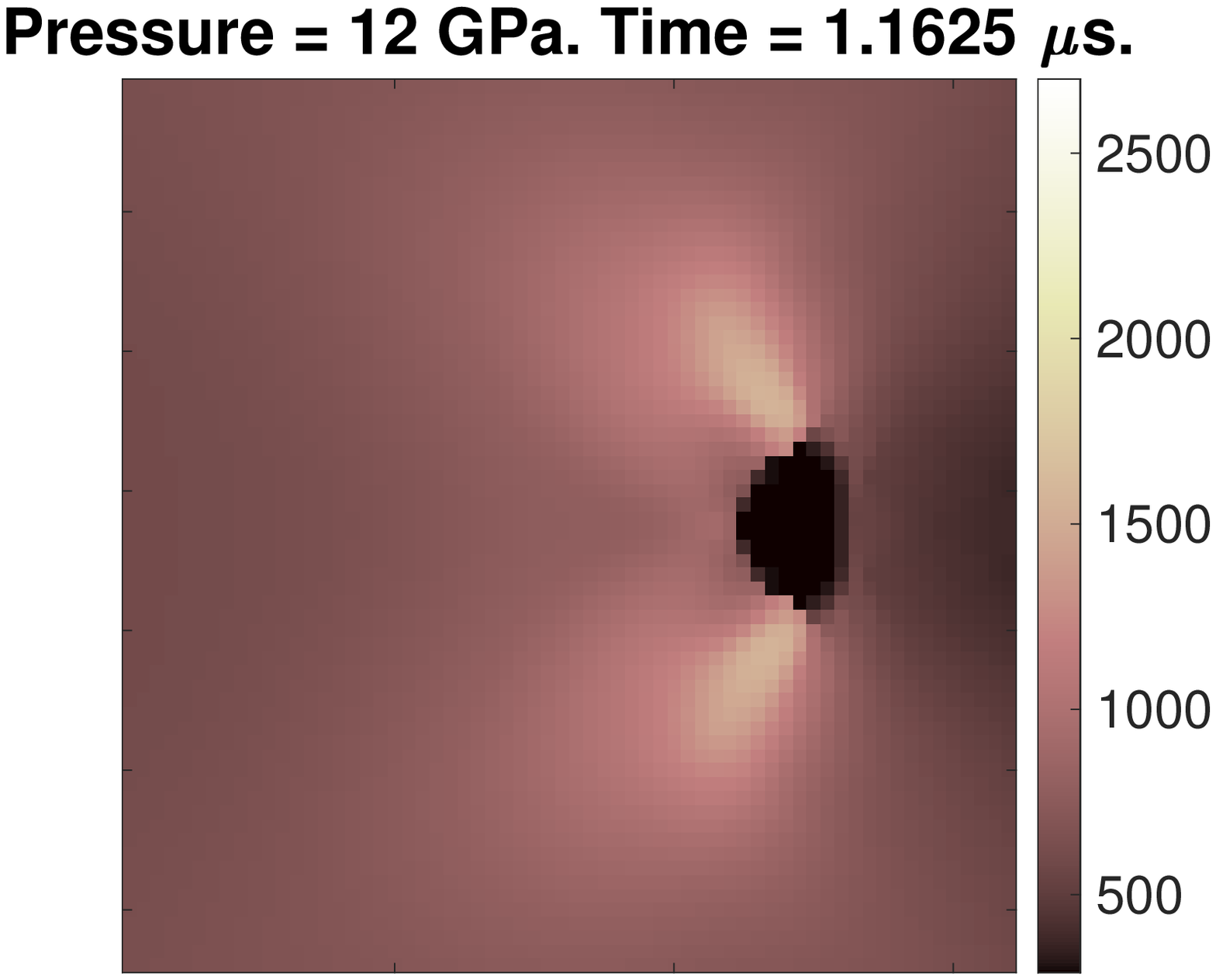}
\includegraphics[width=0.19\linewidth]{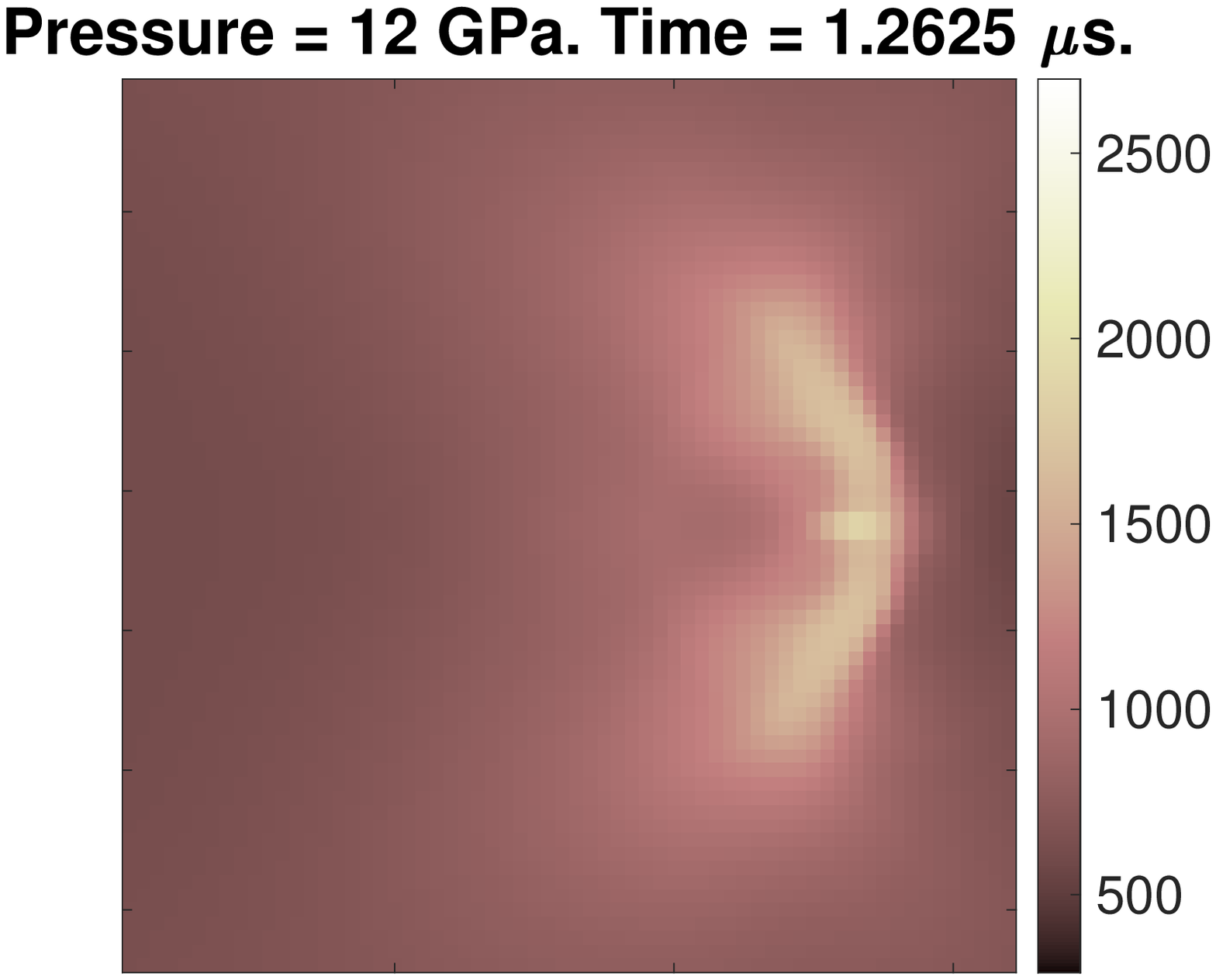}\\
\includegraphics[width=0.19\linewidth]{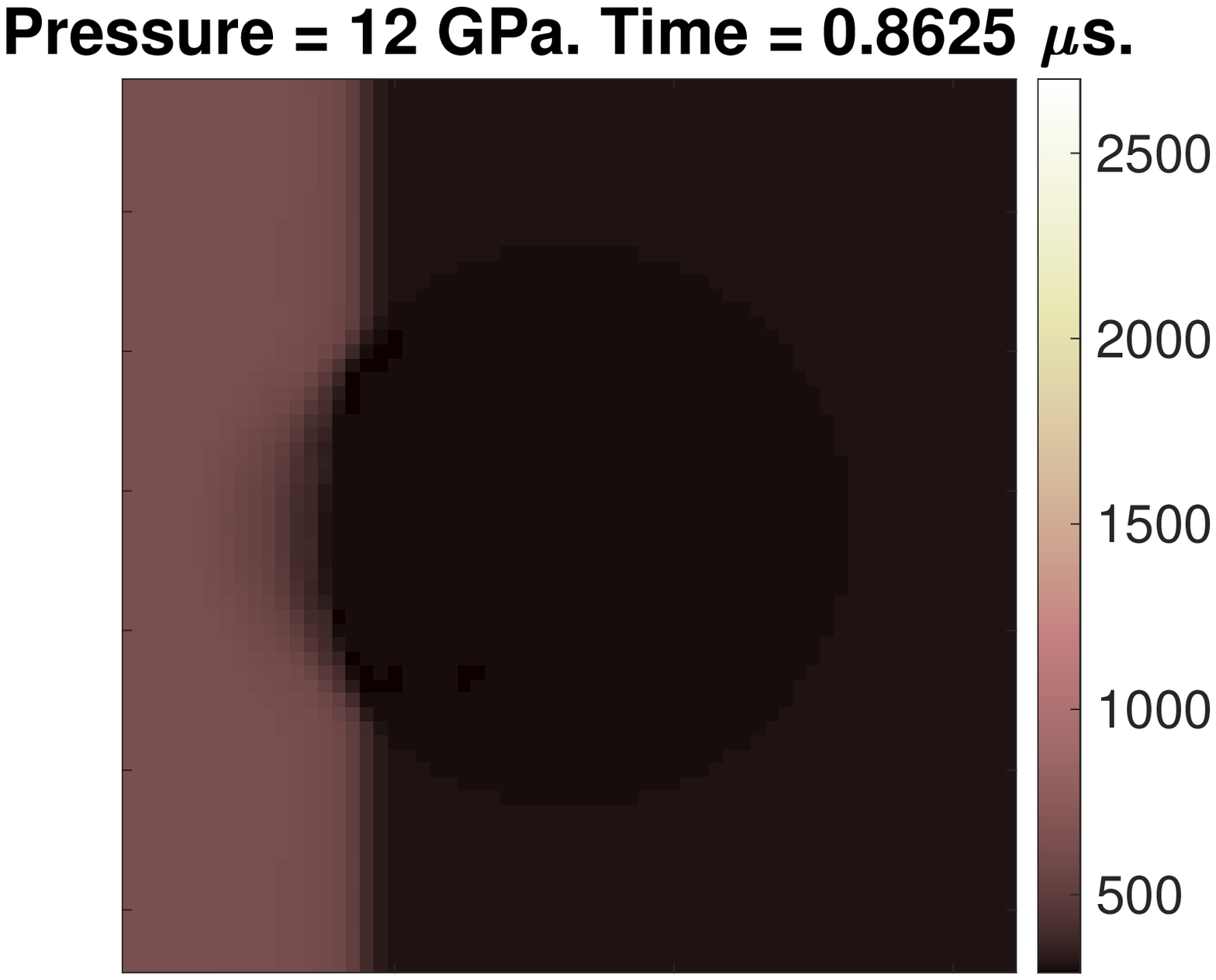}
\includegraphics[width=0.19\linewidth]{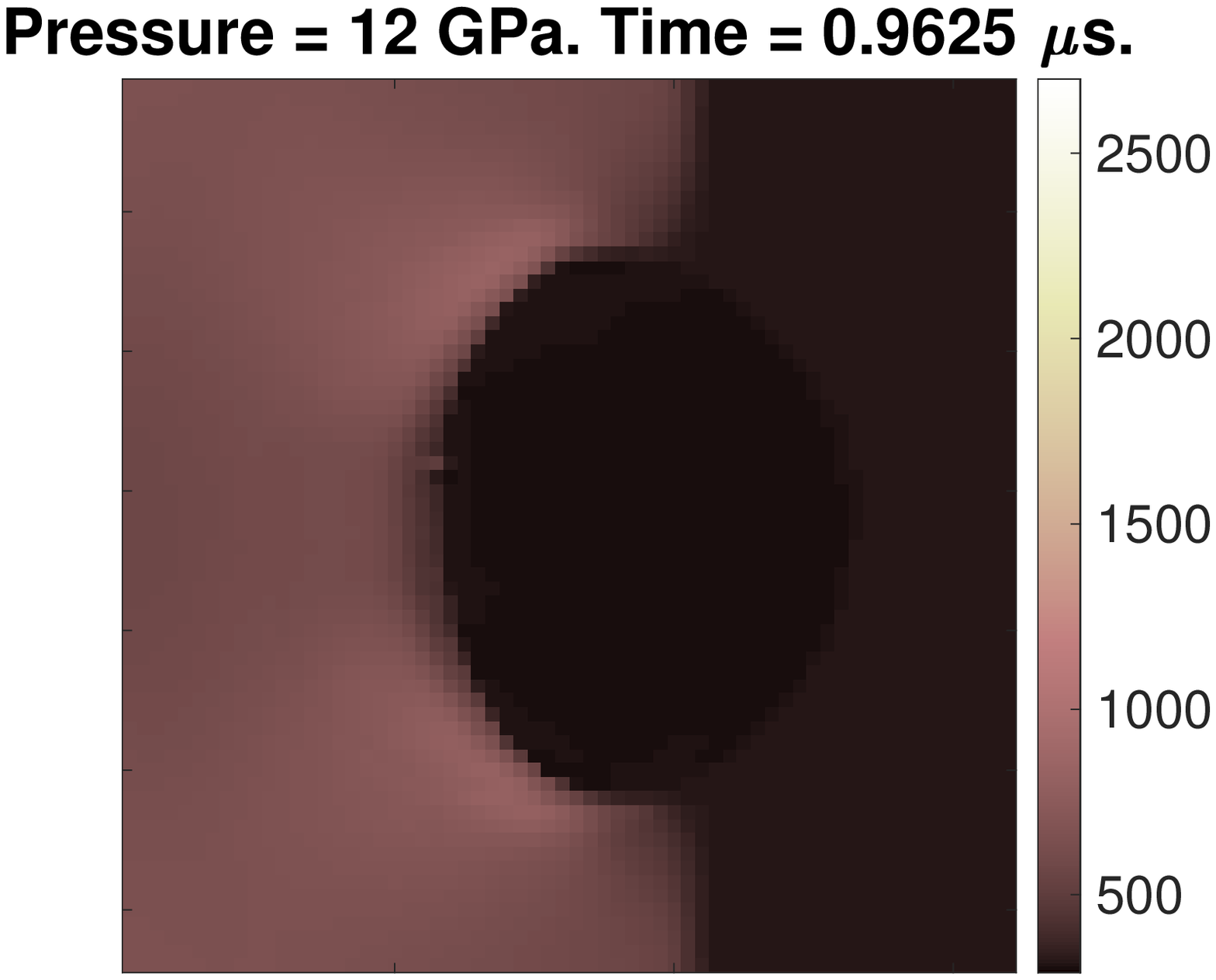}
\includegraphics[width=0.19\linewidth]{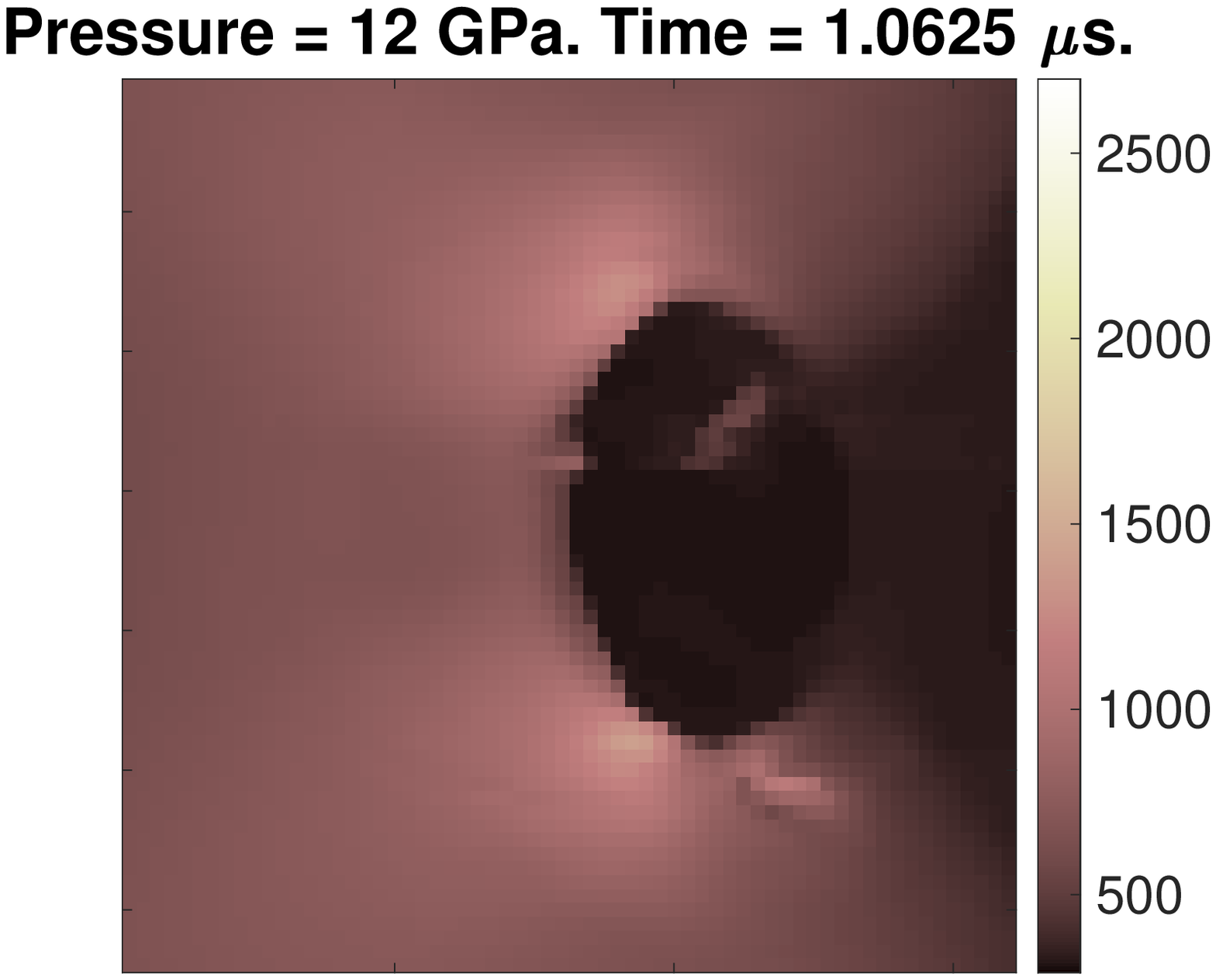}
\includegraphics[width=0.19\linewidth]{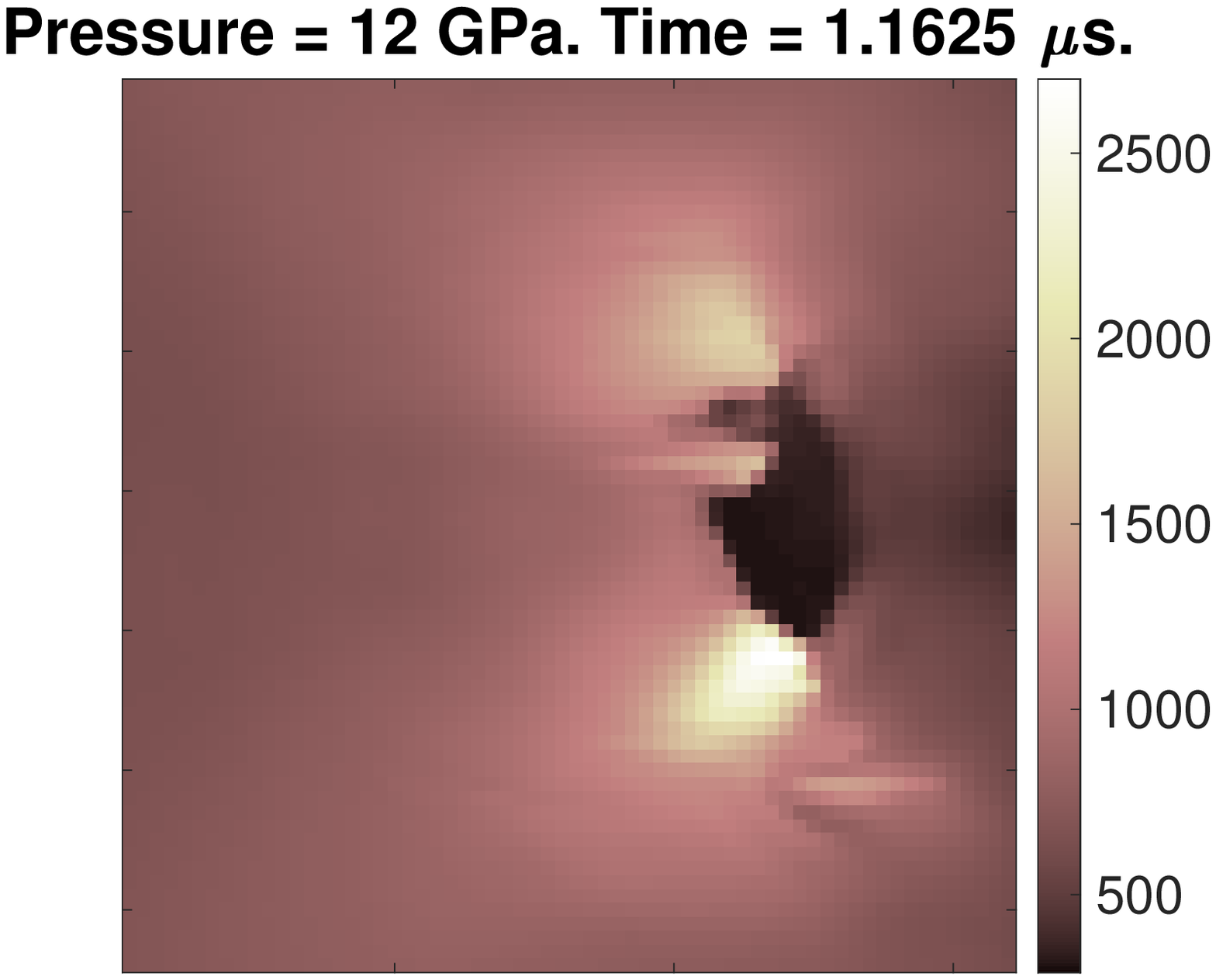}
\includegraphics[width=0.19\linewidth]{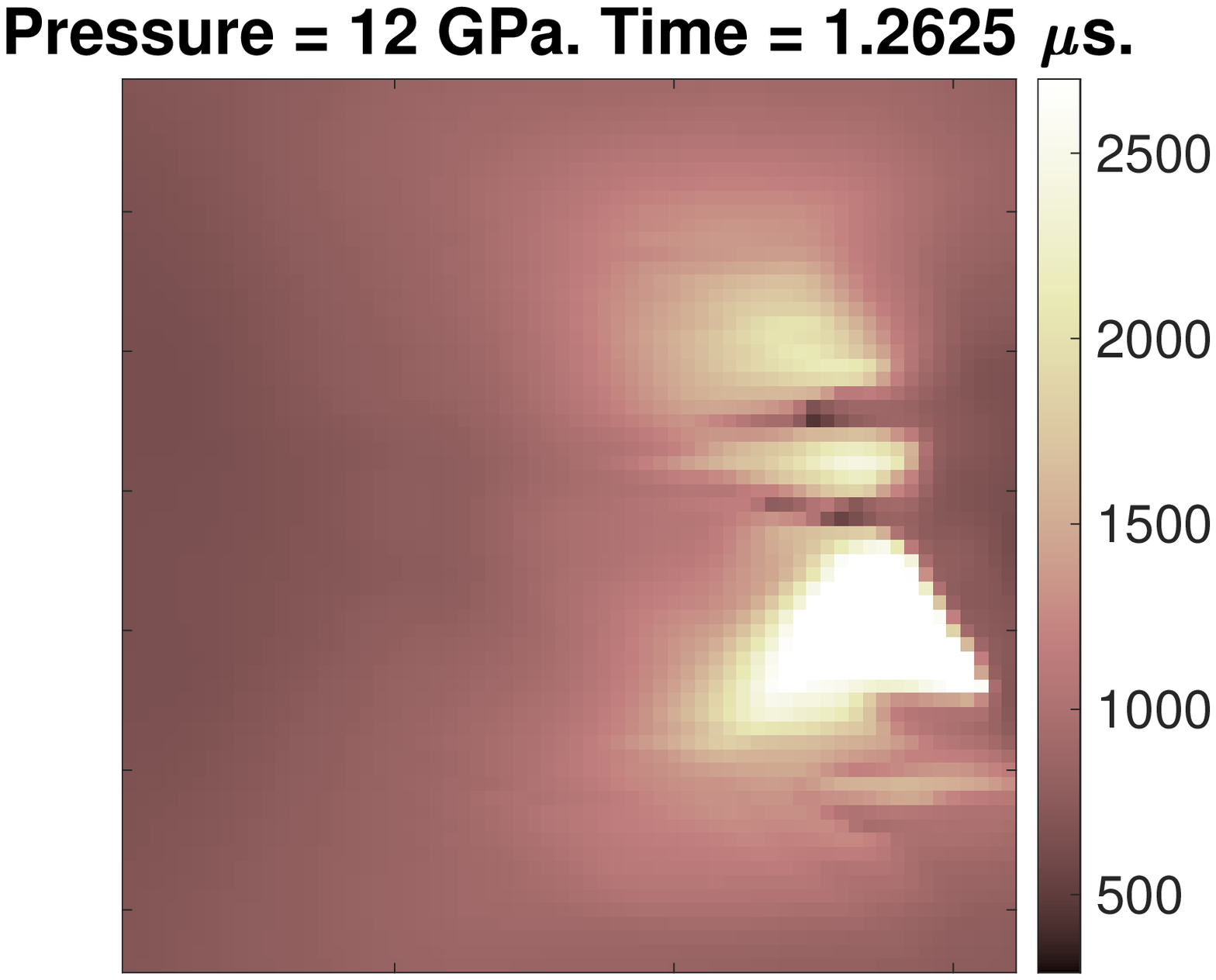}\\
\includegraphics[width=0.19\linewidth]{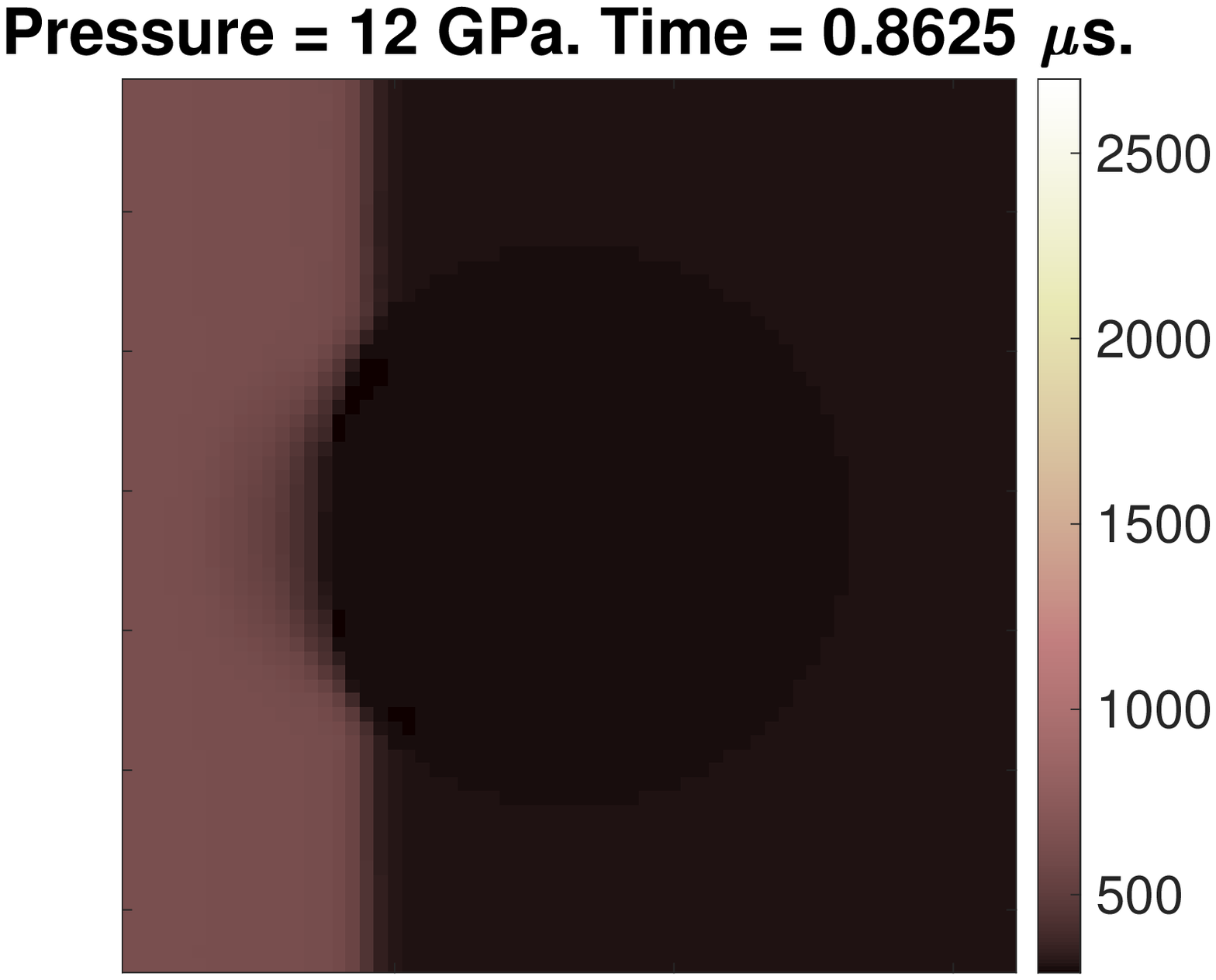}
\includegraphics[width=0.19\linewidth]{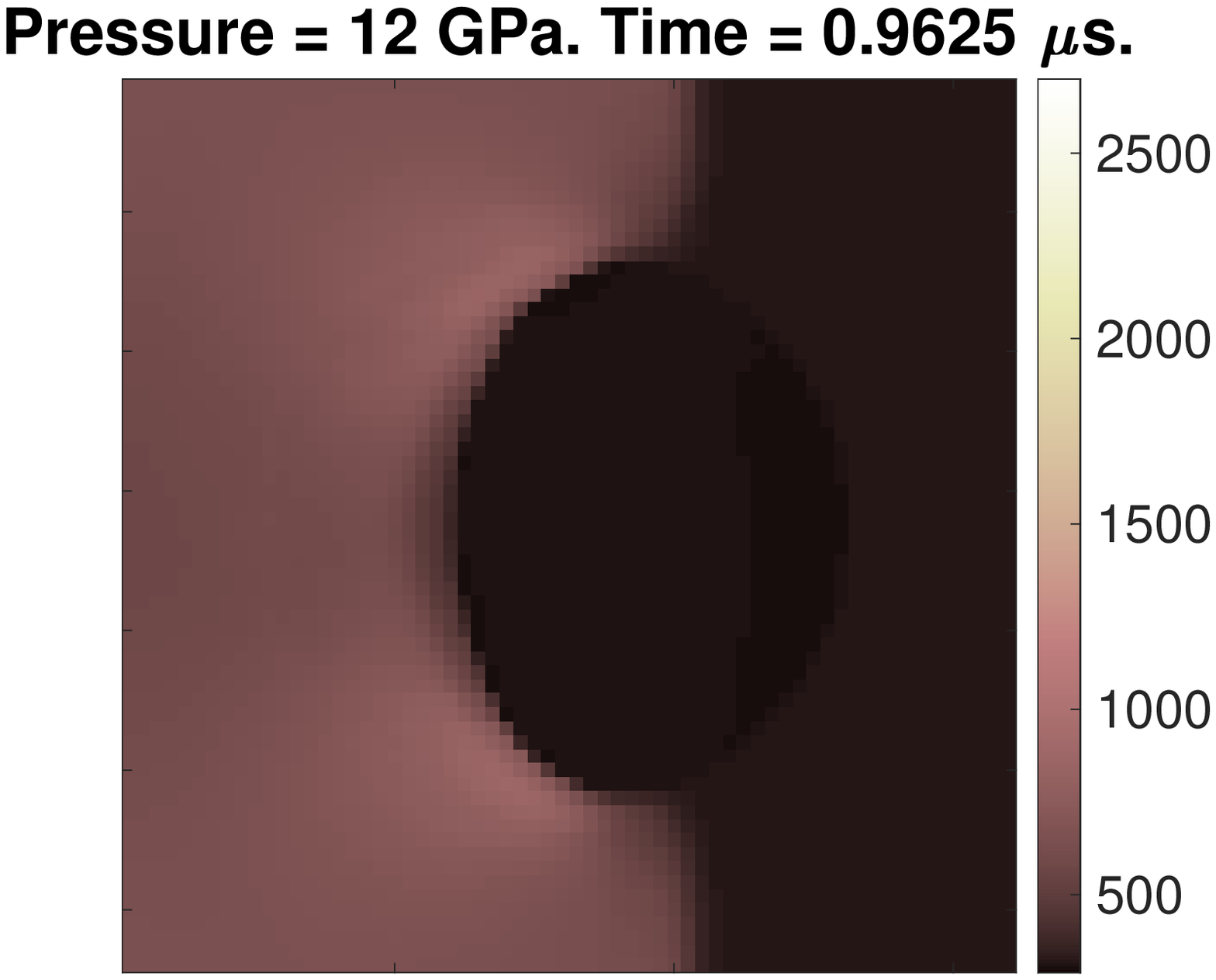}
\includegraphics[width=0.19\linewidth]{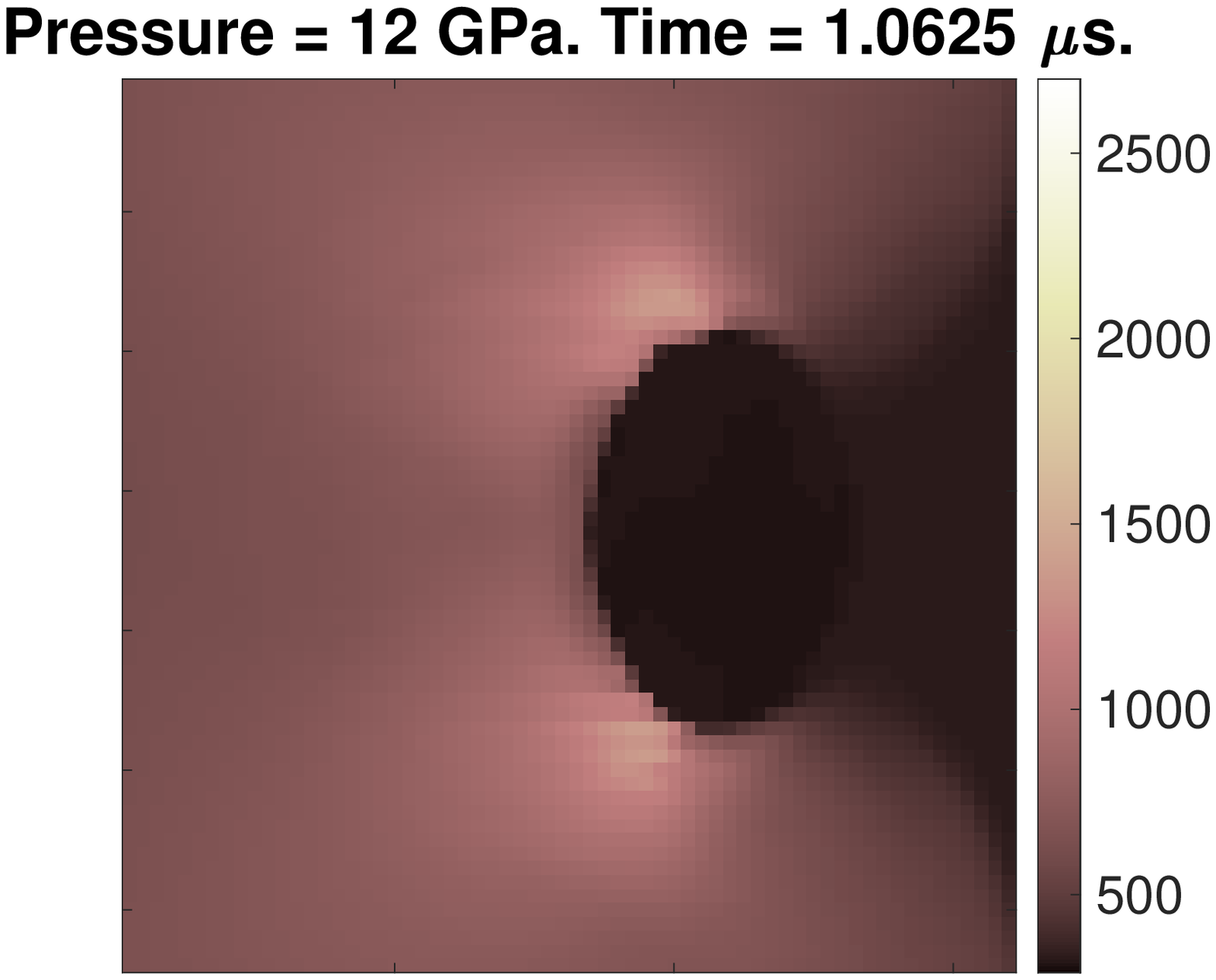}
\includegraphics[width=0.19\linewidth]{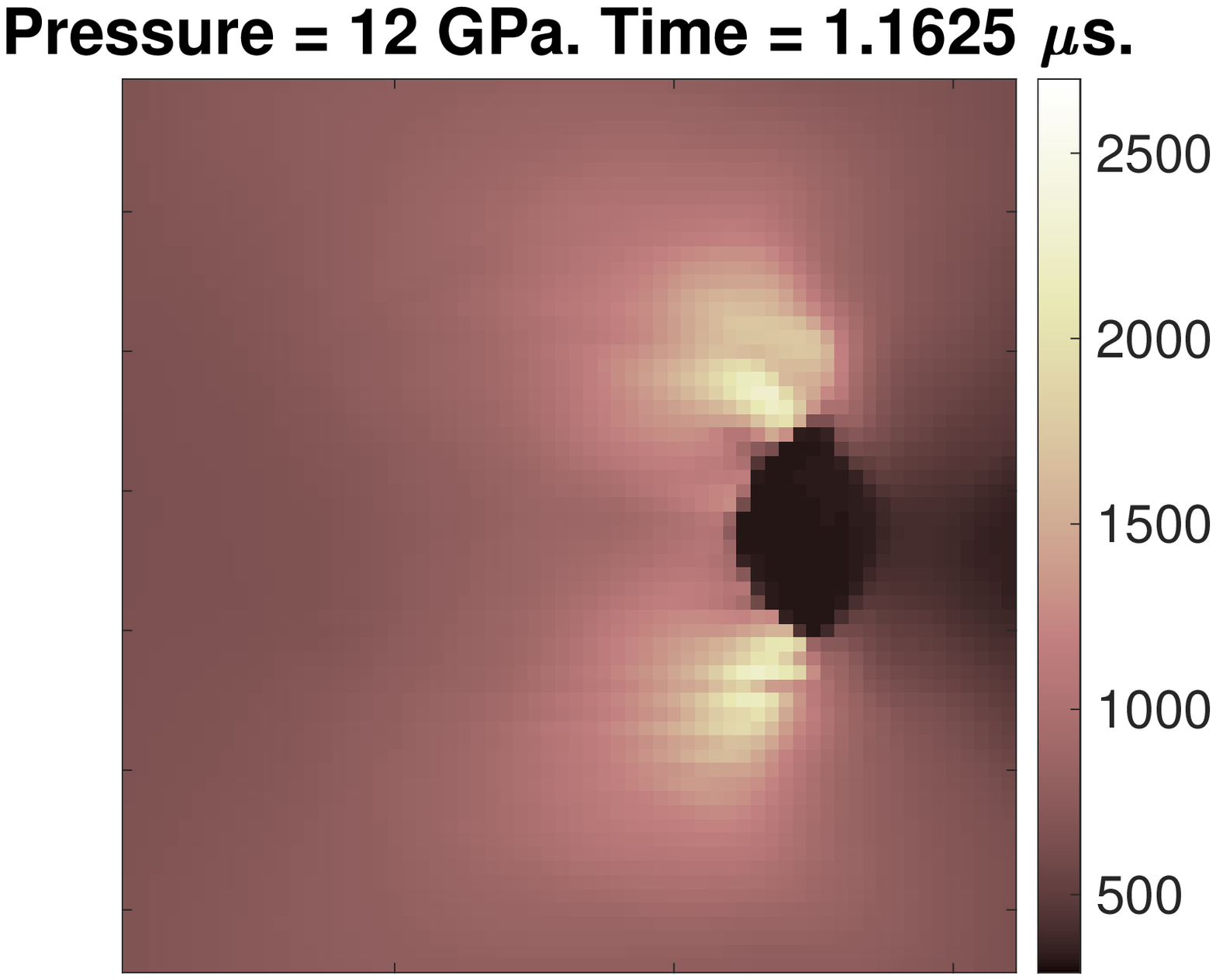}
\includegraphics[width=0.19\linewidth]{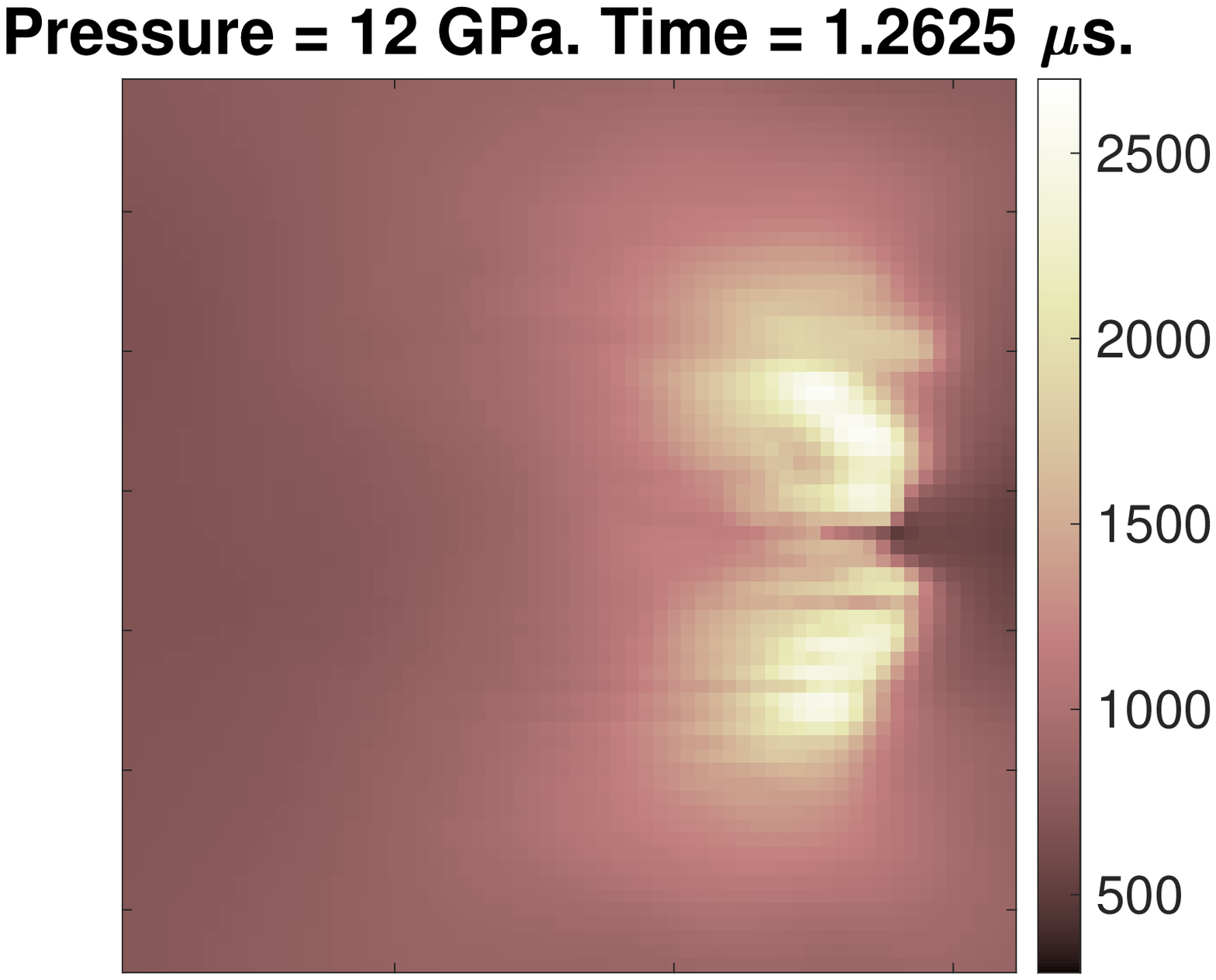}\\
\includegraphics[width=0.19\linewidth]{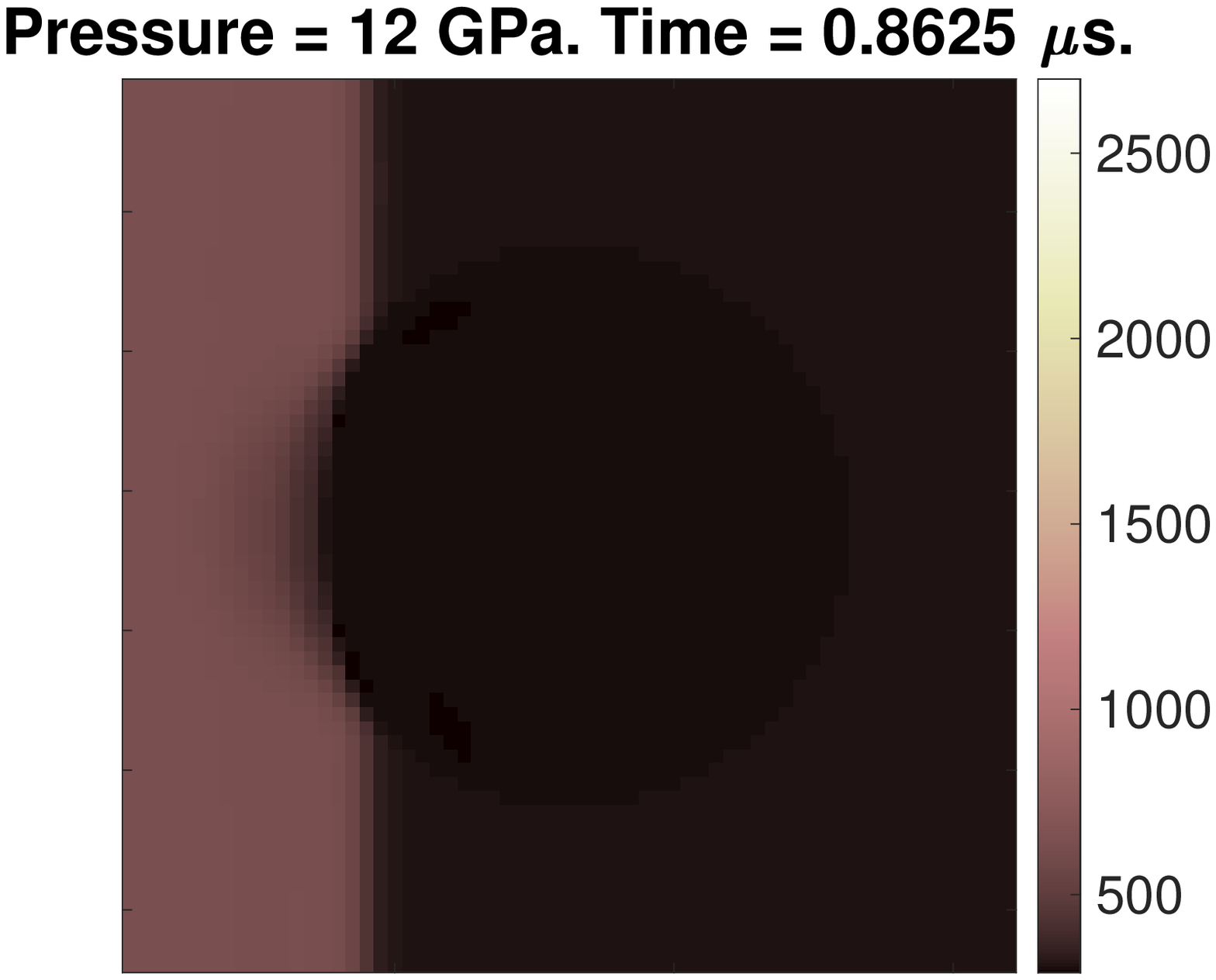}
\includegraphics[width=0.19\linewidth]{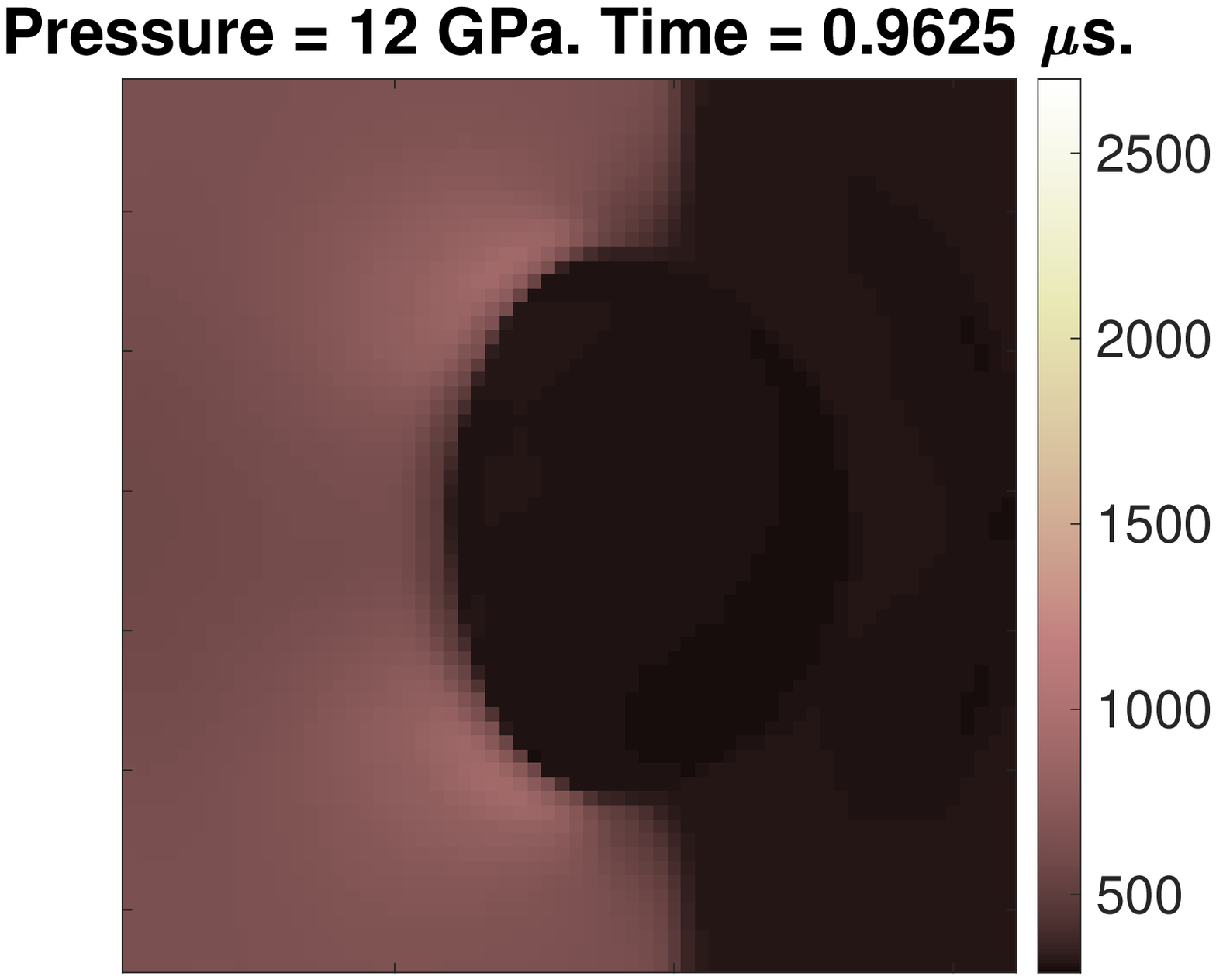}
\includegraphics[width=0.19\linewidth]{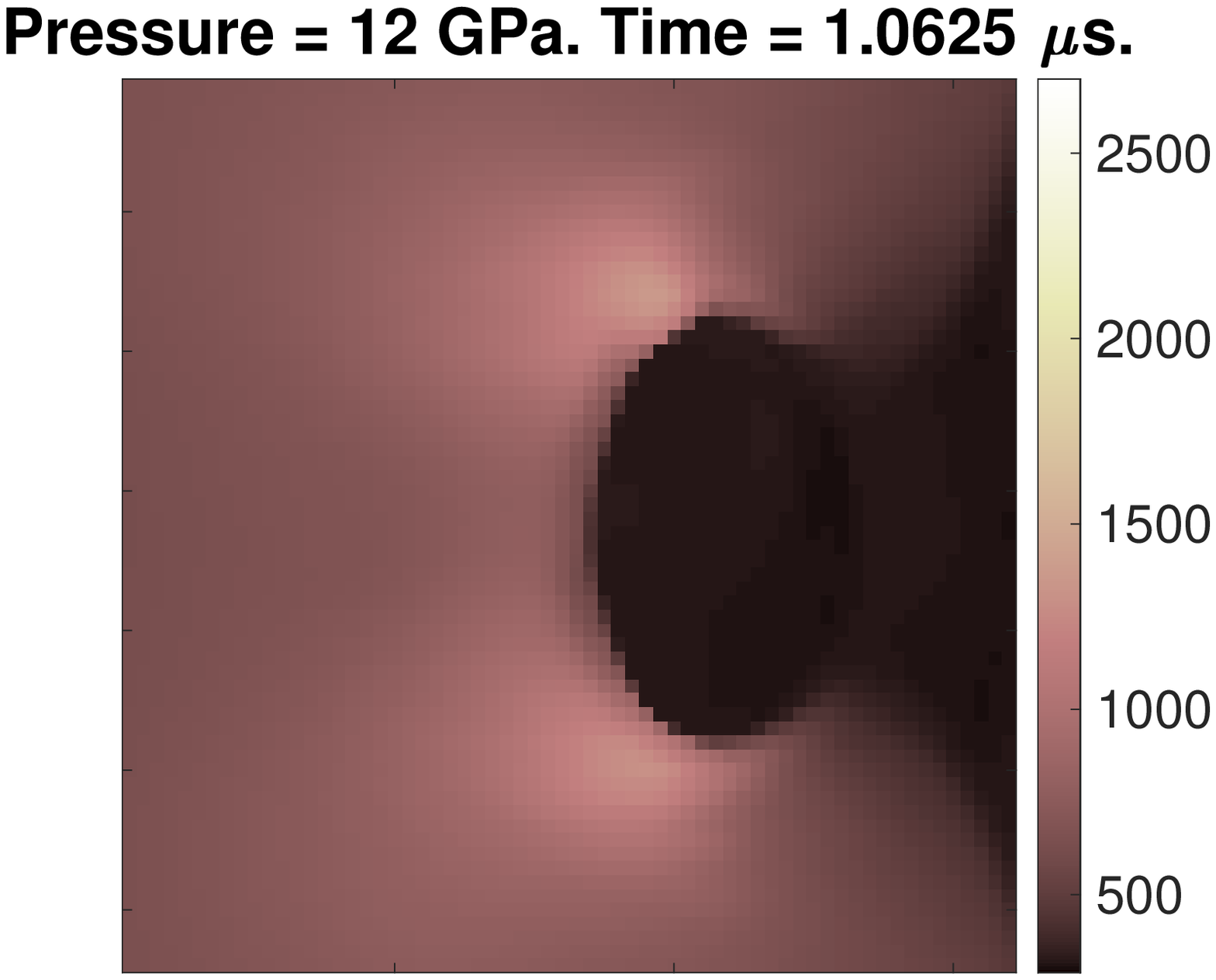}
\includegraphics[width=0.19\linewidth]{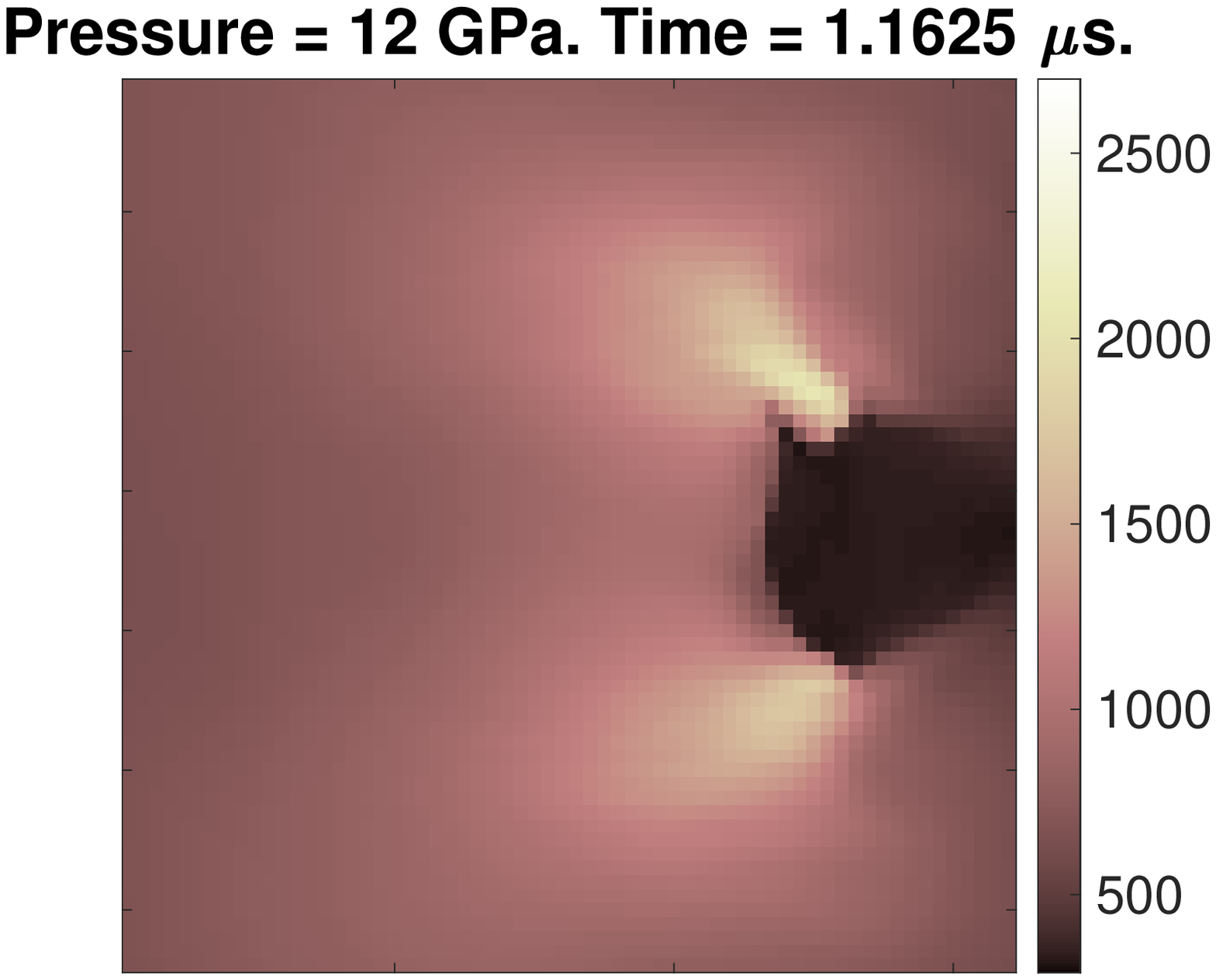}
\includegraphics[width=0.19\linewidth]{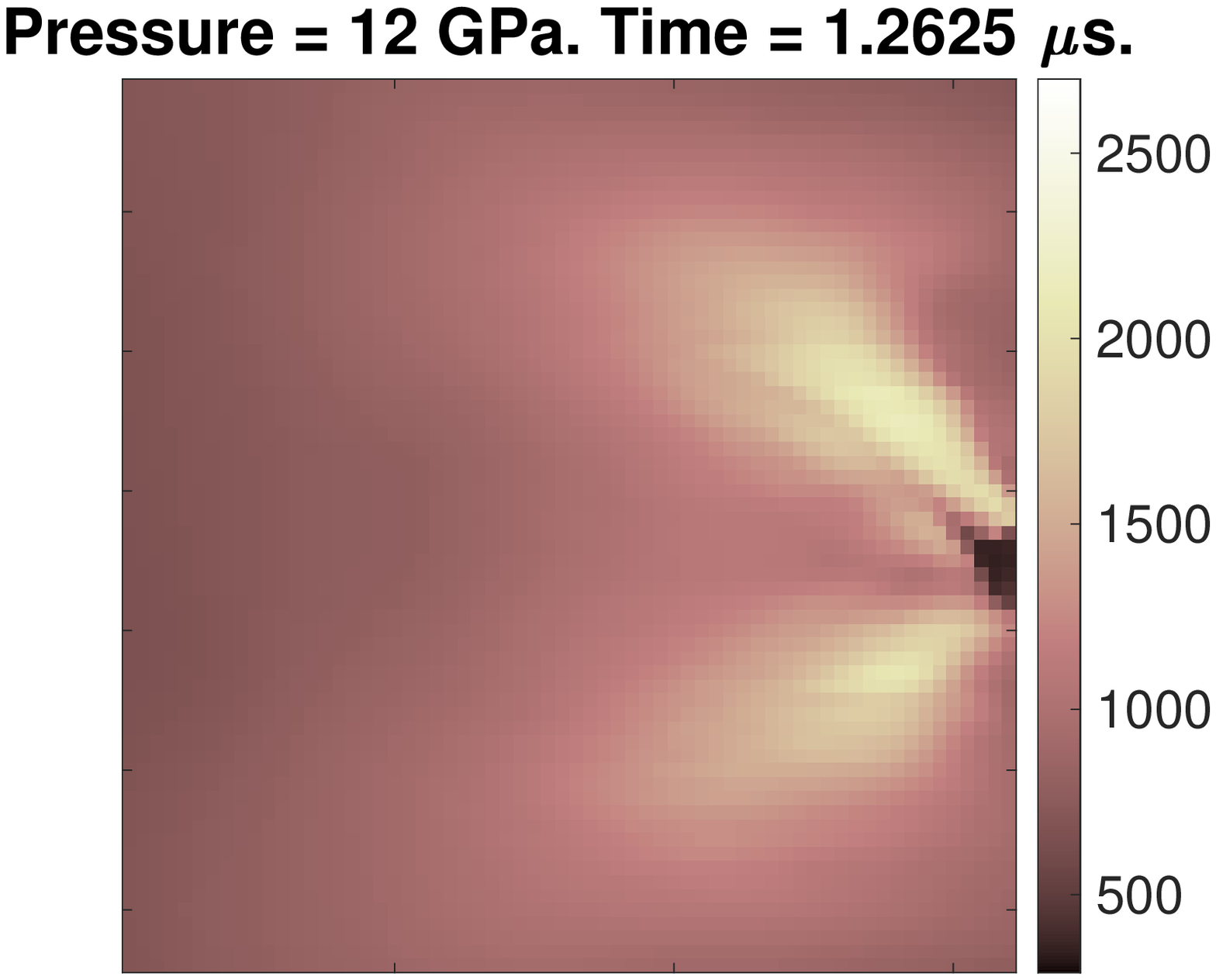}\\
\includegraphics[width=0.19\linewidth]{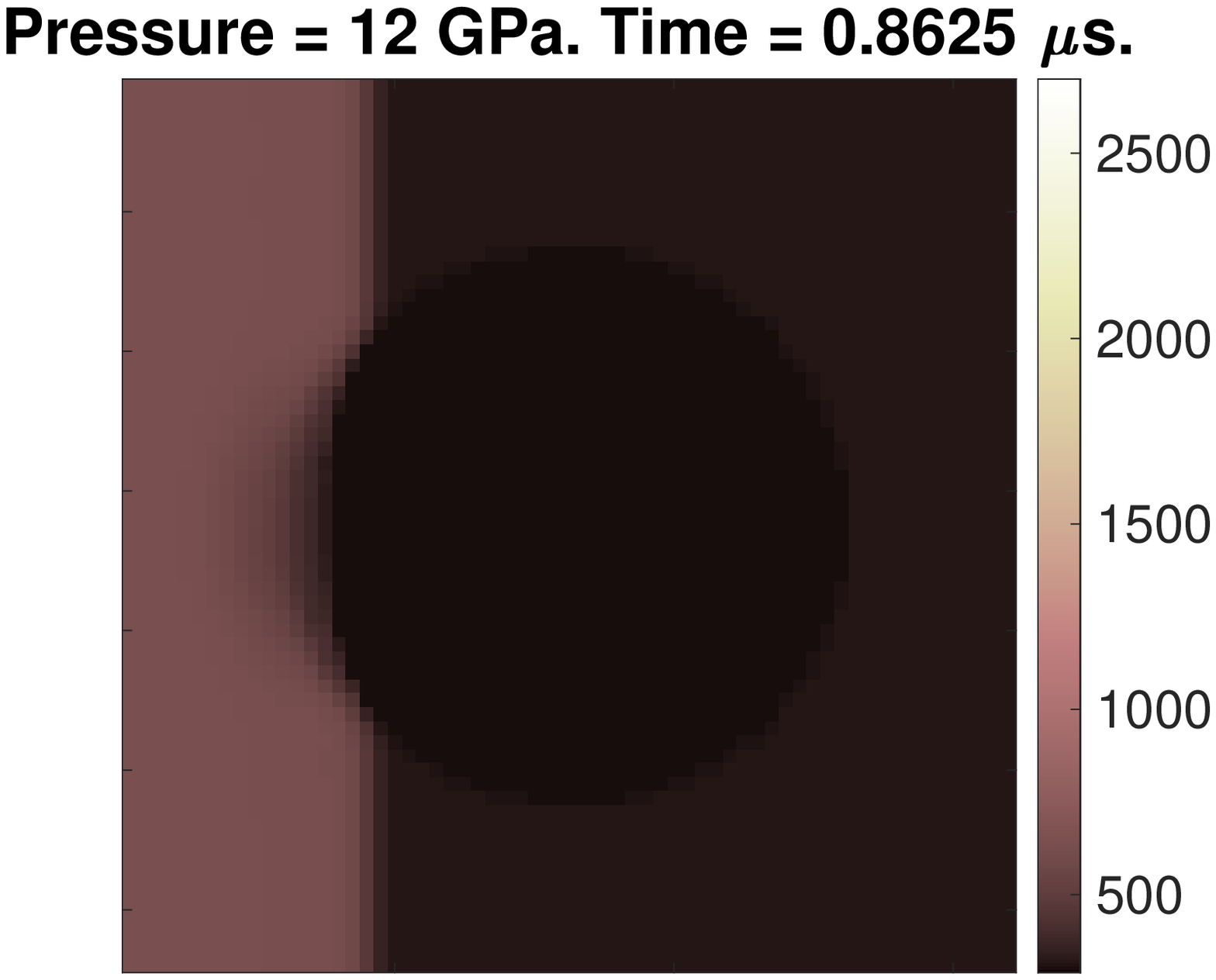}
\includegraphics[width=0.19\linewidth]{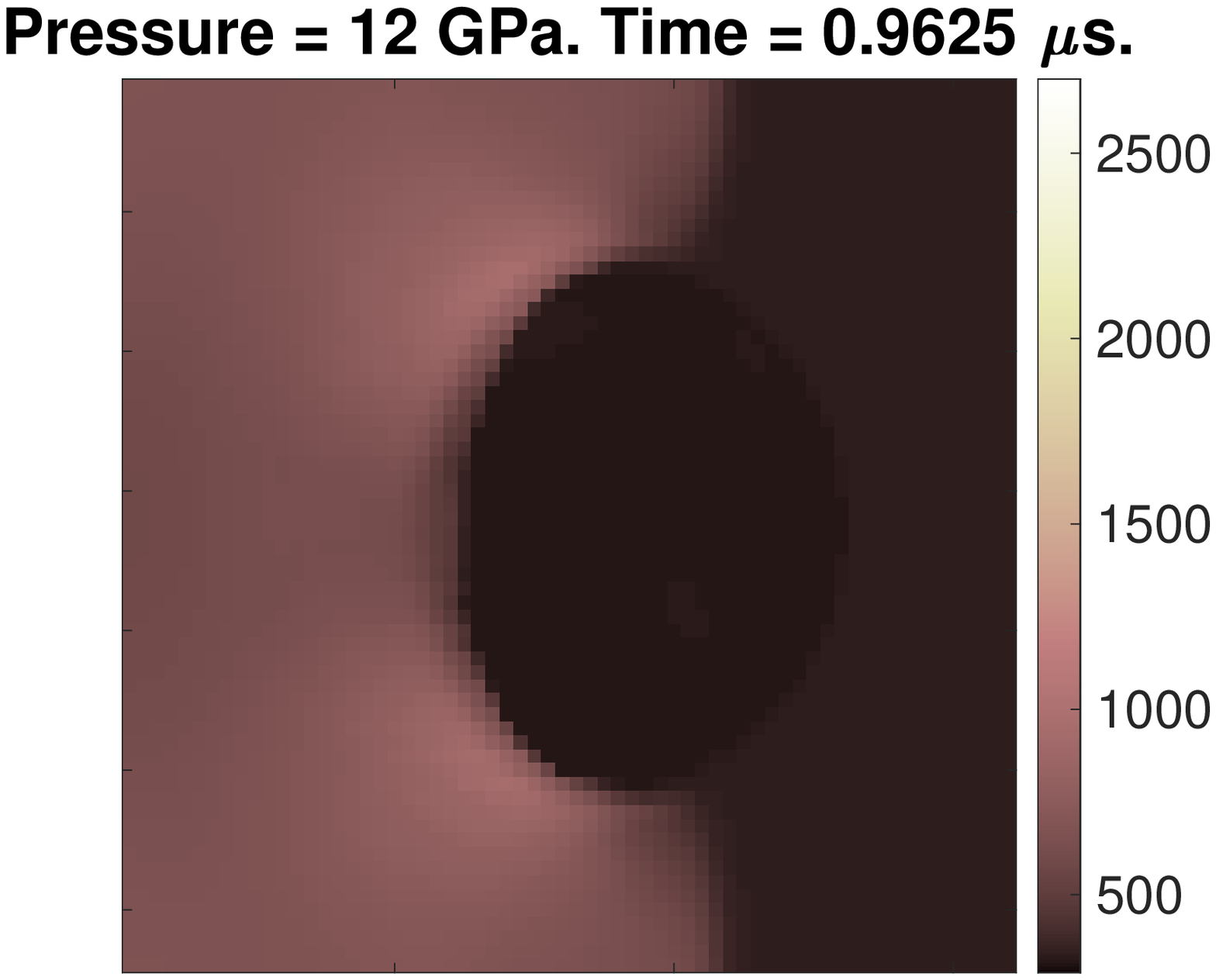}
\includegraphics[width=0.19\linewidth]{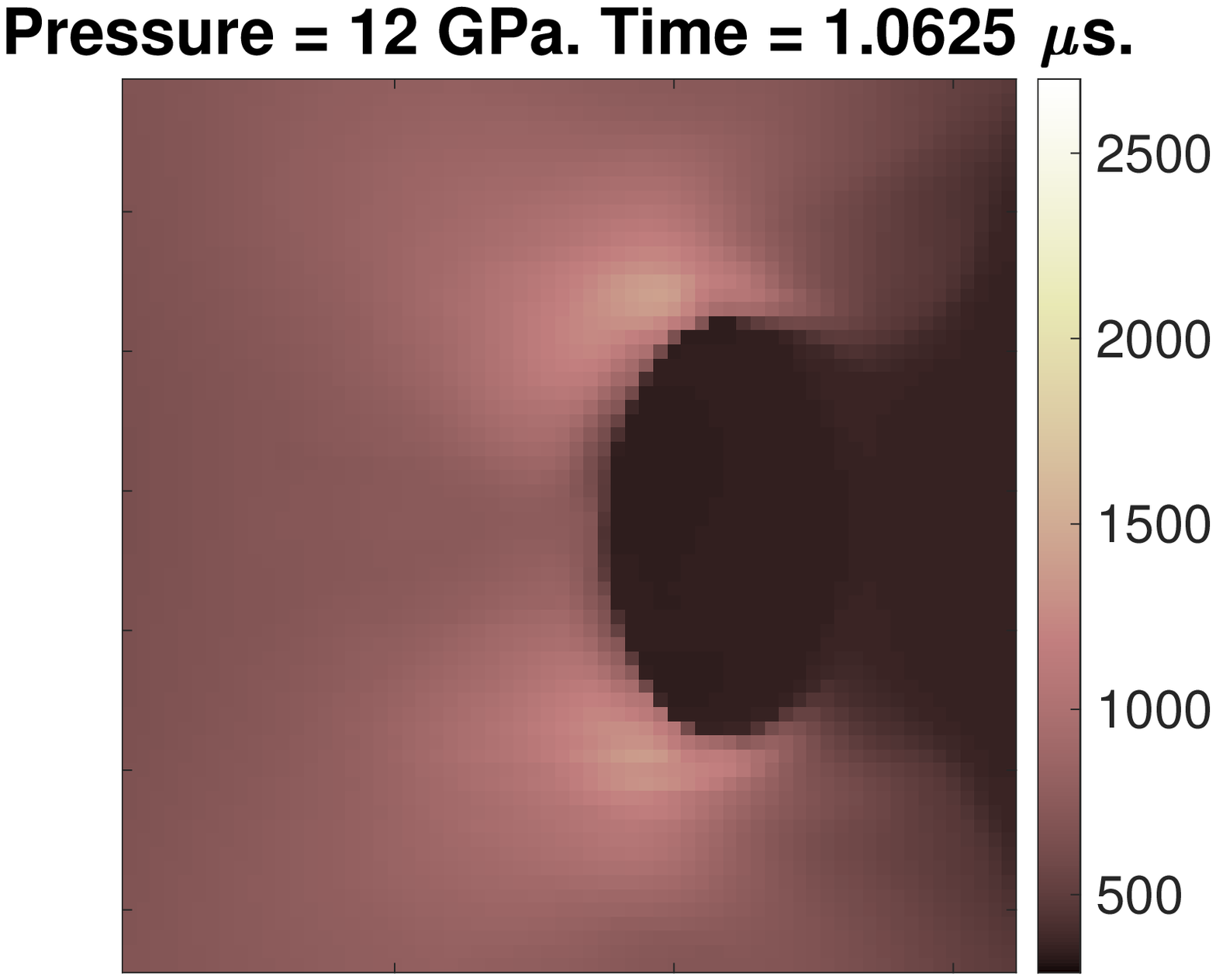}
\includegraphics[width=0.19\linewidth]{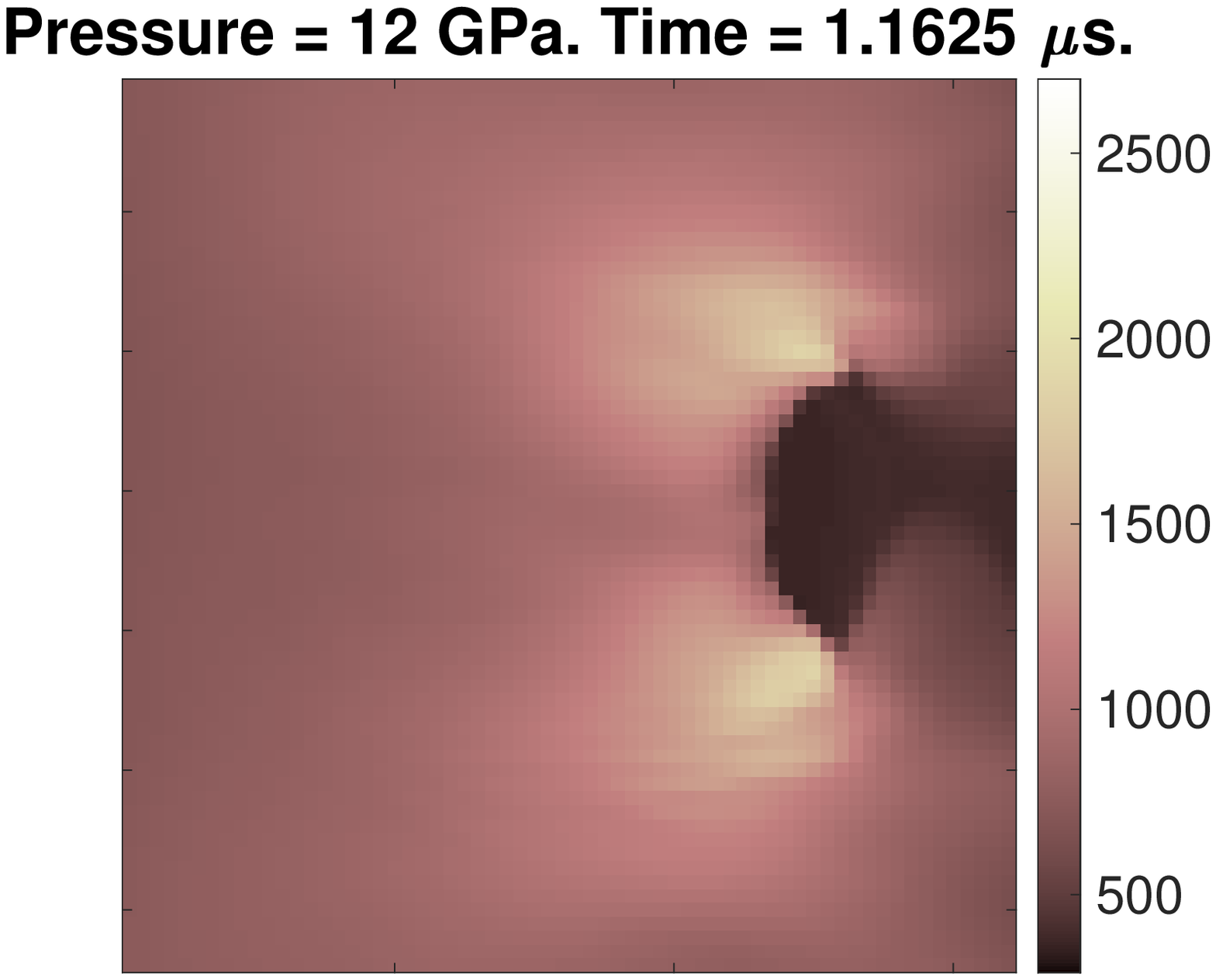}
\includegraphics[width=0.19\linewidth]{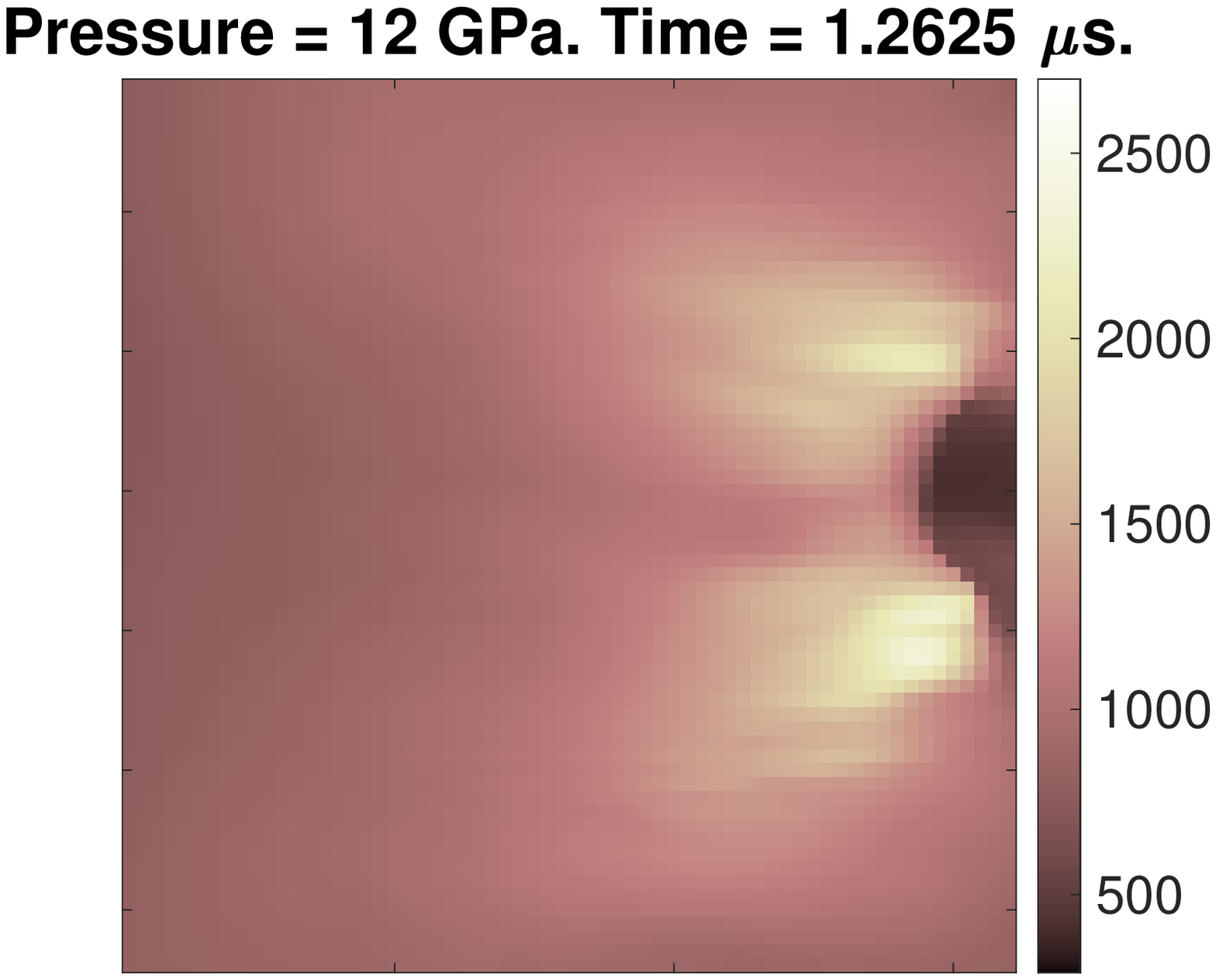}
\caption{Selected snapshots and predictions
of temperature fields at 12GPa. Each row corresponds to:
1. groundtruth snapshots from simulation data,
2. local DW-DMD and $\paramDomain_\text{train} = \{12\}$,
3. parametric DW-DMD and $\paramDomain_\text{train} = \{11,13,15\}$,
4. local DW-DMD and $\paramDomain_\text{train} = \{13\}$,
5. local CcGAN and $\paramDomain_\text{train} = \{12\}$,
6. global CcGAN and $\paramDomain_\text{train} = \{12,14\}$,
7. global CcGAN and $\paramDomain_\text{train} = \{11,13,15\}$, and
8. local CcGAN and $\paramDomain_\text{train} = \{13\}$.}
\label{fig:predict_12}
\end{figure}

In the remaining of this section, we will present numerical results
with various training combinations of surrogate modeling approaches
and training shock pressures $\paramDomain$.
In Figure~\ref{fig:predict_12}, we show the comparison of some selected groundtruth snapshots
and the corresponding surrogate model approximations at $\param = 12$,
with each row corresponds to:
\begin{enumerate}
\item groundtruth snapshots from simulation data,
\item reproductive predictions with local DW-DMD and $\paramDomain_\text{train} = \{12\}$,
\item interpolatory predictions with parametric DW-DMD and $\paramDomain_\text{train} = \{11,13,15\}$,
\item extrapolatory predictions with local DW-DMD and $\paramDomain_\text{train} = \{13\}$,
\item reproductive predictions with local CcGAN and $\paramDomain_\text{train} = \{12\}$,
\item reproductive predictions with global CcGAN and $\paramDomain_\text{train} = \{12,14\}$,
\item interpolatory predictions with global CcGAN and $\paramDomain_\text{train} = \{11,13,15\}$, and
\item extrapolatory predictions with local CcGAN and $\paramDomain_\text{train} = \{13\}$,
\end{enumerate}
and each column corresponds to a time instance, with
$\timeIndex \in \{10, 50, 90, 130, 170\}$,
in the time interval of query $\widetilde{\mathcal{T}}(\param)$,
The reproductive cases will be further explained in Section~\ref{sec:num-reproductive},
and the interpolatory and extrapolatory cases will be further explained in Section~\ref{sec:num-predictive}.
It can be seen that the approximations from DMD,
in the second row to the fourth row,
in general better captures the pore collapse process and
resembles the simulation data in the first row.

Next, we will introduce some performance metric which
allows us to investigate and compare the methods and training combination further.
To evaluate the accuracy of the prediction, we compute
the relative error between the high-fidelity simulation data $\state$ and the
reduced order model approximation $\widetilde{\state}$, i.e.
$\widetilde{\state}_{\text{DMD}}$ or $\widetilde{\state}_{\text{GAN}}$,
at testing shock pressure $\param \in \paramDomain$ and time instance
$\timeSymbol \in \widetilde{\mathcal{T}}(\param)$ by:
$$
\relError(\timeSymbol; \param) = \dfrac{\| \state(\timeSymbol; \param) -
\widetilde{\state}(\timeSymbol; \param) \|}{\| \state(\timeSymbol; \param) \|},
$$
where $\| \cdot \|$ denotes the vector Euclidean norm in $\mathbb{R}^{\sizeFOMsymbol^2}$, or equivalently the matrix Frobenius norm in $\mathbb{R}^{\sizeFOMsymbol \times \sizeFOMsymbol}$.

\subsection{Reproductive cases}
\label{sec:num-reproductive}

As a first experiment, we test the accuracy of surrogate modeling
approaches in reproductive cases,
where the testing shock pressure is identical to that
used in one of the training shock pressures,
i.e. $\param \in \paramDomain_\text{train}$.

\begin{figure}[htp!]
\centering
\includegraphics[width=0.45\linewidth]{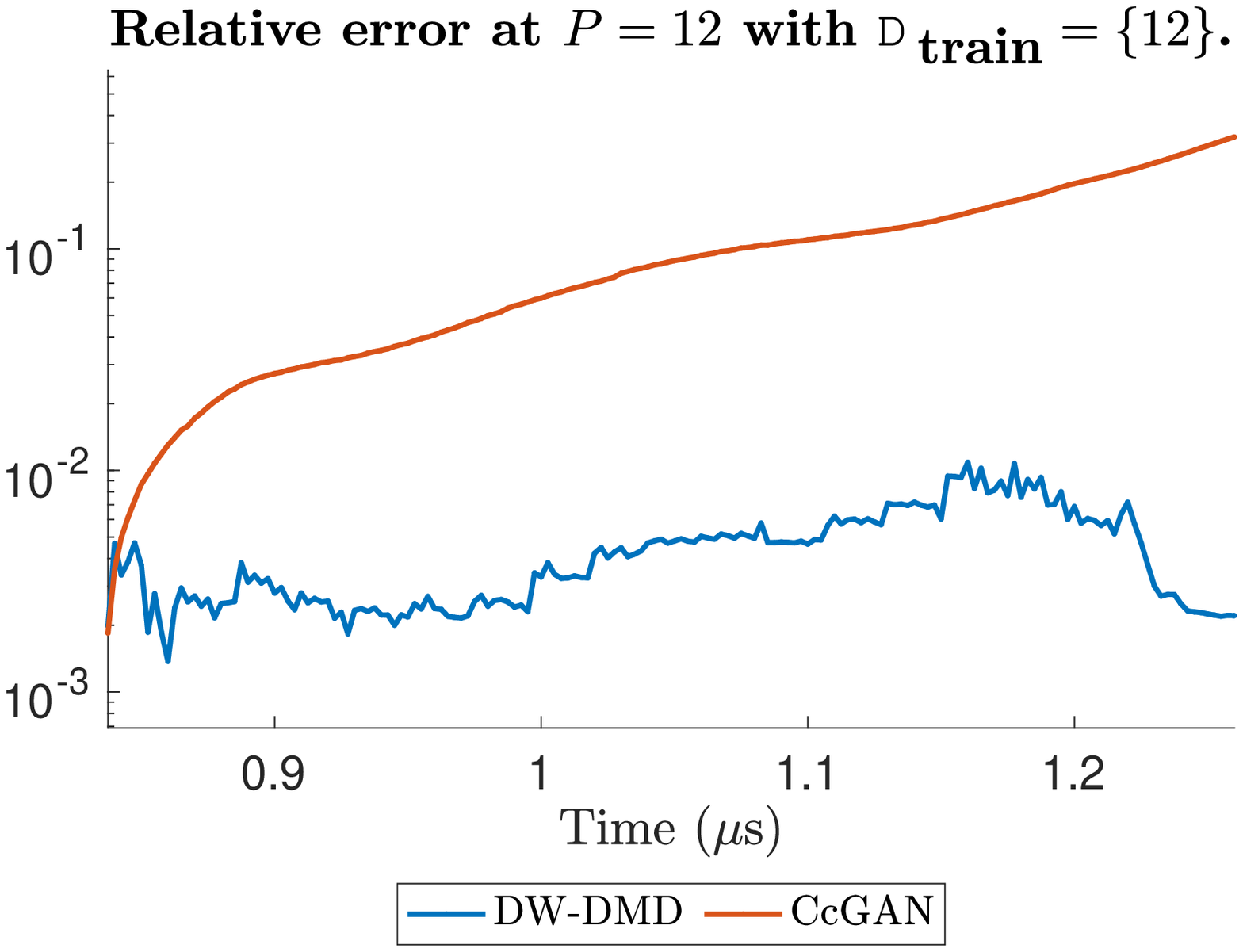}
\includegraphics[width=0.45\linewidth]{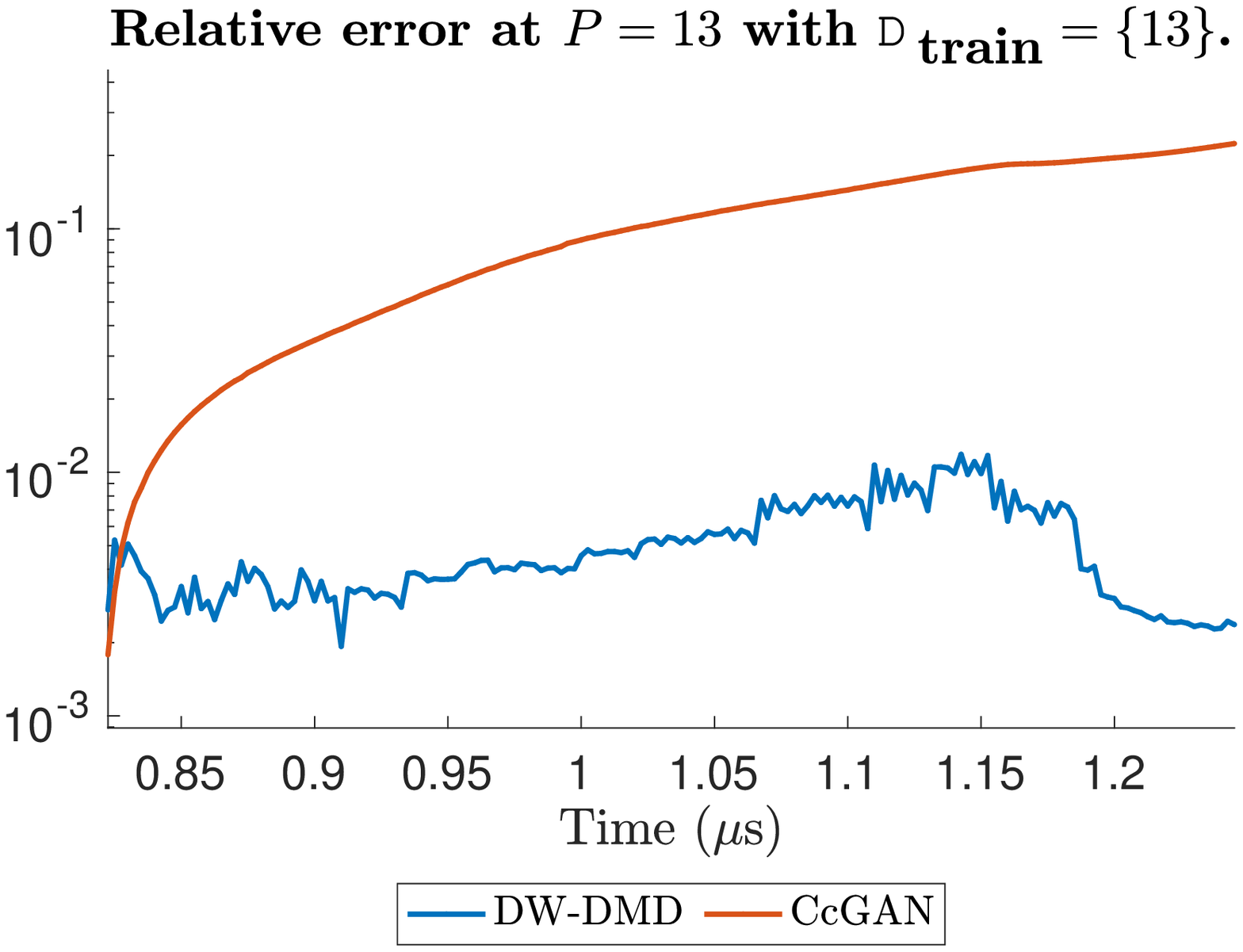}
\caption{Relative error (in logarithmic scale)
of reproductive case
with $\paramDomain_\text{train} = \{12\}$ (left)
and $\paramDomain_\text{train} = \{13\}$ (right),
using local DW-DMD (in blue) and local CcGAN (in red).}
\label{fig:compare-reproductive-loc}
\end{figure}

Figure~\ref{fig:compare-reproductive-loc} shows the comparison
of reproductive accuracy using local DW-DMD and local CcGAN,
in terms of the evolution of relative error (in logarithmic scale),
with $\paramDomain_\text{train} = \{12\}$ and
$\paramDomain_\text{train} = \{13\}$ respectively.
In both cases, local DW-DMD produces more stable reproductive results,
where the relative error stays below 1.2\% in the whole time interval of query,
and terminates at around 0.3\% at final time.
On the other hand, although local CcGAN is able to produce around 0.2\%
error in each time step, the error accumulates quickly
and rises to 32\% and 22\% at the final time of query
with $\paramDomain_\text{train} = \{12\}$ and
$\paramDomain_\text{train} = \{13\}$ respectively.

\begin{figure}[htp!]
\centering
\includegraphics[width=0.45\linewidth]{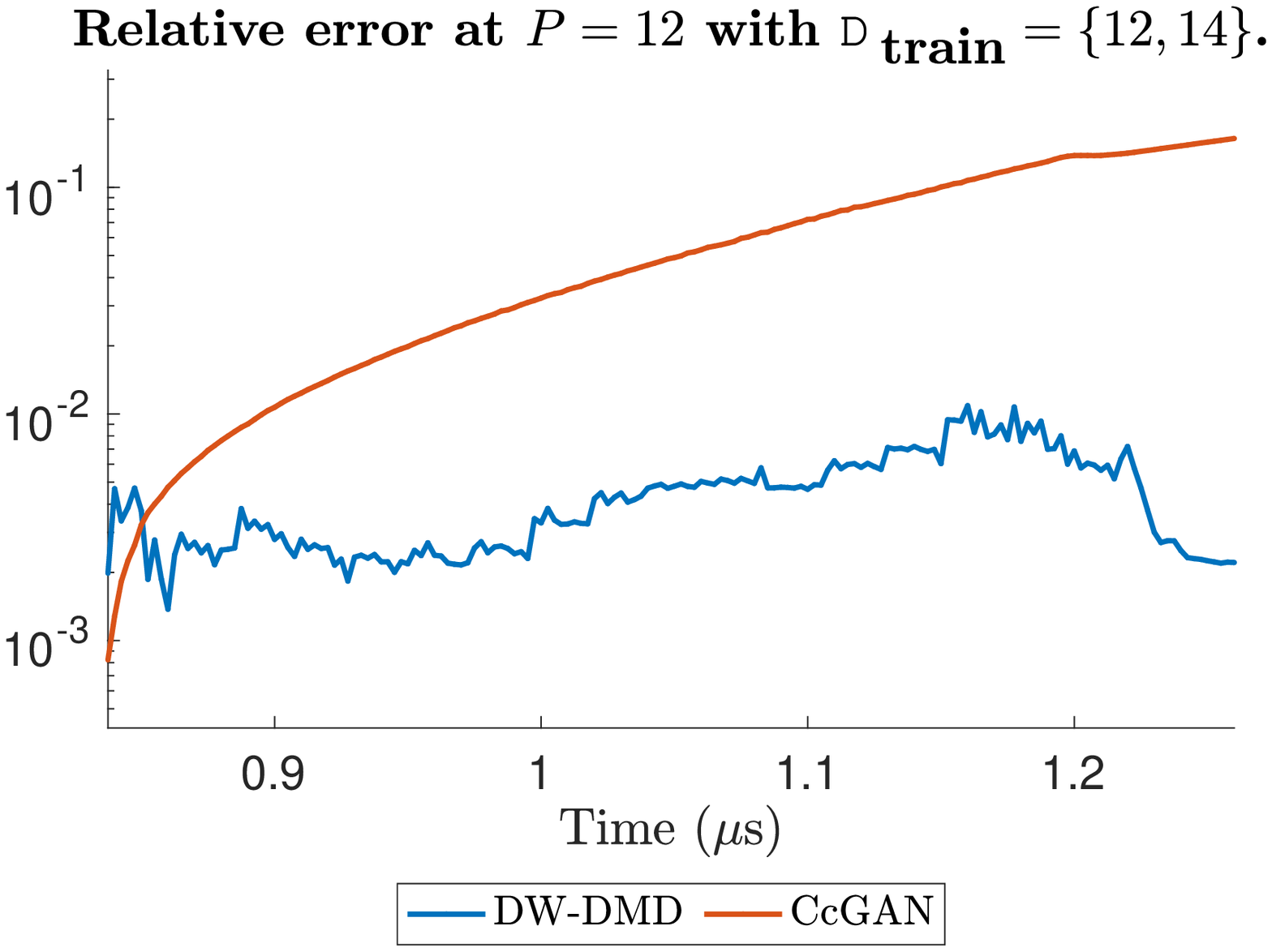}
\includegraphics[width=0.45\linewidth]{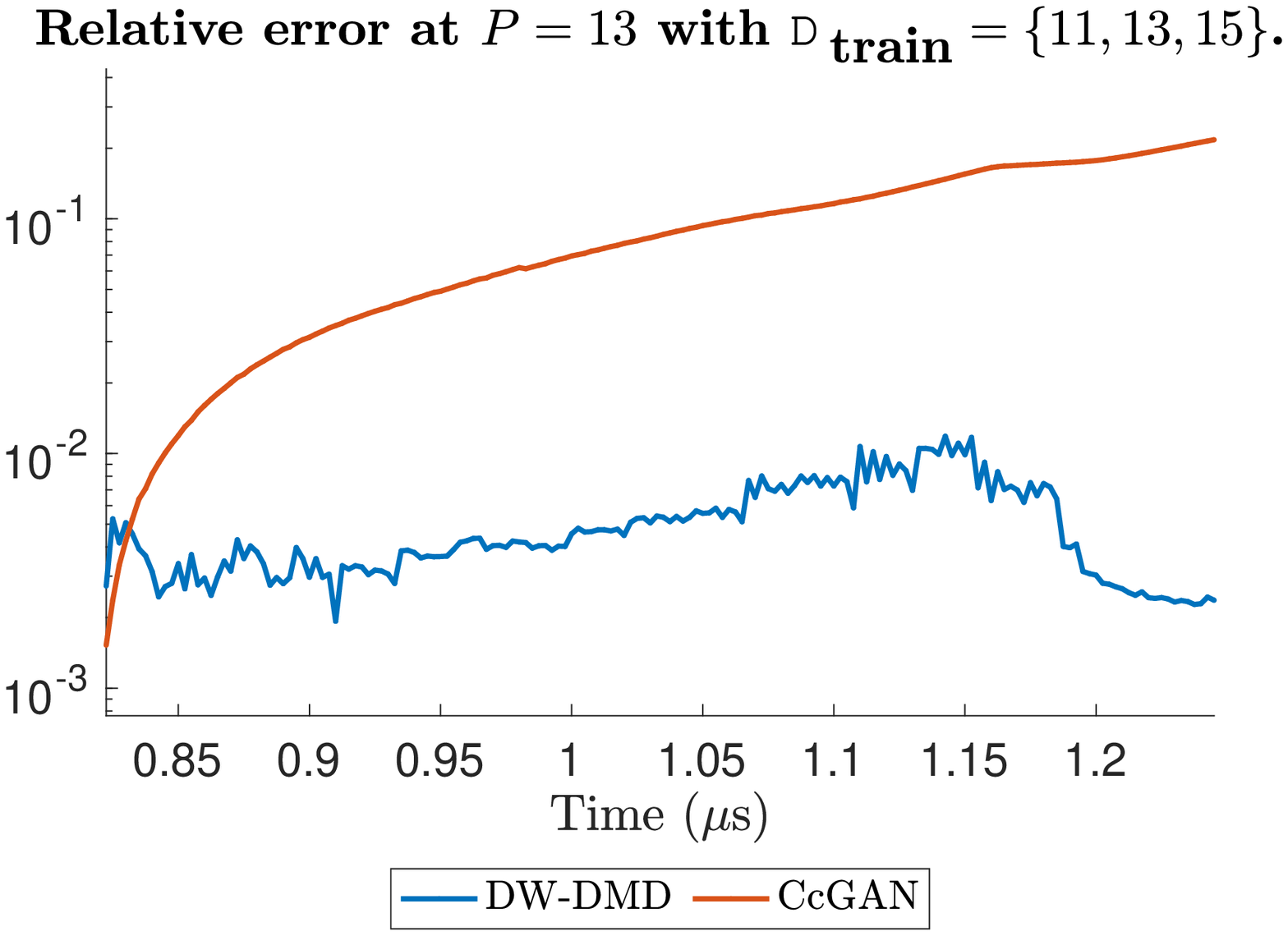}
\caption{Relative error (in logarithmic scale)
of reproductive case
with $\paramDomain_\text{train} = \{12,14\}$ (left)
and $\paramDomain_\text{train} = \{11,13,15\}$ (right),
using parametric DW-DMD (in blue) and global CcGAN (in red).}
\label{fig:compare-reproductive-par}
\end{figure}

Figure~\ref{fig:compare-reproductive-par} shows a similar comparison
with $\paramDomain_\text{train} = \{12,14\}$ and
$\paramDomain_\text{train} = \{11,13,15\}$ respectively.
We remark that the final-time error of DMD at the reproductive cases
remains unchanged at around 0.3\% when adding more training shock pressures,
as explained in Section~\ref{sec:prediction-dmd}.
On the other hand, the final-time error of global CcGAN improves to
16\% with $\paramDomain_\text{train} = \{12,14\}$ and
remains at 22\% with $\paramDomain_\text{train} = \{11,13,15\}$ respectively.

\subsection{Predictive cases}
\label{sec:num-predictive}

In this subsection, we test the accuracy of surrogate modeling
approaches in predictive cases,
where the testing shock pressure is not one of the training shock pressures, i.e.
$\param \in \paramDomain \setminus \paramDomain_\text{train}$.

We begin with some results in the interpolatory cases, i.e.
$\param \in (\min \paramDomain_\text{train}, \max \paramDomain_\text{train})
\setminus \paramDomain_\text{train}$.
Similar to Figure~\ref{fig:compare-reproductive-par},
we compare the relative error at $\param = 12$
with $\paramDomain_\text{train} = \{11,13,15\}$, and
at $\param = 13$ with $\paramDomain_\text{train} = \{11,13,15\}$ respectively.
In the former case, the relative error of parametric DW-DMD is higher than
that of global CcGAN in an earlier stage, but eventually becomes lower.
Throughout the whole time interval of query,
the relative error of parametric DW-DMD stays below 9\% and 4.3\%
and terminates at around 4.7\% and 1.3 \% at final time,
in the former and the latter case respectively.
Meanwhile, the relative error of global CcGAN accumulates to
20\% at the final time of query in both cases.

\begin{figure}[htp!]
\centering
\includegraphics[width=0.45\linewidth]{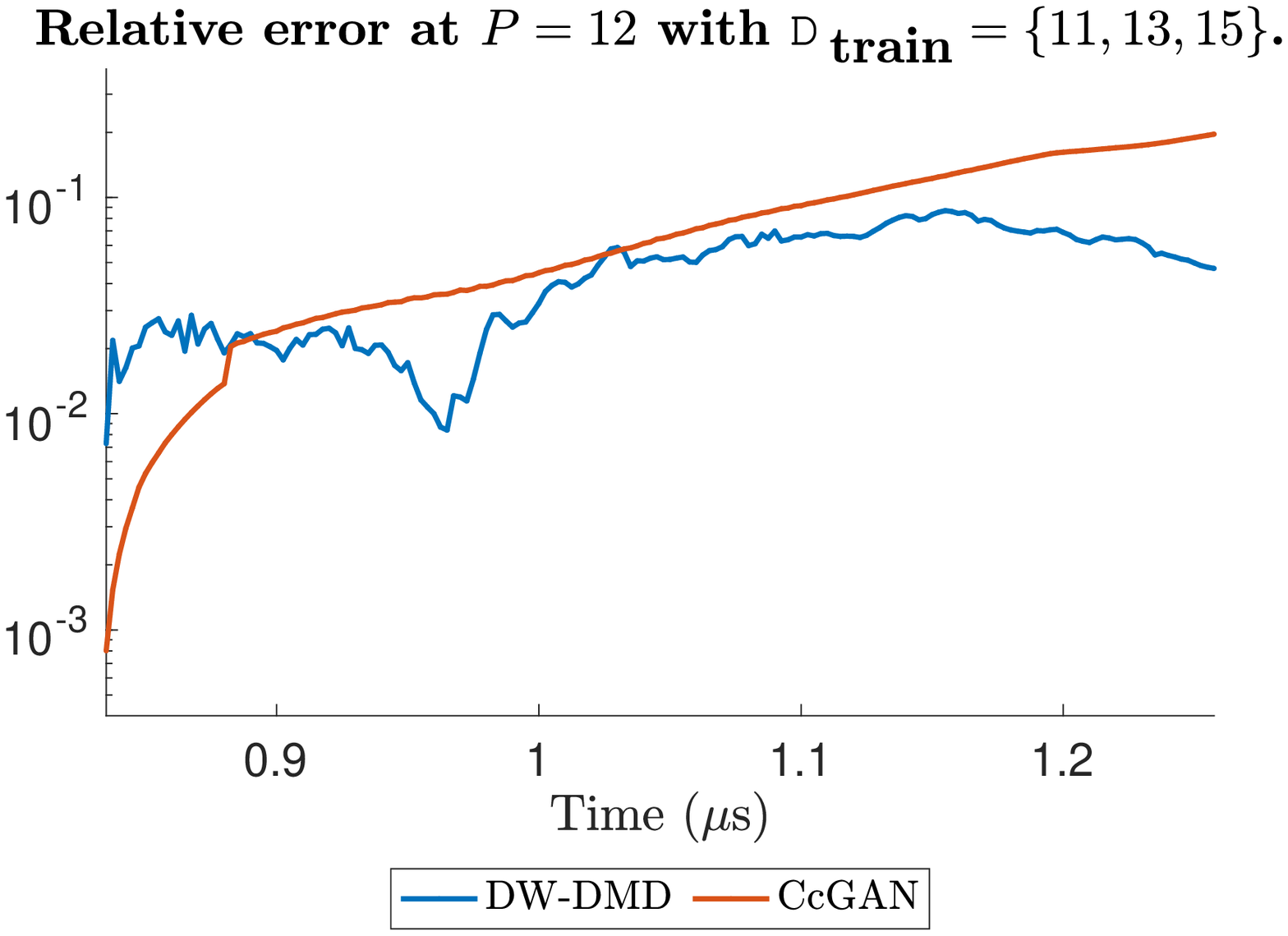}
\includegraphics[width=0.45\linewidth]{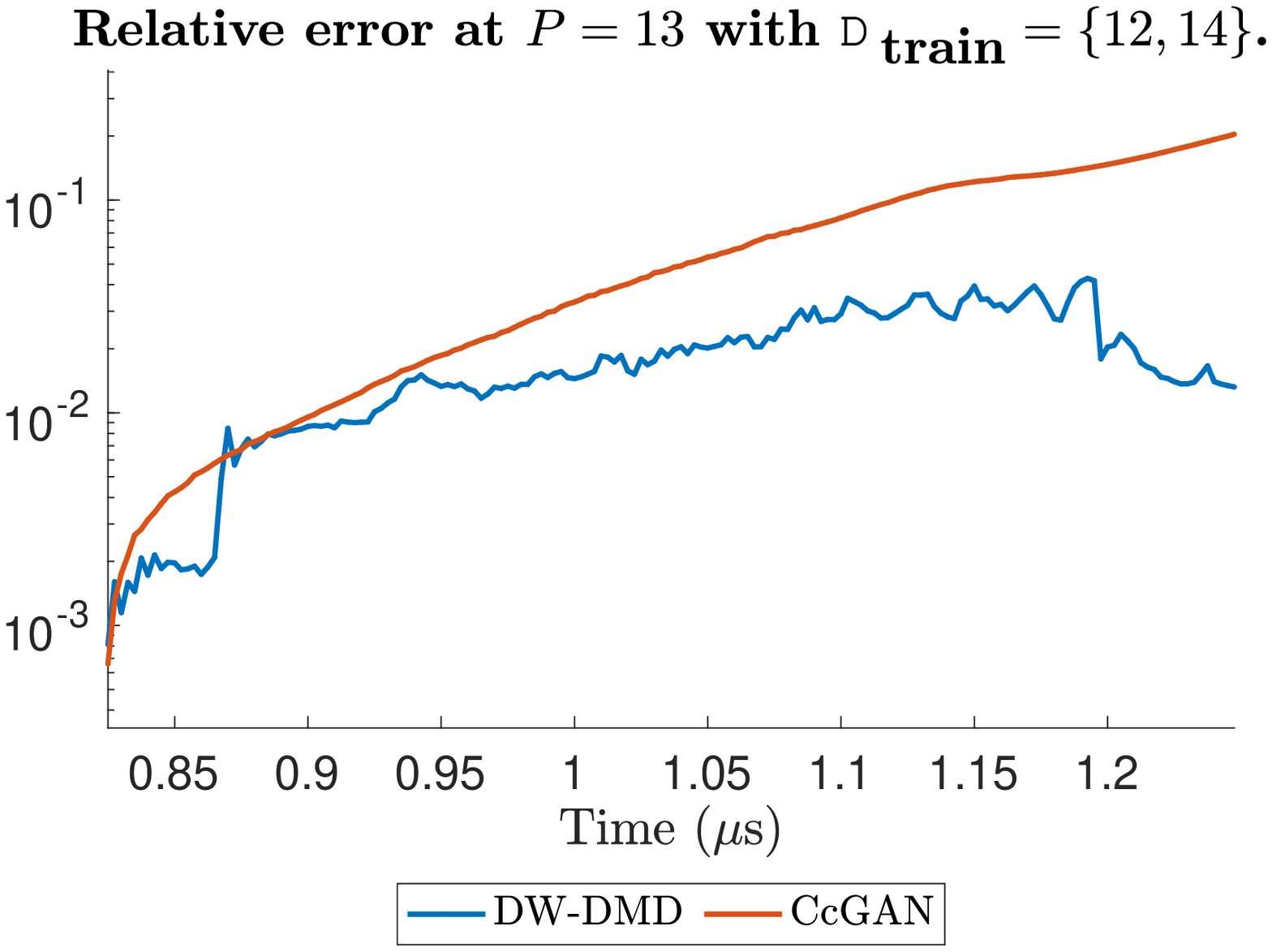}
\caption{Relative error (in logarithmic scale) of interpolatory case
with $\param = 12$ and $\paramDomain_\text{train} = \{11,13,15\}$ (left)
and $\param = 13$ and $\paramDomain_\text{train} = \{12,14\}$ (right),
using parametric DW-DMD (in blue) and global CcGAN (in red).}
\label{fig:compare-interpolatory-par}
\end{figure}

\begin{figure}[htp!]
\centering
\includegraphics[width=0.45\linewidth]{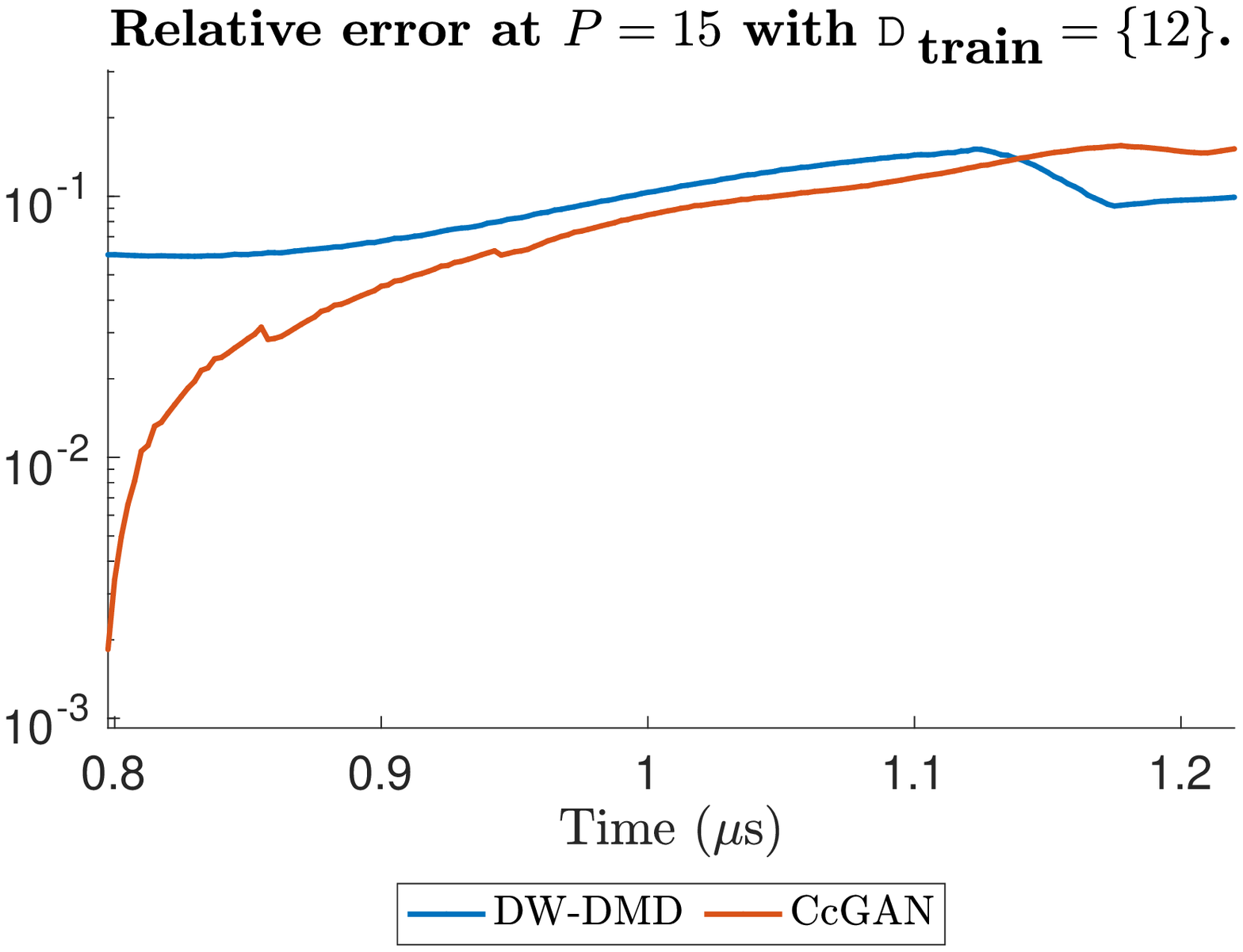}
\includegraphics[width=0.45\linewidth]{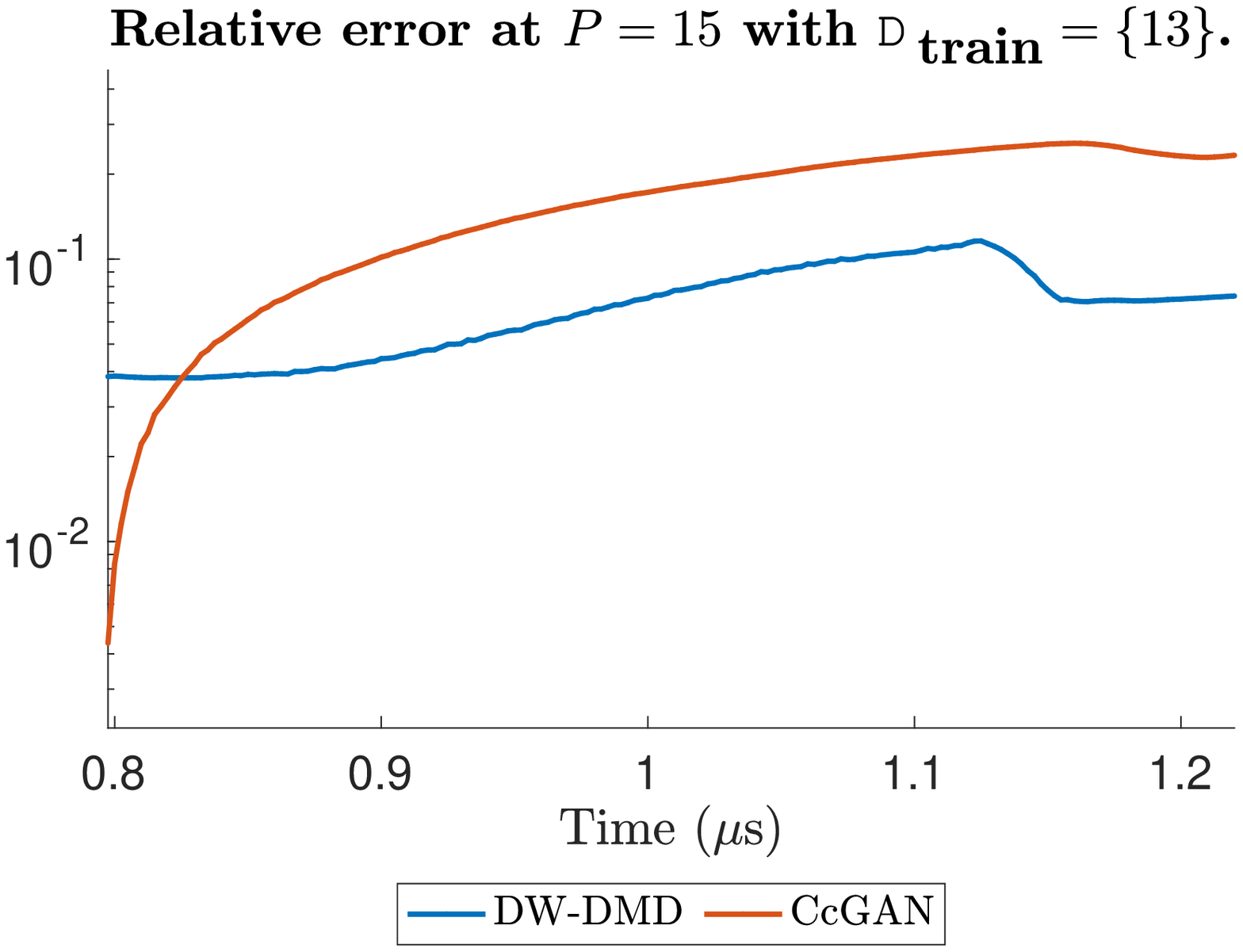}
\caption{Relative error (in logarithmic scale)
of extrapolatory case at $\param = 15$
with $\paramDomain_\text{train} = \{12\}$ (left)
and $\paramDomain_\text{train} = \{13\}$ (right),
using local DW-DMD (in blue) and CcGAN (in red).}
\label{fig:compare-extrapolatory-loc-15}
\end{figure}

Next, we will present some results in extrapolatory cases,
i.e. $\param \in \paramDomain \setminus
(\min \paramDomain_\text{train}, \max \paramDomain_\text{train})$.
Figure~\ref{fig:compare-extrapolatory-loc-15}
shows the comparison of extrapolatory accuracy at
$\param = 15$ using local DW-DMD and local CcGAN,
in terms of the relative error of the temperature field over time,
with $\paramDomain_\text{train} = \{12\}$ and
$\paramDomain_\text{train} = \{13\}$ respectively.
The relative error of DW-DMD
attains a maximum of 15\% and 12\% over time and
terminates at 10\% and 7\% at final time
with $\paramDomain_\text{train} = \{12\}$ and
$\paramDomain_\text{train} = \{13\}$ respectively.
Meanwhile, the relative error of CcGAN attains the maximum
15\% and 23\% at the final time,
with $\paramDomain_\text{train} = \{12\}$ and
$\paramDomain_\text{train} = \{13\}$ respectively.
Unlike the DW-DMD results which shows the extrapolatory accuracy deteriorates
as the testing shock pressure is farther away from the training shock pressure,
the extrapolatory accuracy of CcGAN is unstable with the distance between
testing shock pressure $\param$ and the training shock pressure $\param_1$.

\begin{figure}[htp!]
\centering
\includegraphics[width=0.45\linewidth]{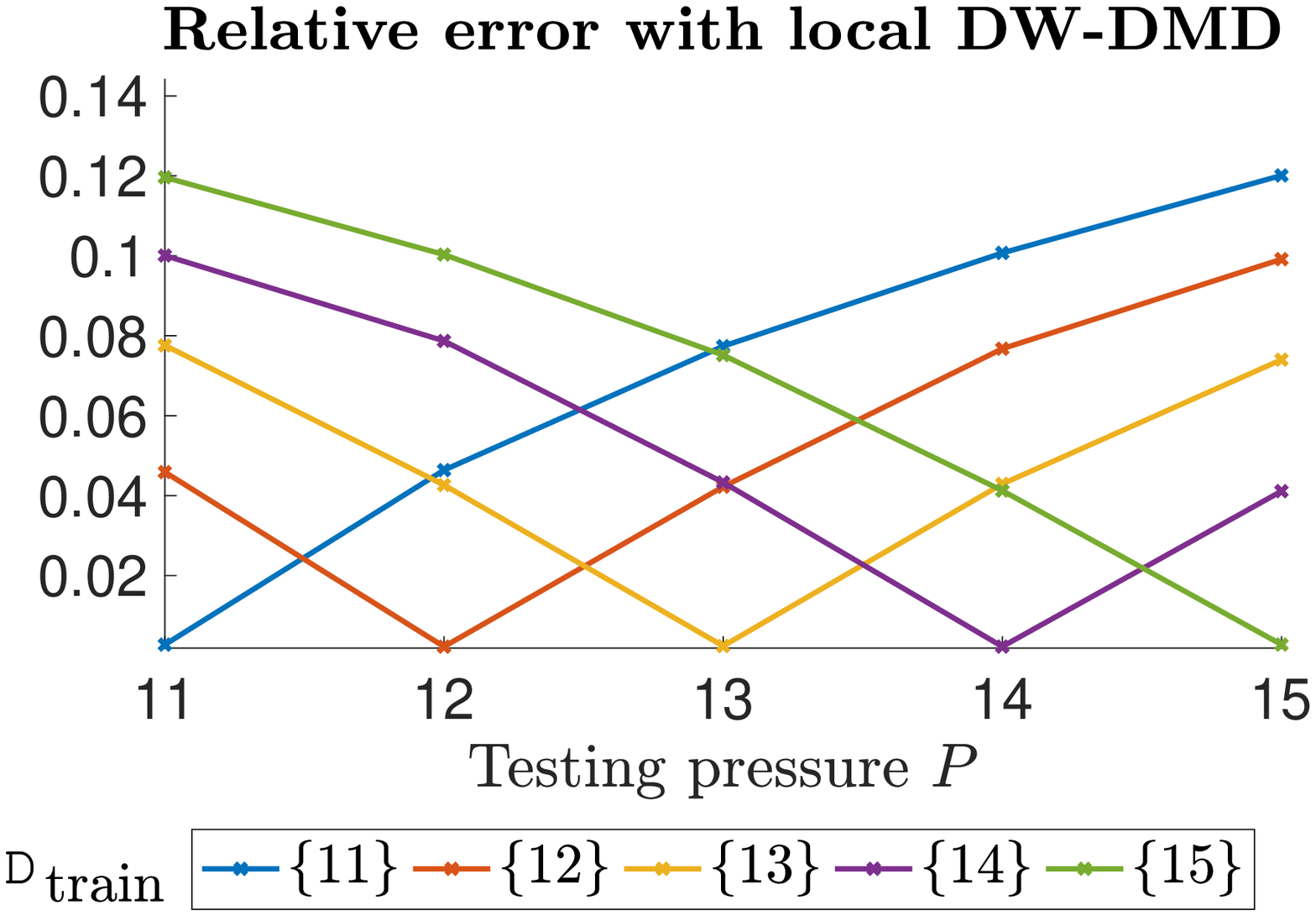}
\includegraphics[width=0.45\linewidth]{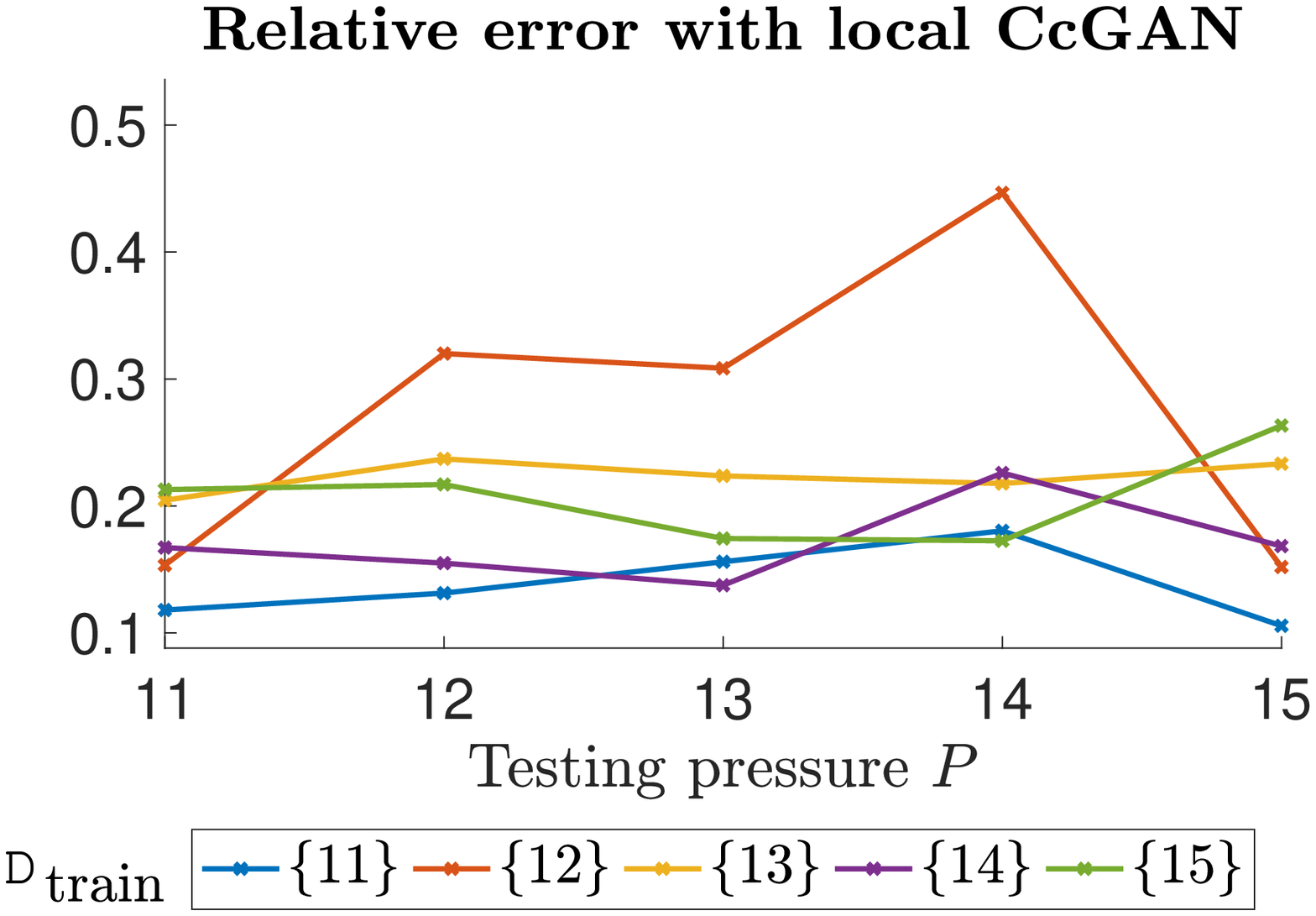}
\caption{Relative error at various testing shock pressures,
using local DW-DMD (left) and CcGAN (right)
with various training shock pressure in $\paramDomain_\text{train}$.}
\label{fig:compare-extrapolatory-loc}
\end{figure}

Figure~\ref{fig:compare-extrapolatory-loc}
shows the comparison of reproductive and extrapolatory accuracy at
various testing shock pressure $\param \in \{11, 12, 13, 14, 15\}$
in terms of the relative error of the temperature field
at the final time of query,
using local DW-DMD and local CcGAN
with respect to different training shock pressure $\param_1 \in \paramDomain_\text{train}$.
It can be observed that with DW-DMD, the relative error at the reproductive case
is always around 0.3\%, while the error at the extrapolatory case increases
as the testing shock pressure is farther away from the training shock pressure, which is
a common phenomenon for parametric reduced order models.
The relative error attains the maximum of 12\%,
when $\vert \param - \param_1 \vert = 4$, in our testing cases.
Meanwhile, the error with CcGAN is always above 10\%
and unstable with the distance between
testing shock pressure $\param$ and the training shock pressure $\param_1$.
With $\paramDomain_\text{train} = \{12\}$, the relative error goes up
to 45\% at $\param = 14$.

\begin{figure}[htp!]
\centering
\includegraphics[width=0.45\linewidth]{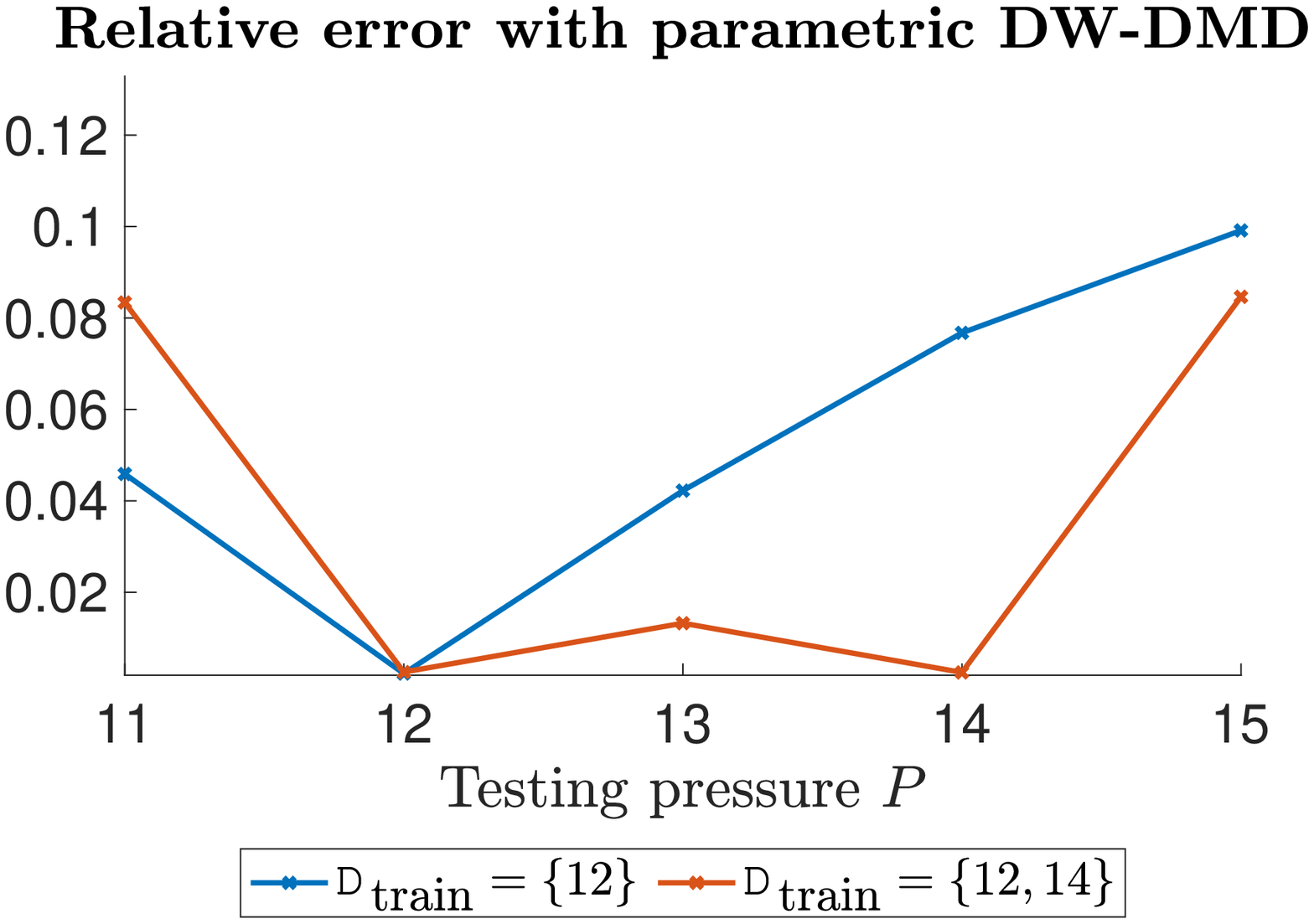}
\includegraphics[width=0.45\linewidth]{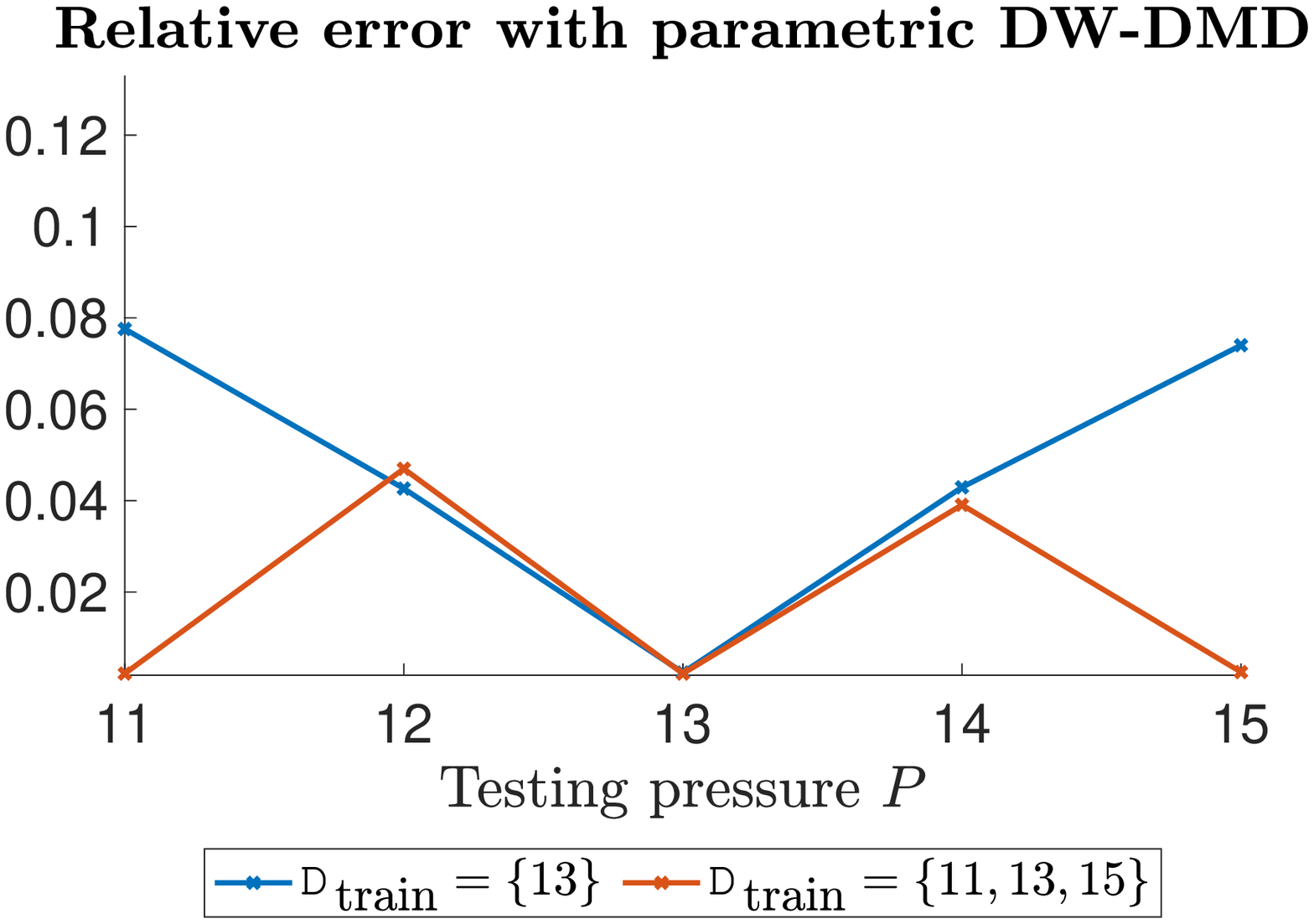}
\caption{Relative error at various testing shock pressures,
using parametric DW-DMD
with $12 \in \paramDomain_\text{train}$ (left)
and $13 \in \paramDomain_\text{train}$ (right).}
\label{fig:compare-predictive-parametric-dmd}
\end{figure}

Figure~\ref{fig:compare-predictive-parametric-dmd}
shows the comparison of reproductive, interpolatory and extrapolatory accuracy at
various testing shock pressure $\param \in \{11, 12, 13, 14, 15\}$
in terms of the relative error of the temperature field
at the final time of query,
using parametric DW-DMD
with $12 \in \paramDomain_\text{train}$
and $13 \in \paramDomain_\text{train}$ respectively.
The error at the newly added training shock pressures is also reduced to around 0.3\%,
and the error at the predictive cases are also reduced in general,
which ranges from 1.3\% to 5\% in the interpolatory cases
and 8\% to 9\% in extrapolatory cases.

\begin{figure}[htp!]
\centering
\includegraphics[width=0.45\linewidth]{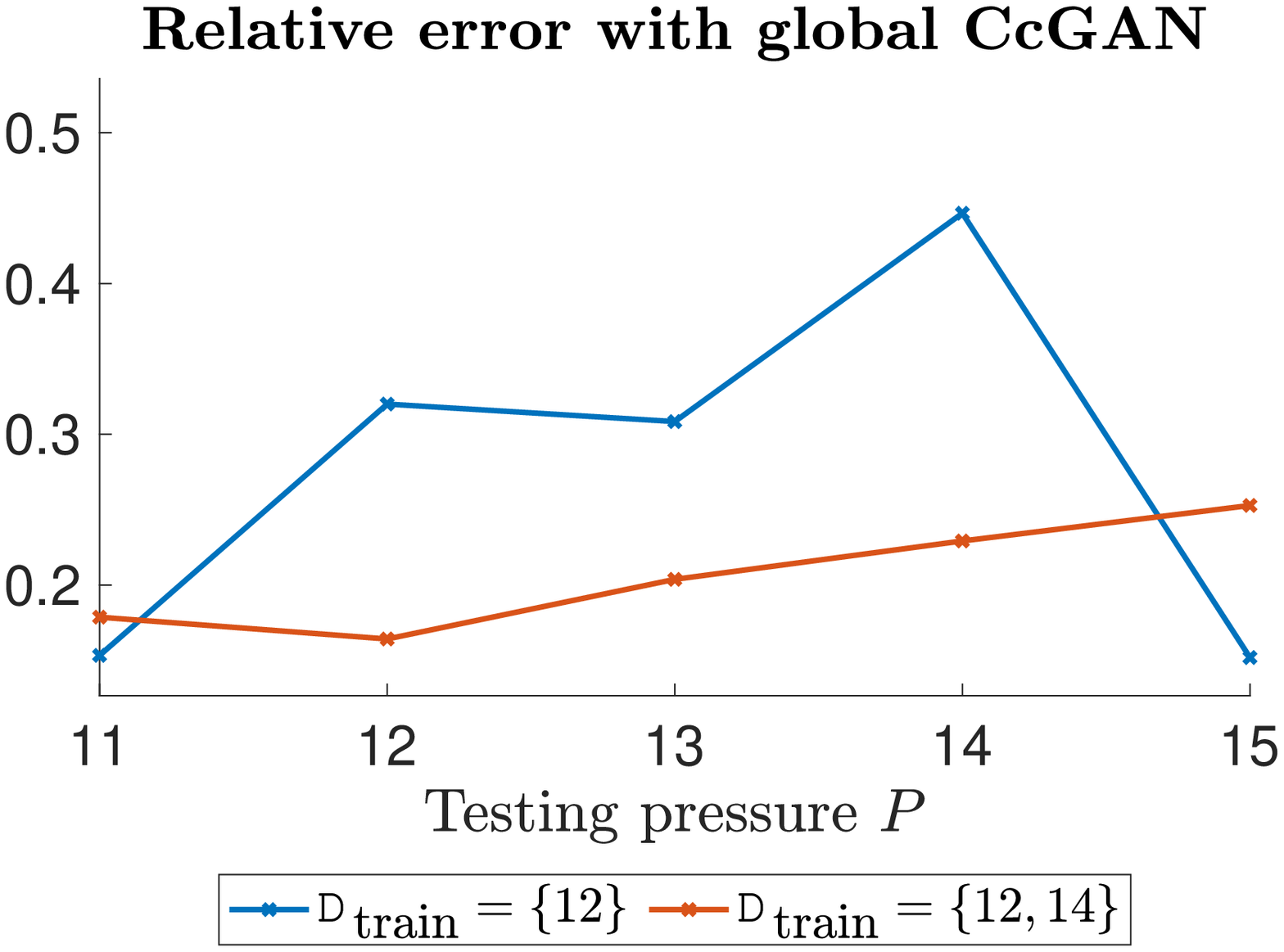}
\includegraphics[width=0.45\linewidth]{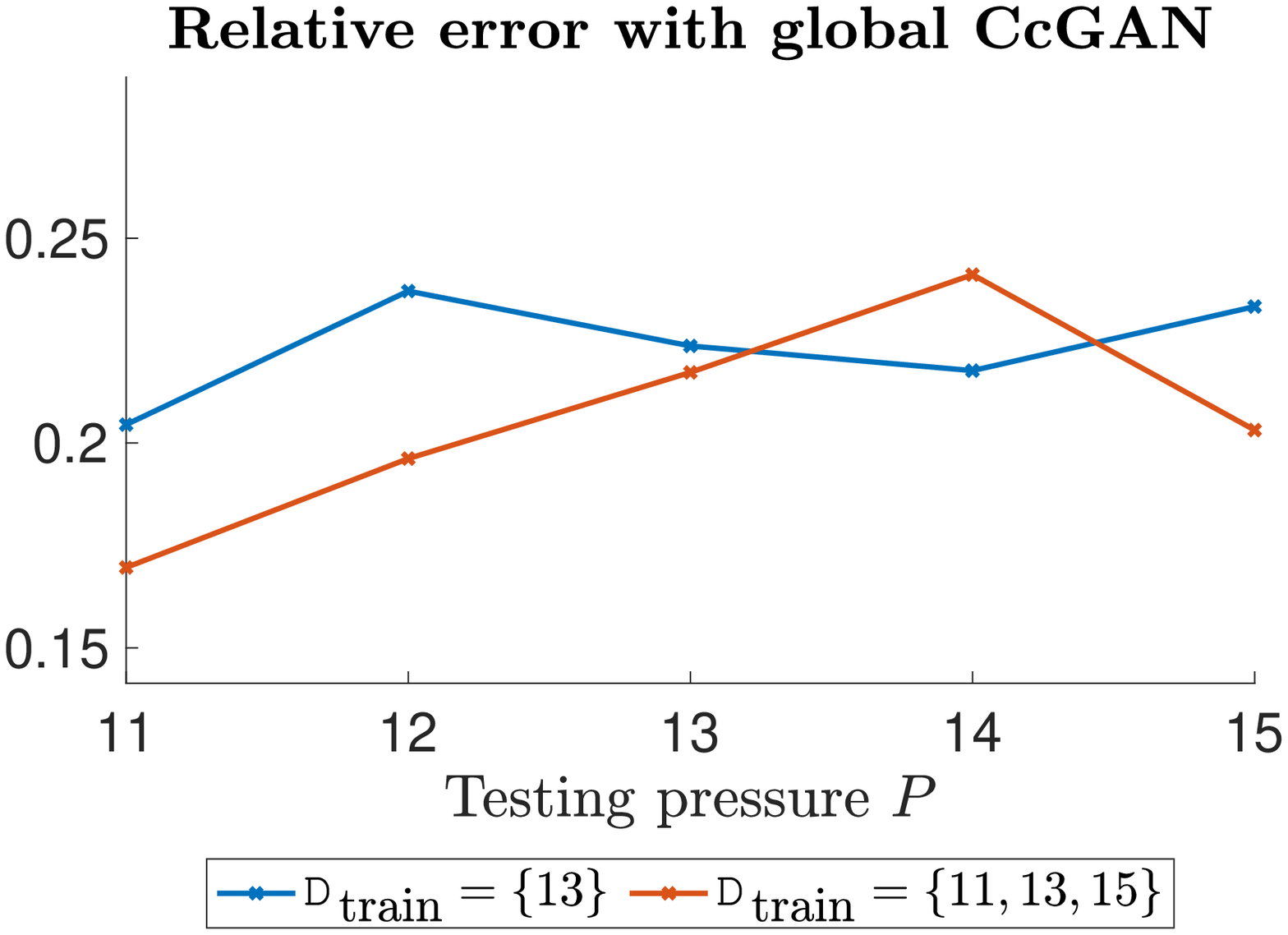}
\caption{Relative error at various testing shock pressures,
using global CcGAN
with $12 \in \paramDomain_\text{train}$ (left)
and $13 \in \paramDomain_\text{train}$ (right).}
\label{fig:compare-predictive-global-gan}
\end{figure}

Figure~\ref{fig:compare-predictive-global-gan}
shows a similar comparison using global CcGAN.
While adding more training shock pressures and
enriching the training datasets in global CcGAN
makes an improvement in the overall solution accuracy,
the error is always around 20\%,
which is still a lot higher than the parametric DW-DMD
by comparing to the same case in Figure~\ref{fig:compare-predictive-parametric-dmd}.

\section{Conclusion}
\label{sec:conclusion}
In this paper, we propose two data-driven surrogate modeling approaches
for computationally economical prediction of complex physics phenomena in shock-induced pore collapse processes.
The surrogate models are built based on dynamic mode decomposition and U-Net generative adversarial networks,
and modified to overcome the challenges of data scarcity and pressure-dependent advective and transport dynamics.
The shock pressure is incorporated in the construction of the surrogate models,
by means of parametric interpolation in dynamic mode decomposition and
conditional input in generative adversarial networks, respectively.
Moreover, windowing is used in dynamic mode decomposition for efficient dimensionality reduction
by further localizing reduced order models in time.

In our numerical realization of these surrogate models,
the training of dynamic mode composition is much more efficient than generative adversarial network.
Moreover, dynamic mode decomposition produces more stable approximation
and accurate prediction for the whole pore collapse processes at unseen shock pressures.
It will be interesting to see how improvements in efficiency and accuracy can be made to neural networks approaches
for dynamic surrogate modeling of data-scarce large-scale applications with
advective and transport phenomena like pore collapse processes.
In the meantime, some physics-guided data-driven approach with simpler machine learning methods,
like the local distance windowing dynamic mode decomposition, will serve as a powerful tool for these applications.

\section*{Acknowledgments}
This work was performed at Lawrence Livermore National Laboratory.
Lawrence Livermore National Laboratory is operated by Lawrence
Livermore National Security, LLC, for the U.S. Department of Energy,
National Nuclear Security Administration under Contract DE-AC52-07NA27344
and LLNL-JRNL-849281.

\section*{Disclaimer}
This document was prepared as an account of work sponsored by an agency of the
United States government.  Neither the United States government nor Lawrence
Livermore National Security, LLC, nor any of their employees makes any warranty,
expressed or implied, or assumes any legal liability or responsibility for the
accuracy, completeness, or usefulness of any information, apparatus, product, or
process disclosed, or represents that its use would not infringe privately owned
rights.  Reference herein to any specific commercial product, process, or
service by trade name, trademark, manufacturer, or otherwise does not
necessarily constitute or imply its endorsement, recommendation, or favoring by
the United States government or Lawrence Livermore National Security, LLC.  The
views and opinions of authors expressed herein do not necessarily state or
reflect those of the United States government or Lawrence Livermore National
Security, LLC, and shall not be used for advertising or product endorsement
purposes.

This article has been authored by an employee of National Technology \& Engineering Solutions of Sandia,
LLC under Contract No. DE-NA0003525 with the U.S. Department of Energy (DOE).
The employee owns all right, title and interest in and to the article and is solely responsible for its contents. 
Sandia National Laboratories is a multimission
laboratory managed and operated by National Technology and Engineering Solutions of Sandia LLC,
a wholly owned subsidiary of Honeywell International Inc. for the U.S. Department of Energy’s National Nuclear
Security Administration under contract DE-NA0003525. 
This paper describes objective technical results and analysis.
Any subjective views or opinions that might be expressed in the paper do not necessarily represent the views of the
U.S. Department of Energy or the United States Government.

\bibliographystyle{unsrt}
\bibliography{reference-pore-coll,reference-rom}

\end{document}